\documentclass[tighten,nolinenumbers, notrackchanges, onecolumn]{aastex62}

\usepackage{amsmath}
\usepackage{amssymb}
\usepackage{amssymb}
\usepackage{mathtools}
\usepackage{mathrsfs}

\usepackage{morefloats}
\usepackage{longtable}
\usepackage{afterpage}

\hypersetup{linkcolor=red,citecolor=blue,urlcolor=magenta}

\usepackage{soul}

\usepackage[makeroom]{cancel}

\usepackage{xspace}
\newcommand{\project}[1]{\textsl{#1}\xspace}
\newcommand*{\XPSI}{\project{X-PSI}}
\newcommand*{\NICER}{\project{NICER}}
\newcommand*{\MultiNest}{\textsc{MultiNest}\xspace}
\newcommand*{\TEMPO}{\textsc{tempo}2\xspace}
\newcommand{\msol}{M$_\odot$}

\newcommand{\TT}[1]{\texttt{#1}}

\received{2019 July 12}
\revised{2019 September 24}
\accepted{2019 September 25}
\published{2019 December 12}

\submitjournal{The Astrophysical Journal Letters}

\shorttitle{A \NICER VIEW OF PSR~J0030$+$0451: PARAMETER ESTIMATION}
\shortauthors{Riley~et~al.}

\begin{document}

\title{A \NICER VIEW OF PSR~J0030$+$0451: MILLISECOND PULSAR PARAMETER ESTIMATION}

\correspondingauthor{T.~E.~Riley}
\email{T.E.Riley@uva.nl}

\author[0000-0001-9313-0493]{T.~E.~Riley}
\author[0000-0002-1009-2354]{A.~L.~Watts}
\email{A.L.Watts@uva.nl}
\affil{Anton Pannekoek Institute for Astronomy, University of Amsterdam, Science Park 904, 1090GE Amsterdam, the Netherlands}

\author[0000-0002-9870-2742]{S.~Bogdanov}
\affil{Columbia Astrophysics Laboratory, Columbia University, 550 West 120th Street, New York, NY 10027, USA}

\author[0000-0002-5297-5278]{P.~S.~Ray}
\affil{Space Science Division, U.S. Naval Research Laboratory, Washington, DC 20375, USA}

\author[0000-0002-8961-939X]{R.~M.~Ludlam}
\affil{Cahill Center for Astronomy and Astrophysics, California Institute of Technology, Pasadena, CA 91125, USA}
\affil{Einstein Fellow}

\author[0000-0002-6449-106X]{S.~Guillot}
\affil{IRAP, CNRS, 9 avenue du Colonel Roche, BP 44346, F-31028 Toulouse Cedex 4, France}
\affil{Universit\'{e} de Toulouse, CNES, UPS-OMP, F-31028 Toulouse, France.}

\author{Z.~Arzoumanian}
\affil{X-Ray Astrophysics Laboratory, NASA Goddard Space Flight Center, Code 662, Greenbelt, MD 20771, USA}

\author{C.~L.~Baker}
\affil{Applied Engineering and Technology Directorate, NASA Goddard Space Flight Center, Code 592, Greenbelt, MD 20771, USA}

\author[0000-0002-7177-6987]{A.~V.~Bilous}
\affil{Anton Pannekoek Institute for Astronomy, University of Amsterdam, Science Park 904, 1090GE Amsterdam, the Netherlands}

\author[0000-0001-8804-8946]{D.~Chakrabarty}
\affil{MIT Kavli Institute for Astrophysics and Space Research, Massachusetts Institute of Technology, Cambridge, MA 02139, USA}

\author{K.~C.~Gendreau}
\affil{X-Ray Astrophysics Laboratory, NASA Goddard Space Flight Center, Code 662, Greenbelt, MD 20771, USA}

\author{A.~K.~Harding}
\affil{Astrophysics Science Division, NASA Goddard Space Flight Center, Greenbelt, MD 20771, USA}

\author[0000-0002-6089-6836]{W.~C.~G.~Ho}
\affil{Department of Physics and Astronomy, Haverford College, 370 Lancaster Avenue, Haverford, PA 19041, USA}
\affil{Mathematical Sciences, Physics and Astronomy, and STAG Research
Centre, University of Southampton, Southampton, SO17 1BJ, UK}

\author[0000-0002-5907-4552]{J.~M.~Lattimer}
\affil{Department of Physics and Astronomy, Stony Brook University, Stony Brook, NY 11794-3800, USA}

\author[0000-0003-4357-0575]{S.~M.~Morsink}
\affil{Department of Physics, University of Alberta, 4-183 CCIS, Edmonton, AB T6G 2E1, Canada}

\author[0000-0001-7681-5845]{T.~E.~Strohmayer}
\affil{Astrophysics Science Division and Joint Space-Science Institute,
  NASA's Goddard Space Flight Center, Greenbelt, MD 20771, USA}

\begin{abstract}
We report on Bayesian parameter estimation of the mass and equatorial radius of the millisecond pulsar PSR~J0030$+$0451, conditional on pulse-profile modeling of \project{Neutron Star Interior Composition Explorer} X-ray spectral-timing event data. We perform relativistic ray-tracing of thermal emission from hot regions of the pulsar's surface. We assume two distinct hot regions \added{based on two clear pulsed components in the phase-folded pulse-profile data}\deleted{, each containing radiating material}; we explore a number of forms (morphologies and topologies) for each hot region, inferring their parameters in addition to the stellar mass and radius. For the family of models considered, the evidence (prior predictive probability of the data) strongly favors a model \replaced{with}{that permits} both hot regions \added{to be} located in the same rotational hemisphere. \explain{The next two sentences were swapped in order as suggested} Models wherein both hot regions are assumed to be simply-connected circular single-temperature spots, in particular those where the spots are assumed to be reflection-symmetric with respect to the stellar origin, are strongly disfavored. For the inferred configuration, one hot region subtends an angular extent of only a few degrees (in spherical coordinates with origin at the stellar center) and we are insensitive to other structural details; the second hot region is far more azimuthally extended in the form of a narrow arc, thus requiring a larger number of parameters to describe. The inferred mass $M$ and equatorial radius $R_\mathrm{eq}$ are, respectively, $1.34_{-0.16}^{+0.15}$~\msol~and $12.71_{-1.19}^{+1.14}$~km, whilst the compactness $GM/R_\mathrm{eq}c^2 = 0.156_{-0.010}^{+0.008}$ is more tightly constrained; the credible interval bounds reported here are approximately the $16\%$ and $84\%$ quantiles in marginal posterior mass. \deleted{Due to the relative size of the marginal credible intervals and the rotational deformations induced by the spin frequency of PSR~J0030$+$0451, $M$ and $R_{\rm eq}$ can, individually, be safely identified as those associated with a nonrotating star with an equivalent number of baryons.}
\end{abstract}


\section{Introduction}\label{sec:intro}
Neutron star (NS) cores are thought to harbor nucleonic matter under extreme conditions: high in density, neutron rich, and potentially strange. Stable states of strange matter may either be bound in the form of hyperons, or deconfined as a mixture of up, down, and strange quarks. The density in the stellar core may reach up to several times the nuclear saturation density $\rho_\mathrm{sat} = 2.8\times 10^{14}$~g/cm$^3$, a range for which we cannot yet calculate the state of nuclear matter from first principles.\footnote{Although note that calculations at sub-saturation densities \citep[see, e.g.,][]{Hebeler10}, and perturbative quantum chromodynamics (QCD) modeling \citep[valid for densities that are several times higher than the maximum expected in NS cores; see][]{Kurkela14} can impose some constraints on the low- and high-density limits of NS core parameter space.} Instead, theorists develop phenomenological models of particle interactions and phase transitions, which must be tested by experiment and observation. Heavy ion collision experiments explore the high-temperature and lower-density parts of the nuclear matter phase diagram; but NSs are unique laboratories for the study of strong and weak force physics in cold, dense matter \citep[for recent reviews see][]{Lattimer16,Oertel17,Baym18}.

The particle interactions on a microphysical scale emerge macroscopically as an equation of state (EOS)---in the context of cold dense matter, a relationship between pressure and (energy) density. The EOS forms part of the relativistic stellar structure equations that enable us, given a central density and a spin rate, to compute model NSs \citep[e.g.,][]{Hartle1967,HT1968}. An EOS function thus maps to a sequence of stable global spacetime solutions, each controlled in the exterior domain by parameters such as the mass and equatorial radius \citep[at low orders in a small dimensionless spin parameter;][]{Hartle1967}. In this work we constrain the total mass (sometimes referred to as the gravitational mass) and the equatorial radius of the star, respectively defined as the mass and coordinate radius in the Schwarzschild metric. If we can statistically estimate \replaced{stellar}{the} masses and radii\replaced{---extending over a range of $\gtrsim1$~\msol---}{ of a set of stars whose central densities span some sufficiently broad range, }we can in principle map out the EOS and hence make inferential statements about the microphysics \citep{Ozel09,Steiner10,Steiner13,Nattila16,Ozel16,Raithel17,Raaijmakers18,Riley18,Greif19}.

The strongest statistical constraints\footnote{Or \textit{measurements}, or \textit{estimates}; in any case, this means some probabilistic measure that is a function of, or otherwise pertains to, model parameters. See Section~\ref{subsec: generative model space} for the probabilistic measures that we consider in this present work.} on NS masses are derived by timing radio pulsars in (compact) binaries, and rely on our well-established understanding of relativistic orbital dynamics. Every EOS function (corresponding to a parameter vector for a parameterized model) permits stable spacetime solutions with a maximum gravitational mass---associated with a specific central density---beyond which no stable solutions can exist. High-mass NSs with tight constraints can therefore effectively exclude\footnote{In a Bayesian context, by truncating the mass likelihood function only far in the tails, leading to a finite but small marginal posterior density for EOSs that do have substantially smaller maximum supported masses.} a subset of EOS function space, barring strong contention with future analyses of independently acquired data. The most informative pulsars in this regard are PSR~J0348$+$0432, with mass $2.01 \pm 0.04$~\msol~\citep[][where the mass is derived by combining pulsar timing and models of the white dwarf companion]{Antoniadis13}, and PSR~J1614$-$2230 with mass $1.908 \pm 0.016$~\msol~\citep[][where the mass comes from Shapiro delay estimation]{nanograv11}.\footnote{Note that PSR~J1614$-$2230 was initially reported as having mass $1.97 \pm 0.04$ \citep{Demorest10}; the inferences have since been updated via analysis of newly acquired data.} More recently, \citet{Cromartie19} have reported a higher---but at present more uncertain---mass of $2.14^{+0.10}_{-0.09}$~\msol~for PSR~J0740$+$6620. We note, however, that the \citet{Cromartie19} measurement is not subject to the systematic uncertainty that should be added to the formal uncertainty on the mass of PSR~J0348+0432 due to the latter's dependence on theoretical models of white dwarf evolution.  

The radio pulsar timing of compact binaries has yet to deliver a radius constraint, although this is feasible and indeed anticipated, via moment of inertia estimation \citep{Kramer09}. There are, however, constraints on radius via X-ray spectral modeling of transiently accreting and bursting NSs \citep[see, e.g.,][]{Steiner13, Ozel16, Nattila17, Shaw18, Baillot19}; we refer the reader to \citet{Miller13} and \citet{Ozel16b} for detailed reviews that include an explanation of these X-ray modeling techniques and associated uncertainties. NS mass and tidal deformability estimates are now also being reported based on the first binary NS merger gravitational wave event, GW170817 \citep{Abbott17,Abbott18,De18}. These can be translated, usually by means of universal relations or EOS model assumptions, into constraints on mass and radius \citep[see, e.g.,][]{Abbott18,Annala18,De18,Most18,Tews18}. Generally, assuming that both NSs have the same EOS, it is found that their radii are nearly equal and (for the 68\% credible interval) have the common value $11.9\pm1.1$ km.

NASA's \project{Neutron Star Interior Composition Explorer} \citep[\NICER;][]{Gendreau16}, a soft X-ray telescope installed on the \textsl{International Space Station} (\textsl{ISS}) in 2017, was developed in part to estimate masses and radii of NSs using pulse-profile modeling of nearby rotation-powered millisecond pulsars \added{(MSPs)}. Pulse-profile modeling is a technique that probes (approximations to) general relativistic effects on thermal emission from hot regions on the stellar surface \citep{bogdanov19b}; these effects are, predominantly, local radiation beaming\footnote{Note that local effective gravity in local comoving frames (instantaneously inertial during rotation) also enters calculation of \textit{atmospheric} beaming of radiation emergent from the local comoving photosphere.} \textit{due to bulk motion} of material on the rotationally-deformed surface,\footnote{Where for statistical applications the surface is either self-consistently computed via matching to a numerical interior solution to the field equations, or is embedded via a quasi-universal relation in an ambient spacetime solution \citep[for an overview see, e.g.,][and the references therein]{Riley18}.} and subsequent ray propagation on the exterior spacetime. Ray propagation includes the canonical \textit{bending} of light, gravitational redshift, and the increasingly small imprints of rotational metric deformation: frame-dragging, a finite mass quadrupole moment, and higher-order (mass and current) multipole moments. As the star spins, the flux and spectrum of X-ray emission registered by a distant observer is modulated in a periodic manner: we can determine the rotational phase evolution of pulsars precisely and build up a pulse-profile (X-ray counts per rotational phase bin per detector channel) by phase-folding X-ray events according to an ephemeris.\footnote{Note a key difference to the X-ray spectral modeling mentioned two paragraphs earlier: pulse-profile modeling involves phase-resolved spectroscopy; spectral modeling is phase-averaged, and does not fully leverage the temporal dimension of information provided by the star's rotation.} The mapping of surface emission into the pulse-profile detected by a distant observer, via relativistic ray-tracing through the spacetime of a rapidly rotating (and hence oblate) star, is well-understood \citep{Pechenick83,Miller98,Poutanen03,Poutanen06,Cadeau07,Morsink2007,Baubock13,AlGendy2014,Psaltis14a,Nattila18,Vincent18}. Thus, given a model for the surface emission (e.g., a geometrically thin atmosphere of some chemical and ionic composition together with a local comoving effective temperature field as a function of surface coordinates), one calculates the expected pulse-profile for a given exterior spacetime solution and a given instrument. By coupling such light-curve models to statistical sampling software via efficient software implementations, we can use Bayesian inference to derive posterior probability distributions for spacetime parameters such as mass and equatorial radius directly from pulse-profile data.

For the pulse-profile modeling technique to deliver tight constraints on mass and radius, rapid spin ($\gtrsim 100$\,Hz) is desirable \citep{Psaltis14b,Miller15,Stevens16}, and one needs high-quality phase- and energy-resolved pulse profiles with time resolution $\le 10\mu$s and a large number of photons. The precise number of photons needed to deliver constraints on mass and radius at levels of a few percent---and by extension tight constraints on EOS models---depends on the geometry of a given source, but is $\sim\!10^6$ pulsed photons \citep{Lo13,Psaltis14b,Miller15}. For the brightest of the rotation-powered \replaced{millisecond X-ray pulsars}{MSPs} targeted by \NICER, it is feasible to collect sufficient data with observation times $\sim\!1$~Ms. The hot regions on rotation-powered \replaced{millisecond pulsars}{MSPs} in theory arise as magnetospheric currents---including return currents---deposit energy in the surface layers of the star; the resulting surface radiation field is \textit{a priori} highly uncertain \citep{Harding01,Gralla17,Baubock19}. \NICER pulse-profile modeling can therefore also help to constrain the characteristics of the hot regions.

In this paper we undertake pulse-profile modeling of \NICER \project{X-ray Timing Instrument} (\project{XTI}) observations of the rotation-powered \replaced{millisecond pulsar}{MSP} PSR~J0030$+$0451. Discovered as a radio pulsar by \citet{Lommen00} and then identified as an X-ray pulsar \citep{Becker00}, PSR~J0030$+$0451 has a spin frequency of $205$ Hz and lies at a distance of $325 \pm 9$\,pc \citep{nanograv11}. There are no independent prior constraints on either mass or radius. Our analysis uses the \project{X-ray Pulsation Simulation and Inference} package \citep[\XPSI\footnote{\url{https://github.com/ThomasEdwardRiley/xpsi}}~\texttt{v0.1};][]{riley19b}. \XPSI is a software package for Bayesian modeling of astrophysical X-ray pulsations generated by the rotating, radiating surfaces of relativistic compact stars. \XPSI couples X-ray pulsation likelihood functionality to open-source statistical sampling software for use on high-performance computing systems; we apply nested sampling \citep[][]{skilling2006} in our analysis (refer to Appendix~\ref{sec:posterior computation}). The work presented here is based on usage of a $500,000$ core-hour grant on the Dutch national supercomputer Cartesius.\footnote{\url{https://userinfo.surfsara.nl/systems/cartesius}}

Section \ref{sec:models} outlines the modeling choices and introduces details specific to this analysis, including issues associated with the PSR~J0030$+$0451 surface radiation field parameterization, instrument response, prior definition and implementation, and the consequences for computational efficiency of posterior sampling. In particular we restrict this analysis to models with two distinct hot regions with various structures. While our choices are physically motivated, it is important to emphasize that our inferences are conditional upon these choices. Posterior\footnote{Appendix~\ref{sec:posterior computation} provides an overview of the methodology used for posterior computation, and of the format used to present the posterior information.} inferences for the models are presented in Section \ref{sec:inferences}, including the inferred posterior probability distributions for the spacetime parameters (mass, radius, and their combination into compactness), surface radiation field parameters (e.g., heating distribution and resulting temperature field), and instrument parameters. The Bayesian evidence for each model is reported, and we also summarize the computational resources required for each \added{parameter estimation} run. We also compare our inference to predictions derived via earlier study of PSR~J0030$+$0451 using \project{XMM-Newton} \citep{Bogdanov09}. \added{Sections \ref{sec:models} and \ref{sec:inferences} are long and detailed; in Section \ref{sec:execsum} we provide a brief overview of some of the key aspects from those sections, to help orient the reader.} In Section \ref{sec:discussion} we discuss the implications of our results for our understanding of dense matter, pulsar emission mechanisms, and stellar evolution. We conclude with a discussion of future work for PSR~J0030$+$0451: variations in the model that should be considered, tests and cross-checks, and the potential for improving the constraints for this source via longer observations or more in-depth analysis.

\section{Modeling procedure}\label{sec:models}
\subsection{Executive summary of modeling procedure and inferences}\label{sec:execsum}
\explain{Added section.}

This paper (and its companions) are the first pulse-profile modeling analyses to emerge from the \NICER mission. We have therefore provided (in Sections \ref{sec:models} and \ref{sec:inferences}) a very detailed description of the methodology, the flow of the analysis, and the results.  Since this is lengthy, we summarize the key aspects in this subsection.

The pulse-profile modeling technique requires us to define a model for the data-generating process, incorporating the physics that we initially assume to be most important. For a given choice of parameters, this model can be used to generate synthetic pulse-profile data sets. The model in part defines the likelihood function (the probability of the data as a function of parameters); the model also defines the prior probability distribution of parameters entering in the likelihood function. The posterior probability distribution of the model parameters (conditional on the observed data) is then sampled during the inference process. We must also pay attention to model complexity, in order to keep the computational load tractable.

In this paper we assume that there are two separate hot regions on the stellar surface; this choice was motivated by the presence of two distinct pulses in the observed (phase-folded) pulse-profile.  However, we considered a number of different possible configurations for the shapes and temperature functions of the hot regions: circular spots, annuli (rings, both centered and off-centered), and crescents; with one or two temperature components. These choices were motivated by contemporary theories for pulsar surface heating distributions as a result of magnetospheric return currents.  We tested configurations where we insisted the two hot regions were antipodal and identical; and where the hot regions were completely independent and potentially non-antipodal.  We then assumed a geometrically-thin fully-ionized hydrogen atmosphere model \citep[\texttt{NSX},][]{Ho09} which characterizes the beaming and spectrum of the emergent thermal radiation \citep[see, e.g.,][for details]{Zavlin96}.

To propagate the emergent radiation towards the observer via relativistic ray-tracing, we use the Oblate Schwarzschild plus Doppler approximation of \citet{Morsink2007} for the NS spacetime. This is sufficiently accurate for our analysis, given the rotation rate of PSR~J0030$+$0451. We define a joint prior distribution of mass and radius (the key parameters specifying the spacetime) that facilitates the subsequent inference of EOS model parameters \citep{Riley18, raaijmakers19}.  For the distance to the source, we use the (Gaussian-distributed) value inferred from radio observations \citep{nanograv11} as a prior in our modeling.  We then need to model the instrument response matrix (which includes both the effective area and the way in which incident photons of a given energy are assigned to specific detector energy channels).  We develop a parameterized model that includes both energy-independent and energy-dependent components. The former attempts to capture absolute calibration uncertainty; for the latter we base our parameterization on residuals derived from \NICER observations of the Crab nebula and pulsar \citep{Ludlam18}.  We also assume a non-source background component, which we treat as a rotational phase--independent channel-by-channel contribution, rather than invoking a specific physics-driven spectral model.  There are no prior constraints on either observer inclination or interstellar absorption for PSR~J0030$+$0451, so we adopt a wide and diffuse prior for both parameters.  

Note that \citet{miller19} have made an independent analysis of the same data set using different modeling choices and methodology. The choices we have made in this paper differ in several regards from those made by \citet{miller19}; some of the most notable differences are in the models for the hot region configurations and the instrumental response, and the specification of the prior on distance. 

During the inference analysis reported in Section \ref{sec:inferences} we considered a sequence of increasingly complex models for the shape and temperature function of the hot regions. All of the other aspects of the modeling described above are shared between models.  Model assessment and comparison then enabled the identification of a superior configuration.  We use a combination of performance measures: the evidence (the prior predictive probability of the data); graphical posterior predictive checking (to verify whether or not a  model generates synthetic data without obvious residual systematic structure in comparison to the real data); visualization of the combined signals from the hot regions; KL-divergences (a measure of the parameter-by-parameter information gain of the posterior over the prior); background-marginalized likelihood functions (useful in combination with evidence to assess whether additional model complexity is helpful); model tractability (posterior computational accuracy being higher for less complex models); and cross-checking of the inferred background against earlier analysis of PSR~J0030$+$0451 with \project{XMM-Newton}.   

Before beginning our analysis, we had mapped out an initial route through the model space of different heating configurations. This was modified as we progressed, informed by the results of each stage.  We began with the simplest model, with single-temperature circular spots. Having the spots be antipodal and identical was quickly ruled out due to large residuals between model and data. Relaxing the requirement that the spots be identical and antipodal largely resolved this issue.  We then nevertheless moved to a more complex model where each hot region consisted of a circular spot---a \textit{core}---and a surrounding annulus with an independently-determined temperature. This model was superior to the simpler one, based on the evidence, but appeared to be overly complex: one hot region was dominated by a small hot circular spot, with negligible emission in the \NICER waveband from the cooler annulus; for the other, emission was dominated by a hot annulus, with a much cooler core making almost no contribution. Simplifying the model such that one hot region was a single-temperature circular spot and the other a single-temperature annulus (with a centered, non-emitting core) produced congruent inferences at lower computational cost. At this point (after assessing the contribution from this component, and our remaining computational resources), we elected to restrict the model for one of the hot regions to be a single-temperature circular spot. The other was restricted to a single temperature, but we increased the complexity of the shape, testing two additional models: an annulus with an off-centered non-emitting core; and a crescent.

The superior configuration to emerge from this sequence of models, in terms of the performance measures listed above, was the final one: one hot region a small circular spot (sufficiently small that we would be insensitive to shape changes); and the other an extended thin crescent. The results that we report in the abstract for the mass and radius of PSR~J0030$+$0451 are those associated with this configuration.

\subsection{Data pre-processing}\label{subsec:data}

\begin{figure*}[t!]
\centering
\includegraphics[clip, trim=0cm 0cm 0cm 0cm, width=0.75\textwidth]{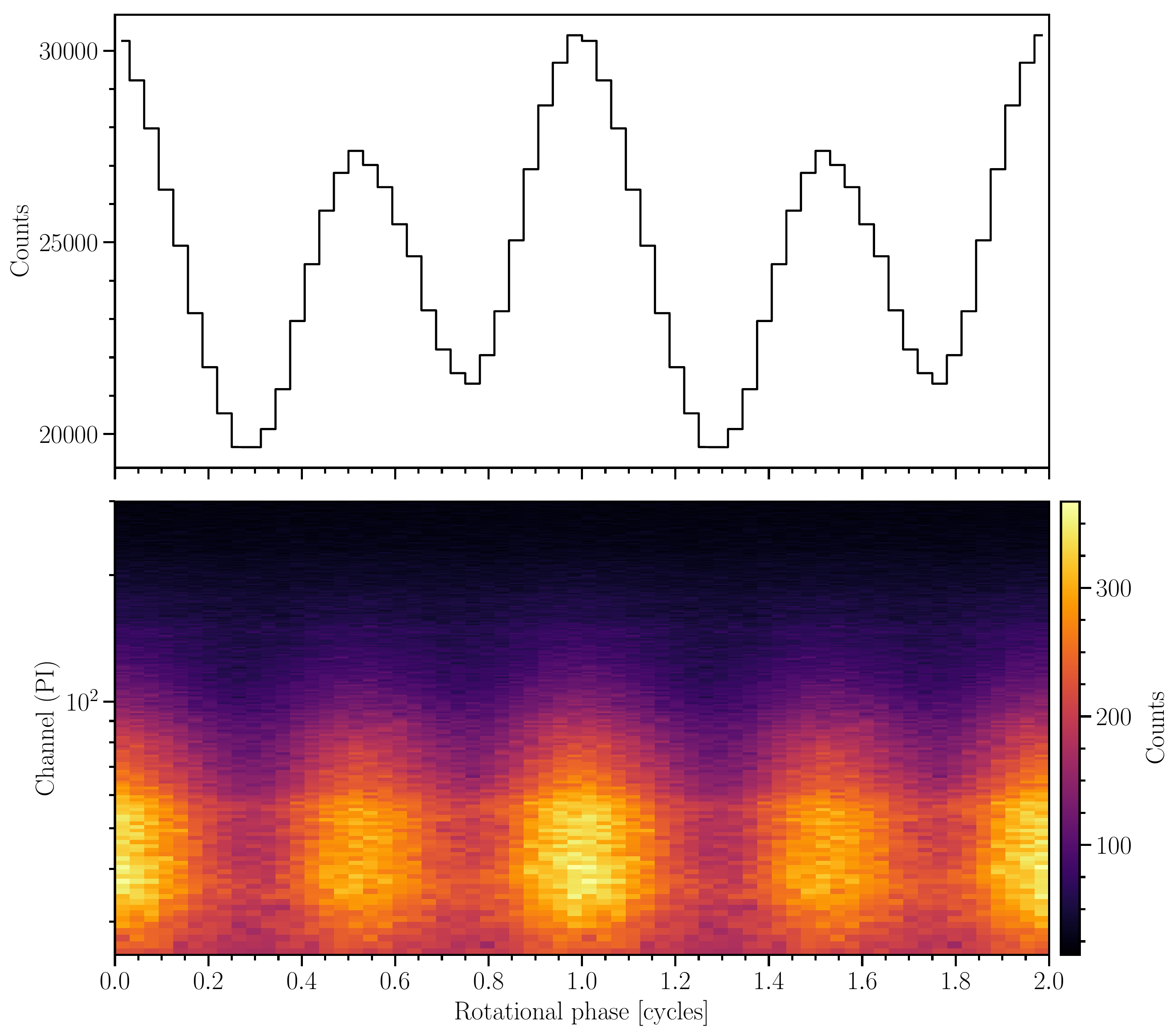}
\caption{\small{Phase-folded PSR~J0030$+$0451 event data split over two rotational cycles for clarity: we use $32$ phase intervals (bins) per cycle and the count numbers in bins separated by one cycle---in a given channel---are identical. The total number of counts is given by the sum over all phase-channel pairs. The \textit{top} panel displays the pulse-profile summed over the contiguous subset of channels $[25,300)$. The \textit{bottom} panel displays the phase-channel resolved count numbers for channel subset $[25,300)$. For likelihood function evaluation (see Section~\ref{subsubsec:likelihood functions}) we group all event data registered in a given channel into phase intervals spanning a single rotational cycle. Moreover, we do not indicate the count-number noise in the \textit{top} panel to avoid confusion: the observed events are viewed as fixed random variates that do not have errors, and whose parameterized joint sampling distribution is to be modeled (see Section~\ref{subsec: generative model space}).}}
\label{fig:J0030 count data}
\end{figure*}

The \NICER~\project{XTI} data set $\boldsymbol{d}$ associated with the rotation-powered \replaced{millisecond pulsar}{MSP} PSR~J0030$+$0451 is necessarily a product of pre-processing, and its curation is largely described in \citet{bogdanov19a}. However, there are a number of details that are specific to each data set and so we record a summary of these details below. The raw \project{XTI} event data are publicly available via the \NICER archive (accessible via HEASARC\footnote{\url{https://heasarc.gsfc.nasa.gov/}}). The processed data set $\boldsymbol{d}$ may be found in the persistent repository of \citet{bogdanov_slavko_2019_3524457}.

For this analysis we consider only the detector channel\footnote{Conventionally termed a pulse-invariant (PI) channel, or alternatively a charge pulse amplitude (or height) channel. Note the distinction between \textit{channel}, which is only indicative of photon energy in units of $10$~eV, and incident photon \textit{energy intervals} that map probabilistically to the channel set via the redistribution matrix, a standard calibration product provided by the instrument team.} subset $[25,300)$---meaning channels $25$ through $299$ inclusive---nominally corresponding to $0.25$--$3$~keV. Below channel $25$, there is increased ``optical loading'' contamination (electronic noise due to ambient light), and there is greater uncertainty in the detector readout triggering efficiency for valid X-ray events. Above channel $300$ the soft thermal emission from PSR~J0030$+$0451 becomes negligible relative to the non-source background.

For PSR~J0030$+$0451 we use $49$ out of the $52$ active detectors (excluding the three detectors that are frequently ``hot,'' i.e., prone to excess electronic noise) and only \NICER pointings subtending an angle $>80^{\circ}$ to the direction of the Sun. In addition to the standard filtering criteria applied to \project{XTI} events, we excluded times where the planetary K-index $K_P \ge 5$, and excluded regions in the \NICER orbit where the cutoff rigidity (the COR\_SAX parameter) was less than 1.5. We further screened for occurrences of elevated background and eliminated all $16$-second time intervals in which the $0.25$--$8.00$~keV count rate exceeded $3$~counts/s, yielding an integrated exposure time of $T_{\textrm{exp}}\coloneqq\mathop{\sum}_{\ell}\Delta t_{\ell}=1,936,864$~s, where each $\Delta t_{\ell}$ is the time interval for the $\ell^{\mathrm{th}}$ exposure.

Harnessing the flux and timing stability of PSR~J0030$+$0451, the entire event data set in each detector channel is phase-folded coherently according to the best available radio pulsar timing solution from \citet{nanograv11} and using two approaches to check for consistency: using the \TEMPO~\citep{Hobbs06}~\TT{photons} plug-in and the PINT\footnote{\url{https://github.com/nanograv/PINT}}~\TT{photonphase} tool. Differences of $\lesssim1\,\mu{\rm s}$ as a phase-offset are observed, but are deemed negligible for the present analysis as such an offset corresponds to $\lesssim 0.02\%$ of the total phase, or less than $0.7\%$ of a bin width. The resulting folded event list, obtained with the PINT \TT{photonphase} tool, is summarized as count data in a set of rotational phase intervals and detector channels. We display in Figure~\ref{fig:J0030 count data} the count data for channels in the interval $[25,300)$, nominally corresponding to the energy range $0.25$--$3.00$~keV. We reserve the remaining details of data-space definition for Section~\ref{subsubsec:likelihood functions}, wherein we formalize the likelihood function applied in this specific work.

\subsection{Generative modeling}\label{subsec: generative model space}

We begin the description of our modeling process by outlining useful mathematical objects that are assigned statistical meaning, and which can interface with open-source computational machinery. In Section \ref{subsec:overarching definitions}, we build upon this conceptual groundwork and assign astrophysical meaning to such objects.

There exists an inherent degree of freedom as to the precise definition of a \textit{model}. We consider a \textit{generative} model for the X-ray data set $\boldsymbol{d}$ curated for PSR~J0030$+$0451. Moreover, we opt to define a generative model as the union of the following components: (i) a data space $\mathscr{D}$ in which a data set $\boldsymbol{d}\in\mathscr{D}$ exists as a fixed vector of numbers; (ii) an abstract model space $\mathscr{M}$ of elements, such that each element completely specifies a joint sampling distribution on space $\mathscr{D}$ (i.e., the sampling distribution can be evaluated conditional on the element); (iii) a joint probability distribution defined on space $\mathscr{M}$ that is not conditional on data vector $\boldsymbol{d}$. Such a model is considered as generative because one can define a Bayesian joint distribution on the joint space of $\mathscr{D}$ and $\mathscr{M}$: the data vector and a model vector are both interpreted as (finite-dimensional) multivariate random variables, and in order to simulate (or generate) data sets, one can jointly draw random variates to populate the vector elements.

The model space $\mathscr{M}$ is in general a discrete-continuous mixed space, meaning that its elements form a discrete-continuous mixed set: it follows that one could in principle identify each element as a model. Ultimately, however, one can distinguish continuous subsets of $\mathscr{M}$ as models, which is perhaps more common. To formalize the framework we are working in further, let us define a discrete-continuous mixed space $\mathscr{M}$ as a union $\bigcup_{m\in\mathscr{F}}\mathcal{M}_{m}$, where $\mathscr{F}\subset\mathbb{N}$ is the space of a discrete parameter $m$---a flag or label---and where each $\mathcal{M}_{m}$ may be considered as a model within $\mathscr{M}$. Let each model $\mathcal{M}_{m}$ define a continuous space $\varTheta_{m}\subseteq\mathbb{R}^{n_{m}}$ such that $\mathscr{M}$ defines a space given by the union of Cartesian products
\begin{equation}
\varTheta\coloneqq\bigcup_{m\in\mathscr{F}}\left(\{m\}\times\varTheta_{m}\right).
\end{equation}
The number of dimensions of the space $\varTheta_{m}$ is given by $n_{m}$; a parameter vector $\boldsymbol{\theta}\in\varTheta_{m}\subseteq\mathbb{R}^{n_{m}}$ then has $n_{m}$ elements. We thus consider each space in the union $\mathscr{M}$ as a model: unless explicitly stated otherwise, we hereafter consider a model as an element in a discrete set, where each element has an associated continuous parameter space $\varTheta_{m}$ (among other constructs, e.g., probability measures such as a joint prior density distribution).

Working within the scope of a given model $\mathcal{M}_{m}$ with continuous parameter space $\varTheta_{m}$, the prior support $\mathcal{S}_{m}\subseteq\varTheta_{m}$ is a (compact) subset of $\mathbb{R}^{n_{m}}$ on which the joint prior density is finite. The target distribution is the joint posterior density distribution, denoted by $\pi(\boldsymbol{\theta}\,|\,\boldsymbol{d},\mathcal{M}_{m})$, which is related to the joint prior density distribution, $p(\boldsymbol{\theta}\,|\,\mathcal{M}_{m})$, via the probability identity (Bayes' theorem)
\begin{equation}
\pi(\boldsymbol{\theta}\,|\,\boldsymbol{d},\mathcal{M}_{m})p(\boldsymbol{d}\,|\,\mathcal{M}_{m})=p(\boldsymbol{\theta},\boldsymbol{d}\,|\,\mathcal{M}_{m})=p(\boldsymbol{d}\,|\,\boldsymbol{\theta},\mathcal{M}_{m})p(\boldsymbol{\theta}\,|\,\mathcal{M}_{m}),
\end{equation}
where $p(\boldsymbol{d}\,|\,\boldsymbol{\theta},\mathcal{M}_{m})$ is the likelihood function for model $\mathcal{M}_{m}$. The likelihood function is the sampling distribution on the space $\mathscr{D}$ evaluated at the fixed data vector $\boldsymbol{d}$, as a function of $\boldsymbol{\theta}$, conditional on model $\mathcal{M}_{m}$; the normalization of the posterior is the prior predictive probability\footnote{The scalar expectation of the likelihood function with respect to the prior $p(\boldsymbol{\theta}\,|\,\mathcal{M}_{m})$, commonly referred to as the evidence or fully-marginal likelihood.} $p(\boldsymbol{d}\,|\,\mathcal{M}_{m})$ of the data conditional on model $\mathcal{M}_{m}$. We document the techniques implemented for posterior computation in Appendix~\ref{sec:posterior computation}. It is necessarily the case that for construction of a generative model, a joint prior distribution injects a finite quantity of information about the elements of the model space. From a Bayesian perspective, models and parameters are random variables, whose joint prior distribution strictly encodes the information available before acquisition of---and computation given---data. Hereafter we do not need to typeset the symbol $\varTheta$ (with or without a subscript) to represent a space, and thus we explicitly free it for a different use.

The model space we consider is some mixture of $\mathscr{M}$-complete and $\mathscr{M}$-open \citep[][]{vehtari2012}: we do not believe the true data-generating process exists in $\mathscr{M}$, nor do we believe that an element in $\mathscr{M}$ is the closest approximation (to the true data-generating process) that is achievable and tractable (in the future) given more resources (i.e., both cognition and computing time). However, we proceed as though we can imagine a true data-generating process to physically exist---i.e., is plausibly the product of simple physical laws and initial conditions---and that parameterized approximations can in principle approach the real process arbitrarily closely, at least in terms of predictive performance. Moreover, we view the model space for this specific work as effectively the best available to us given current resources and consider it plausible that it is sufficiently rich to predict features in the procured X-ray event data. A subset of continuous parameters are shared by all models in the discrete set: these are considered to be of an (effectively) fundamental physical nature, such as the distance to a source system and parameters controlling the exterior spacetime solution of a compact star in general relativistic gravity. On the other hand, parameters are also defined that are of an overtly phenomenological nature and are intended to approximate reality only insofar as: (i) predicting---via interplay with the (almost) fundamental parameters---features observed in data; and (ii) supporting future development of physics models that are considered more realistic but crucially remain tractable for statistical falsification.

When we increment model complexity, it is intended to be in an intuitive and natural manner: we generally do so by breaking a form of symmetry, forming nested relationships between models (see Section~\ref{subsec:definitions} and then Section~\ref{subsubsec: model relationships}). Nevertheless, the form of the simplest model and the increments in complexity are ultimately of a subjective nature and exhibit a degree of arbitrariness---different practitioners would have defined their model spaces differently.

The discrete set of models we condition on was determined partly out of consideration for available resources. Given that posterior computation cost generally increases with prior predictive complexity,\footnote{Which is easily proven for (non-dynamic) nested sampling processes (see Appendix~\ref{sec:posterior computation} for an outline of the sampling procedure).} we consider it justified to organize a set of increasingly costly problems wherein each is an extension of the former. The salient advantage in this respect is facilitating robustness: by monitoring and analyzing sampling processes operating on gradually more difficult problems, tractability can be gauged for subsequent problems given available computing resources. Moreover, if a pair of models form an exact or approximate nested relationship we expect the posterior parameter inferences to be consistent with that of the simpler model if the additional complexity is unhelpful.

Aside from resource management, and conditioned on the assumption that posterior computation is sufficiently accurate, one can pose the question of how much complexity is useful to capture structure in observational data. In a Bayesian framework one can in principle estimate prior predictive probabilities of data conditional on a model. It is generally argued that Occam's razor is inherent to prior predictive probability integrals: predictive complexity is penalized if predictions are not \textit{expected} to be at least as commensurate with the data as a simpler (or nested) model. The interpretation of prior predictive probabilities is often fraught with problems, principally sensitivity to prior definition. In this work the joint prior distribution defined for the continuous parameters of each model is not rigorously chosen according to an information-theoretic criterion, nor to accurately quantify belief for overtly phenomenological parameters. However, the prior choices are viewed as being weakly informative\footnote{Also known as \textit{vague} or \textit{diffuse}.} for most parameters of interest in the absence of existing constraints, and are viewed as being consistent between models.

As acknowledged above, a widely held view is that evidence estimation\footnote{For calculation of Bayes' factors.} does not solve the problem of model comparison. In order to evaluate model performance we thus employ both a form of graphical posterior-predictive checking, and prior predictive probabilities that hereafter we will refer to simply as \textit{evidences}---a less accurate but canonical descriptor of $p(\boldsymbol{d}\,|\,\mathcal{M})$. If the evidence increases, it is generally accurate to conclude that additional complexity is warranted; if evidence does not increase, however, graphical posterior predictive checking on local modes is useful for determining whether or not facets of the higher-complexity model are a promising avenue to pursue in model development---i.e., if the likelihood function maxima are larger.

\subsection{Overarching definitions}\label{subsec:overarching definitions}
In this section we describe model aspects that are generally shared between all models in the discrete set $\mathscr{M}$. These model facets are in some cases described in detail elsewhere in the \NICER literature (\citealt{bogdanov19a}; \citealt{bogdanov19b}; \citealt{bogdanov19c}) and thus we are brief where possible. Due to the large number of symbols required to describe the models in this paper, the symbols used to describe geometric variables shared with \citet{bogdanov19b} are different; Table~\ref{table:symbol diffs} provides symbol translation from \citet{bogdanov19b}, the theory in which underpins the present work.

\begin{table*}[t!]
\centering
  \caption{Translation of symbols for angle variables typeset in \textit{both} this work and \citet{bogdanov19b}.}
   \label{table:symbol diffs}
  \begin{tabular}{ll|l}
    \hline
    \hline
    Symbol & Description & \citet{bogdanov19b} \\
    \hline
    $i$ & Earth inclination to pulsar rotation axis & $\zeta$ \\
    $\Theta$ & colatitude of center of a circular hot spot\footnote{We also use the symbol $\Theta$, with subscripts, to denote the colatitudes of hot regions whose shapes are more complex than circular spots.} & $\theta_{c}$ \\
    $\zeta$, $\psi$ & angular radius of a circular hot spot\footnote{We also use these symbols to parameterize hot regions with more complex shapes, such as rings with outer angular radius $\zeta$ and inner angular radius $\psi$.} & $\theta_{\rm spot}$ \\
    $\phi$ & pulsar rotational phase\footnote{We also use the symbol $\phi$, with subscripts, to denote the azimuthal coordinates of hot regions.} & $\phi$ \\
    \hline
    \hline
  \end{tabular}
\end{table*}

\subsubsection{Source}\label{subsubsec:source}

\replaced{We usually refer to \textit{the source} in the context of our simplistic modeling as the}{The pulsed sources are assumed to be thermally-emitting,} \added{rotating} hot surface regions of \deleted{a millisecond pulsar} PSR~J0030$+$0451.\deleted{; such emission is explicitly defined in mathematical terms at the surface and the corresponding signal incident on \NICER is explicitly computed. However, the intended meaning of the word \textit{source} is also contextual, and is in some instances used more generally to mean a source of radiation---even a background source---when discussing physical reality, as distinct from our simple approximative models.}

\textbf{Parameterization.} The exterior spacetime solution is approximated as follows: we embed in each temporal hyperslice of an ambient Schwarzschild spacetime, a (quasi-universal) oblate $2$-surface, such that the geometric center coincides with the origin of the Schwarzschild coordinate chart \citep{Morsink2007}. The coordinate equatorial radius is denoted by $R_{\textrm{eq}}$, and the circumference of the equator is $2\pi R_{\rm eq}$; the total mass in the ambient spacetime is denoted by $M$. The polar axis of the Schwarzschild chart is defined as the pulsar rotational axis.

\added{Rotational deformation of the metric away from spherical symmetry is neglected; the current dipole and mass quadrupole moments of the exterior metric enter at first- and second-order in dimensionless angular velocity $\bar{\Omega}$ \citep{Hartle1967} but are sufficiently small---regardless of the EOS---that they can be neglected in pursuit of a tractable likelihood function \citep{Morsink2007}, especially during initial modeling and in a context where we do not expect the likelihood function to be sensitive to the finiteness of these moments. The perturbations, at constant baryon number, to both the total mass and circumferential radius of a nonrotating star are second-order in $\bar{\Omega}$ \citep{Hartle1967}, and are thus small for a spin of $205$\,Hz (see also Section~\ref{sec:mrcontext}) but are implicitly accounted for. The perturbation to the polar coordinate radius is second-order in $\bar{\Omega}$, and the surface oblateness is controlled by an EOS-insensitive constraint equation \citep[as is the effective gravity along the surface;][]{Morsink2007}. These small changes in the shape of the stellar cross-section induce tilt to the surface, which affects the rays that connect spacetime events at the surface to an observer; together with the change in the projected surface area of a tilted surface, and the change in effective gravity, the effect on light-curves manifests at first-order in $\bar{\Omega}$. Crucially, there is a performance floor for light-curve integration demarcated by a spherically symmetric exterior spacetime solution: embedding an oblate surface in such an ambient spacetime results in negligible increase in computation time per call to a light-curve integrator. We therefore are not concerned about quantifying the difference in our statistical inferences due to inclusion of oblateness over a spherical surface---more resources are required to quantify this rigorously than to simply account for oblateness.}

A distant, static, and fictitious\footnote{A notion borne from the nature of the event-data pre-processing \citep[see, e.g.,][for details pertaining to \XPSI]{riley19b}.} instrument (see Section~\ref{subsubsec:instrument general}) is located at radial coordinate $D$, and subtends colatitude---hereafter termed inclination---denoted by $i$. Interstellar light-matter interaction is described by absorption within a column of material. The attenuation factor is parameterized solely by the column density $N_{\textrm{H}}$ of neutral hydrogen and we assume relative abundances for the interstellar medium from \citet{Wilms2000}; we implement \added{the} \TT{tbnew}\footnote{\added{\url{https://pulsar.sternwarte.uni-erlangen.de/wilms/research/tbabs/}}} \added{model} \replaced{through a set of precomputed}{to precompute a set of} lookup tables \added{for attenuation as a function of photon energy}.

In each model the spatial dependence of the surface radiation field is of a phenomenological nature: the aim is to introduce sufficient complexity so as to represent the basic notion of pulsar surface heating due to energy deposition by magnetospheric currents (in the vicinity of the magnetic poles). We are largely ignorant of spatial structure in the surface radiation field because the star is not spatially resolved; moreover, it is intractable for us to consider more self-consistent numerical models of the surface radiation field, in part due to the expense of statistical computation. In the simplest case the radiation field is constructed by filling two closed simply-connected regions on the surface---which do not mutually overlap---with radiating material;\footnote{\added{Note that although a cooler radiating hydrogen atmosphere should exist globally over the stellar surface (as observed for PSR~J0437$-$4715, \citealt{Durant12,Guillot16b,Gonzalez19}, and PSR~J2124$-$3358, \citealt{Rangelov17}), we make no explicit reference to it when defining our likelihood function---i.e., we do not compute any radiative signal from the atmosphere exterior to the closed regions. The atmosphere cannot be globally uniform because local heating by magnetospheric currents will affect the local temperature and ionization degree; effective gravity also varies due to rotation. Reference to the global atmosphere is implicit due to the fluid properties required for containment of hot material in the closed regions.}} these regions may be interpreted to each result from magnetospheric polar cap heating. We only compute a radiative signal from these hot regions, and therefore in the context of each of our models, \textit{hot region} can be viewed as synonymous with \textit{radiating region}.

For all models a geometrically-thin (and thus plane-parallel) fully-ionized hydrogen \TT{NSX} atmosphere is invoked for the radiating material \citep[][]{Ho01,Ho09}. The radiation field is precomputed and represented as a lookup table for cubic polynomial interpolation of specific intensity, $I_{E}/k_{\rm B}T_{\rm eff}^{3}$, with respect to four variables defined in a surface local comoving frame: effective temperature, $T_{\rm eff}$; effective gravity; photon energy, $E/k_{\rm B}T_{\rm eff}$; and the cosine of the ray zenith angle (to surface normal). A quasi-universal relation for surface effective gravity\footnote{Equatorially reflection-symmetric.} is adopted from \citet[][]{AlGendy2014} in order to evaluate local radiation field intensities. We do not explicitly compute emission from the stellar surface exterior to the closed regions (i.e., as a function of source parameters controlling the exterior radiation field),\footnote{Such computation is supported by \XPSI with specialization to ensure (almost) exact areas as described above \citep[see also][]{bogdanov19b}, but requires a choice of surface radiation field (e.g., atmosphere ionization degree and chemical composition). We therefore opt to capture the \textit{non-pulsed} fraction of emission from the stellar surface exterior to the regions via our default background treatment. If evidence for unmodeled soft pulsed emission arises \textit{a posteriori} one could then consider explicit computation of such emission.} but the phase-invariant background model we invoke in all cases can capture non-pulsed components of surface and off-surface emission (see Section~\ref{subsubsec:background general}).

Let us hereafter refer to the geometric configuration of the infinitesimal radiating surface elements simply as the \textit{shape} of a hot region---including both exterior and interior boundaries. In more complex models each hot region is constructed using additional shape parameters and in some cases a second temperature component. We consider the shape and temperature of a hot region to result from the interaction of \textit{two} closed regions---hereafter \textit{members}. The members partially or wholly\footnote{Such that one member is a superset of the other.} overlap, and one member takes precedence when evaluating local radiation intensities along rays (null geodesics) that connect spacetime events on the rotating surface to a distant observer. We term one member as \textit{ceding}, and the other as \textit{superseding}. As discussed in the appendix of \citet{bogdanov19b}, the \XPSI implementation \citep[referred to as the AMS code in][]{bogdanov19b} of a radiating region is specialized for fast likelihood function evaluation when said function is a callback for sampling processes: whilst numerical approximations are necessary in general, it is relatively inexpensive to ensure that the proper area of each finite-element (discretised) radiating region is computed to a precision that (almost) exactly matches their mathematical definition.\footnote{The overall numerical accuracy remains implementation dependent.} When two members overlap to form a hot region: (i) the area of a discretely-represented superseding member is (almost) exact; and (ii) the area the discretely-represented, \textit{non-superseded subset} of a ceding member is also (almost) exact. The subset that is not superseded can itself be simply-connected or non-simply-connected depending on the model (where the set of configurations assigned finite prior density is model-dependent). For the precise details of the hot regions, refer forward to Sections~\ref{subsubsec:ST-S and ST-U} through \ref{subsubsec: model relationships}.

We opt for two disjoint hot regions: the two distinct pulses visible in the phase-folded event data (Figure~\ref{fig:J0030 count data}) are suggestive of two such regions being widely separated. Initially, we impose parity in the complexity of each hot region---i.e., an equal number of shape parameters and temperature components. We then consider models in which the hot regions have unequal complexities because it becomes clear that increasing the complexity of a particular hot region is unwarranted. In all cases we define the support of the joint prior so as to exclude limiting configurations in which the hot regions overlap; the reason being that extension of scope to such configurations requires specification and implementation\footnote{Efficient finite-element representations.} of additional logical conditions for a complete order of precedence in local radiation intensity evaluation. For a subset of models we impose antipodal symmetry of the hot regions, in order to crudely represent a heating distribution that is consistent with symmetry in the physical mechanisms driving surface X-ray emission, such as a dominantly or perfectly centered-dipolar field configuration. A magnetic field with finite higher-order structure is viewed as a closer approximation of physical reality; we represent this case crudely by breaking antipodal symmetry and defining additional parameters for a secondary hot region that are not derived from parameters of a primary hot region. However, self-consistent coupling of the magnetosphere to the surface radiation field is beyond the scope of this work, partly due to the associated increase in complexity of efficient model implementation for posterior computation.

We consider models with three or more mutually disjoint and separated radiating regions to be a logical extension of the model space if it is deemed that incrementally increasing the complexity of only two such regions is yielding insufficient advancement in posterior predictive performance---i.e., is not satisfactorily capturing observed structure in data for the resource expenditure---when approaching or extrapolating to the limit of what is considered computationally tractable by a group executing posterior computation. A salient advantage of such an approach is that it is more exhaustive with ideas for two hot regions and incrementally breaks symmetries; an apparent disadvantage is that some small set of conceivable closely-related models with equal (continuous parameter) dimensionality are not applied.\footnote{For example, configurations in which three single-temperature regions are disjoint.} It is however necessary to be selective---inherent to which is subjectivity and arbitrariness.

The above choices for surface radiation field configuration are somewhat consistent with the notion of the source being a rotation-powered X-ray pulsar with two relatively small hot regions that are disjoint. Therefore we consider the proposed model space to be logically structured and a reasonable representation of widely-held conceptions of such stars that are yet to be falsified statistically.

Whether or not such a model space is tractable given algorithm properties and computing resources is highly sensitive to the choice of parameterization for posterior computation, especially in phenomenological contexts. If degeneracies plague the problem at hand, it may be considered an indicator that either: (i) the model is simply ill-defined, leading to forms of invariance of the parameterized sampling distribution on the space of the data; (ii) the model is needlessly or at least \textit{unhelpfully} complex for describing observations, because despite (physical) parameters having a finite effect on forward data-generation, one is ultimately insensitive to such model structure. Usually this equates---at least in-part---to transforming away nonlinear likelihood function degeneracies where possible. A number of sophisticated open-source sampling software packages efficiently handle linear degeneracies, even in multi-modal contexts, but nonlinear degeneracy remains fiendish: certain sampling algorithms can perform accurately,\footnote{At reduced efficiency relative to simpler contexts---see the \MultiNest sampler cited in Appendix~\ref{sec:posterior computation}.} but coupled with an expensive (numerical) likelihood function and moderately high-dimensional sampling spaces, still require massive computing resources.

As we highlight in Section~\ref{subsec:definitions} (where we provide more precise definitions of surface radiation field structure), the choice of parameterization of a (largely phenomenological) hot region plays a crucial role in sampling-space definition. In Appendix~\ref{sec:posterior computation} we summarize the techniques adopted for posterior computation: we opt to perform nested sampling \citep[][]{skilling2006}. The natural space for nested sampling is usually that of a unit hypercube, which maps to an equal-dimensional physical parameter space according to an inverse transformation derived from a joint (prior) probability distribution on the physical space \citep[e.g.,][]{MultiNest_2009}. Given that nested sampling algorithms tend to operate in such a native space, a parameterization that approaches optimality involves both the physical parameterization \textit{and} the inverse transformation from the native to physical space, and is such that continuous likelihood function degeneracies, if existent, manifest effectively linearly in the native space.

Usually a joint prior distribution is chosen to be weakly informative---or ``flat''---in the context of the likelihood function, and in some cases is defined as an absolutely flat density function with respect to a joint space. It follows that the mapping from the parameter space to the native sampling space will then approximately preserve the linear degeneracy of a posterior mode. An exception to this occurs if the boundary of the support of the joint prior satisfies some set of non-trivial constraint equations.\footnote{An example of trivial constraint equations here are those that generate a (hyper-)rectangular support boundary in parameter space.} 

\textbf{Priors}. We define the joint prior density distribution $p(M,R_{\textrm{eq}})$ to be \textit{jointly} flat with compact support: a prophylactic choice that eases future use of samples on the $(M,R_{\textrm{eq}})$-subspace for computing an approximative marginal likelihood function, which in turn can be used for estimation of interior source-matter properties---principally EOS parameters \citep[see][]{Riley18,riley19b}. We also choose the boundary of prior support to be close to maximally inclusive in regards to theoretical EOS predictions: we impose hard bounds $M\in[1,3]\;M_{\odot}$, and impose that $R_{\textrm{eq}}\in[3r_{g},16]$ km, where $r_{g}=r_{g}(M) = GM/c^2$ is the gravitational radius.

Although we allow the boundary of the prior support to extend down to the photon sphere of the ambient spacetime solution, when computing pulse-profiles we only integrate over the \textit{primary} images (along rays with angular deflection $\leq\pi$) of radiating elements subtended on the sky of the instrument. For a spherical star of radius $R<3.52r_{g}(M)$, multiple images of parts of the star will be visible \citep[][]{Pechenick83,bogdanov19b} requiring that light from the primary, secondary, and higher-order images be included---at additional computational expense---in an exact calculation of the flux; this issue, and how it pertains to oblate stars, is discussed in more detail elsewhere \citep{bogdanov19b}. From a computational statistics perspective, one could view the inclusion of one or more higher-order images as a modeling refinement to be made if, \textit{a posteriori}, a rotating star is favored to be sufficiently compact: e.g., a substantial fraction of posterior mass lies at $R_{\rm eq}/r_{g}(M)\lesssim3.6$. The parameter inference reported in this work favors much less compact stars \textit{a posteriori} (refer forward to Figure~\ref{fig:STpPST spacetime marginal posterior}), so multiple imaging is not deemed important. However, when images are neglected, the issue of choosing the most appropriate support for a joint prior distribution of $M$ and $R_{\rm eq}$ remains an open problem for statistical modeling.

A typical likelihood function for pulse-profile modeling will express many modes of dependence on the compactness $r_{g}(M)/R_{\textrm{eq}}$, and will generally be more sensitive to this combination than to $M$ (or $R_{\textrm{eq}}$) individually. We ensure the mapping from the parameter space to the native space preserves such linear degeneracy between $M$ and $R_{\textrm{eq}}$. In Appendix~\ref{app:prior transforms} we provide implementation details for the joint density $p(M,R_{\textrm{eq}})$.

Finally, there exists a constraint on the distance of the PSR~J0030$+$0451 system \citep[][]{nanograv11} that we adopt---in approximation---as an informative prior.

\subsubsection{Instrument}\label{subsubsec:instrument general}

In defining a generative model, the data space is constructed by phase-folding X-ray events in each detector channel and grouping those events into a uniform set of phase intervals (bins) to curate a set \textit{count} numbers, typically with cardinality $\mathcal{O}(10^{3})$. The conditional joint sampling distribution of these count numbers is always constructed in terms of a phase-energy-resolved signal that is generated during a \textit{single} rotation of the source. Together with an appropriate nuisance background model (see Section~\ref{subsubsec:background general}), it follows that the instrument in such an analysis represents the temporal-mean operation of all detectors collectively in response to the incident radiation field from the source during the observation time intervals. We reserve a more elaborate discussion on these modeling facets for Section~\ref{subsubsec:likelihood functions} \citep[and also refer the reader to][]{riley19b}.

For every model we invoke the instrument response model: the on-axis \TT{v1.02} \textsl{ancillary response function} (ARF) and an updated version of the \TT{v1.02} \textsl{redistribution matrix file} \citep[RMF; private communication from James Steiner, see][for details of the updates]{Hamaguchi19} to generate a reference (or nominal) response matrix $\mathcal{R}^{\star}$ derived from microphysical knowledge. Let detector channels increment with row number $i$, and energy intervals increment with column number $j$, such that an element of the reference response matrix is denoted by $\mathcal{R}_{ij}^{\star}$. We use this reference matrix as a basis for a parameterized family of response matrices, and aim to compute (for each model) a joint posterior density distribution of continuous source parameters and continuous instrument (response matrix) parameters. It is well-founded to parameterize the instrument because despite its synthetic nature, we do not consider its microphysical operation to be sufficiently \textit{known};\footnote{In-flight astrophysical calibration sources, for instance, are in practice far brighter than science targets, and operation is conditional on the radiation field incident on the detectors.} nevertheless, we define far fewer parameters for the instrument than for the source. Whilst the following model ensures that operational uncertainty is included, the continuum of response models and the associated prior density distribution does not attempt to rigorously represent uncertainty in microphysical knowledge; given a close approximation to the radiation field incident on the telescope, we would expect the model to be conservative in terms of \textit{prior} predictive performance.

We parameterize the response matrix using a calibration product derived from observations of a calibration source.  For this work we use instrumental residuals derived from \NICER observations of the Crab. These residuals are derived using the observations and following the procedure outlined in \citet{Ludlam18}, modified to use the appropriate number of detectors, ARF, RMF, and sun-angle cut consistent with the PSR~J0030$+$0451 data set (refer to Section~\ref{subsec:data}). We acknowledge that the Crab is a remarkably different source to PSR~J0030$+$0451: the expected operation of \NICER (and X-ray instruments in general) in response to incident radiation fields is a function of its properties. The Crab exhibits a very different spectrum to PSR~J0030$+$0451, being harder, more absorbed, and subject to astrophysical features; the Crab is also an extended source, not a point source, and is much brighter than the rotation-powered MSPs targeted by \NICER.

The calibration product is a channel-by-channel vector $\mathscr{R}$ of ratios of observed Crab count numbers to count numbers derived using a theoretical incident spectrum and the reference response matrix $\mathcal{R}^{\star}$. Let the \replaced{vector $\mathscr{R}\coloneqq\mathcal{C}\oslash\left(\mathcal{R}^{\star}\cdot\mathcal{F}\right)$: in words, a Hadamard operation between an \textit{observed} count vector $\mathcal{C}$ from some calibration source, and}{vector elements $\mathscr{R}_{i}\coloneqq\mathcal{C}_{i}/\left(\mathcal{R}^{\star}\cdot\mathcal{F}\right)_{i}$: in words, the element-wise division of an \textit{observed} count vector $\mathcal{C}$ from some calibration source, by} a vector $\mathcal{R}^{\star}\cdot\mathcal{F}$ where $\mathcal{F}$ is a vector of photon \textit{fluences} (in the set of energy intervals inherent to the definition of $\mathcal{R}^{\star}$) computed given some theoretical model of said calibration source during the calibration observations. Note that the calibration product is derived from observations of a single chromatic source and thus is not resolved over elements of the matrix, only over the set of channels; therefore we apply the ratio $\mathscr{R}_{i}$ for the $i^{th}$ channel to all elements $\mathcal{R}^{\star}_{ij}$. The theoretical model is uncertain in the lowest ten channels we consider: the Crab is highly absorbed so that there is less data at low energies, and the telescopes used to generate the reference spectra for the residuals also perform poorly in this regime.  In this work we therefore assume that $\mathscr{R}_{i}$ for $i\in[25,35)$ is equal to $\mathscr{R}_{35}$.

\begin{figure*}[t!]
\centering
\includegraphics[clip, trim=0cm 0cm 0cm 0cm, width=\textwidth]{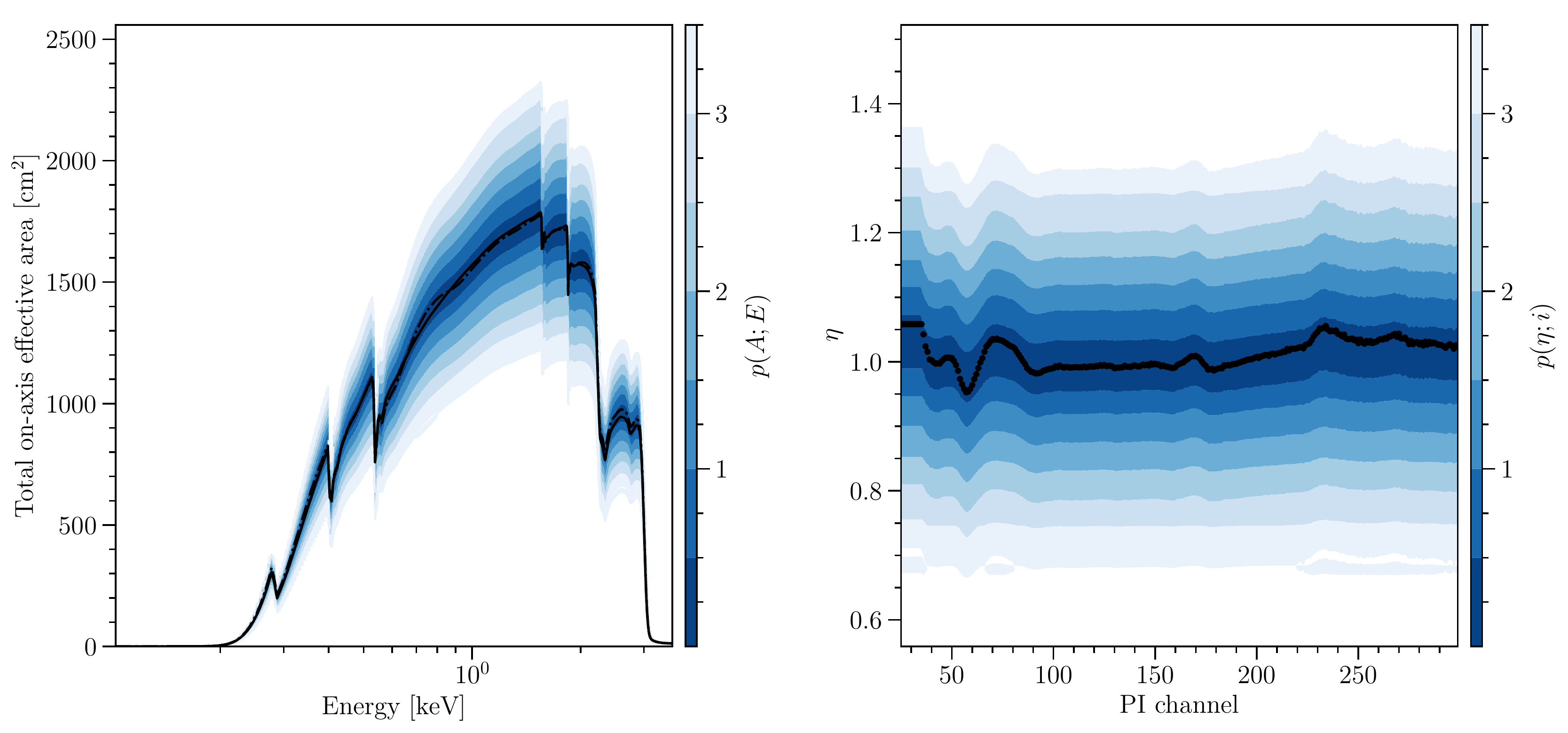}
\caption{\small{In the \textit{left} panel we display marginal conditional prior probability density distributions of total on-axis effective area $A$ as a function of energy (integrated over uniform intervals of width $5\times10^{-3}$ keV), summed over the contiguous channel subset $[25,300)$. Note that the total \NICER effective area at energies $\gtrsim3$~keV in particular is far greater than shown here. The prior $p(A;E)$ is represented by the \textit{blue} contours. At energy $E$ a set of (one-dimensional) \textit{highest-density} credible intervals are estimated for $A=A(\alpha,\beta,\gamma)$; the credible intervals are connected as a function of energy, such that, e.g., the second-darkest band encodes the energy-dependence of the estimated $68.3\%$ highest-density credible interval. We give the marginal prior distributions for the parameters $\alpha$, $\beta$, and $\gamma$ in Table~\ref{table: STpPST}. Note that the posterior information shown is \textit{not} that of the joint distribution of effective areas over energy intervals: the effective areas are coupled by a functional form with three parameters. The range of the energy intervals is determined based on the RMF of the reference matrix $\mathcal{R}^{\star}$ and on the curated data set in the contiguous set of channels $[25,300)$, where these channels are summed over. The \textit{solid} curve is that of the reference matrix $\mathcal{R}^{\star}$ (with $\gamma=1$); the \textit{dash-dot} curve is that of the calibrated matrix $\mathcal{R}_{ij}=\mathscr{R}_{i}\mathcal{R}^{\star}_{ij}$ (with $\alpha=1$). In the \textit{right} panel we display as \textit{black} points the elements of the vector $\mathscr{R}$ of multipliers supplied for instrument parameterization based on the Crab as a calibration source. The prior $p(\eta;i)$, where $i$ enumerates channels and $\eta$ is defined in Equation~(\ref{eqn: instrument parameterization}), is represented by the \textit{blue} contours; the probabilistic information is otherwise congruent in nature to that described for the \textit{left} panel.}}
\label{fig:NICER instrument effective area prior}\label{fig:NICER instrument multiplier prior}
\end{figure*}

We choose to construct the response matrix as a continuous three-parameter family, where the parameters are denoted by $\NICER\;\alpha$, $\NICER\;\beta$, and $\NICER\;\gamma$ where possible, but reduced to the aliases $\alpha$, $\beta$, and $\gamma$ respectively for clarity of mathematical expressions. First we give the definition, and then we offer an interpretation in words. The parameterized matrix is defined as
\begin{equation}
\begin{aligned}
    \mathcal{R}_{ij}(\alpha,\beta,\gamma;\mathscr{R},\mathcal{R}^{\star})
    &\coloneqq
    (1-\beta)\gamma\mathcal{R}^{\star}_{ij}
    +
    \beta\alpha\mathscr{R}_{i}\mathcal{R}^{\star}_{ij}\\
    &\equiv
    \gamma\mathcal{R}^{\star}_{ij}
    +
    \beta
    \underbrace{\left[\alpha\mathscr{R}_{i}\mathcal{R}^{\star}_{ij}
    -
    \gamma\mathcal{R}^{\star}_{ij}\right]}_{\textrm{calibration shift}}
    \equiv
    \mathcal{R}^{\star}_{ij}
    \underbrace{\left[\gamma + \beta \left(\alpha\mathscr{R}_{i} - \gamma\right)\right]}_{\eta_{i}}.
\end{aligned}
\label{eqn: instrument parameterization}
\end{equation}
In general an additional condition must be invoked: $\mathcal{R}_{ij}(\alpha, \beta,\gamma)\coloneqq0$ if $\left[(1-\beta)\gamma\mathcal{R}^{\star}_{ij}+\beta\alpha\mathscr{R}_{i}\mathcal{R}^{\star}_{ij}\right]\leq0$ for $(\alpha,\beta,\gamma)\in\mathcal{S}_{(\alpha,\beta,\gamma)}$, where $\mathcal{S}_{(\alpha,\beta,\gamma)}$ is the joint prior support on the $(\alpha, \beta,\gamma)$-subspace.\footnote{Alternatively, one might define the joint prior support such that if for $(\alpha, \beta)=(\alpha^{\prime},\beta^{\prime},\gamma^{\prime})$ the inequality is true for any $(i,j)$, then the joint density is locally zero at $(\alpha^{\prime},\beta^{\prime},\gamma^{\prime})$.} We note, however, that this parameterization exhibits a finite degree of degeneracy.

The parameter $\alpha$ scales the calibration vector $\mathscr{R}$, and manifests to target the assumption that the vectors $\mathcal{C}$ and $\mathcal{F}$ are \textit{known}: in regards to $\mathcal{C}$, the assumption that the temporal-mean operation of the instrument between calibration and science observations is invariant (including, e.g., pointing vector relative to source line-of-sight and the flux-dependent effects); and in regards to $\mathcal{F}$, the assumption that the expectation of the incident radiation field from the Crab during the calibration observations is known. Thus, the parameter $\alpha$ represents the product of dimensionless element-invariant scaling factors applied, respectively, to vectors $\mathcal{C}$ and $\mathcal{F}$. The support of $\alpha$ is such that $\mathcal{S}_{\alpha}=\{\alpha\colon\alpha\in\mathbb{R}^{>0}\wedge1-\epsilon_{\alpha}\leq\alpha\leq1+\epsilon_{\alpha}\}$ where $0<\epsilon_{\alpha}<1$; the prior density function allocates mass mostly to the near vicinity of unity.

The parameter $\gamma$ is an element-invariant scaling factor applied to the reference matrix $\mathcal{R}^{\star}$ alone. The reason that we choose $\alpha\neq\gamma$ is that $\gamma$ operates on $\mathcal{R}^{\star}$, and thus because $\mathscr{R}$ is a function of $\mathcal{R}^{\star}$ (via the \replaced{Hadamard operation}{element-wise vector division} written above), $\gamma$ cancels. The support of $\gamma$ is such that $\mathcal{S}_{\gamma}=\{\gamma\colon\gamma\in\mathbb{R}^{>0}\wedge1-\epsilon_{\gamma}\leq\gamma\leq1+\epsilon_{\gamma}\}$ where $0<\epsilon_{\gamma}<1$; the prior density function allocates mass mostly to the near vicinity of unity.

The parameter $\beta$ is a weighting factor between matrices with elements $\gamma\mathcal{R}^{\star}_{ij}$ and $\alpha\mathscr{R}_{i}\mathcal{R}^{\star}_{ij}$---i.e., the element-invariant coefficient of the element-dependent calibration shift away from the $\gamma\mathcal{R}^{\star}$ matrix. Thus $\beta$ may be interpreted as the degree to which the calibration vector $\mathscr{R}$ is used to modify the reference matrix $\mathcal{R}^{\star}$. Note that the prior support for a weighting parameter such as $\beta$ is the unit interval $\beta\in[0,1]$. We consider the limit $\beta\to0$ as a useful safeguard against erroneous calibration---e.g., artifacts may be introduced by invoking a calibration source to which the instrument responds appreciably differently than it does to PSR~J0030$+$0451. As $\beta\to0$, $\mathcal{R}_{ij}(\alpha,\beta,\gamma)\to\gamma\mathcal{R}_{ij}^{\star}$, capturing a simple element-invariant scaling of $\mathcal{R}^{\star}$. On the other hand, in the limit $\beta\to1$, we have $\mathcal{R}_{ij}(\alpha, \beta,\gamma)\to\alpha\mathscr{R}_{i}\mathcal{R}_{ij}^{\star}$.

\textbf{Priors.} To demonstrate the properties of the family of response matrices defined above, we: (i) define a joint probability density distribution $p(\alpha,\beta,\gamma)$ that could be plausibly viewed as a \textit{weak} prior on its compact support given that the instrument is an artificial system that has been closely studied; and (ii) then visualize the distribution in terms of derived properties (such as total effective area as a function on energy interval). Let: $\alpha\sim N(1,0.1)$ truncated such that $\alpha\in[0.5,1.5]$; $\beta\sim U(0,1)$; and $\gamma\sim N(1,0.1)$, truncated such that $\gamma\in[0.5,1.5]$. In Figure~\ref{fig:NICER instrument effective area prior} we display the prior distribution of the total on-axis effective area as a function of energy.

\subsubsection{Likelihood functions and background}\label{subsubsec:likelihood functions}\label{subsubsec:background general}

In the \XPSI documentation \citep{riley19b} we offer a more complete overview of the supported class of generative models than is appropriate for this work---instead we adapt the \XPSI documentation to provide a summary. A generative model for raw on-board event data is eschewed by subsuming a non-parameterized portion of the modeling within a data pre-processing phase. In this work, a pulsar radio timing solution is invoked to transform events into a simpler time domain: a fictitious instrument, which is static (or Eulerian) and distant in the (Schwarzschild) spacetime of the source, is implicitly constructed to register events against the elapsed natural number rotations of the star, which is a clock related to the Schwarzschild coordinate time simply by an affine transformation. In all models we condition on the phase-resolved specific flux signal (incident on the instrument) generated by precisely \textit{one} rotation of the star: accurate computation of such a signal, even when invoking spacetime spherical symmetry, is approaching the limit of what we consider tractable at present in terms of likelihood function callback cost for sampling processes in $\mathcal{O}(10)$-dimensional spaces.

In the context of a model for the joint probability distribution of observed events \citep[see][]{riley19b},\footnote{Leading to an unbinned likelihood function.} phase-folding said events \deleted{can be mathematically (and numerically) is equivalent to \textit{tessellation}} \footnote{\deleted{One can also visualize the operation as the cloning---or duplication---of the signal many times to span the subset of the time domain of the instrument over which events are distributed.}} \replaced{of the computed signal}{is equivalent to computing one (average) rotational pulse and replicating it} over the many rotational cycles in order to evaluate the likelihood function. It follows that, in this limit, no information is lost by transforming events to the unit interval \added{because the underlying information content in the model is not summarized for comparison to data}. In each instrument channel we choose to group events into a set of uniform-width phase intervals (bins) that are subsets of the unit interval, and define the data space as $\mathscr{D}\coloneqq\mathbb{N}^{I\times K}$, where $I\in\mathbb{N}$ is the number of channels over which the folded events are distributed, and $K\in\mathbb{N}$ is the number of phase intervals; we choose $K=32$ and $I=275$ corresponding to channel subset $[25,300)$. Information loss is an inherent consequence of compression of events into a smaller set of summary quantities, but the phase resolution is sufficiently high here to mitigate our concern about the use of a binned likelihood function instead of an unbinned likelihood function. The conditional joint sampling distribution on the space $\mathscr{D}$ is assumed to be purely Poissonian and separable over channels: the \NICER instrument exhibits sufficiently high-resolution event-timing capabilities for the Poissonian nature of the incident radiation field to be effectively conserved as an event arrival process in the on-board time domain \citep[see][for a more explicit set of arguments pertaining to this matter]{riley19b}.

Let the likelihood function for phase-folded and binned events be defined by \citep[see also][]{Miller15}
\begin{equation}
L(\boldsymbol{\theta},\boldsymbol{B})
\coloneqq
p(\{\boldsymbol{d}_{i}\}_{i=1,\ldots,I}\,|\,\boldsymbol{\theta},\{B_{i}\})
=
\mathop{\prod}_{i,k}p(d_{ik}\,|\,\boldsymbol{\theta},B_{i}),
\end{equation}
where: $i\in[1,I]$ enumerates channels of the instrument; each $\boldsymbol{d}_{i}$ is a data vector associated with the $i^{th}$ channel, constituted by count numbers $\{d_{ik}\}_{k=1,\ldots,K}$ where $k$ enumerates phase intervals $\phi_{k}\subset[0,1]$; $B_{i}$ is the (nuisance) background count-rate parameter in the $i^{th}$ channel; and $\boldsymbol{\theta}$ are the continuous source parameters that constitute the sampling space, and on which source expected count numbers $s_{ik}(\boldsymbol{\theta})$ are dependent. Each $B_{i}$ is defined as the expectation of a homogeneous---i.e., time-invariant---Poisson arrival process: up to a known constant,
\begin{equation}
\ln L(\boldsymbol{\theta},\boldsymbol{B})
=
-2\mathop{\sum}_{i,k}d_{ik}\ln\left[s_{ik}(\boldsymbol{\theta})+B_{i}\left(\mathop{\sum}_{\ell}\Delta t_{\ell}\right)\mathop{\int}_{\phi_{k}}d\phi\right]
-
s_{ik}(\boldsymbol{\theta})-B_{i}\left(\mathop{\sum}_{\ell}\Delta t_{\ell}\right)\mathop{\int}_{\phi_{k}}d\phi.
\label{eqn:channel-by-channel backgrounds likelihood function}
\end{equation}
Such background treatment is the default for \NICER parameter estimation work and \XPSI \citep[for implementation details see appendix~B of][]{riley19b}, and is based on \citet[]{Miller15}. In reality, the statistical properties of backgrounds do not exhibit time invariance, but exhibit long-term variation over the $\mathcal{O}(1)$~year observation epoch of MSPs target by the \NICER mission, especially due to factors such as dynamical space weather \citep{bogdanov19a}. However, any background emission processes---and dynamical emission processes in the local vicinity of PSR~J0030$+$0451---that are not harmonically coupled to the surface X-ray emission will \textit{decohere} over the unit phase interval whose boundary is periodic. It follows that event phase-folding enables invocation of phase-invariant channel-by-channel (background) count-rate terms.

The background below $\sim\!3$~keV for \NICER observations of PSR~J0030$+$0451 observations consists of \citep{bogdanov19a}: (i) cosmic energetic particle events and \deleted{unresolved} diffuse X-ray emission over the $\sim\!30$~arcmin$^{2}$ field of view \citep[][]{Arzoumanian14}; (ii) many nearby X-ray point sources in the field that make a small total contribution relative to the targeted MSP; and (iii) solar system contamination, including optical loading (pointing sun-angle dependent), and high-energy non-cosmic particles and radiation. Considering a proper subset of detector channels and filtering background events during the pre-processing phase acts to reduce background contribution, but some subset of background events survive and must be modeled \citep{bogdanov19a}.

The source terms $s_{ik}(\boldsymbol{\theta})$ in Equation~(\ref{eqn:channel-by-channel backgrounds likelihood function}) are then derived as follows. Let $F(\phi,E;\boldsymbol{\theta})$ denote the incident specific photon flux from the source as a function of rotational phase $\phi$. The function $F(\phi,E;\boldsymbol{\theta})$ is evaluated numerically at a regular discrete set of points in the joint space of energy and phase, as an approximating two-dimensional integral over the solid angle of the image of the source subtended on the sky of a point in the vicinity of the distant static instrument; given the discrete representation, a continuous representation is constructed via spline interpolation in \XPSI.

Let the symbol $\mathcal{R}_{\ell ij}(\hat{r})$ denote a temporal-mean point-source response matrix invoked for the $\ell^{th}$ observing interval, which is dependent on the radial coordinate unit vector $\hat{r}$ in the Schwarzschild chart at the location of the instrument---i.e., the pointing of the telescope relative to the source. The response matrix with elements
\begin{equation}
\mathcal{R}_{ij}(\hat{r})
=
\mathop{\sum}_{\ell}
\frac{\Delta t_{\ell}}{\left(\mathop{\sum}_{\ell}\Delta t_{\ell}\right)}
\mathcal{R}_{\ell ij}(\hat{r})
\coloneqq
\mathcal{R}_{ij}(\alpha,\beta,\gamma)
\end{equation}
is the exposure-time-weighted mean response matrix that is modeled as the matrix defined in Equation~(\ref{eqn: instrument parameterization}) and Section~\ref{subsubsec:instrument general}.

The source contribution to the expected number of counts is given by
\begin{equation}
\begin{aligned}
s_{ik}(\boldsymbol{\theta})
&=
\mathop{\sum}_{\ell,j}
\mathcal{R}_{\ell ij}(\hat{r})
\Delta t_{\ell}
\mathop{\int}_{\phi_{k}\times\boldsymbol{E}_{j}}F(\phi,E;\boldsymbol{\theta})dEd\phi\\
&=
\mathop{\sum}_{j}
\left[\left(\mathop{\sum}_{\ell}\Delta t_{\ell}\right)
\underbrace{\mathop{\int}_{\phi_{k}\times\boldsymbol{E}_{j}}F(\phi,E;\boldsymbol{\theta})dEd\phi}_{\boldsymbol{F}_{jk}(\boldsymbol{\theta})}\right]
\mathop{\sum}_{\ell}
\frac{\Delta t_{\ell}}{\left(\mathop{\sum}_{\ell}\Delta t_{\ell}\right)}
\mathcal{R}_{\ell ij}(\hat{r}).
\end{aligned}
\label{eqn:phase-folded source count numbers}
\end{equation}
A matrix of source count numbers may thus be evaluated as the dot-product\footnote{An approximation in many ways, one being the discrete representation of an instrument that responds in a continuous manner to input.}
\begin{equation}
\boldsymbol{s}(\boldsymbol{\theta})
=
\left(\mathop{\sum}_{\ell}\Delta t_{\ell}\right)
\mathcal{R}(\alpha,\beta,\gamma)
\cdot
\boldsymbol{F}(\boldsymbol{\theta}),
\end{equation}
where $\boldsymbol{F}(\boldsymbol{\theta})$ is a matrix of phase-integrated incident photon fluxes.\footnote{In the \XPSI implementation phase integration is performed using splines after a dot-product operation on a matrix of instantaneous incident photon fluxes.} Note that the elements of the matrix $\boldsymbol{F}(\boldsymbol{\theta})$ may be approximated using instantaneous fluxes at points \textit{within} the finite phase intervals instead of explicitly integrating over those intervals, provided that the intervals are determined to be sufficiently small.

We numerically marginalize the likelihood function given by Equation~(\ref{eqn:channel-by-channel backgrounds likelihood function}) over the subspace of nuisance background parameters $\boldsymbol{B}$ in order to improve tractability of the sampling process. The target distribution (the posterior) for sampling is written conditional on model $\mathcal{M}\subset\mathscr{M}$ as
\begin{equation}
\pi(\boldsymbol{\theta}\,|\,\boldsymbol{d},\mathcal{M})
=
\mathop{\int}\pi(\boldsymbol{\theta},\boldsymbol{B}\,|\,\boldsymbol{d},\mathcal{M})d\boldsymbol{B}
\propto
p(\boldsymbol{\theta}\,|\,\mathcal{M})
\underbrace{\mathop{\int}
L(\boldsymbol{\theta},\boldsymbol{B})
p(\boldsymbol{B}\,|\,\mathcal{M})
d\boldsymbol{B}}_{L(\boldsymbol{\theta})},
\end{equation}
where $L(\boldsymbol{\theta})=p(\boldsymbol{d}\,|\,\boldsymbol{\theta},\mathcal{M})$ is the marginal likelihood function supplied as a callback for a sampling process, and the joint prior density distribution $p(\boldsymbol{\theta},\boldsymbol{B}\,|\,\mathcal{M})$ is separable with respect to $\boldsymbol{\theta}$ and $\boldsymbol{B}$. The joint prior distribution $p(\boldsymbol{B}\,|\,\mathcal{M})$ is equivalent for all models: jointly flat and separable. Crucially, such a phenomenological background model exhibits a large prior complexity; for instance, the joint density at a background count rate vector $\boldsymbol{B}$ where the variation between channels is always small relative to the limiting instrument count rate, is equivalent to the joint density at a vector $\boldsymbol{B}$ whose elements exhibit vast channel-to-channel variations. It follows that a joint flat, separable prior is not considered representative of our prior belief. However, we consider---without proof---the prior to be weakly informative because: (i) the conditional likelihood function (given a fixed source vector) exhibits a large curvature relative to the prior density function; (ii) the source photon flux signal always has few extrema in the joint space of energy and phase; and (iii) \textit{a posteriori} the conditional likelihood function maxima do not wildly fluctuate as a function of channel because such structure does not exist in the data set. Nevertheless, if one were to compare, based on prior predictive performance, the models that we consider in this work with a model invoking a background component with far lower complexity, one should not be surprised if the former are strongly disfavored.

In general, the support of $p(\boldsymbol{B}\,|\,\mathcal{M})$ is compact and bounds can be specified on a channel-by-channel basis to truncate the marginalization integrals. Lower-bounds may be derived, for example, from calibration observations of nearby fields that exclude the PSR~J0030$+$0451 and are otherwise devoid of bright sources. Upper-limits may, for example, be based on distinct \NICER observations of the field containing PSR~J0030$+$0451.

For this work, however, we define the lower-bound as zero for each channel, and we eschew definition of an upper-bound in each channel because the posterior is considered integrable: non-diverging on joint compact support, and the conditional likelihood function---$L(\boldsymbol{\theta},\boldsymbol{B})$ for fixed $\boldsymbol{\theta}$---asymptotes to zero at large background count rates. If a set of sufficiently high upper-bounds \textit{were} specified (e.g., based on \NICER count rate limits), the associated normalizing constant for the joint prior, equal to the reciprocal of the products of those bounds, would not modulate relative probability measures defined on $\mathscr{M}$.\footnote{The raw event rate during the exposures used to curate the data set for this work is known to be far below limiting and thus the conditional likelihood function is always relatively small for near-limiting background count rates. A set of upper-bounds defined in this limit therefore truncates the evidence integral in a regime where model-dependent sensitivity is negligible.} We therefore do not view the improperness of the above prior as a misdemeanor, but it does mean that we should not describe our model as \textit{generative} in the strictest sense.

The numerical marginalization operation implemented is described in Appendix~B of the \XPSI documentation \citep{riley19b}.

\subsection{Model-specific definitions}\label{subsec:definitions}

The surface heating distribution by realistic magnetospheric (return) currents remains uncertain. The global magnetic field may be more complex than a simple dipole (at least in the near vicinity of the surface), whilst the mapping between currents and surface temperature field is not well-determined by existing theoretical models \citep{Harding01,Harding11,Timokhin13,Philippov15, Gralla17,Lockhart19}. We thus consider a set of simplified models that are representative of the various theoretical possibilities, albeit restricting our analysis to models with two distinct hot regions. We allow for the possibility of the hot regions being non-antipodal and non-identical \citep{Pavlov97,Bogdanov07,Bogdanov08,Bogdanov13}; we also consider various hot-region shapes, including circles, rings, and crescents filled with material of uniform local comoving temperature.

While our choices are physically motivated, it is important to emphasize that our inferences are conditional upon these choices. However, posterior computation is computationally intensive and scales with model complexity. The work presented here is based on usage of a $500,000$ core-hour grant on the Dutch national supercomputer Cartesius; we thus find it pragmatic to disseminate information to the community at this point. Further exploration of model variants or execution of higher-resolution calculations requires additional resource allocation on high-performance systems, building upon the information offered here as guidance. In particular, it will be important to explore the sensitivity of marginal posterior estimates of fundamental physical parameters of interest---i.e., exterior and interior spacetime parameters---to expansion of the space of models that we have had the resources to consider.

In this section we detail the properties that distinguish the models in our model space. For each model we give the parameterization details and any remarkable prior details; notes on the \textit{support} of the joint prior distribution are given where appropriate. We also discuss the existence of (continuous linear and nonlinear) degeneracy in posterior modes for a given parameterization, which we interpret as one indicator of unnecessary complexity for a surface radiation field with phenomenological spatial structure. For each model whose associated posterior distribution we compute, we provide a summary table containing a more precise definition of the joint prior distribution; these tables may be found in Section~\ref{sec:inferences} and in Appendix~\ref{app:model summaries}.

\subsubsection{Single-temperature regions with antipodal symmetry (\TT{ST-S}) and with unshared parameters (\TT{ST-U})}\label{subsubsec:ST-S and ST-U}

\begin{figure*}[t!]
\centering
\includegraphics[clip, trim=0cm 0cm 0cm 0cm, width=0.75\textwidth]{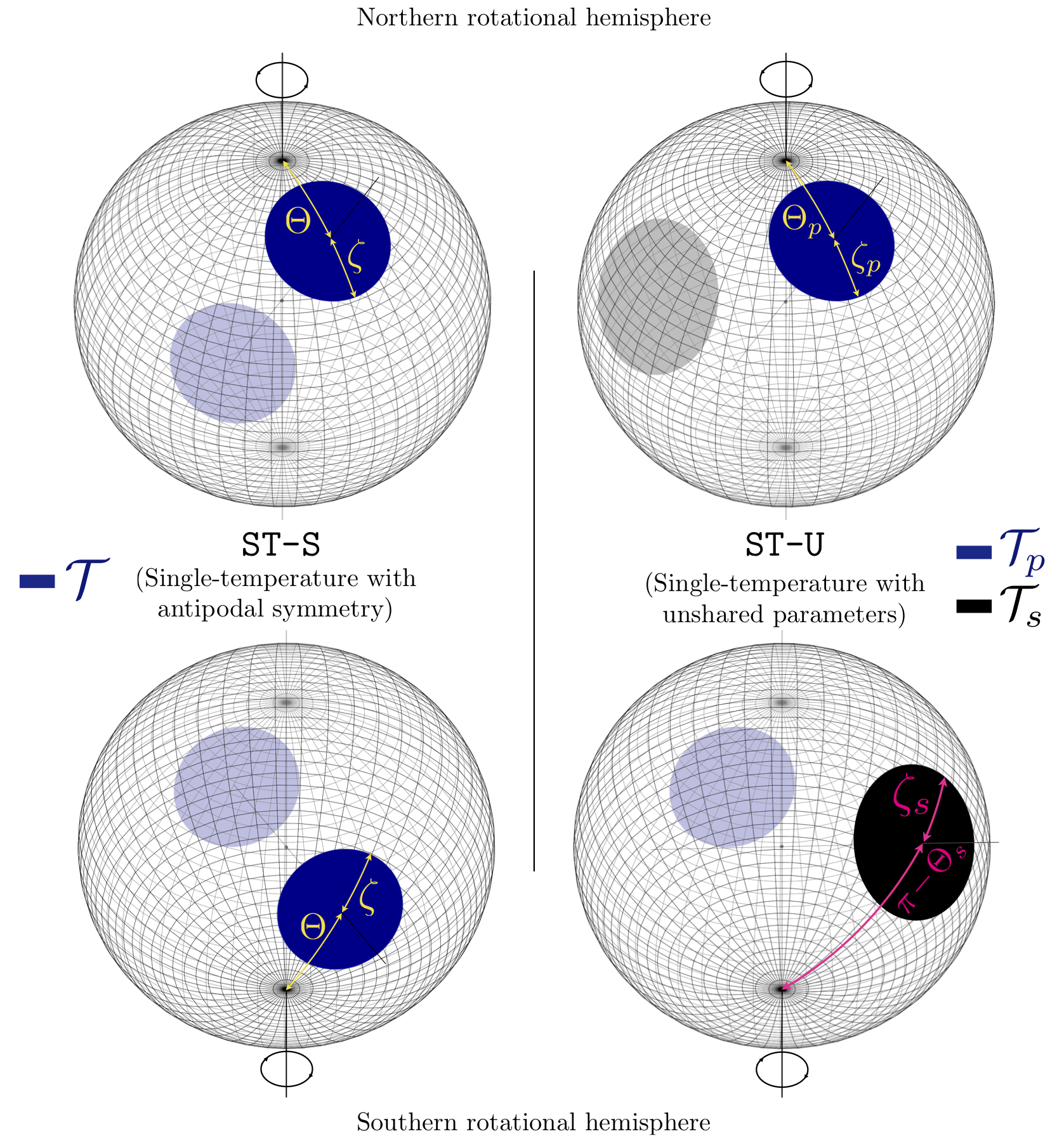}
\caption{\small{Schematic diagrams of models with \textit{single-temperature} regions: \TT{ST-S} defined by antipodal symmetry of the primary and secondary regions, and \TT{ST-U} defined by the primary and secondary regions \textit{not} sharing any parameters. Note that the $2$-surface shown is spherical: the hot regions are defined in spherical coordinates on a (unit) sphere and projected onto the rotationally-distorted (oblate) spheroid representing the $2$-surface of the star, such that the proper areas of the regions are only implicitly defined. To construct the \TT{ST-U} diagram from the \TT{ST-S} diagram we: displaced the secondary region (defined by having a center colatitude $\Theta_{s}\geq\Theta_{p}$) away from the antipode of the center of the primary region (with center colatitude $\Theta_{p}$); decreased $\zeta_{s}$ such that $\zeta_{s}<\zeta_{p}$; and assigned the secondary region a distinct temperature parameter. Note that no numeric relationship is implied between the effective temperatures of material within each member---indeed, the prior is separable with respect to the temperatures ($\mathcal{T}_{p}$ and $\mathcal{T}_{s}$), and the support includes combinations $\mathcal{T}_{s}\leq\mathcal{T}_{p}$ and $\mathcal{T}_{s}>\mathcal{T}_{p}$. For clarity we display for \TT{ST-S} a projection showing the southern rotational hemisphere; in subsequent diagrams we omit such a projection when antipodal symmetry applies.}}
\label{fig: schematic ST diagrams}
\end{figure*}

\textbf{Parameterization.} The primary hot-region (refer to the leftmost panel of Figure~\ref{fig: schematic ST diagrams}) is simply-connected and encloses radiating material---a fully-ionized hydrogen \TT{NSX} atmosphere with effective temperature $\mathcal{T}$. The boundary of the region is \textit{circular}: i.e., given a \textit{center} point on the surface with colatitude $\Theta_{p}$ and azimuth $\phi_{p}$ (in a spherical coordinate basis whose polar axis is defined as the stellar rotation axis), the boundary is the locus of points that are equidistant\footnote{The ambient spacetime is static, and with respect to a Schwarzschild chart, the points are equidistant in angular coordinates. However, when projected from a spherical $2$-surface onto that of a rotationally deformed (oblate) spheroid, the spacelike separation---on a Schwarzschild temporal hyperslice---between the center point and boundary points is \textit{not} invariant.} in angular space from the center point.

Hereafter we use the alias \TT{ST-S}, parsed as \textit{Single-Temperature-Shared}. The surface radiation field associated with the secondary hot-region is derived exactly by applying antipodal symmetry to the primary region: there are no free parameters associated with the secondary region.

Similarly, we use the alias \TT{ST-U}, parsed as \textit{Single-Temperature-Unshared}. The primary region (refer to the rightmost panel of Figure~\ref{fig: schematic ST diagrams}) definition is retained from \TT{ST-S} as defined above. The secondary region, however, is now endowed with distinct parameters---i.e., the region is \textit{not} derived from the primary region under antipodal symmetry. The parameters of the secondary region have an otherwise equivalent meaning---in terms of surface radiation field specification---to their primary-region counterparts. 

\textbf{Degeneracy.} We note that a discrete degeneracy---multi-modality---can in principle arise for a source such as PSR~J0030$+$0451, but may only be weak when there is detectable asymmetry between the two component pulses over the course of one rotational cycle. There may exist two phase solutions, each corresponding to a distinct mapping between hot regions (distinguished by colatitude) and the pulse components in the event data. For instance, the primary (lower-colatitude) region could in principle generate either of the component pulses, whilst the secondary region generates the other. Fortunately, a number of open-source sampling software packages are designed to handle multi-modality efficiently (at least in the absence of nonlinear degeneracy). If the asymmetry between the component pulses is clear, the posterior mass in one mode may be entirely dominant.

\textbf{Priors.} For \TT{ST-S}, we eliminate a region-exchange degeneracy by imposing a constraint $\Theta_{p}\leq\pi/2$ on the prior support. The primary region is uniquely defined as the region whose center subtends the smallest colatitude, $\Theta_{p}$, to the rotational axis, if the region colatitudes are different.

For \TT{ST-U}, we eliminate a region-exchange degeneracy by imposing a constraint $\Theta_{p}\leq\Theta_{s}$ on the support of the joint prior distribution. The primary region is uniquely defined as the region whose center subtends the smallest colatitude, $\Theta_{p}$, to the rotational axis, if the region colatitudes are different; the regions are distinguishable when $\Theta_{p}=\Theta_{s}$ according to the subset of parameters that controls their physical manifestation. The joint prior support is such that the two regions cannot overlap but otherwise are not restricted to be antipodally symmetric.

\subsubsection{Concentric single-temperature regions with antipodal symmetry (\TT{CST-S}) and with unshared parameters (\TT{CST-U})}\label{subsubsec:CST}

\begin{figure*}[t!]
\centering
\includegraphics[clip, trim=0cm 0cm 0cm 0cm, width=0.75\textwidth]{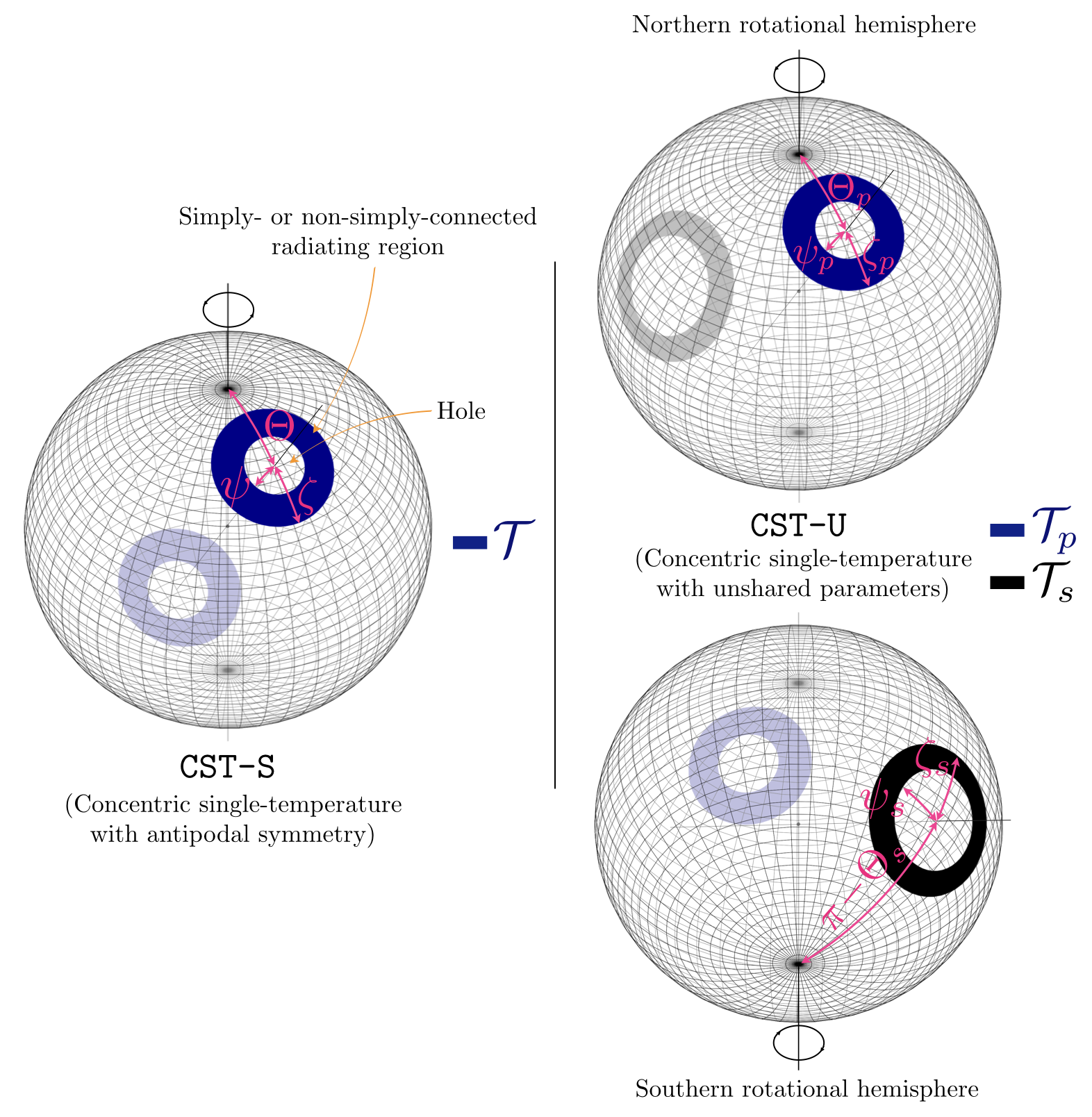}
\caption{\small{Schematic diagrams of models with \textit{concentric single-temperature} regions: \TT{CST-S} defined by antipodal symmetry of the primary and secondary regions, and \TT{CST-U} defined by the primary and secondary regions \textit{not} sharing any parameters.}}
\label{fig: CST schematic}
\end{figure*}

\textbf{Parameterization.} The primary hot-region (see the leftmost panel of Figure~\ref{fig: CST schematic}) is a non-simply-connected annulus (or ring) with outer angular radius $\zeta$, which contains material with effective temperature $\mathcal{T}$. The non-radiating \textit{hole} with angular radius $\psi$ is concentric (in angular coordinates) with the radiating annulus. We thus recover the shape defined for the \TT{ST-S} and \TT{ST-U} variants in the limit $\psi\to0$ (which is at the boundary of the prior support).

\begin{figure*}[t!]
\centering
\includegraphics[clip, trim=0cm 0cm 0cm 0cm, width=0.75\textwidth]{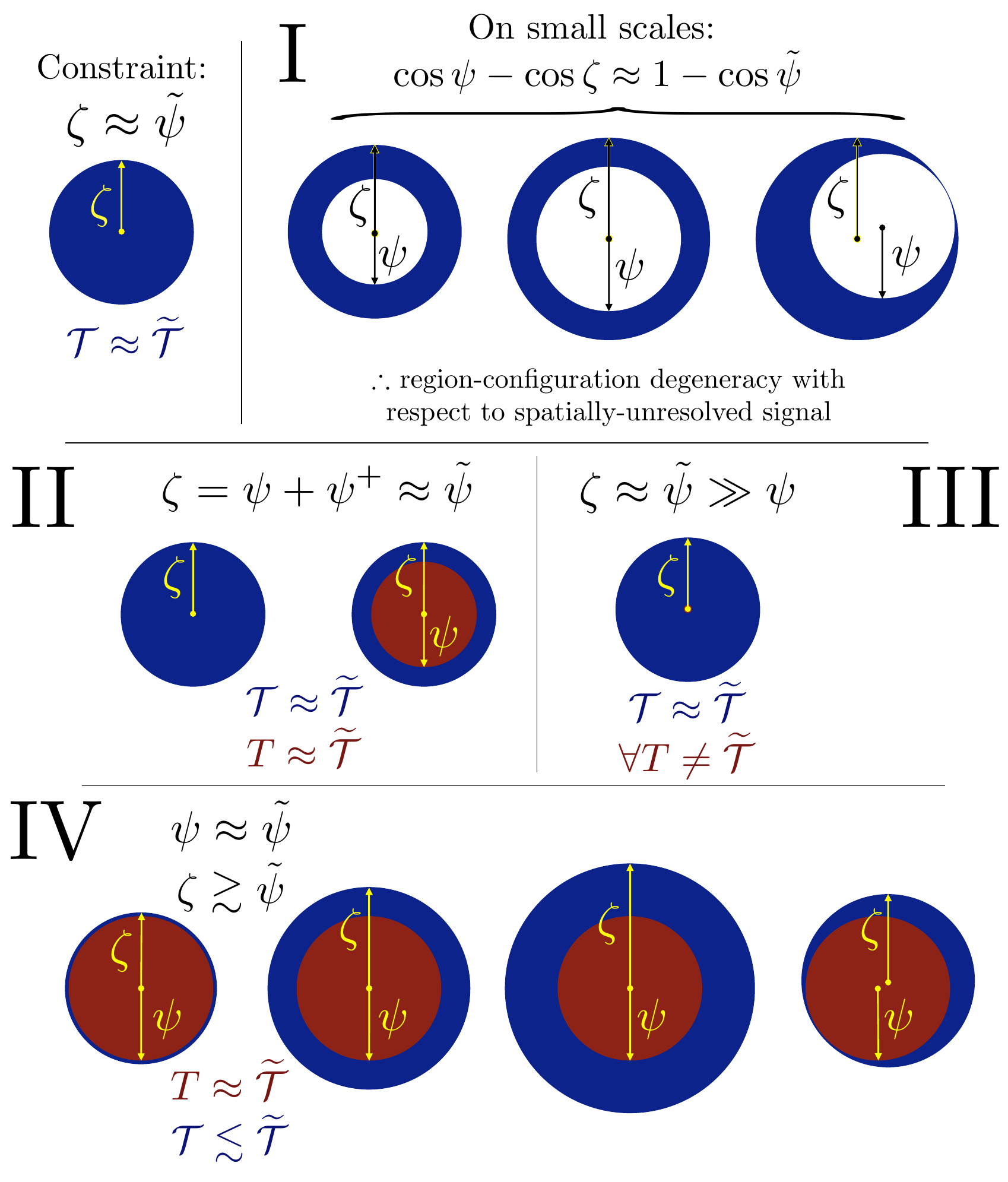}
\caption{\small{Schematic diagram of (approximate and exact) continuous degeneracies between hot-region structure parameters with respect to the energy-phase-resolved signal incident on a distant telescope. Degeneracy type I is discussed in detail in Section \ref{subsubsec:CST}, whilst types II through IV are discussed in Section \ref{subsubsec:CDT-S and CDT-U}; here we focus on defining the components of the diagram. In all cases, let us suppose that there exists a strong posterior constraint on the (coordinate) solid angle $\varpi$ subtended by a component of a hot region (at the stellar origin) with temperature $\widetilde{\mathcal{T}}$. \textit{Type I}.---We consider a hot region with a single-temperature component, and represent such a region as simply-connected with a strongly constrained radius $\zeta=\tilde{\psi}$. If the angular extent of the headed region is sufficiently small that we are insensitive (in the absence of spatial imaging power) to structural details other than solid angle $\varpi$, then we can introduce a hole of angular radius $\psi$ such that the hot region has the topology of a ring and a likelihood function degeneracy $\varpi=\cos\psi-\cos\zeta=\textrm{const.}$ manifests (on small angular scales). Similar arguments can be made for other parameterized structural modifications to the hot region. \textit{Type II}.---Let us consider a hot region constructed from an annulus and a hole (concentric or eccentric), each filled with material of local comoving temperature $\mathcal{T}$ and $T$ respectively. Let us assume a strong constraint on the outer angular radius $\zeta=\tilde{\psi}$ of the annulus, where $\tilde{\psi}$ is now \textit{not} restricted to small angular scales; let us further assume a strong constraint on the structure of the hot region as simply-connected circular with uniform temperature $\widetilde{\mathcal{T}}$, where only one temperature component is useful. It follows that degeneracy manifests for $\mathcal{T}\approx T\approx\widetilde{\mathcal{T}}$ and $\psi+\psi^{+}\approx\tilde{\psi}$, where $\psi^{+}\coloneqq\zeta-\psi$. \textit{Type III}.---As type I, but in the limit that $\zeta\gg\psi$, such that the signal generated by the hole, whose temperature is degenerate over a range dependent on $\psi$, is effectively turned off. \textit{Type IV}.---The signal generated by the \textit{hole} now satisfies the constraints assumed for type II---i.e., $T\approx\widetilde{\mathcal{T}}$ and $\psi\approx\tilde{\psi}$---whilst the solid angle, temperature, and coordinates of the annulus exhibit degeneracy with respect to configurations that effectively turn off the signal generated by the annulus.}}
\label{fig: continuous region degeneracies}
\end{figure*}

A generally useful way to distinguish the hole and the annulus---in particular for further increments in complexity---is as follows. Recall the term \textit{member} from Section~\ref{subsubsec:source}: consider two simply-connected partially-overlapping member regions, each wholly filled with radiating material, but impose the logical condition that when evaluating radiating intensities at a spacetime event on the stellar surface, one member---the hole---takes precedence if the event falls within its boundary. In this case (for the models here described), let the temperature of the material in the hole be (effectively) zero so that no signal need be computed for the hole. The statements in Section~\ref{subsubsec:source} pertaining to the proper areas in finite-element representations of radiating regions apply here: the annulus is a subset of a (simply-connected circular) ceding member that is \textit{not superseded} by the hole when evaluating local radiation intensities, and its proper area is computed (almost) exactly. These constructions are useful for further extension of the model.

Hereafter we use the alias \TT{CST-S}, parsed as \textit{Concentric-Single-Temperature-Shared}. For \TT{CST-S}, the surface radiation field associated with the secondary region is derived exactly by applying antipodal symmetry to the primary region: there are no free parameters associated with the secondary region. The annuli share an outer angular radius $\zeta$, and the holes share a fractional angular radius $f$ such that the hole angular radii are $\psi\coloneqq f\zeta$. 

Similarly, we use the alias \TT{CST-U}, parsed as \textit{Concentric-Single-Temperature-Unshared}. For \TT{CST-U}, the primary region (refer to the rightmost panel of Figure~\ref{fig: CST schematic}) definition is retained from \TT{CST-S} as defined above. The secondary region, however, is now endowed with distinct parameters---i.e., it is \textit{not} derived from the primary region under antipodal symmetry. The parameters of the secondary region have an otherwise equivalent meaning---in terms of surface radiation field specification---to their primary-region counterparts.

\textbf{Degeneracy.} We now consider the continuous degeneracy labeled I in Figure~\ref{fig: continuous region degeneracies}. When the angular extent of a radiating region is small, the signal generated by that region---as registered by a distant detector that does not spatially resolve (image) the star---is insensitive to its shape (detailed spatial structure). Sensitivity is here a measure in terms of the likelihood: i.e., the \textit{total} variation of the parameterized joint sampling distribution of a set of random variables,\footnote{And whose number usually exceeds the number of parameters.} in response to motion along a certain set of curves\footnote{Where those curves may more generally together generate $m$-dimensional surfaces in an $n$-dimensional space where usually $n>m$---if $m=n$ then no facet of the model is constrainable and the model is arguably not useful unless it can meaningfully tested in some other manner.} in parameter space, is small as summarized by the scalar likelihood. It follows that if the shape is parameterized with more than a single degree of freedom, the likelihood function is degenerate with respect to shapes that satisfy a constraint on the solid angle subtended by the radiating region at the center of the star, and thus which satisfy a constraint on the proper area of the radiating region.

The degeneracy is not in this instance exact, but holds approximately. Consider a ceding member of angular extent $\cos\zeta\lesssim1$ and hole with angular extent $\cos\psi>\cos\zeta$: the degeneracy is such that the solid angle, $\varpi(\zeta,\psi)=\cos\psi-\cos\zeta$, of the hot region (the annulus), is approximately $\varpi=1-\cos\tilde\psi$, where $\zeta=\tilde\psi$ is the angular radius in the limit $\psi\to0$. In other words, the relative size---and indeed existence---of the hole is at most weakly constrained on small angular scales. This degeneracy in the $(\zeta,\psi)$-subspace is nonlinear for $\zeta\to\tilde\psi$, but linearizes for increasing $\zeta>\tilde\psi$; however, with increasing $\zeta$ the signal generated by the region evolves away from that generated in the limit $\zeta\to\tilde\psi$ with $\psi\to0$. Note that if the superseding member is not concentric with the ceding member, and the overlap is only partial, the form of the degeneracy---the constraint equation satisfied---has additional dependence on the coordinates of the center of the hole relative to the center of the ceding member.

It is important to be aware of such degeneracy for the purpose of efficient posterior computation---inherent to which is accuracy. In our case, if the type of signal that is superior for describing the event data is generated by radiation from a localized region on the star, the constraint on the solid angle of the region can be viewed as dragging a posterior mode through parameter space along the type I degeneracy direction. Thus, with this parameterization, nonlinearity will exist that will reduce efficiency to some degree.\footnote{Which can only be robustly learned during computation.}

If the posterior predictive performance is maximal for signals that are generated by localized emission, and degenerate posterior structure is observed, then clearly the most effective manner in which to achieve efficiency increase is to \textit{reduce} the complexity of the structure of the radiating region. One thus inserts a simpler model into the model space, with the caveat that whilst estimation of ulterior model parameters should be insensitive to this reduction in complexity, the evidence may not be.

Alternatively, working with an integral summary variable such as the solid angle $\varpi$ of the (uniform temperature) radiating region is useful from the perspective of eliminating degeneracy by parametrising directly in terms of variables to which we are statistically sensitive. On the other hand, the mapping from $\varpi$ to variables that directly control the shape of the radiating regions can behave undesirably, and can thus complicate the action of extending models.

One potential avenue for efficiency improvement by linearizing the degeneracy, is to sample in the space of $(\cos\zeta,\cos\psi)$, which will eliminate the emergent small-scale nonlinear degeneracy. The cost is complication of the joint prior definition and implementation: in this case, a singularity\footnote{Note that Taylor-expanding on small angular scales to work in the joint space of $(\zeta^{2},\psi^{2})$ by definition cannot bypass the singularity.} exists in the mapping for $\psi\to0$, which is the boundary at which the radiating region reduces to being simply-connected as for both \TT{ST-S} and \TT{ST-U}. If we define finite joint prior density $p(\zeta,\psi)$ at points where $\psi=0$, the joint density $p(\cos\zeta,\cos\psi)$ is divergent (although integrable, possibly in closed form).

\textbf{Priors.} An issue to be aware of is that the mapping from the native sampling space---that of the unit hypercube---would need to avoid introducing nonlinearity, otherwise the effort to improve efficiency may be in vain (see the discussion in Section~\ref{subsubsec:source} regarding implementation of the joint prior distribution of $M$ and $R_{\textrm{eq}}$).

We define the prior as separable on the joint space of $f$ and $\zeta$. Specifically, we condition on $f\sim U(0,1)$ and $\zeta\sim U(\epsilon_{\zeta},\pi/2-\epsilon_{\zeta})$ where $\epsilon_{\zeta}\ll\pi/2$. For posterior computation, in order to ease prior implementation, we opt to transform the prior onto the joint space of $\psi$ and $\zeta$ and accept the nonlinear degeneracy in the limit $\zeta\to\tilde\psi$ for small $\tilde\psi$. The chosen joint prior is not separable on this chosen space, but remains straightforwardly implementable.

For \TT{CDT-U}, we eliminate a region-exchange degeneracy by imposing a constraint $\Theta_{p}\leq\Theta_{s}$ on the support of the joint prior distribution. The primary region is uniquely defined as the region whose center subtends the smallest colatitude, $\Theta_{p}$, to the rotational axis; the regions are distinguishable when $\Theta_{p}=\Theta_{s}$ according to the subset of parameters that controls their physical manifestation. The joint prior support is such that the two regions cannot overlap but otherwise are not restricted by antipodal symmetry. Note that imposing that the regions are non-overlapping modifies the marginal prior density $p(\zeta)$---and thus $p(\psi)$---by redistributing prior mass to lower angular radii.

\subsubsection{Concentric dual-temperature regions with antipodal symmetry (\TT{CDT-S}) and with unshared parameters (\TT{CDT-U})}\label{subsubsec:CDT-S and CDT-U}

\begin{figure*}[t!]
\centering
\includegraphics[clip, trim=0cm 0cm 0cm 0cm, width=0.75\textwidth]{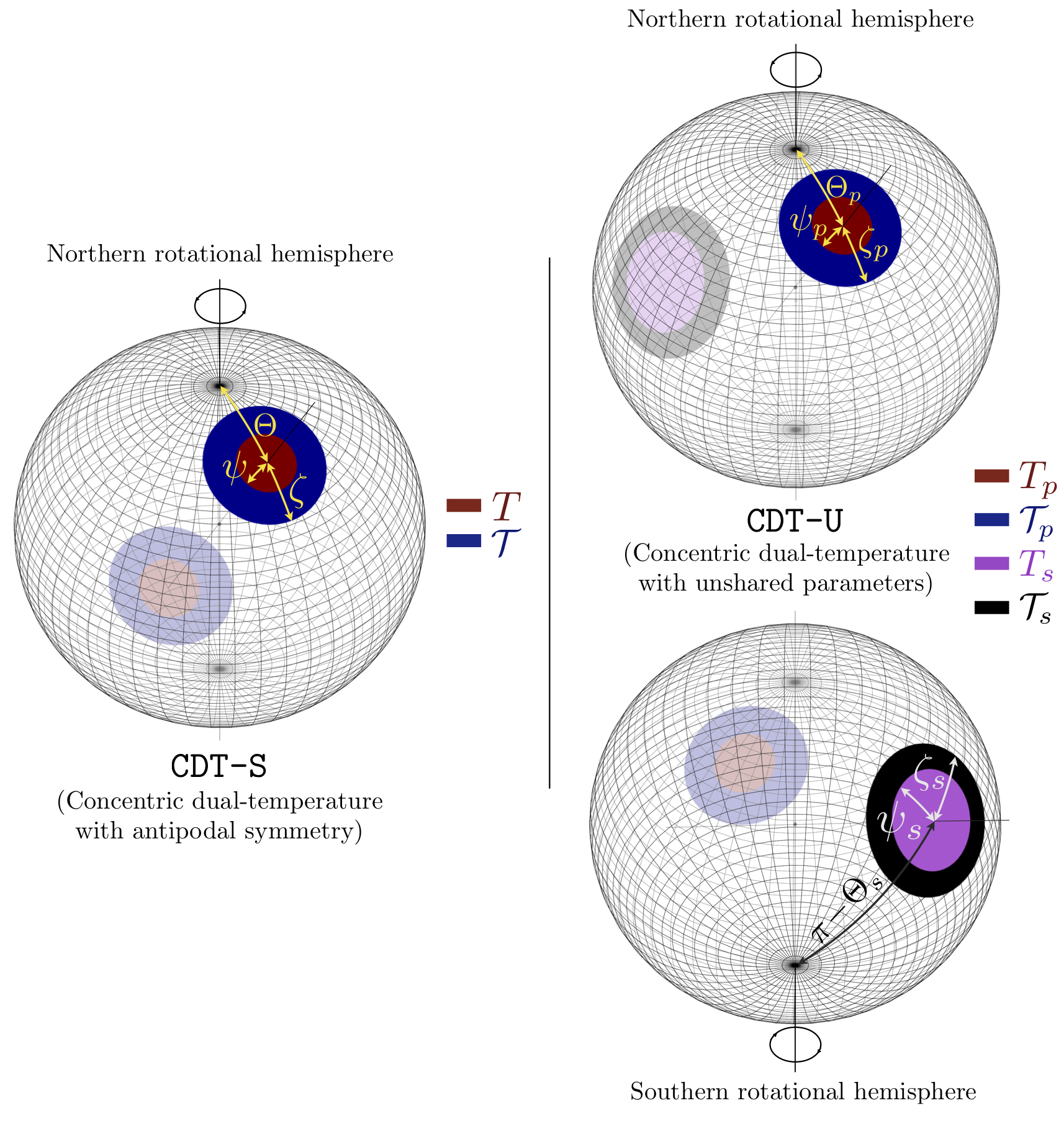}
\caption{\small{Schematic diagrams of models with \textit{concentric dual-temperature} regions: \TT{CDT-S} defined by antipodal symmetry of the primary and secondary regions, and \TT{CDT-U} defined by the primary and secondary regions \textit{not} sharing any parameters. To construct the \TT{CDT-U} diagram from the \TT{CDT-S} diagram we: displaced the secondary region (defined by having a center colatitude $\Theta_{s}\geq\Theta_{p}$) away from the antipode of the center of the primary region (with center colatitude $\Theta_{p}$); decreased $\zeta_{s}$ such that $\zeta_{s}<\zeta_{p}$ whilst leaving $\psi_{s}=\psi_{p}$, equivalently meaning $f_{s}>f_{p}$; and assigned distinct colors to the hole and annulus of the secondary region. We retained the caveat that no numeric meaning is implied regarding the effective temperature of the material in the members---indeed, the prior is separable with respect to all temperatures ($T_{p}$, $\mathcal{T}_{p}$, $T_{s}$, $\mathcal{T}_{s}$), and the support includes combinations $\mathcal{T}_{s}\leq T_{s}$ and $\mathcal{T}_{s}>T_{s}$.}}
\label{fig: schematic CDT diagrams}
\end{figure*}

\textbf{Parameterization.} We extend the \TT{CST} models by filling the \textit{holes} (the superseding members) with radiating material at finite temperature and evaluating the signal generated by both the superseding and ceding members. The hot region is simply the union of the members together with an order of precedence, and can be considered to have two heated \textit{subregions} with distinct temperatures, each of which generates a \textit{component} of the signal.

Hereafter we use the alias \TT{CDT-S}, parsed as \textit{Concentric-Dual-Temperature-Shared}. For \TT{CDT-S}, the surface radiation field associated with the secondary region is derived exactly by applying antipodal symmetry to the primary region: there are no free parameters associated with the secondary region. The annuli share an outer angular radius $\zeta$, and the holes share a fractional angular radius $f$ such that the hole angular radii are given by $\psi\coloneqq f\zeta$.

Similarly, we use the alias \TT{CDT-U}, parsed as \textit{Concentric-Dual-Temperature-Unshared}. For \TT{CDT-U}, the primary region (see the rightmost panel of Figure~\ref{fig: schematic CDT diagrams}) definition is retained from \TT{CDT-S}. The secondary region, however, is now endowed with distinct parameters---i.e., it is \textit{not} derived from the primary region under antipodal symmetry. The parameters of the secondary region have an otherwise equivalent meaning---in terms of surface radiation field specification---to their primary-region counterparts. It follows that parameter vectors that correspond to, e.g., the primary region having a hotter annulus whilst the secondary region has a hotter hole, are assigned finite local joint prior density.

\textbf{Degeneracy.} Introducing a second temperature component also introduces degenerate structure in the likelihood function---a fundamental problem with such phenomenological descriptions of the surface radiation field. We now consider the continuous degeneracies labeled I, II, III, and IV in Figure~\ref{fig: continuous region degeneracies}.

Degeneracy of type I manifests on small angular scales when the temperature $T$ of the superseding member is small relative to the temperature $\mathcal{T}$ of the ceding member. In this case, the signal from the superseding member is dominated and we effectively recover the structure from the \TT{CST} models; the reader should imagine the hole in panel I of Figure~\ref{fig: continuous region degeneracies} as colored and labeled $T\ll\mathcal{T}$.

Degeneracy of type II manifests because the temperatures of the ceding and superseding members can be approximately and exactly equal. Thus, given a strong constraint on the temperature, the solid angle ($\varpi\approx1-\cos\zeta$) constraint on the radiating region can be satisfied $\forall\psi\leq\zeta$. Note that unlike degeneracy of type I, degeneracy of type II can be exact (for $T=\mathcal{T}$) and exists invariantly of angular scale. The smaller the solid angle constraint, the weaker the constraint on topology of the hotter subregion (annulus or hole), and thus degeneracy of type II should be present. On larger angular scales, the signal offers a much stronger constraint on topology: if the hottest subregion is ring-like,\footnote{Formed by imposing a relatively large, cooler superseding member (effectively a hole).} degeneracy of type II cannot be present \textit{a posteriori}.\footnote{Here we assume a unimodal posterior distribution, but more generally, the degeneracy will not be present for a \textit{local} mode in which the topology is strongly constrained as ring-like.}

Degeneracy of type III manifests because the smaller the superseding member is in relation to the ceding member, the greater the dominance of the signal from the ceding member, and thus the greater the range of values $T$ can assume whilst leaving the signal approximately invariant. Thus, given a strong constraint on the temperature and solid angle ($\varpi\approx1-\cos\zeta$) of the dominant component, the constraint can be satisfied $\forall\psi\ll\zeta$. Again note that unlike degeneracy of type I, the existence of degeneracy of type II is insensitive to angular scale. However, if there is a strong constraint on the topology of the subregion that dominates the joint signal, and that subregion is ring-like, degeneracy of type II will not be present \textit{a posteriori}.

We now consider degeneracy IV and its implications for sampling. Suppose that the signal can be dominated by that from a simply-connected subregion: if the superseding member satisfies both a temperature constraint and a solid angle constraint ($\varpi\approx1-\cos\psi$), the ceding member then forms an annulus whose width $\psi^{+}$ and whose temperature $\mathcal{T}$ can together assume a wide range of values. In Figure~\ref{fig: continuous region degeneracies} we illustrate the case of $\mathcal{T}\ll T$ for varying $\zeta$ (and, in the last cartoon, for a shift of the center of the ceding member in angular space).\footnote{Note that a less important degeneracy occurs, which we do not illustrate: the annulus is arbitrarily hotter ($\mathcal{T}>T$) but $\psi^{+}\to0$ so that the superseding member also dominates the signal. In the $T\approx\mathcal{T}$ transition zone---which coincides with the type II degeneracy---$\zeta$ approximately satisfies the solid angle constraint.} The superseding member need only be described by four parameters, whilst the non-superseded subset of the ceding member is described by five parameters, three of which are shared with the superseding member. If the ceding member is relatively cool, then the signal it generates is dominated by that of a much hotter superseding member; the signal is then degenerate with respect to $\mathcal{T}$. Crucially, if one writes the properties of the superseding member in terms of that of the ceding member, nonlinear degeneracy can arise in what otherwise may appear to be a natural parameterization, as we show below.

For posterior computation (see Appendix~\ref{sec:posterior computation} for methodology) we explicitly consider a parameter space in which the superseding member is constructed using the minimal number of parameters necessary, which eliminates needless \textit{nonlinear} degeneracy. For example, if we work in the joint space of $f$ and $\zeta$, the local direction of approximate signal invariance---which emerges when the superseding member with angular radius $\psi$ is dominant---is \textit{not} everywhere a basis vector, but is given by the gradient of $f=\psi/\zeta$ for constant $\psi\in\mathbb{R}^{+}$: $\partial f/\partial\zeta=-\psi/\zeta^{2}$, so the direction is $(1,-\psi/\zeta^{2})$ at point $(\zeta, f)$. This degeneracy is linearized by working in the joint space of $\psi$ and $\zeta$, or alternatively, in the joint space of $\psi$ and the angular annular width $\psi^{+}\coloneqq\zeta-\psi$.

We note that unlike degeneracy of type I, the existence of degeneracy of type IV is insensitive to angular scale. However, if there exists a strong constraint on the topology of the heated subregion that dominates the signal, and that subregion is ring-like, degeneracy of type IV will not be present \textit{a posteriori}---in the same vein that degeneracy of types II and III are absent. This does not mean that the posterior modes are devoid of degeneracy if the hot regions are both constrained to be ring-like: the temperature of a superseding member can assume a wide range of values (in logarithmic space) lower than that of the ring, because the additional complexity beyond that of a non-radiating hole (i.e., devoid of heated material) is unwarranted. However, the degeneracy is linear and thus handled straightforwardly.

In many cases, degeneracies are characterized by allowing one temperature component (a heated subregion) to dominate the signal, whilst one temperature---or the solid angle subtended at the stellar origin by a subregion---assumes relatively low values. In these cases the degeneracy is not expected to be particularly problematic because it will be effectively linear, parallel to basis vectors of the parameter space, and truncated by the boundary of the prior support. Thus, provided that the mapping from the native sampling space to the parameter space preserves this behavior, nested sampling efficiency may not be affected as adversely as when posterior modes exhibit strong nonlinear degeneracy.

For instance, consider degeneracy such as type IV: let the angular scale be sufficiently large for the dominant component to be constrained to be simply-connected. The degenerate subsets of the $(\psi,\zeta)$- and $(\psi,\mathcal{T})$-subspaces will be linear and orthogonal to the $\hat{\psi}$ basis vector. The notable source of efficiency reduction for type IV will be the boundary of the degenerate subset of the $(\zeta,\mathcal{T})$-subspace: whilst there will be a hard lower-limit on $\mathcal{T}$, the boundary is otherwise dependent on combinations of parameters and will not conform trivially to common nested-sampling active-point bounding algorithms.

On the other hand, if degeneracy of type I arises and the hole is filled with material of temperature $T$, we incur a nonlinear degeneracy in the $(\zeta,\psi)$-subspace, \textit{and} non-trivial boundaries of the degenerate subsets of the $(\zeta,T)$ and $(\psi,T)$-subspaces. In this case the potential for efficiency reduction is greater.

It is interesting to note that degeneracy can also arise when one temperature component does \textit{not} dominate the signal. Let us suppose that two components with distinct temperatures are favored \textit{a posteriori}, where those components both contribute non-negligibly to the total signal. We implement a numerical geometrically thin atmosphere where the effective temperature \added{and effective gravity} control the local comoving specific intensity as a function of photon energy and direction \replaced{(beaming)}{(i.e., control the spectrum and beaming)}; it follows that if a hot region exhibits small angular extent, two components (subregions) can in principle generate signals that are commensurate in total count rate, provided that the solid angles subtended by the subregions at the stellar origin have appropriate relative sizes. For the \TT{CDT} models, a \textit{discrete} degeneracy---multi-modality---can then arise on small angular scales: pairs of configurations of the components generate approximately the same total signal. In one configuration the hole hosts a particular component, whilst in the alternate configuration that component is hosted by the annulus; the relative solid angles subtended by the hole and annulus depend on which component is hosted. However, for a more complex model where the superseding and ceding members are not defined as concentric (see Figure~\ref{fig: continuous region degeneracies}, and Section~\ref{subsubsec:EST and EDT} and beyond), it is clear that (approximate) continuous degeneracies would also arise and thus complicate matters.

We should also be aware of the relative prior masses associated with subsets of parameter space over which the signal is approximately invariant, governed roughly by the dimensionality of the degeneracy. Degeneracy of types II and III generally occupy a subset of parameter space with smaller prior mass than that of type I and IV, the former two being effectively to one-dimensional,\footnote{An  unillustrated degeneracy, characterized by a hotter ring in the limit $\psi^{+}\to0$, is also effectively one-dimensional} and the latter two being effectively two-dimensional. Note that the type I degeneracy is effectively two-dimensional \textit{because} we fill the hole from the \TT{CST} models with material whose temperature is finite and a model parameter.

\textbf{Priors.} We retain almost all prior definitions from the \TT{CST} models (Section~\ref{subsubsec:CST}). We again transform the prior onto the joint space of $\psi$ and $\zeta$ in order ensure a more optimal parameterization for our sampling algorithm of choice, as highlighted above. Crucially, we need to transform from the native space to the parameter space in a manner that preserves linearity of degeneracy IV. We achieve this by inverse sampling the marginal prior density $p(\psi)$ and the condition prior density $p(\zeta\,|\,\psi)$; we give these transforms in Appendix~\ref{app:prior transforms}. In some cases we opt to transform to the joint space of $\psi$ and $\psi^{+}$ space in post-processing for optimality of kernel density estimation when $\psi^{+}$ approaches a boundary---e.g., due to degeneracy of type IV.

\subsubsection{Interlude: On degeneracy and complexity}\label{subsubsec:degeneracy and complexity}

\begin{figure*}[t!]
\centering
\includegraphics[clip, trim=0cm 0cm 0cm 0cm, width=\textwidth]{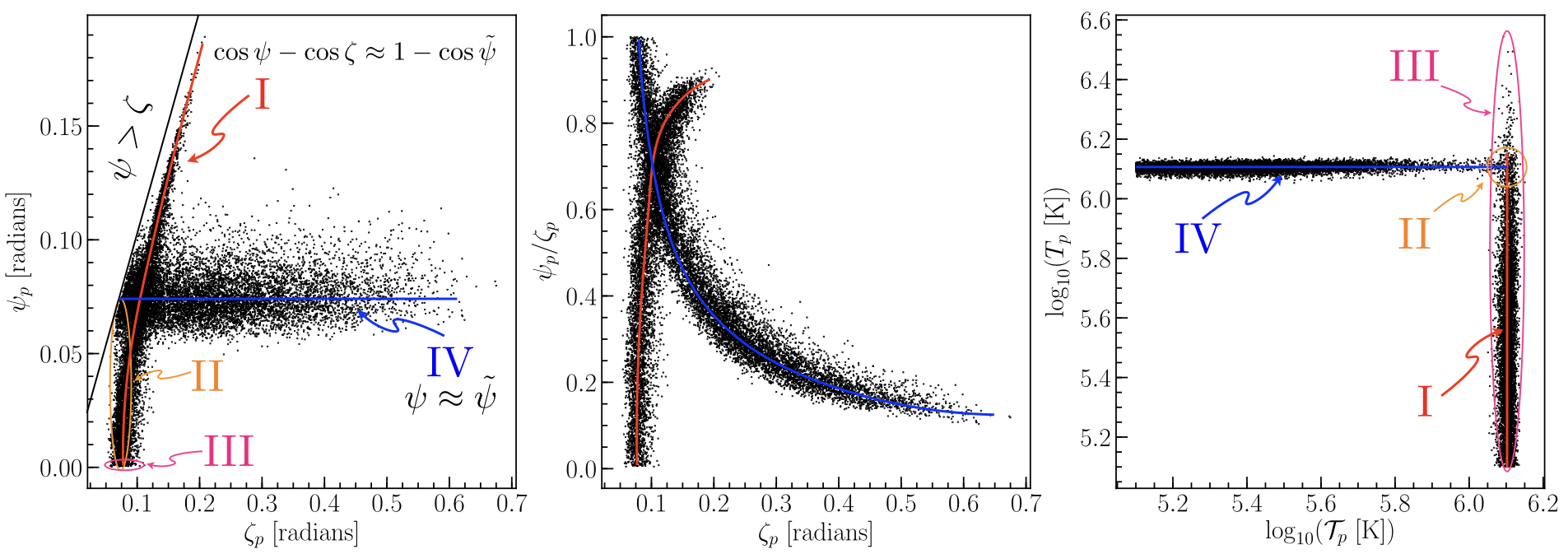}\caption{\small{Conditional on \TT{CDT-U}, we display points (nested samples) $\{\boldsymbol{\theta}_{k}\}$ that reported (background-marginalized) likelihood function values $L(\boldsymbol{\theta}_{k})>L^{\prime}$, where $\max_{k}\ln L(\boldsymbol{\theta}_{k})-\ln L^{\prime}\lesssim13$; the choice of threshold $L^{\prime}$ is somewhat arbitrary beyond needing a large enough number of points to clearly resolve the form of the degeneracies. We display these points in a selection of two-dimensional spaces to focus on the degeneracies that emerge upon application of \TT{CDT-U}. These degeneracies pertain to the structure of the \textit{primary} (lower-colatitude) hot region, which generates a signal describing a particular pulse component visible in Figure~\ref{fig:J0030 count data}; the hot region exhibits a small angular extent on the stellar surface, and we are statistically insensitive to the additional complexity offered by a \TT{CDT} cap over a \TT{ST} cap. In the \textit{left} panel we show the points in the joint space $(\zeta_{p},\psi_{p})$---i.e., in the joint space of the angular radii of the ceding and superseding members. We indicate which type of degeneracy illustrated in Figure~\ref{fig: continuous region degeneracies} corresponds to which structure in parameter space; the curves (\textit{blue} and \textit{red}) each correspond to a constraint equation. Types II and III occupy far smaller prior masses than types I and IV, so we choose to indicate the structures associated with the former types with ellipses; in a finite-sample context these structures are allocated points more sparsely, in proportion to the associated prior mass in the full $n$-dimensional parameter space. Note that the region $\psi_{p}>\zeta_{p}$ is not a subset of the \TT{CDT-U} prior support. In the \textit{center} panel we display the points in the joint space of $(\zeta_{p},f_{p})$}, where $f_{p}\coloneqq\psi_{p}/\zeta_{p}$; our sampling processes (refer to Appendix~\ref{sec:posterior computation} for details) were executed in a space in which likelihood function isosurfaces are structured nonlinearly as suggested in this center panel, leading to reduced posterior computation efficiency and higher nested-sampling error. In Sections~\ref{subsubsec:CST} and \ref{subsubsec:degeneracy and complexity} we discuss the interpretation and treatment of degeneracy: in hindsight, higher efficiency would have been achieved if we had worked in a joint space that linearized degeneracy of type IV. In the \textit{right} panel we display the corresponding structures in a joint space trivially related to $(\mathcal{T}_{p},T_{p})$---i.e., in the joint space of the effective temperatures of the material enclosed by the ceding and superseding members. The points displayed here are a composite set from two sampling processes, where due to reduced efficiency and increased nested-sampling error, neither process individually resolved all of the degenerate structure present.}
\label{fig: CDTU degeneracies primary}
\end{figure*}

\begin{figure*}[t!]
\centering
\includegraphics[clip, trim=0cm 0cm 0cm 0cm, width=\textwidth]{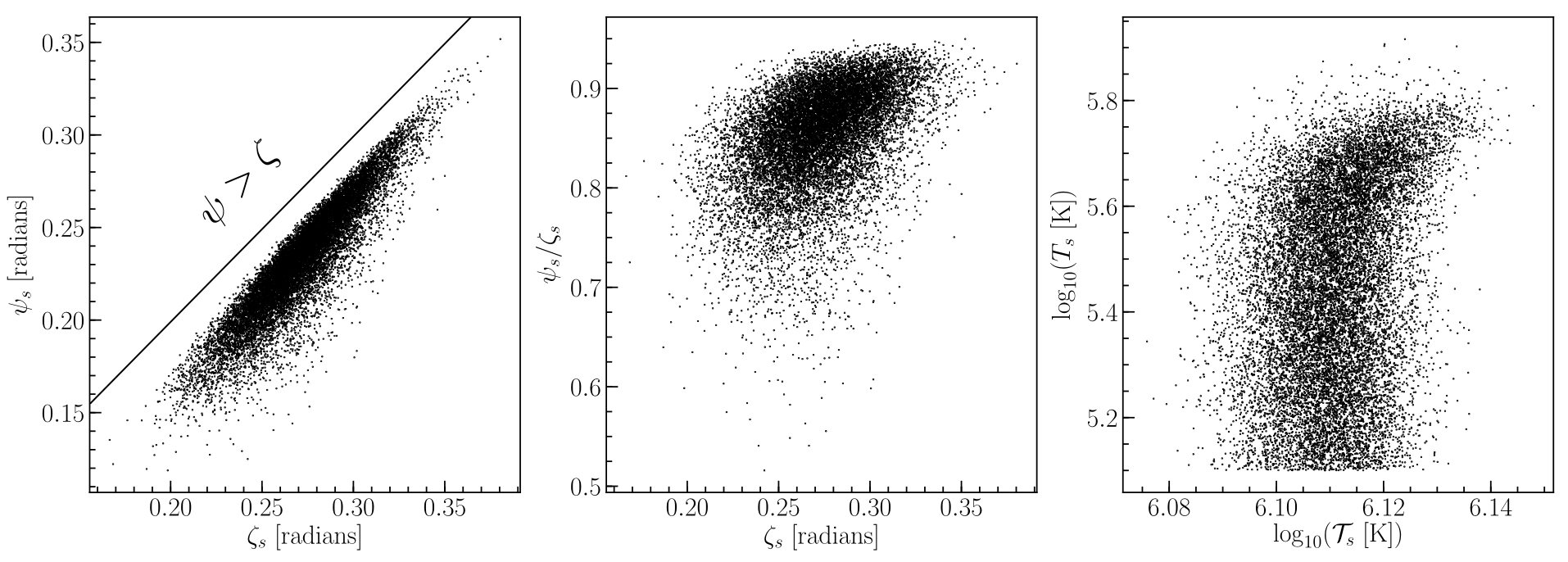}
\caption{\small{As for Figure~\ref{fig: CDTU degeneracies primary}, conditional on \TT{CDT-U}, we display points (nested samples) $\{\boldsymbol{\theta}_{k}\}$ that reported (background-marginalized) likelihood function values $L(\boldsymbol{\theta}_{k})>L^{\prime}$, where $\max_{k}\ln L(\boldsymbol{\theta}_{k})-\ln L^{\prime}\lesssim13$. We display these points in a selection of two-dimensional spaces to focus on the \textit{secondary} (higher-colatitude) hot region. The secondary region exhibits a larger angular extent on the stellar surface than the primary region and, in contrast, we are statistically sensitive to the topology of the secondary region whilst being statistically insensitive to the additional complexity offered by a \TT{CDT} cap over a \TT{CST} cap. In other words, we do not strongly require the material in the \textit{hole} to radiate when considering the \NICER waveband.}}
\label{fig: CDTU degeneracy secondary}
\end{figure*}

\textbf{General.} The continuous degeneracies discussed in Sections~\ref{subsubsec:CST} and \ref{subsubsec:CDT-S and CDT-U}, when in mutual existence, generally form a connected structure in parameter space. These degeneracies could be characterized in more detail, and also with reference to ulterior parameters (including those of the other hot region). We do not study the degeneracies further here for lack of a clear way to simultaneously linearize each of them via transformation.

We instead observe a fundamental aspect of the modeling process \textit{on a source-by-source basis}: It is judicious to evaluate the utility of additional complexity \textit{on a region-by-region basis}, if \textit{a posteriori} one infers localized emission and/or that a solitary temperature component per hot region suffices in describing a particular pulse component in the phase-folded event data. In practice: (i) design a model space with increments in complexity, moving toward more sophisticated region topologies and/or boundaries, and/or an additional temperature component as warranted by the data; (ii) insert models into one's model space that reduce region complexity, if it is apparent \textit{a posteriori} that a model with unhelpful complexity was first applied; or (iii) reduce region complexity as in point (ii), whilst increasing the complexity of another facet of the model, based for example on graphical posterior predictive checking.

To provide an example, suppose that \textit{a posteriori}, given \TT{CDT-U}, hot-region $\mathcal{A}$ is degenerate with respect to multiple parameters controlling its surface radiation field, whilst hot-region $\mathcal{B}$ is only degenerate with respect to temperature of one relatively cool component. Specifically, suppose $\mathcal{A}$ is constrained to have a small angular extent and a single temperature component, whilst offering satisfactory posterior predictive performance in comparison to a particular pulse component. Continuous degeneracies of type I and IV are then clear in the posterior distribution. Additionally, suppose that the secondary region has a strong constraint on its topology, with its dominant component being ring-like with moderate angular extent. On the basis of this inference, one might argue that it is justified to modify the model space: one proceeds to define a model wherein the complexities of  $\mathcal{A}$ and $\mathcal{B}$ are unequal. In doing so one can reduce or redistribute complexity whilst improving computational efficiency. For the aforementioned example, one may define a model \TT{ST+CST}, parsed as \textit{Single-Temperature + Concentric-Single-Temperature}, which is intermediary in relation to \TT{ST-U} and \TT{CST-U}: $\mathcal{A}$ has one simply-connected component, whilst $\mathcal{B}$ has one non-simply-connected component (which can reduce to a simply-connected component). The dimensionality of the parameterization is then $n=9$, reduced from the \TT{CDT-U} value of $n=12$. One could also define additional models that increase the complexity of $\mathcal{B}$ (see Sections~\ref{subsubsec:EST and EDT} and \ref{subsubsec:PST and PDT}). 

As a brief second example: if, given \TT{CDT-U}, $\mathcal{B}$ was constrained to have \textit{two} temperature components, one could instead define a model \TT{ST+CDT} with dimensionality $n=10$. One could then proceed to increment the complexity of $\mathcal{B}$.

An important remark here is that the performance of $\mathcal{A}$ cannot strictly be decoupled from the performance of $\mathcal{B}$ in relation to their respective target pulse components: the regions exist on the same rotating $2$-surface and share an ambient spacetime solution. It follows that the complexity of one region can in principle affect our conclusions about the level of complexity of the other. Such conclusions cannot be based solely on the existence of degenerate structure in the joint posterior distribution of the region parameters. If $\mathcal{A}$ exhibits degenerate structure whilst $\mathcal{B}$ does not, the structure may manifest---at least in part---\textit{because} the complexity of $\mathcal{B}$ is insufficient.

A more robust basis for breaking parity between the hot-region complexities would be the existence of degenerate structure, \textit{and} satisfactory posterior predictive performance of \textit{both} regions in comparison to their respective pulse components. A practical basis for breaking parity, given the existence of efficiency-reducing degenerate structure \textit{a posteriori}, may be resource availability and management.

\textbf{Modeling PSR~J0030$\mathbf{+}$0451.} For the source we focus on in this work, degeneracy emerged upon application of \TT{CDT-U}. We display the corresponding structures, in parameter space, in Figure~\ref{fig: CDTU degeneracies primary} and Figure~\ref{fig: CDTU degeneracy secondary}. In summary, the primary region was constrained to have a small angular extent and one dominant temperature component, and could generate data structurally consistent with a pulse component \deleted{pulse} visible in the phase-folded event data (Figure~\ref{fig:J0030 count data}); the primary region thus here corresponds to region $\mathcal{A}$ above. Degeneracy of types I and IV arose and dominated in prior mass (and by extension, in posterior mass). The nonlinear degeneracies in the native nested-sampling space also suppressed the efficiency of the sampling processes. We therefore now have grounds to design and apply models that break the parity in complexity between hot regions. We redistribute complexity from the primary region to the secondary region, such that the dimensionality of the parameter space does not exceed that of \TT{CDT-U}. In Appendix~\ref{app: supplementary models} we define models that preserve parity between regions---a brief continuation of the above scheme for model extension.

In Sections~\ref{subsubsec:EST and EDT} and \ref{subsubsec:PST and PDT} we explain how the complexity of a solitary hot region is extended. In Section~\ref{subsubsec:unequal complexities} we briefly summarize and provide diagrams of the relevant models formed from two disjoint hot regions with unequal complexities. In Section~\ref{subsubsec: model relationships} we then provide diagrams summarizing the relationships between all models considered in the scope of this work.

\subsubsection{Eccentric single- and dual-temperature regions (\TT{EST} and \TT{EDT})}\label{subsubsec:EST and EDT}

\textbf{Parameterization.} We extend the \TT{CST} and \TT{CDT} hot-region types by not requiring the ceding and superseding members to be concentric. For a \TT{CST} region, the superseding member was a hole in a ceding member, devoid of radiating material; for a \TT{CDT} region, the superseding member was similarly a hole, but filled with radiating material. For both \TT{CST} and \TT{CDT} regions, the \textit{non-superseded} subset of the ceding member was a radiating annulus with concentric inner- and outer-boundaries; now the radiating annulus is \textit{eccentric}. Further we require---via the joint prior support---that the ceding member is strictly a superset of the superseding member; in other words, the superseding member is a hole in the ceding member, again forming an annulus or ring.\footnote{The hole reduces to a point at the boundary of the prior support for an \TT{EST} region, whilst for a \TT{EDT} region the hole is not permitted to reduce to a point. Also note that because the ceding and superseding members are circular, their boundaries may intersect at a maximum of one point.}

\begin{figure*}[t!]
\centering
\includegraphics[clip, trim=0cm 0cm 0cm 0cm, width=0.7\textwidth]{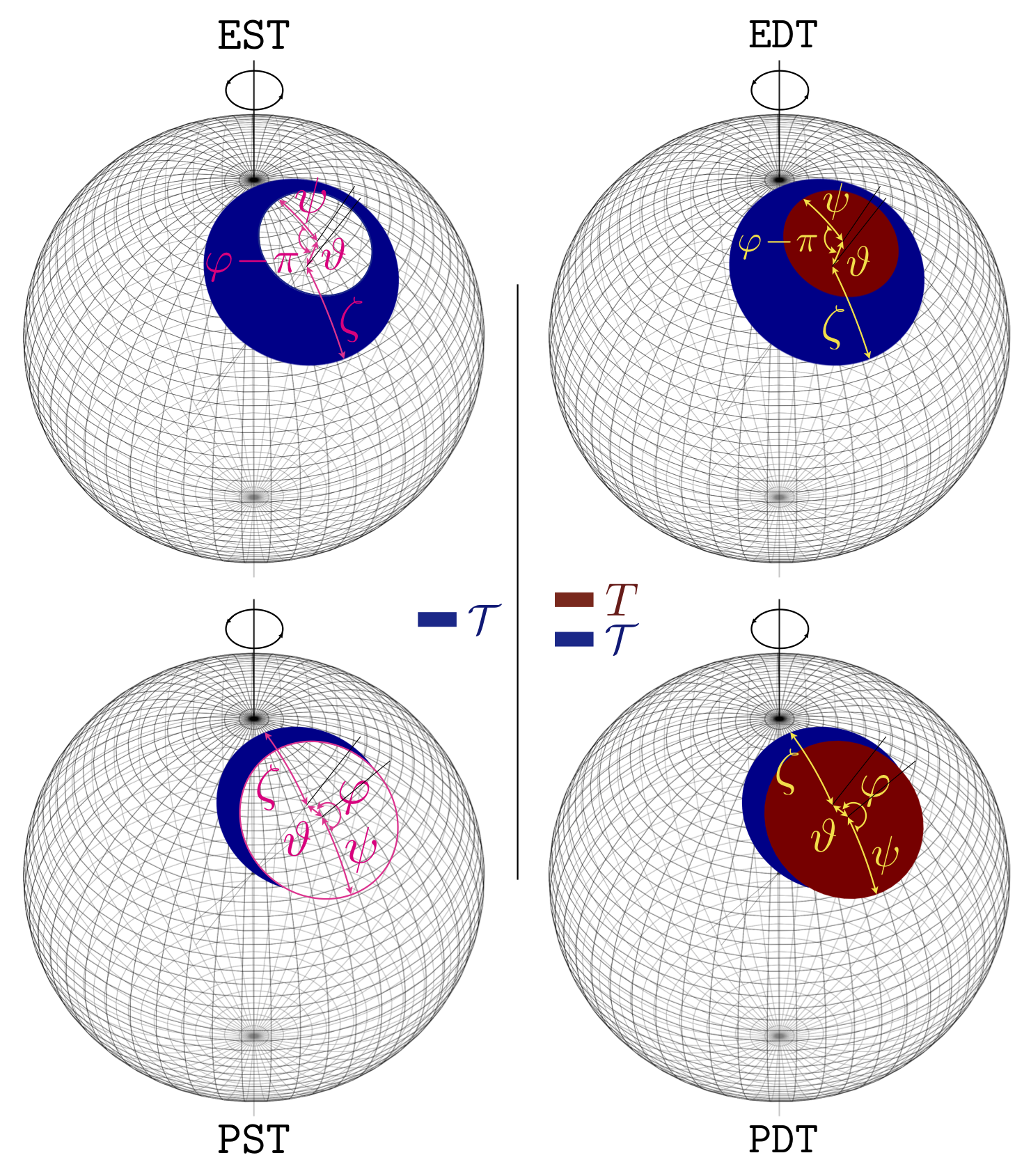}
\caption{\small{Schematic diagrams of solitary hot regions beyond the \TT{CST} and \TT{CDT} complexity levels, as described in Sections~\ref{subsubsec:EST and EDT}~and~\ref{subsubsec:PST and PDT}. We delineate the boundary of the superseding member for the \TT{PST} region.}}
\label{fig: schematic region diagrams}
\end{figure*}

We illustrate these hot-region structures in the topmost panels of Figure~\ref{fig: schematic region diagrams}; we also alluded to such configurations in Figure~\ref{fig: continuous region degeneracies} and in Sections~\ref{subsubsec:CST} and \ref{subsubsec:CDT-S and CDT-U}. Hereafter we use the aliases \TT{EST} and \TT{EDT}, respectively parsed as \textit{Eccentric-Single-Temperature} and \textit{Eccentric-Dual-Temperature}.

An additional two parameters must be defined to specify the eccentricity as the great-circle segment separating the centers of the members: the magnitude and direction of the offset. To do so, we consider a spherical coordinate basis such that: (i) the polar axis subtends an angle $\Theta$ to the stellar rotation axis; (ii) the southern rotational pole subtends angle $\pi-\Theta$ to the polar axis, and has azimuth \textit{zero} (such that the northern rotational pole has azimuth $\pi$ radians). The azimuth of the annulus center, $\varphi$, in the rotated basis uniquely selects the great circle and direction of the offset (depicted in Figure~\ref{fig: schematic region diagrams}). We choose define the magnitude of the angular separation as a fraction $\varepsilon$ of the \textit{difference} $\psi^{+}\coloneqq\zeta-\psi\equiv \zeta(1-f)$ between the angular radii of the hole and annulus (see Section~\ref{subsubsec:CDT-S and CDT-U}). The magnitude of the angular offset between the centers---equal to the colatitude of the annulus center in the rotated spherical coordinate basis defined above---is thus $\vartheta\coloneqq \varepsilon\psi^{+}$.\footnote{An alternative is to parameterize in terms of the fraction of the sum of their angular radii such that $\vartheta\coloneqq \varepsilon(1+f)\zeta$; it is this combination that is the limit of zero partial overlap when the superseding member lies partially exterior to the ceding member (see Section~\ref{subsubsec:PST and PDT}). In order to impose that the superseding member is a hole in the ceding member, we would then require the support constraint that $\varepsilon(1+f)\zeta+f\zeta\leq\zeta$ and thus that $\varepsilon\leq(1-f)/(1+f)$.}

\textbf{Degeneracy.} In a similar vein to the \TT{CDT} parameterization, optimality of the parameterization for an \TT{EDT} region should be considered. By this point in the modeling process, we have a handle on a region-by-region basis whether the additional complexity offered by an \TT{EST} or \TT{EDT} region is a justified modeling step. Nevertheless, one can construct the sampling space using the minimum number of parameters need to fully control the superseding member. The parameter vector for the regions is thus of the form $\boldsymbol{v}=(\psi,\Theta,\phi,\zeta,\varepsilon,\varphi)$, and the mapping from the native sampling space to the parameter space is formed in part using the marginal density $p(\psi)$ and the conditional density $p(\zeta\,|\,\psi)$.

\textbf{Priors.} As for the \TT{CDT} type of region (Sections~\ref{subsubsec:CDT-S and CDT-U}), joint prior support is such that the material within the annulus may exhibit an effective temperature $\mathcal{T}<T$ or $\mathcal{T}\geq T$. Moreover, when an \TT{EST}/\TT{EDT} region shares the stellar surface with another hot region $\mathcal{C}$, the joint prior support does not include configurations wherein $\mathcal{C}$ overlaps with the ceding (annular) region of the \TT{EST}/\TT{EDT} region. Note that a continuous set of coordinate singularities exist in the mapping $(\varepsilon,\varphi)\mapsto(\Theta^{\prime},\phi^{\prime})$, where the coordinates $(\Theta^{\prime},\phi^{\prime})$ are the colatitude and azimuth of the center of the annulus in a spherical coordinate system with polar axis defined as the stellar rotation axis. The mapping is singular for all points $(\varepsilon,\varphi)=(0,\varphi)$ at the boundary of the prior support, such that by defining a finite joint prior density at such singular points the joint density at $(\Theta^{\prime},\phi^{\prime})=(\Theta,\phi)$ diverges. We consider the spherical coordinates $(\vartheta,\varphi)$ to be more natural for increasing complexity and thus we display density functions with respect to $(\varepsilon,\varphi)$ to avoid difficulty in representing density functions accurately.

\subsubsection{Protruding single- and dual-temperature regions (\TT{PST} and \TT{PDT})}\label{subsubsec:PST and PDT}

\textbf{Parameterization.} We extend the \TT{EST} and \TT{EDT} hot-region types by not requiring the ceding member be a strict superset of the superseding member. In other words, the superseding member need not be a hole in the ceding member, leading to radiating component with the topology of a ring. If the superseding member is \textit{not} a hole nor a point, and the boundaries of the ceding and superseding members intersect at two points, then the \textit{non-superseded} subset of the ceding member is simply-connected \textit{and} has a non-circular boundary: it cannot therefore always be considered an annulus, and can assume a crescent- or arc-like morphology.\footnote{It is now particularly clear why we supplant the \textit{hole} and \textit{annulus} descriptors with those in terms of simply-connected \textit{members} and evaluation precedence as noted in Section~\ref{subsubsec:source}: one member is distinguished from the other simply by which takes precedence when evaluating radiation intensities along a ray (null geodesic) connecting the stellar surface to a distant observer.} Note that we do not increase complexity here by incrementing the dimensionality of parameter spaces defined for the \TT{EST} and \TT{EDT} regions. We nevertheless choose to distinguish between \TT{EST}/\TT{EDT} and \TT{PST}/\TT{PDT} in view of a symmetry being broken: that of the circularity of the outer boundary of the region. The superseding member can also now subtend the larger (coordinate) solid angle at the stellar origin, and in the limit that the member boundaries touch at a point, the hot region is simply-connected, with a boundary that is either circular or union of two circular subregion boundaries. 

We illustrate these hot-region structures in the bottommost panels of Figure~\ref{fig: schematic region diagrams}; we also alluded to such configurations in Sections~\ref{subsubsec:CST} and \ref{subsubsec:CDT-S and CDT-U}. Hereafter we use the aliases \TT{PST} and \TT{PDT}, respectively parsed as \textit{Protruding-Single-Temperature} and \textit{Protruding-Dual-Temperature}: the superseding member can protrude from the ceding member, for parameter vectors within the joint prior support.

We modify the \TT{EST} (and thus \TT{EDT}) parameterization in order to permit the superseding member to: (i) protrude\footnote{Or partially overlap with the ceding member.} from the ceding member; and (ii) subtend a larger (coordinate) solid angle at the stellar origin than the ceding member. Let us denote the angular radius of the \textit{largest} member as $\xi$. Let us then consider the interval $f\in[\epsilon_{f},2-\epsilon_{f}]$, where $\epsilon_{f}$ is some small number: when $f\leq1$ the angular radii of the superseding and ceding members are, respectively, $\psi=f\xi$ and $\zeta=\xi$, as in the \TT{EDT} variants. However, when $f>1$, the angular radii of the superseding and ceding members are respectively $\psi=\xi$ and $\zeta=(2-f)\xi$. The angular radii are thus piecewise in $f$, and are continuous at the transition point $f=1$. It follows that the interval for the angular radius $\xi$ of the largest member considered in the previous variants can be maintained, whilst the member that is largest switches. If one varies $f$ through the interval $f\in[\epsilon_{f},2-\epsilon_{f}]$ whilst all other parameters are fixed, the solid angle subtended by the superseding member increases to match that of the ceding member; at the $f=1$ transition the superseding member stops expanding, and for $f$ increasing beyond unity, the solid angle subtended by the ceding member decreases.

We choose to define the magnitude of the angular separation as a fraction\footnote{Equal to the colatitude of the center of the ceding member in the rotated spherical coordinate basis defined above.} $\varkappa$ of a combination of the angular radii of the superseding and ceding members. For $f\leq1$, the combination is the \textit{sum} of the angular radii of the superseding and ceding members, $\psi=f\xi$ and $\zeta=\xi$, respectively, for $f\leq1$: $\vartheta(\xi,f\leq1,\varkappa)\coloneqq \varkappa(1+f)\xi=\varkappa(\psi+\zeta)$. It is this combination that is the limit of zero partial overlap between the members when the superseding member lies partially exterior to the ceding member. We define the prior support for $\varkappa$ simply as the unit interval: we permit only configurations wherein the members partially overlap (including at a point), with maximum separation $\vartheta=\psi+\zeta$.

We require the angular separation $\vartheta$ as piecewise in $f$:  for $f>1$, we let $\vartheta(\xi,f>1,\varkappa)\coloneqq \psi-\zeta+2\varkappa\zeta=(f-1)\xi + 2\varkappa(2-f)\xi$. Note that the minimum separation here with respect to $\varkappa$, for fixed $\xi$ and fixed $f$, is $\vartheta=\psi-\zeta$, equivalent to the transition from the ceding member being \textit{partially superseded} to being \textit{wholly superseded}; the minimum separation for the subinterval $f\leq1$ is zero. The maximum separation (with respect to $\varkappa$ for fixed $\xi$ and fixed $f$) is $\vartheta=\psi+\zeta$, which is equivalent to the expression for the maximum separation for $f\leq1$. Note that at the transition $f=1$, the piecewise components of $\vartheta(\xi,f,\varkappa)$ continuously match at a value of $2\varkappa\xi$.

The coordinates of the center of the ceding member remain written in terms of those of the center of the superseding member. We cannot define a pair of coordinates as being associated with the \textit{larger} member (in the same vein that $\xi$ is directly associated with the larger constituent) without generally introducing a discontinuous transition with respect to the configuration of the members at $f=1$.

Note that the above prescription is equivalent in principle to defining a binary discrete parameter (with associated uniform prior probability mass function) that controls which member $\xi$ in turn directly controls, and thus which member $f$ directly controls, where the upper-bound of the prior support of $f$ remains as unity. In this equivalent alternative, the piecewise definitions of $\vartheta$ and the angular radii are required, but are recast with respect to the binary parameter. However, extension of the support of a continuous parameter to include these configurations eases posterior computation.

\textbf{Degeneracy.} The two clear exact degeneracies beyond those considered above\footnote{The degeneracies considered above should at this stage be of little to no concern if computation of the simpler (nested) models has \textit{not} indicated that the complexity thus far introduced is unhelpful.} are: (i) equal temperatures where the superseding member is wholly enclosed by the ceding member, leading to three-dimensional degeneracy in the subspace of $(f,\varkappa,\varphi)$; and (ii) the superseding member \textit{wholly supersedes} the ceding member ($\varkappa=0$ for $f\geq1$), leading to a two-dimensional degeneracy in the $(f,\varphi)$-subspace. There are also discrete member-exchange degeneracies for a given region when the temperatures are equivalent.

\textbf{Priors.} The joint prior support is again such that the material within the ceding member may exhibit an effective temperature $\mathcal{T}<T$ or $\mathcal{T}\geq T$. Moreover, when a \TT{PST}/\TT{PDT} region shares the stellar surface with another hot region $\mathcal{C}$, the joint prior support could be defined to exclude configurations wherein $\mathcal{C}$ overlaps with the \TT{PST}/\TT{PDT} region. We note that due to choice of parameterization and prior support, the prior density function for the angular separation between the ceding and superseding members has changed from the corresponding \TT{EST}/\TT{EDT} density function.

\subsubsection{Overview of the model space}\label{subsubsec:unequal complexities}\label{subsubsec: model relationships}

\begin{figure*}[t!]
\centering
\includegraphics[clip, trim=0cm 0cm 0cm 0cm, width=0.7\textwidth]{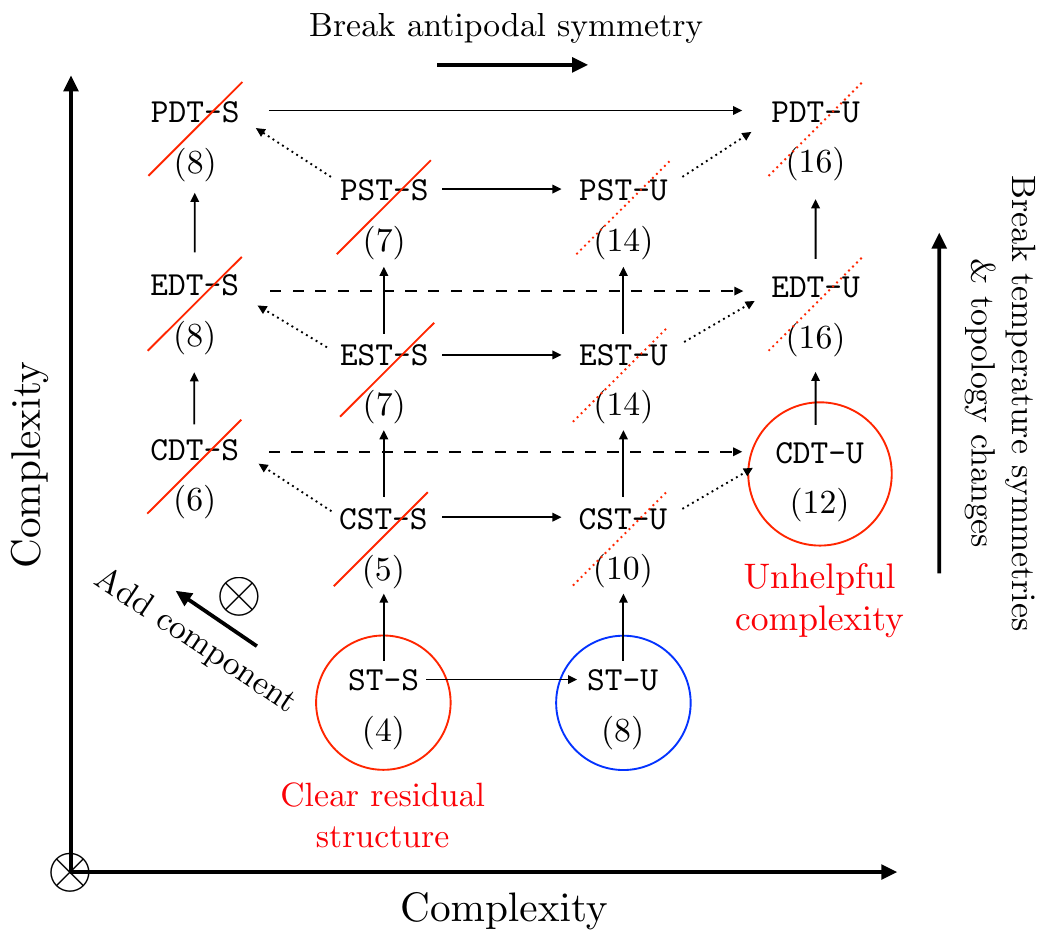}
\caption{\small{Diagram of relationship between a subset of models $\{\mathcal{M}\}$, where each $\mathcal{M}\subset\mathscr{M}$ and $\mathscr{M}$ is the model space as defined in Section~\ref{subsec: generative model space}. The models are: \TT{ST-S} and \TT{ST-U} (Section~\ref{subsubsec:ST-S and ST-U}); \TT{CST-S} and \TT{CST-U} (Section~\ref{subsubsec:CST}); \TT{CDT-S} and \TT{CDT-U} (Section~\ref{subsubsec:CDT-S and CDT-U}); \TT{EDT-S} and \TT{EDT-U} (Appendix~\ref{app: supplementary models}); \TT{PDT-S} and \TT{PDT-U} (Appendix~\ref{app: supplementary models}). The integers in parentheses are the number of continuous parameters controlling the hot regions; to obtain the total number of parameters constituting the sampling space for each model, add $8$ to these numbers---corresponding to the vector $(M,R_{\rm eq},i,D,N_{\rm H},\alpha,\beta,\gamma)$ shared between all models. Prior predictive complexity increases with dimensionality and/or prior support expansion; complexity increments are achieved by breaking symmetries, allowing the morphology and topology of the radiating regions to change, and adding a second component with parameterized temperature. Solid arrows between model nodes delineate a nested relationship between models: the model at the tail is nested within the model at the head, the latter of which has greater complexity. Adding arrows head-to-tail at a node conserves such relationships---i.e., the model at the tail of the resultant arrow is nested within the more complex model at the head. The \textit{dashed} arrows are in the background (visualize a third dimension of the graph as indicated by $\otimes$). The \textit{dotted} arrows between single- and dual-temperature models indicate that the nested relationship is weaker: in the single-temperature models no material (or material with zero temperature) fills the superseding member and thus no signal is generated, whereas for the dual-temperature models we fill the superseding member with material of finite temperature $T\gtrsim\mathcal{O}(10^5)$ K, and a signal is thus physically generated and computed. The \textit{blue} ring indicates the simplest model that can generate data that is visually comparable to the real data set (refer to Section~\ref{sec:inferences} for discussion). The \textit{red} annotations are to denote where we learn that a given model is not performing adequately or has unhelpful complexity. A strikethrough denotes that a model was not applied: (i) a \textit{solid} strikethrough denotes that the model is considered as being incapable of generating synthetic event data that resembles the real \project{XTI} event data, based on performance of a simpler model with which some degree of symmetry is shared;  (ii) a \textit{dotted} strikethrough denotes that the model is considered to include unhelpful complexity for at least one hot region.}}
\label{fig: discrete model space diagram}
\end{figure*}

\begin{figure*}[t!]
\centering
\includegraphics[clip, trim=0cm 0cm 0cm 0cm, width=0.7\textwidth]{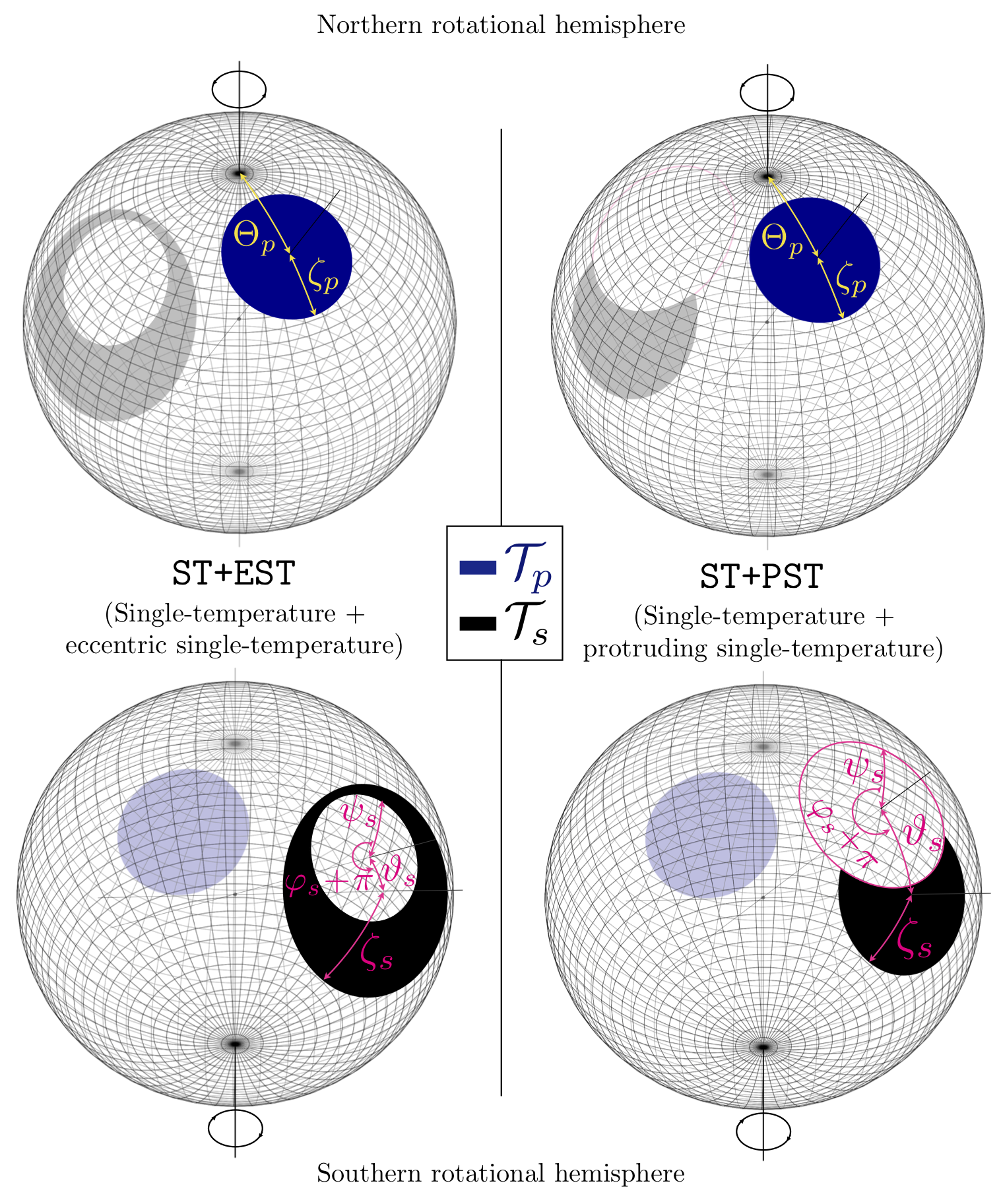}
\caption{\small{Schematic diagrams of models wherein an \TT{ST} region shares the stellar surface with a higher-complexity \TT{EST} or \TT{PST} region.}}
\label{fig: schematic ST+EST and ST+PST diagrams}
\end{figure*}

\begin{figure*}[t!]
\centering
\includegraphics[clip, trim=0cm 0cm 0cm 0cm, width=0.4\textwidth]{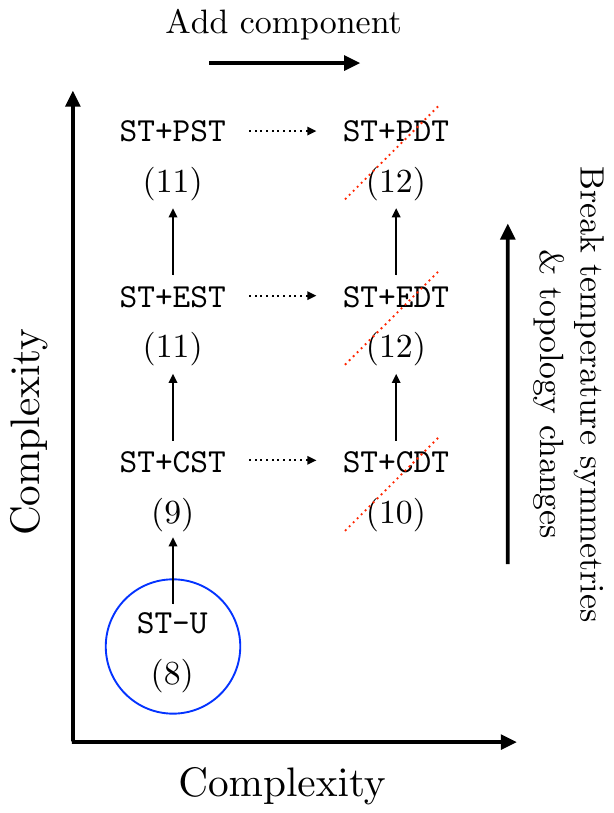}
\caption{\small{Diagram of relationship between dual-hot-region models with unequal complexities, constructed via extension of \TT{ST-U}. These models exist within the scope of those displayed in Figure~\ref{fig: discrete model space diagram}.}}
\label{fig: discrete model space diagram 2}
\end{figure*}

In Figure~\ref{fig: discrete model space diagram} we provide a diagram of the relationships between the models---spanning the discrete model space---which impose equal complexities. A subset of these models are based on the extensions defined in Sections~\ref{subsubsec:EST and EDT} and \ref{subsubsec:PST and PDT}; diagrams of \TT{EDT-S}, \TT{EDT-U}, \TT{PDT-S}, and \TT{PDT-U} are given in Appendix~\ref{app: supplementary models} because we do not compute posterior distributions that are jointly conditional on these models and the event data.

We also define models wherein an \TT{ST} region shares the stellar surface with a higher-complexity hot region. We provide diagrams of \TT{ST+EST} and \TT{ST+PST} in Figure~\ref{fig: schematic ST+EST and ST+PST diagrams}. We provide diagrams of \TT{ST+EDT} and \TT{ST+PDT} in the online figure set associated with Figure~\ref{fig: schematic ST+EST and ST+PST diagrams}. In Figure~\ref{fig: discrete model space diagram 2} we illustrate the relationships between these models. Note that these models with unequal complexities exist within the scope of those models present in Figure~\ref{fig: discrete model space diagram}.

When the hot regions are distinguishable by complexity level we can label the regions as \textit{primary} and \textit{secondary} according to that complexity: let the higher-complexity region be \textit{secondary}. Our application of \TT{CDT-U} suggested that evolving the lower-colatitude (primary) region from \TT{ST} to \TT{CDT} did not yield any improvement in describing the event data; the lower-colatitude region in this case describes the pulse component in the event data (Figure~\ref{fig:J0030 count data}) that peaks at $\sim\!0.5$ rotational cycles. Moreover, for both \TT{ST+CST} and \TT{ST+CDT}, the secondary region is formed (not necessarily by trivial union) from two members---one or both of which radiate---that are concentric and thus share a single colatitude parameter. Therefore, for \TT{ST+CST} and \TT{ST+CDT} it is reasonable to retain the constraint on the prior support that $\Theta_{p}\leq\Theta_{s}$---i.e., that the colatitude of the \textit{center} of the \TT{ST} region is at most the colatitude of the \textit{center} of the \TT{CST}/\TT{CDT} region.\footnote{A model wherein the \TT{CST} or \TT{CDT} region is at \textit{higher} colatitude can be continuously connected to to this model by simply by forgoing the constraint on the prior support that $\Theta_{p}\leq\Theta_{s}$.}

Notably, we cannot consider a model with an \TT{EST}/\TT{EDT} or \TT{PST}/\TT{PDT} region as nested within \TT{CDT-U}, and there is thus not an obvious ordering of regions in colatitude that can be applied based on the \TT{CDT-U} posterior information. We therefore do not impose such a constraint on the prior support, meaning that the order of the regions in model names \TT{ST+EST}, \TT{ST+EDT}, \TT{ST+PST}, and \TT{ST+PDT} does \textit{not} indicate an order in colatitude.

Based on the \TT{CDT-U} posterior information we are now interested in models wherein we couple the lower-complexity (primary) region to the pulse component in the event data (Figure~\ref{fig:J0030 count data}) that peaks at $\sim\!0.5$ rotational cycles. We consider a practical reason to identify a particular hot region with a particular pulse component in the phase-folded event data: to ease posterior computation by focusing sampling resolution in the vicinity of such configurations. If we do not opt for this coupling, a local posterior mode (or modes) can absorb sampling resolution, despite contributing relatively low posterior mass and thus being ultimately uninteresting for both parameter estimation and evidence estimation; such a local mode is characterized by the \TT{ST} region describing the pulse component at approximately zero rotational cycles (Figure~\ref{fig:J0030 count data}). Mitigating such an effect may then require additional active-points and/or activation of the mode-separation \MultiNest sampling variant (refer to Appendix~\ref{sec:posterior computation}).

To intentionally couple the hot regions to pulse components, we restrict the prior support for the \TT{ST} region to an interval $\phi_{p}\in[a,b]$ where $(b-a)<1$ (such that the boundary of the support is not periodic with respect to $\phi_{p}$). Remarkably, whilst such restriction may not affect parameter estimation,\footnote{Provided that: (i) the dominant posterior mode corresponds to a \TT{ST} primary region; and (ii) that the marginal prior density $p(\phi_{p})$ remains weakly informative relative to the likelihood function} the evidence does clearly depend on changes to the prior support, especially for weakly informative priors whose density does \textit{not} fall to negligible values at the changing boundary of the support. It follows that in order to compare models based on evidences, we should obtain a lower-bound on the evidence of, e.g., \TT{ST+EST}, by accounting for the increased prior mass in the posterior mode due to contraction of the \TT{ST} phase support.

We reserve the remaining prior implementation details for Appendix~\ref{app:hotregions}.

\section{Inferences}\label{sec:inferences}

In this section we provide (posterior) summary information about each model applied and compare them. Tables giving numerical information for all but one model are available in Appendix~\ref{app:model summaries}. \added{Numerical files associated with the nested sample sets may be found in the persistent repository of \citet{riley19c}.} \explain{The DOI has been reserved; when we get a DOI from ApJ and all relevant permissions, the samples can be published.}. As regards model comparison, there is no clear maximally optimal measure for relative model performance (and certainly not for performance in an absolute sense). We deem \TT{ST+PST} to be superior considering the following mixture of measures, both quantitative and qualitative.


\figsetstart
\figsetnum{13}
\figsettitle{Data, model, and residuals for graphical posterior-checking.}

\figsetgrpstart
\figsetgrpnum{13.1}
\figsetgrptitle{\texttt{ST+PST}.}
\figsetplot{f13_1.pdf}
\figsetgrpnote{Count data $\{d_{ij}\}$, posterior-expected count numbers $\{\lambda_{ij}\}$, and (Poisson) residuals for \texttt{ST+PST}.}
\figsetgrpend

\figsetgrpstart
\figsetgrpnum{13.2}
\figsetgrptitle{\texttt{ST+EST}.}
\figsetplot{f13_2.pdf}
\figsetgrpnote{Count data $\{d_{ij}\}$, posterior-expected count numbers $\{\lambda_{ij}\}$, and (Poisson) residuals for \texttt{ST+EST}.}
\figsetgrpend

\figsetgrpstart
\figsetgrpnum{13.3}
\figsetgrptitle{\texttt{ST+CST}.}
\figsetplot{f13_3.pdf}
\figsetgrpnote{Count data $\{d_{ij}\}$, posterior-expected count numbers $\{\lambda_{ij}\}$, and (Poisson) residuals for \texttt{ST+CST}.}
\figsetgrpend

\figsetgrpstart
\figsetgrpnum{13.4}
\figsetgrptitle{\texttt{ST-U}.}
\figsetplot{f13_4.pdf}
\figsetgrpnote{Count data $\{d_{ij}\}$, posterior-expected count numbers $\{\lambda_{ij}\}$, and (Poisson) residuals for \texttt{ST-U}.}
\figsetgrpend

\figsetgrpstart
\figsetgrpnum{13.5}
\figsetgrptitle{\texttt{ST-S}.}
\figsetplot{f13_5.pdf}
\figsetgrpnote{Count data $\{d_{ij}\}$, posterior-expected count numbers $\{\lambda_{ij}\}$, and (Poisson) residuals for \texttt{ST-S}.}
\figsetgrpend

\figsetgrpstart
\figsetgrpnum{13.6}
\figsetgrptitle{\texttt{CDT-U}.}
\figsetplot{f13_6.pdf}
\figsetgrpnote{Count data $\{d_{ij}\}$, posterior-expected count numbers $\{\lambda_{ij}\}$, and (Poisson) residuals for \texttt{CDT-U}.}
\figsetgrpend

\figsetend

\begin{figure*}
\centering
\includegraphics[clip, trim=0cm 0cm 0cm 0cm, width=0.7\textwidth]{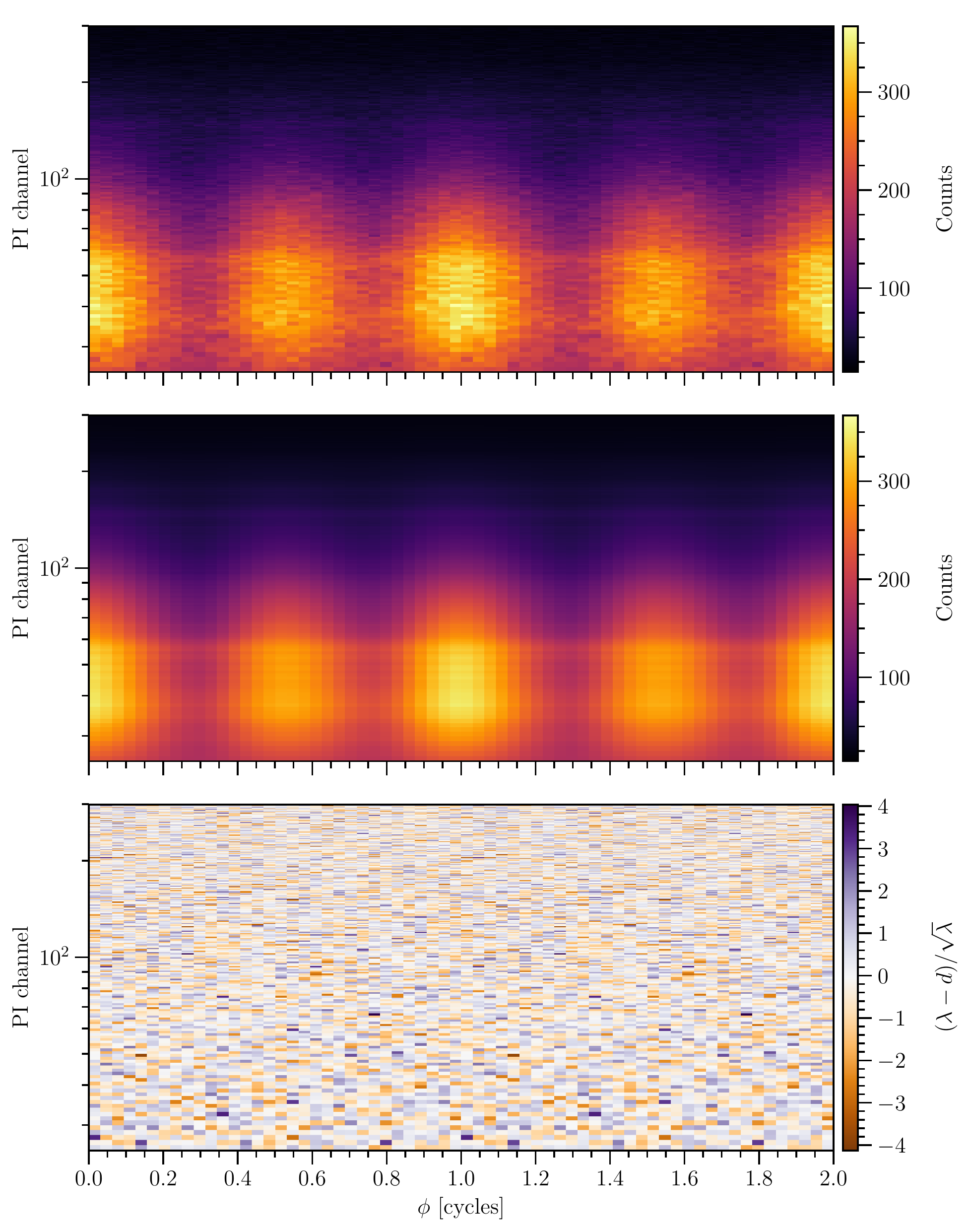}
\caption{\small{Count data $\{d_{ij}\}$, posterior-expected count numbers $\{\lambda_{ij}\}$, and (Poisson) residuals for \TT{ST+PST}. Note that we split the count numbers in the upper two panels over two rotational cycles, such that the information on phase interval $\phi\in[0,1]$ is identical to the information on $\phi\in(0,2]$; our data sampling distribution, however, is defined as the (conditional) joint probability of all event data grouped into phase intervals on $\phi\in[0,1]$. We display the standardized (Poisson) residuals in the \textit{bottom} panel: the residuals for the rotational cycle $\phi\in[0,1]$ were calculated in terms of all event data on that interval (as for likelihood definition), and simply cloned onto the interval $\phi\in(1,2]$. In Appendix~\ref{subsubsec:posterior checking figure} we elaborate on the information displayed here. The complete figure set ($6$ images) is available in the online journal, for the $\TT{ST+PST}$, $\TT{ST+EST}$, $\TT{ST+CST}$, $\TT{ST-U}$, $\TT{ST-S}$, and $\TT{CDT-U}$ models.}}
\label{fig:STpPST residuals}
\end{figure*}

\textbf{Posterior predictive performance.} Our crude graphical posterior predictive checking procedure suggests that even \TT{ST-U} can generate synthetic event data that is structurally commensurate with the \project{XTI} event data---at least in the channel subset $[25,300)$. For reference, see Appendix~\ref{subsubsec:posterior checking figure}, together with Figure~\ref{fig:STpPST residuals} and the associated figure set. Absent are obvious systematic differences in the (Poisson) standardized residuals over phase-channel intervals. Nevertheless one could study the residual differences for modeling background event arrival processes, instrument operation, or noise properties of the event data---e.g., whether the event arrival processes could detectably deviate from being Poissonian (see also Figure~\ref{fig:residuals summary}). For the purpose of constructing a model that can simply generate event data that is similar to PSR~J0030$+$0451 under visual inspection, \TT{ST-U} is competitive with all higher-complexity models.

\begin{figure*}[t!]
\centering
\includegraphics[clip, trim=0cm 0cm 0cm 0cm, width=0.675\textwidth]{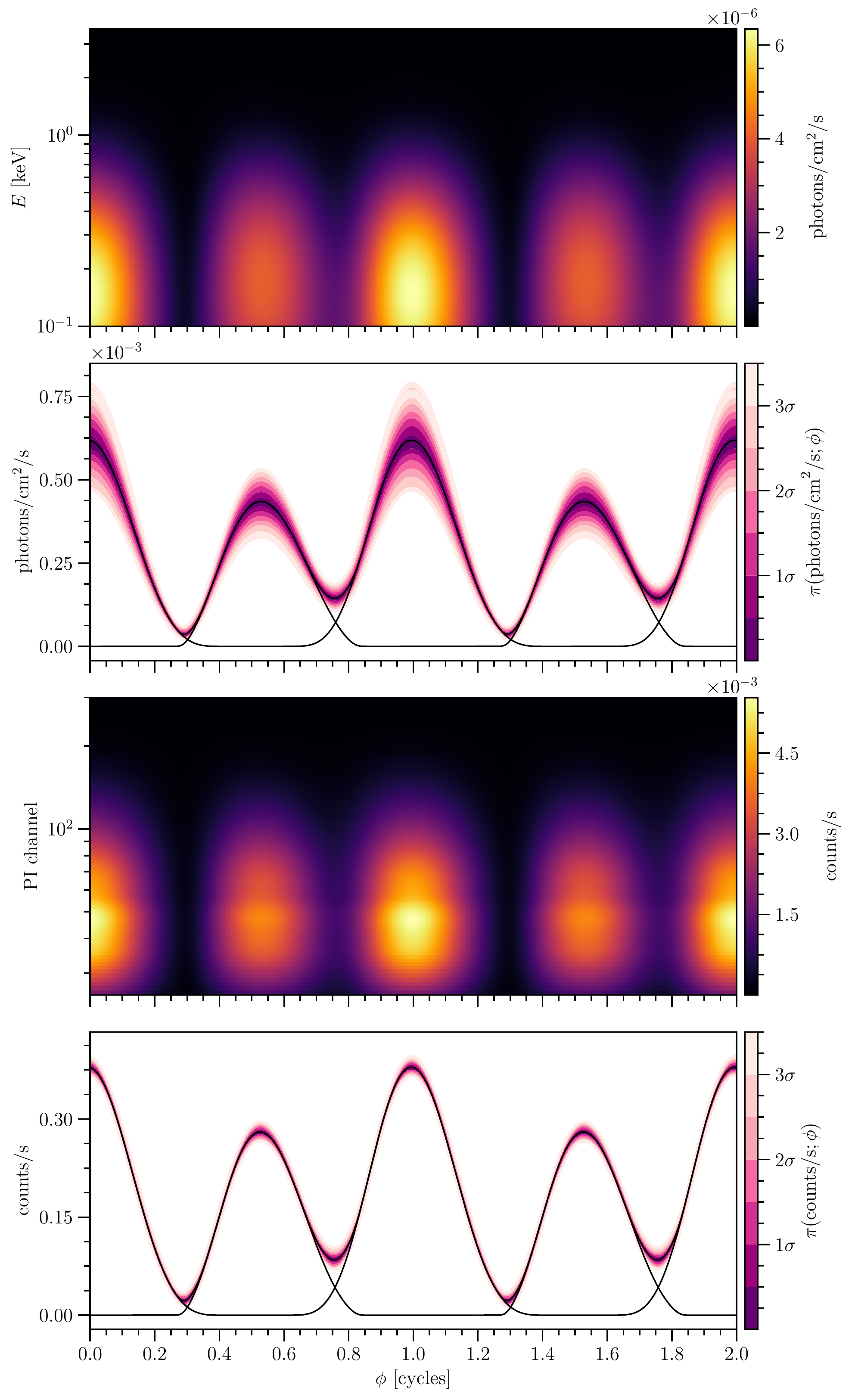}
\caption{\small{The posterior-expected signal for \TT{ST+PST}, both incident on the instrument (\textit{top} and \textit{top-center}) and as registered by the instrument (\textit{bottom-center} and \textit{bottom}). The signal in the \textit{top} panel has been integrated over the linearly-spaced instrument energy intervals, and is effectively proportional to the specific photon flux. The \textit{black} count-rate curves are the posterior-expected signals generated by each hot region separately, and in combination. We also represent the conditional posterior distribution of the incident photon flux (\textit{top-center}) and count-rate (\textit{bottom}) at each phase as a set of one-dimensional highest-density credible intervals, and connect these intervals over phase via the contours; these distributions are denoted by $\pi(\mathrm{photons/cm^{2}/s};\phi)$ and $\pi(\mathrm{counts/s};\phi)$. Note that the fractional width of the credible interval at each phase is usually higher for $\pi(\mathrm{photons/cm^{2}/s};\phi)$ than for $\pi(\mathrm{counts/s};\phi)$ because of the variation permitted for the instrument model; in combination, the signal registered by the instrument is more tightly constrained. To generate the conditional posterior bands we apply the \XPSI package, which in turn wraps the \TT{fgivenx}~\citep{fgivenx} package. The complete figure set ($6$ images) is available in the online journal.}}
\label{fig:STpPST source}
\end{figure*}

\begin{figure*}[t!]
\centering
\includegraphics[clip, trim=0cm 0cm 0cm 0cm, width=0.65\textwidth]{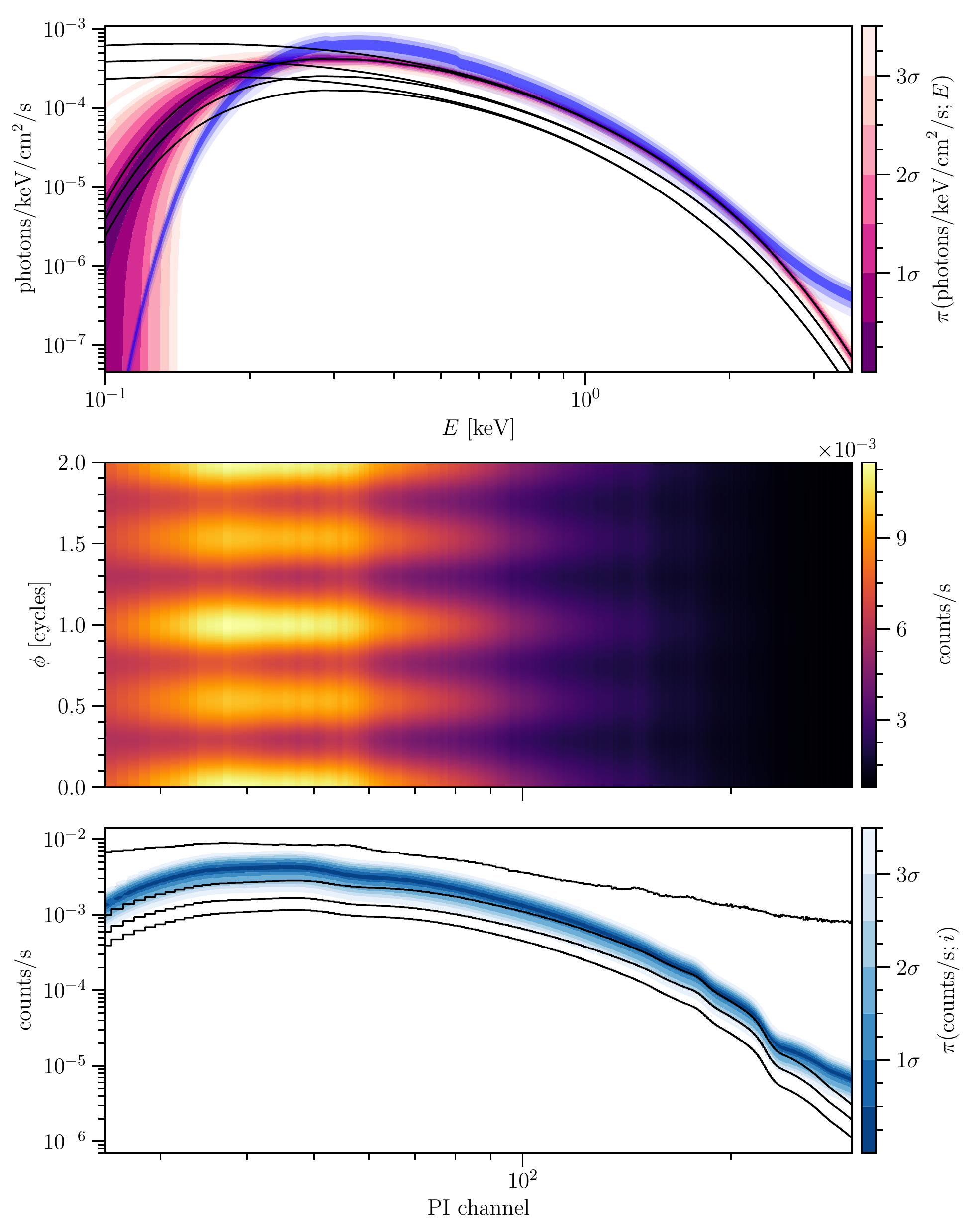}
\caption{\small{The posterior-expected spectrum for \TT{ST+PST}, both incident on the instrument (\textit{top}) and as registered by the instrument (\textit{center} and \textit{bottom}). The \textit{black} count-rate curves are the posterior-expected spectra generated by each hot region separately, and in combination. In the \textit{top} panel we display as \textit{black} curves the incident photon specific flux spectra both with and without interstellar absorption. We represent the conditional posterior distribution $\pi(\mathrm{photons/keV/cm^{2}/s};E)$ of the \textit{absorbed} incident photon specific flux at each energy as a set of one-dimensional highest-density credible intervals, and connect these intervals over phase via the contours (\textit{top}); the energies displayed are those spanning the waveband of channel subset $[25,300)$. The credible intervals fan-out at the lowest energies because: (i) conditional on event data for channel subset $[25,300)$, we are relatively insensitive to the details of the signal for $E\in[0.1,0.2)$; and (ii) the interstellar attenuation factor is stronger the lower the photon energy, and thus the incident signal varies strongly as a function of the neutral hydrogen column density---a free parameter that operates as an exponent. In the \textit{top} panel we overlay the incident absorbed spectrum inferred by \citet[][]{Bogdanov09} based on a phase-averaged analysis of low-background \project{XMM} observations; the \textit{blue} band denotes the estimated fitting uncertainty on this spectrum, at each energy, as a Gaussian with fractional standard deviation $\sigma/\mu=0.15$, and the three opacity levels indicate intervals $1\sigma$ through $3\sigma$. There is additional systematic \project{XMM} flux calibration uncertainty at the $\sim\!10\%$-level that is not included here. In the \textit{center} panel we display the background-marginalized posterior-expectation of the source count-rate signal, \textit{plus} the background count-rate terms that maximize the \textit{conditional} likelihood function; the signal is equivalent to that displayed in the center panel of Figure~\ref{fig:STpPST residuals}. In the \textit{bottom} panel we display the posterior-expected count-rate spectra generated by the hot regions in combination and individually; we opt not to render the conditional posterior count-rate distribution for each channel because it is too narrow about the expected spectrum to be useful. Moreover, the topmost \textit{black} step function is the phase-average of the \textit{center} panel---it is effectively, but not exactly, the observed count-number spectrum divided by the total exposure time $T_{\textrm{exp}}$. We combine the \project{XMM}-derived count-rate spectrum (and its associated uncertainty) with the marginal \NICER instrument posterior on parameters $\alpha$, $\beta$, and $\gamma$ to simulate a conditional probability distribution $\pi(\textrm{counts/s};i)$ for the count-rate in the $i^{th}$ channel; these conditional distributions are connected via the contours in a manner congruent to the \textit{top} panel described above. The complete figure set ($6$ images) is available in the online journal.}}
\label{fig:STpPST spectrum}
\end{figure*}

\begin{figure*}[t!]
\centering
\includegraphics[clip, trim=0cm 0cm 0cm 0cm, width=0.675\textwidth]{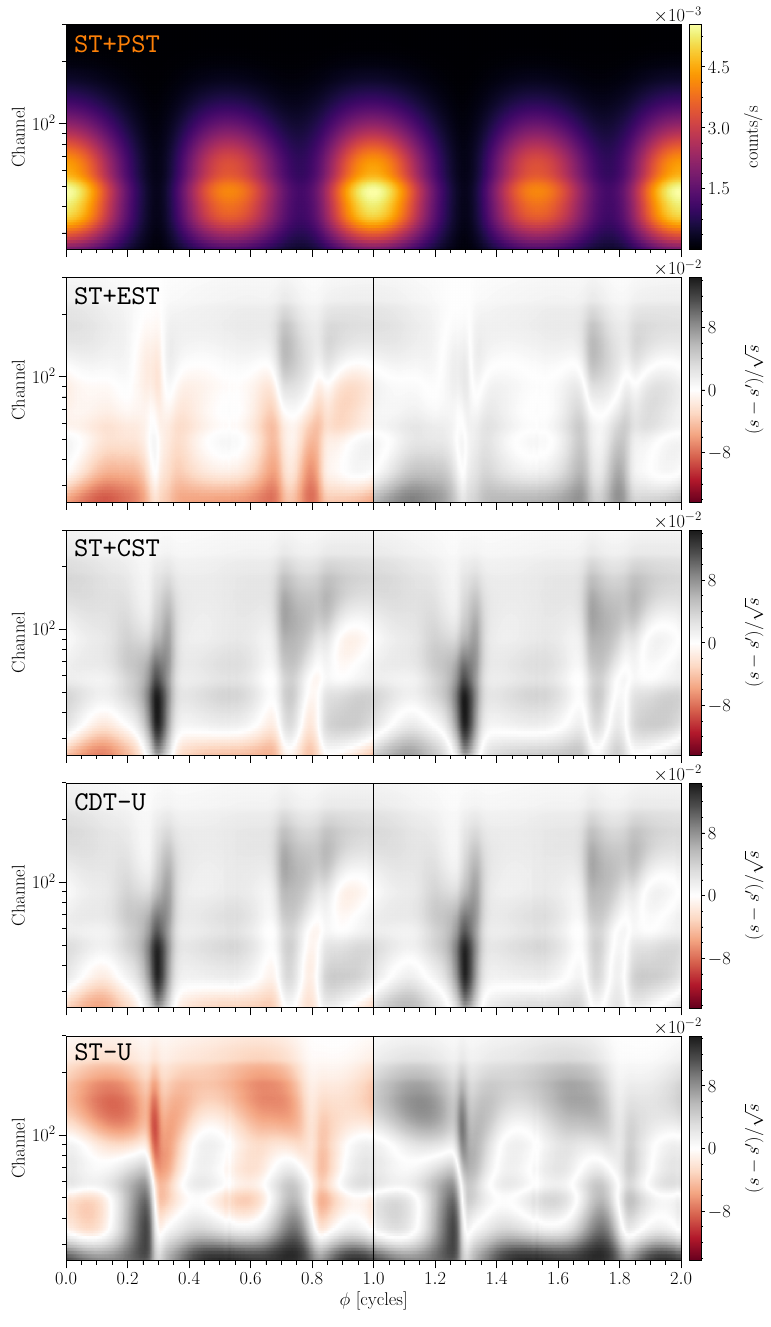}
\caption{\small{\textit{Top panel}: the posterior-expected channel-by-channel source count rate signal generated as a function of rotational phase by \TT{ST+PST} surface emission: $s\coloneqq\mathbb{E}_{\pi(\boldsymbol{\theta})}[g(\boldsymbol{\theta})]$, where $g(\boldsymbol{\theta})$ is a map from parameter space to the corresponding count rate signal. \textit{Other panels}: the difference between signal $s$ (for \TT{ST+PST}) and the posterior-expected signal conditional on some other model: let $s^{\prime}\coloneqq\mathbb{E}_{\pi(\boldsymbol{\theta})}[h(\boldsymbol{\theta})]$, where $h(\boldsymbol{\theta})$ is a map from parameter space to a count-rate signal conditional on some model other than \TT{ST+PST}. The signals $s$ and $s^{\prime}$ represent Poissonian arrival processes; we therefore opt to display the difference as $(s-s^{\prime})/\sqrt{s}$, and scale the signals by $T_{\textrm{exp}}/32$. A grayscale value is then locally representative of the absolute change (shown exclusively on the phase interval $\phi\in[1,2]$), in units of the Poisson standard deviation, within a phase interval of width $1/32$~cycles as defined during event data pre-processing.}}
\label{fig: posterior expected source signal differencing}
\end{figure*}

\begin{figure*}[t!]
\centering
\includegraphics[clip, trim=0cm 0cm 0cm 0cm, width=0.8\textwidth]{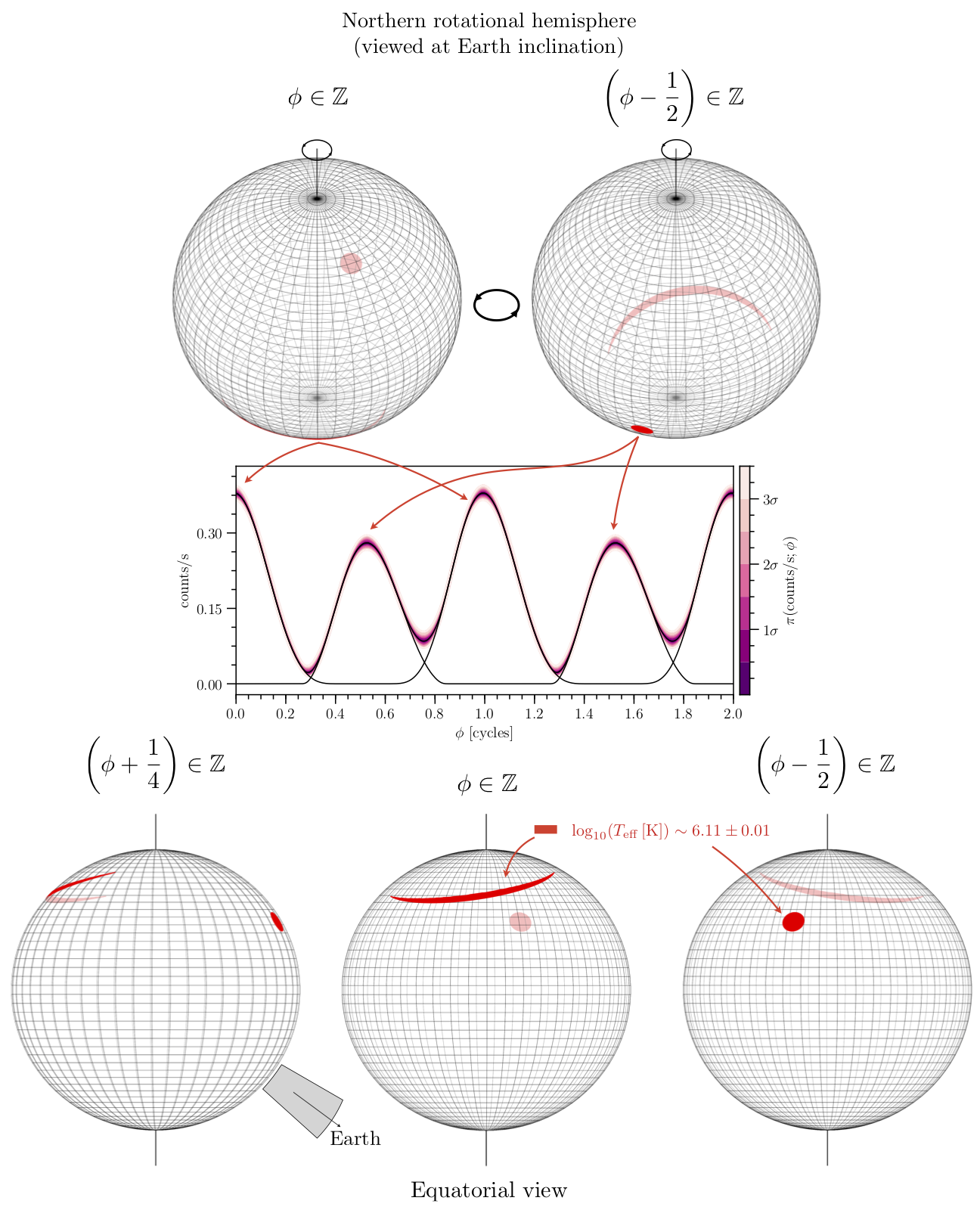}
\caption{\small{\textit{Top panel}: schematic diagram of a surface heating configuration---and Earth inclination---representative of those corresponding to points in the \TT{ST+PST} posterior mode. The configuration rendered here corresponds to the \textit{sample} that reported the highest background-marginalized likelihood function value across all models (amongst the values reported by the set of all nested samples). We project the hot regions onto a (unit) sphere and view from the Earth inclination with no ambient gravitational field. The regions are constrained to exist in the same hemisphere, but with remarkably different morphologies. The hot regions are approximately equal in effective temperature \added{and thus we define a new temperature symbol $T_{\rm eff}$ that is common to both}. We also display the channel-summed count rate pulse generated by the source emission and indicate which region generates which component; we refer the reader to Figure~\ref{fig:STpPST source}, where this signal is also displayed, for more information. \textit{Bottom panel}: note that we impose (via the prior support) that the Earth inclination lies within the northern rotational hemisphere, but an identical configuration (in terms of the physics that we consider and thus signal generation) is given via an equatorial reflection of both the Earth direction and radiating regions; we render this alternative configuration as viewed from the equatorial plane, and display the Earth inclination \added{(but not azimuth)} as the shaded angular interval bounded by the $16\%$ and $84\%$ quantiles in marginal posterior mass (see Figure~\ref{fig:STpPST obs marginal posterior} and Table~\ref{table: STpPST} for the numerical interval).}}
\label{fig:STpPST MML config}
\end{figure*}

\begin{figure*}[t!]
\centering
\includegraphics[clip, trim=0cm 0cm 0cm 0cm, width=0.875\textwidth]{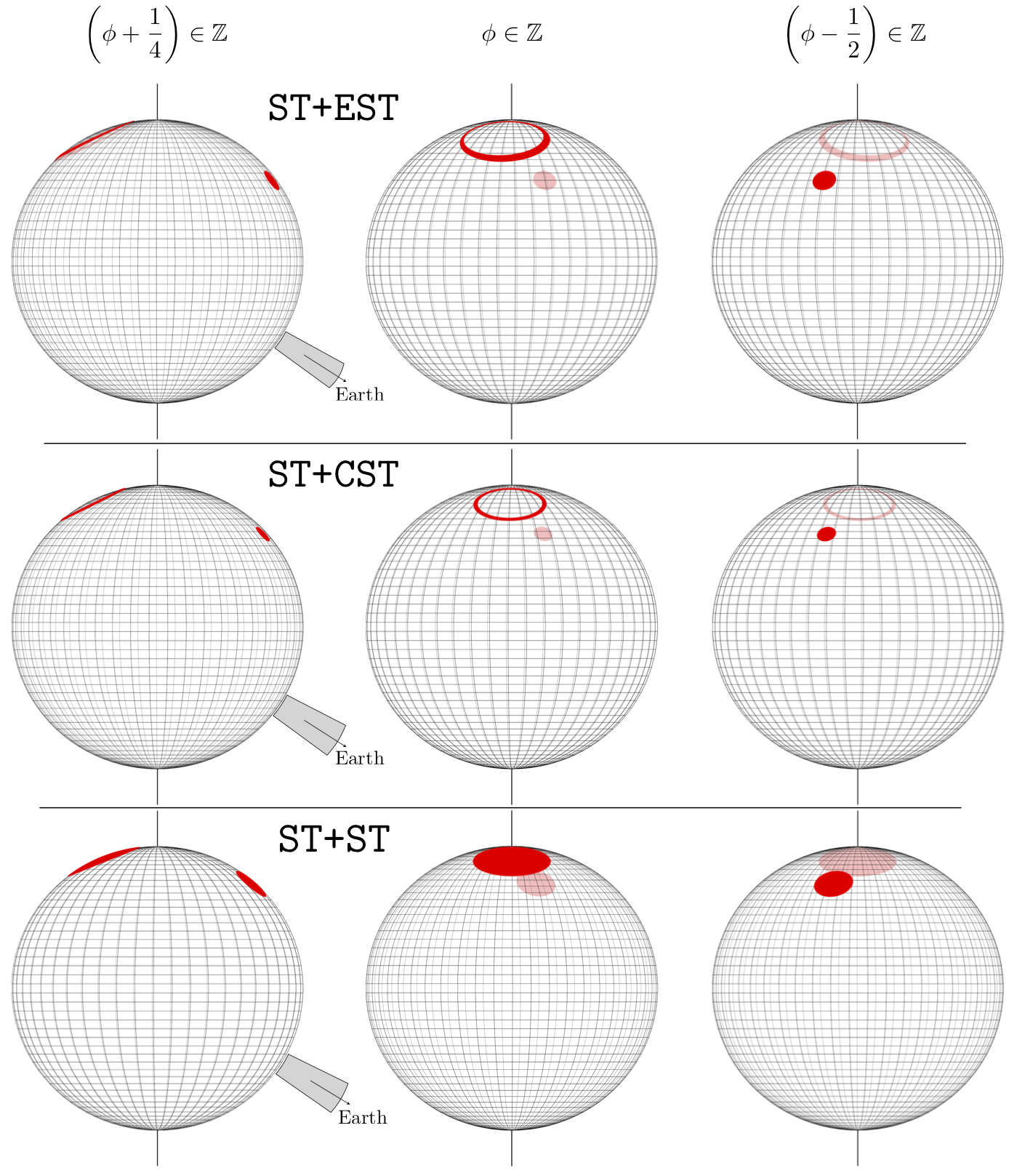}
\caption{\small{Surface heating configurations conditional on \TT{ST+ST} (equivalent to \TT{ST-U}), \TT{ST+CST}, and \TT{ST+EST}. For each model the hot regions correspond to the sample assigned the greatest posterior weight and is thus representative, being a draw from the posterior typical set. For each model we display the Earth inclination \added{(but not azimuth)} as an angular interval bounded by the $16\%$ and $84\%$ quantiles in marginal posterior mass.}}
\label{fig:surface heating schematic}
\end{figure*}

\textbf{Visualization.} In Figures~\ref{fig:STpPST source}~and~\ref{fig:STpPST spectrum} (and the associated online figure sets) we display, in posterior-expected form, various signals (derived quantities) generated by the hot regions; the reader can thus get a handle on the source contribution to the model displayed in Figure~\ref{fig:STpPST residuals} (center panel). In Figure~\ref{fig: posterior expected source signal differencing} we display the model-to-model evolution of the posterior-expected count-rate signals generated. To aid visualization, we provide in Figures~\ref{fig:STpPST MML config}~and~\ref{fig:surface heating schematic}, schematics of the hot regions on the surface that generate the aforementioned signals.


\figsetstart
\figsetnum{14}
\figsettitle{Posterior-expected source signals incident on and registered by the instrument.}

\figsetgrpstart
\figsetgrpnum{14.1}
\figsetgrptitle{\texttt{ST+PST}.}
\figsetplot{f14_1.pdf}
\figsetgrpnote{Posterior-expected source signal for \texttt{ST+PST}.}
\figsetgrpend

\figsetgrpstart
\figsetgrpnum{14.2}
\figsetgrptitle{\texttt{ST+EST}.}
\figsetplot{f14_2.pdf}
\figsetgrpnote{Posterior-expected source signal for \texttt{ST+EST}.}
\figsetgrpend

\figsetgrpstart
\figsetgrpnum{14.3}
\figsetgrptitle{\texttt{ST+CST}.}
\figsetplot{f14_3.pdf}
\figsetgrpnote{Posterior-expected source signal for \texttt{ST+CST}.}
\figsetgrpend

\figsetgrpstart
\figsetgrpnum{14.4}
\figsetgrptitle{\texttt{ST-U}.}
\figsetplot{f14_4.pdf}
\figsetgrpnote{Posterior-expected source signal for \texttt{ST-U}.}
\figsetgrpend

\figsetgrpstart
\figsetgrpnum{14.5}
\figsetgrptitle{\texttt{ST-S}.}
\figsetplot{f14_5.pdf}
\figsetgrpnote{Posterior-expected source signal for \texttt{ST-S}.}
\figsetgrpend

\figsetgrpstart
\figsetgrpnum{14.6}
\figsetgrptitle{\texttt{CDT-U}.}
\figsetplot{f14_6.pdf}
\figsetgrpnote{Posterior-expected source signal for \texttt{CDT-U}.}
\figsetgrpend

\figsetend

\figsetstart
\figsetnum{15}
\figsettitle{Posterior-expected source spectra incident on and registered by the instrument.}

\figsetgrpstart
\figsetgrpnum{15.1}
\figsetgrptitle{\texttt{ST+PST}.}
\figsetplot{f15_1.pdf}
\figsetgrpnote{Posterior-expected source spectra for \texttt{ST+PST}.}
\figsetgrpend

\figsetgrpstart
\figsetgrpnum{15.2}
\figsetgrptitle{\texttt{ST+EST}.}
\figsetplot{f15_2.pdf}
\figsetgrpnote{Posterior-expected source spectra for \texttt{ST+EST}.}
\figsetgrpend

\figsetgrpstart
\figsetgrpnum{15.3}
\figsetgrptitle{\texttt{ST+CST}.}
\figsetplot{f15_3.pdf}
\figsetgrpnote{Posterior-expected source spectra for \texttt{ST+CST}.}
\figsetgrpend

\figsetgrpstart
\figsetgrpnum{15.4}
\figsetgrptitle{\texttt{ST-U}.}
\figsetplot{f15_4.pdf}
\figsetgrpnote{Posterior-expected source spectra for \texttt{ST-U}.}
\figsetgrpend

\figsetgrpstart
\figsetgrpnum{15.5}
\figsetgrptitle{\texttt{ST-S}.}
\figsetplot{f15_5.pdf}
\figsetgrpnote{Posterior-expected source spectra for \texttt{ST-S}.}
\figsetgrpend

\figsetgrpstart
\figsetgrpnum{15.6}
\figsetgrptitle{\texttt{CDT-U}.}
\figsetplot{f15_6.pdf}
\figsetgrpnote{Posterior-expected source spectra for \texttt{CDT-U}.}
\figsetgrpend

\figsetend

\textbf{Evidence.} We report evidence estimates in Tables~\ref{table: STpPST} through \ref{table: ST-S}. For the family of models that we have considered, the evidence strongly favors \citep[adopting the guidelines of][]{KassRaft95} a model wherein antipodally reflection symmetry is \textit{not} imposed; in other words, \TT{ST-S}, and by extrapolation any such model with the \TT{-S} extension (refer to Figure~\ref{fig: discrete model space diagram}), is considered to be strongly disfavored. Moreover, for all computed models that do not impose antipodal reflection symmetry, posterior modes contain configurations characterized by both hot regions being located in the same rotational hemisphere---the opposite rotational hemisphere to the Earth direction. We also conclude that at least one region should be modeled with more complexity than offered by a single-temperature simply-connected circular (\TT{ST}) region; in other words, there is deemed sufficient evidence to disfavor \TT{ST-U}. However, there is insufficient evidence to resolve between \TT{CDT-U}, \TT{ST+CST}, \TT{ST+EST}, and \TT{ST+PST}, especially when considering the evidence not as a scalar estimator, but as a random variable with a (simulated) distribution for estimating error intervals. The estimated error intervals (defined to contain 90\% of evidence estimates based on nested sampling process realizations) typically overlap for these models, whilst the expected log-evidences are within $\sim\!2$ units.

\begin{figure*}[t!]
\centering
\includegraphics[clip, trim=0cm 0cm 0cm 0cm, width=0.85\textwidth]{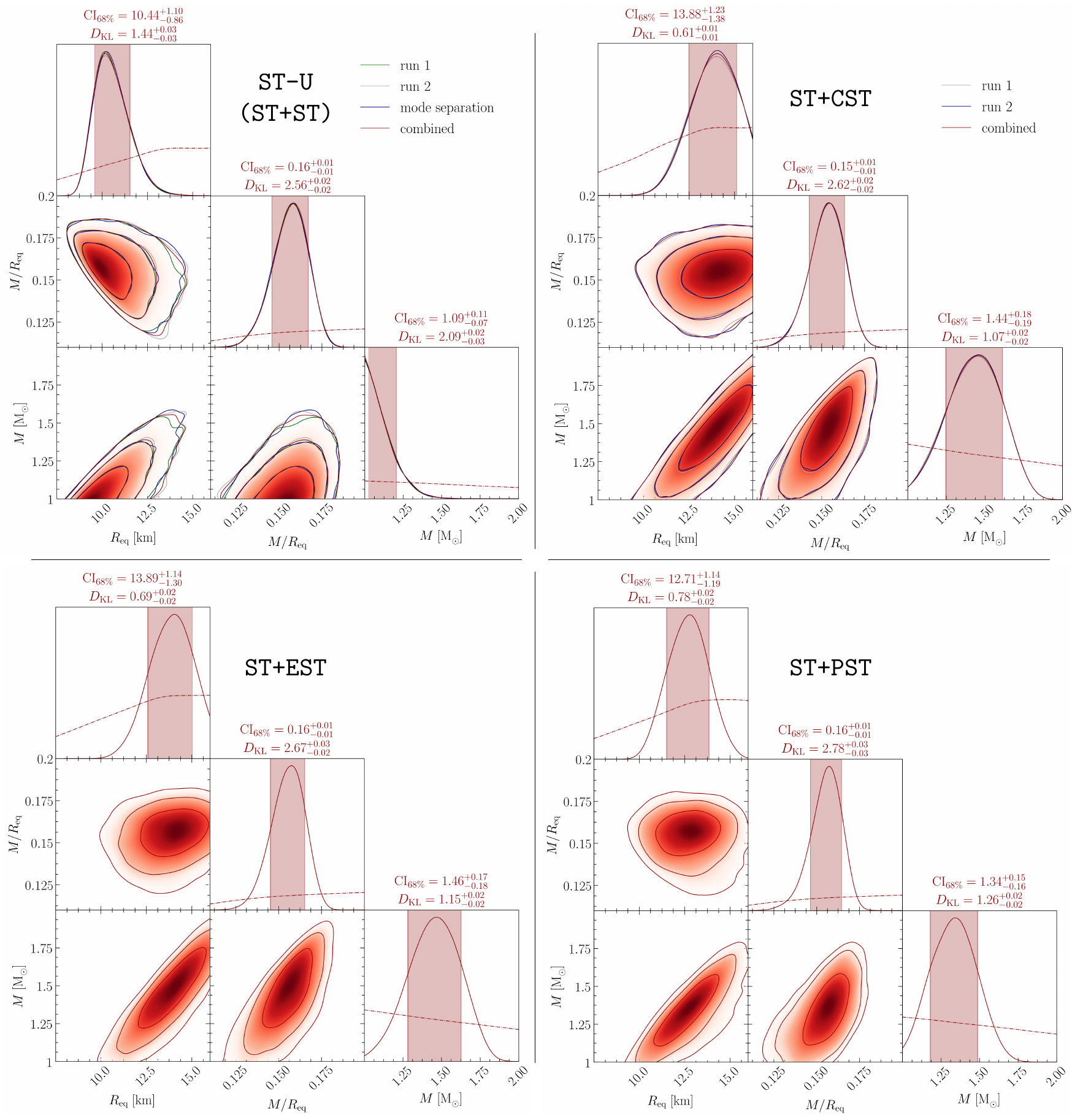}
\caption{\small{One- and two-dimensional marginal posterior density distributions for the MSP spacetime parameters conditional on the each in the sequence \TT{ST-U}, \TT{ST+CST}, \TT{ST+EST}, and \TT{ST+PST}. From leftmost to rightmost in each panel, the parameters are equatorial radius, (equatorial) compactness, and gravitational mass. We display the marginal prior density distributions for each parameter as the \textit{dash-dot} functions. For the less expensive \TT{ST-U} and \TT{ST+CST} models we executed two production runs and combined them, whilst for each of \TT{ST+EST} and \TT{ST+PST} we executed a single production run. We also display an \TT{ST-U} run in which the mode-separation \MultiNest variant was activated, but because neither the theory nor the software exists for combining such a sampling process with runs in default mode (where sampling threads can migrate between posterior modes), it is not included in the combined run. The mode-separation run allocated $\sim\!1/3$ of the sampling resolution (i.e., sampling threads, or active points) to a second posterior mode with negligible local mass; this mode corresponds to a distinct phase configuration, with the hot regions---which are distinguished by their order in colatitude---transposed in their coupling to the pulse components visible in the phase-folded event data. We report the KL-divergence, $D_{\mathrm{KL}}$, from prior to posterior in \textit{bits} for each parameter, together with an error interval containing $68.3\%$ of $\widehat{D}_{\mathrm{KL}}$ estimates based on simulated nested sampling process realizations. The shaded credible intervals $\mathrm{CI}_{68\%}$ for each parameter are symmetric in marginal posterior mass about the median, containing $68.3\%$ of the mass; the (barely discernible) darker intervals at the $\mathrm{CI}_{68\%}$ boundaries contain $68.3\%$, respectively, of the $15.85\%$ and $84.15\%$ quantiles in posterior mass, again based on simulated nested sampling process realizations. The credible regions in the off-diagonal panels, on the other hand, are uniquely the \textit{highest-density}---and thus the smallest possible---credible regions, containing $68.3\%$, $95.4\%$, and $99.7\%$ of the posterior mass. In Appendix~\ref{sec:posterior computation} we provide additional information regarding posterior kernel density estimation, error analysis, and the estimators displayed here. The complete figure set ($7$ images) is available in the online journal.}}
\label{fig:STpPST spacetime marginal posterior}
\end{figure*}

\begin{figure*}[t!]
\centering
\includegraphics[clip, trim=0cm 0cm 0cm 0cm, width=1.0\textwidth]{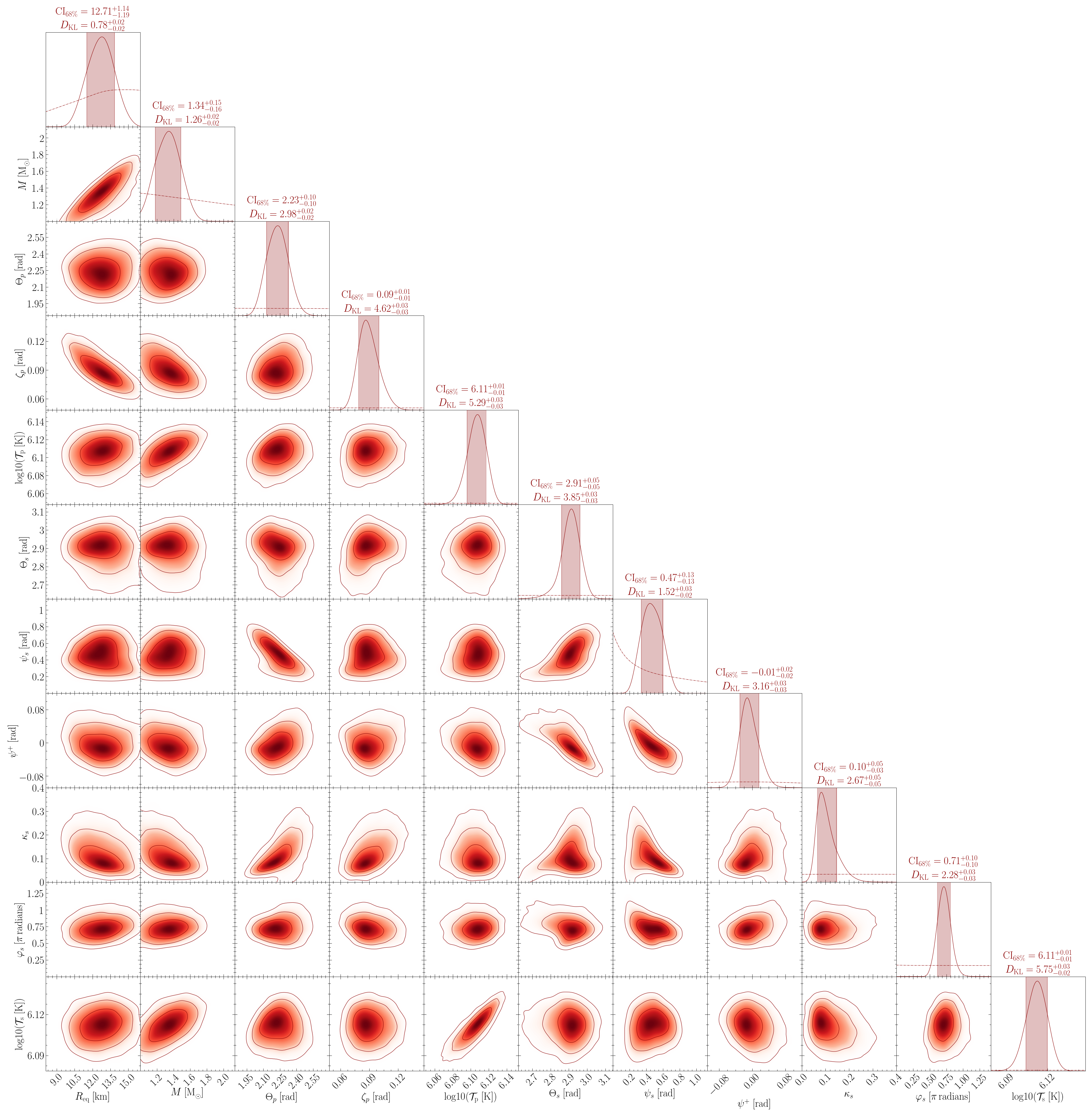}
\caption{\small{One- and two-dimensional marginal posterior density distributions for the MSP parameters conditional on \TT{ST+PST}. From leftmost to rightmost: equatorial radius $R_{\mathrm{eq}}$; gravitational radius $M$; \TT{ST} center colatitude $\Theta_{p}$; \TT{ST} angular radius $\zeta_{p}$; \TT{ST} \TT{NSX} effective temperature $\log_{10}\mathcal{T}_{p}$; \TT{PST} superseding member center colatitude $\Theta_{s}$; \TT{PST} superseding angular radius $\psi_{s}$; \TT{PST} superseding member angular radius difference $\psi_{s}^{+}=\zeta_{s}-\psi_{s}$, where $\zeta_{s}$ is the angular radius of the ceding member; \TT{PST} ceding member fractional angular offset $\varkappa_{s}$ \added{(labeled as $\kappa_{s}$)}; \TT{PST} ceding member azimuthal offset $\varphi_{s}$; and \TT{PST} ceding \TT{NSX} effective temperature $\log_{10}\mathcal{T}_{s}$. For descriptions of the information displayed, refer to Figure~\ref{fig:STpPST spacetime marginal posterior}; note that here we display the marginal posterior density distribution for each parameter as a single solid function due to the number of panels. We choose not to display joint posterior distributions for all pairs of model parameters because the number of panels is prohibitive; moreover, the posterior azimuthal separation of the \TT{ST} and \TT{PST} regions is displayed in Figure~\ref{fig:STpPST obs marginal posterior}. The complete figure set ($6$ images) is available in the online journal.}}
\label{fig:STpPST source marginal posterior}
\end{figure*}

\begin{figure*}[t!]
\centering
\includegraphics[clip, trim=0cm 0cm 0cm 0cm, width=1.0\textwidth]{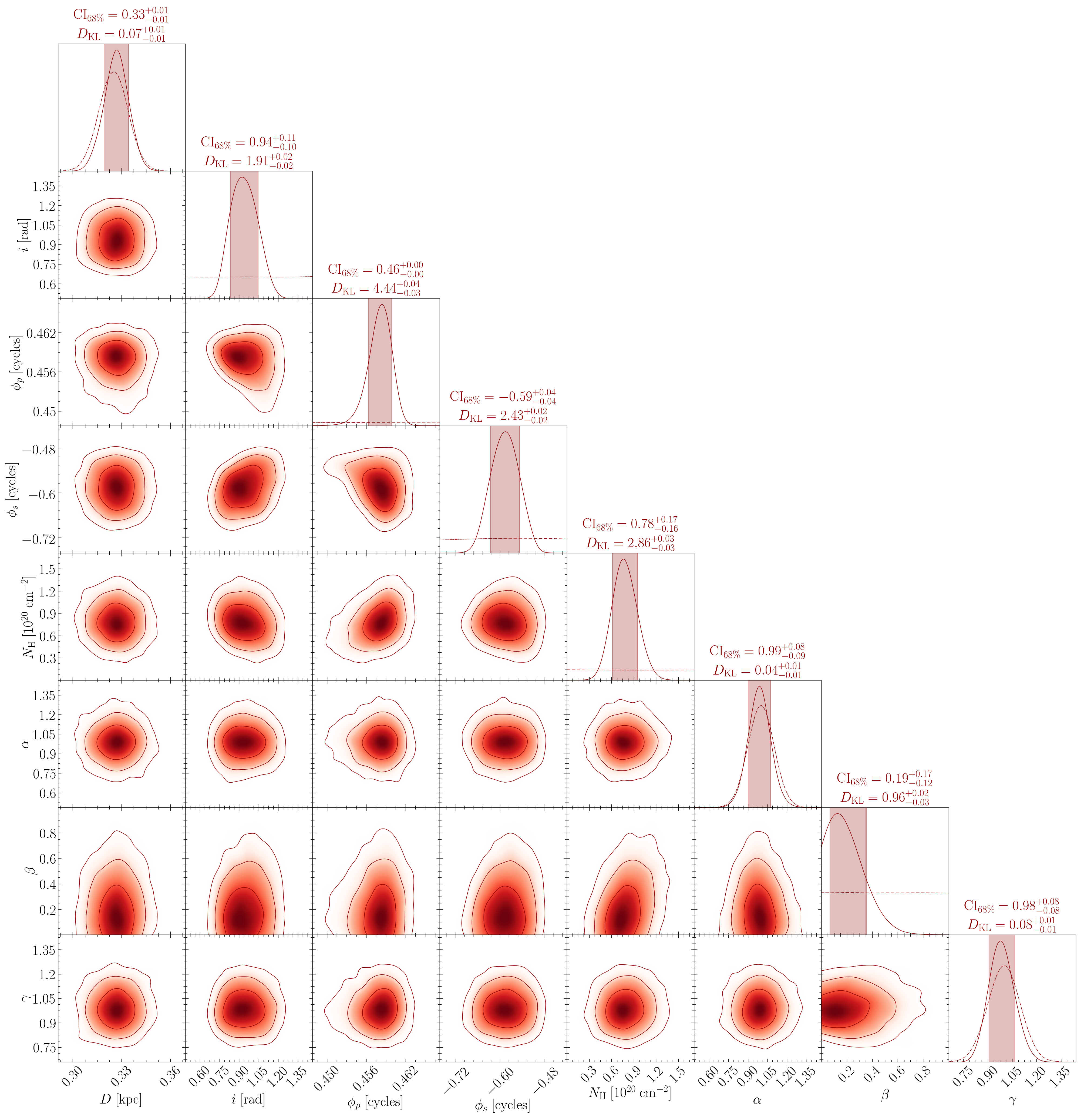}
\caption{\small{One- and two-dimensional marginal posterior density distributions for parameters pertaining mostly to the MSP \textit{observation}, conditional on \TT{ST+PST}. From leftmost to rightmost: distance $D$; Earth inclination $i$; \TT{ST} center azimuth (phase) relative to Earth direction $\phi_{p}$; \TT{PST} superseding member center azimuth (phase) relative to Earth antipode $\phi_{s}$; interstellar neutral hydrogen column density $N_{\mathrm{H}}$; and \NICER instrument parameters $\alpha$, $\beta$, and $\gamma$. For descriptions of the information displayed, refer to Figure~\ref{fig:STpPST spacetime marginal posterior}; note that here we display the marginal posterior density distribution for each parameter as a single solid function due to the number of panels. The complete figure set ($6$ images) is available in the online journal.}}
\label{fig:STpPST obs marginal posterior}
\end{figure*}

\begin{figure*}[t!]
\centering
\includegraphics[clip, trim=0cm 0cm 0cm 0cm, width=\textwidth]{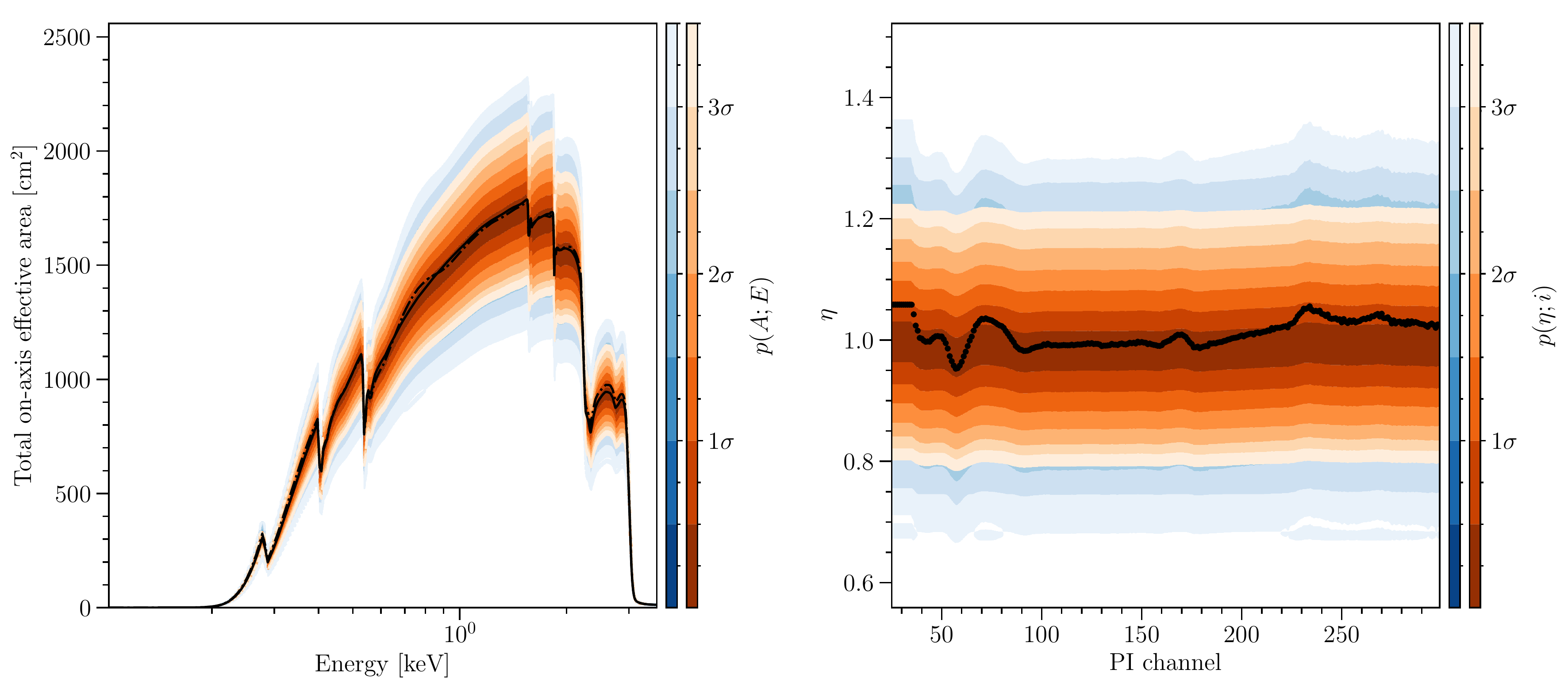}
\caption{\small{We update the instrument prior (displayed in Figure~\ref{fig:NICER instrument effective area prior}) conditional on \TT{ST+PST}. The conditional posterior distributions are represented by the \textit{orange} bands; the conditional prior distributions are displayed in \textit{blue} to indicate information gain about the instrument, and the \textit{black} lines and points are equivalent to those in Figure~\ref{fig:NICER instrument effective area prior}. The complete figure set ($6$ images) is available in the online journal.}}
\label{fig:NICER instrument STpPST posterior}
\end{figure*}

\textbf{Kullback-Leibler divergence.} We report global and marginal divergence estimates in Tables~\ref{table: STpPST} through \ref{table: ST-S}, and in Figures~\ref{fig:STpPST spacetime marginal posterior}, \ref{fig:STpPST source marginal posterior}, and \ref{fig:STpPST obs marginal posterior}. Information gain from prior to posterior is defined as a divergence integral over some subset of parameter dimensions: it is a non-negative real scalar measure of the difference between normalized density functions (see Appendix~\ref{subsubsec:divergence} for a more detailed description). The larger the number, the larger the information gain, whilst a minimum divergence of zero indicates that the density functions are identical. The global information gain---the divergence integral over all dimensions---is comparable for all models. Based on comparison of parameter-by-parameter divergence estimates, and visual comparison of the marginal prior and posterior density functions hence summarized, we consider the joint prior distribution to be weakly informative in the context of the the likelihood function for most source parameters; the main exception is the distance, which is strongly prior-dominated. The other parameters whose marginal posterior distributions are entirely prior-dominated are the two of the instrument parameters, $\alpha$ and $\gamma$ (refer to Figure~\ref{fig:STpPST obs marginal posterior} and Figure~\ref{fig:NICER instrument STpPST posterior}). These informative prior distributions are however shared by all models, as are the improper flat prior density functions described in the background treatment (refer to Section~\ref{subsubsec:likelihood functions}). Note that more information is gained about the instrument parameter $\beta$---the weighting factor between two response matrices---and we can conclude that the instrument manifests \textit{a posteriori} as a mixture weighted appreciably toward the nominal response for all posterior computations; we reserve discussion on why the instrument calibration may not be accurate for Section~\ref{subsubsec: discuss instrument}.


\figsetstart
\figsetnum{19}
\figsettitle{One- and two-dimensional marginal posterior distributions of spacetime parameters.}

\figsetgrpstart
\figsetgrpnum{19.1}
\figsetgrptitle{Four-panel comparison version.}
\figsetplot{f19_1.pdf}
\figsetgrpnote{Posterior information conditional on \texttt{ST-U}, \texttt{ST+CST}, \texttt{ST+EST}, and \texttt{ST+PST}.}
\figsetgrpend

\figsetgrpstart
\figsetgrpnum{19.2}
\figsetgrptitle{\texttt{ST+PST}.}
\figsetplot{f19_2.png}
\figsetgrpnote{Posterior information conditional on \texttt{ST+PST}.}
\figsetgrpend

\figsetgrpstart
\figsetgrpnum{19.3}
\figsetgrptitle{\texttt{ST+EST}.}
\figsetplot{f19_3.png}
\figsetgrpnote{Posterior information conditional on \texttt{ST+EST}.}
\figsetgrpend

\figsetgrpstart
\figsetgrpnum{19.4}
\figsetgrptitle{\texttt{ST+CST}.}
\figsetplot{f19_4.png}
\figsetgrpnote{Posterior information conditional on \texttt{ST+CST}.}
\figsetgrpend

\figsetgrpstart
\figsetgrpnum{19.5}
\figsetgrptitle{\texttt{ST-U}.}
\figsetplot{f19_5.png}
\figsetgrpnote{Posterior information conditional on \texttt{ST-U}.}
\figsetgrpend

\figsetgrpstart
\figsetgrpnum{19.6}
\figsetgrptitle{\texttt{ST-S}.}
\figsetplot{f19_6.png}
\figsetgrpnote{Posterior information conditional on \texttt{ST-S}.}
\figsetgrpend

\figsetgrpstart
\figsetgrpnum{19.7}
\figsetgrptitle{\texttt{CDT-U}.}
\figsetplot{f19_7.png}
\figsetgrpnote{Posterior information conditional on \texttt{CDT-U}.}
\figsetgrpend

\figsetend

\figsetstart
\figsetnum{20}
\figsettitle{One- and two-dimensional marginal posterior distributions of MSP parameters.}

\figsetgrpstart
\figsetgrpnum{20.1}
\figsetgrptitle{\texttt{ST+PST}.}
\figsetplot{f20_1.png}
\figsetgrpnote{Posterior information conditional on \texttt{ST+PST}.}
\figsetgrpend

\figsetgrpstart
\figsetgrpnum{20.2}
\figsetgrptitle{\texttt{ST+EST}.}
\figsetplot{f20_2.png}
\figsetgrpnote{Posterior information conditional on \texttt{ST+EST}.}
\figsetgrpend

\figsetgrpstart
\figsetgrpnum{20.3}
\figsetgrptitle{\texttt{ST+CST}.}
\figsetplot{f20_3.png}
\figsetgrpnote{Posterior information conditional on \texttt{ST+CST}.}
\figsetgrpend

\figsetgrpstart
\figsetgrpnum{20.4}
\figsetgrptitle{\texttt{ST-U}.}
\figsetplot{f20_4.png}
\figsetgrpnote{Posterior information conditional on \texttt{ST-U}.}
\figsetgrpend

\figsetgrpstart
\figsetgrpnum{20.5}
\figsetgrptitle{\texttt{ST-S}.}
\figsetplot{f20_5.png}
\figsetgrpnote{Posterior information conditional on \texttt{ST-S}.}
\figsetgrpend

\figsetgrpstart
\figsetgrpnum{20.6}
\figsetgrptitle{\texttt{CDT-U}.}
\figsetplot{f20_6.png}
\figsetgrpnote{Posterior information conditional on \texttt{CDT-U}.}
\figsetgrpend

\figsetend

\figsetstart
\figsetnum{21}
\figsettitle{One- and two-dimensional marginal posterior distributions of (mainly) observational parameters.}

\figsetgrpstart
\figsetgrpnum{21.1}
\figsetgrptitle{\texttt{ST+PST}.}
\figsetplot{f21_1.png}
\figsetgrpnote{Posterior information conditional on \texttt{ST+PST}.}
\figsetgrpend

\figsetgrpstart
\figsetgrpnum{21.2}
\figsetgrptitle{\texttt{ST+EST}.}
\figsetplot{f21_2.png}
\figsetgrpnote{Posterior information conditional on \texttt{ST+EST}.}
\figsetgrpend

\figsetgrpstart
\figsetgrpnum{21.3}
\figsetgrptitle{\texttt{ST+CST}.}
\figsetplot{f21_3.png}
\figsetgrpnote{Posterior information conditional on \texttt{ST+CST}.}
\figsetgrpend

\figsetgrpstart
\figsetgrpnum{21.4}
\figsetgrptitle{\texttt{ST-U}.}
\figsetplot{f21_4.png}
\figsetgrpnote{Posterior information conditional on \texttt{ST-U}.}
\figsetgrpend

\figsetgrpstart
\figsetgrpnum{21.5}
\figsetgrptitle{\texttt{ST-S}.}
\figsetplot{f21_5.png}
\figsetgrpnote{Posterior information conditional on \texttt{ST-S}.}
\figsetgrpend

\figsetgrpstart
\figsetgrpnum{21.6}
\figsetgrptitle{\texttt{CDT-U}.}
\figsetplot{f21_6.png}
\figsetgrpnote{Posterior information conditional on \texttt{CDT-U}.}
\figsetgrpend

\figsetend

\figsetstart
\figsetnum{22}
\figsettitle{Conditional, marginal posterior distributions of instrument response properties.}

\figsetgrpstart
\figsetgrpnum{22.1}
\figsetgrptitle{\texttt{ST+PST}.}
\figsetplot{f22_1.pdf}
\figsetgrpnote{Posterior information conditional on \texttt{ST+PST}.}
\figsetgrpend

\figsetgrpstart
\figsetgrpnum{22.2}
\figsetgrptitle{\texttt{ST+EST}.}
\figsetplot{f22_2.pdf}
\figsetgrpnote{Posterior information conditional on \texttt{ST+EST}.}
\figsetgrpend

\figsetgrpstart
\figsetgrpnum{22.3}
\figsetgrptitle{\texttt{ST+CST}.}
\figsetplot{f22_3.pdf}
\figsetgrpnote{Posterior information conditional on \texttt{ST+CST}.}
\figsetgrpend

\figsetgrpstart
\figsetgrpnum{22.4}
\figsetgrptitle{\texttt{ST-U}.}
\figsetplot{f22_4.pdf}
\figsetgrpnote{Posterior information conditional on \texttt{ST-U}.}
\figsetgrpend

\figsetgrpstart
\figsetgrpnum{22.5}
\figsetgrptitle{\texttt{ST-S}.}
\figsetplot{f22_5.pdf}
\figsetgrpnote{Posterior information conditional on \texttt{ST-S}.}
\figsetgrpend

\figsetgrpstart
\figsetgrpnum{22.6}
\figsetgrptitle{\texttt{CDT-U}.}
\figsetplot{f22_6.pdf}
\figsetgrpnote{Posterior information conditional on \texttt{CDT-U}.}
\figsetgrpend

\figsetend

\textbf{Likelihood function.} We generally refrain from reporting parameter vector point-measures such as the maximum \textit{a posteriori} (MAP) vector or the maximum likelihood vector for the purpose of formal quantitative model comparison (see Appendix~\ref{subsubsec:posterior checking figure} for reasoning). We note, however, that the background-marginalized likelihood function exhibits the largest values in any posterior \textit{typical set} (by $\sim\!4$ natural logarithmic units), across all models, for \TT{ST+PST}. It follows that a subset of the additional complexity introduced by \TT{ST+PST} is helpful---and suggestive of avenues for future modeling---whilst a subset of the complexity is unhelpful, leading to only commensurate evidence relative to \TT{ST+EST}. In other words, the (background-marginalized) likelihood function, over a subset of the additional prior support, is at most commensurate with the prior expectation of the likelihood function for the other models; globally, the model does not therefore improve as measured solely by the evidence.

\textbf{Tractability and complexity.} Posterior computation accuracy and reproducibility are generally higher the less complex the model. The \TT{ST+CST}, \TT{ST+EST}, and \TT{ST+PST} models redistributed complexity based on the performance of \TT{CDT-U} in pursuit of \textit{helpful} complexity and thus higher-efficiency resource consumption. We demonstrate in Figure~\ref{fig: posterior expected source signal differencing} that effectively all of the improvement of \TT{CDT-U} over \TT{ST-U} is captured by the intermediary \TT{ST+CST}. As a consequence, \TT{ST+PST} in particular exhibited substantial improvement in regard to the (background-marginalized) likelihood function in the posterior mode, whilst reducing the number of parameters by limiting the complexity of one hot region; it follows that it should be clearly preferred over \TT{CDT-U}.\footnote{An implementation-dependent note is that likelihood function evaluation will generally be faster for a temperature \replaced{function}{field} that is azimuthally invariant \added{where it is finite}, as is by definition the case for regions that are of a single temperature.}

We remark that although the complexity of one hot region was ultimately limited at the \TT{ST}-level, this does not mean that the signal generated by that region is fixed or performs maximally in some absolute sense. Indeed, the signal generated by one region cannot be decoupled (as stated in Section~\ref{subsubsec:degeneracy and complexity}) from the signal generated by the other for the purpose of statistical inference: (i) the regions are restricted to exist on the same $2$-surface within an ambient spacetime; and (ii) for some subset of rotational phases, and for the configurations that report higher (background-marginalized) likelihood values, the images of both regions are simultaneously visible (\textit{a posteriori}) and registered by the instrument in combination. As a result, upon examination of Figure~\ref{fig: posterior expected source signal differencing}, we observe that the posterior-expected count-rate signal generated by the \TT{ST} region evolves along the \TT{ST-U}, \TT{ST+CST}, \TT{ST+EST}, and \TT{ST+PST} sequence, both at phases where the non-\TT{ST} region is invisible, and at phases where both regions are simultaneously visible. The reader should not therefore interpret our decision to limit the complexity of one region to \TT{ST} as a statement that an \TT{ST} region is forecasted to be sufficient for the purpose of future modeling efforts nor that exploration of different models will fail to yield improvement; indeed, one could even consider models such as \TT{PST-U} given that our modeling route for one region was based on \TT{CDT-U}.

\textbf{Telescope cross-checking.} In Figure~\ref{fig:STpPST spectrum} we display the posterior information for the phase-integrated spectrum generated by the \TT{ST+PST} hot regions; corresponding figures for the other models may be found in the online figure set. We overlay a model PSR~J0030$+$0451 spectrum, derived via analysis of low-background low-signal-to-noise phase-integrated \project{XMM} observations \citep[see][]{Bogdanov09}; this spectrum is considered as a guiding upper-bound for all emission in the \NICER waveband from the PSR~J0030$+$0451 system. The \TT{ST-U}, \TT{CDT-U}, \TT{ST+CST}, \TT{ST+EST}, and \TT{ST+PST} models do not violate this condition in channels $[25,300)$ spanned by the event data considered, nor at incident photon energies that couple strongly to this channel subset; the \TT{ST-S} model clearly violates this condition. The \project{XMM}-derived model is more uncertain than shown at low energies in Figure~\ref{fig:STpPST spectrum}, due to unresolved likelihood function degeneracy between $N_{\mathrm{H}}$ and the thermal components. The \project{XMM} spectral analysis also included a power-law component, which is subsumed into the phase-invariant (background) terms in our analysis of \NICER data (see the topmost \textit{black} count-rate step function in the \textit{bottom} panel of Figure~\ref{fig:STpPST spectrum}); the discrepancy at higher energies is thus accounted for. Nevertheless, a pertinent question remains regarding how much of the signal generated by surface emission is captured by the phase-invariant count-rate terms; these terms combine linearly with the signals from the hot regions and are intended to parameterize the background contribution. We refer the reader to Section~\ref{sec:assump} for more detailed discussion on this topic.

\begin{longtable}{l|l|l|ll}
\caption{Summary table for \TT{ST+PST}. We provide: (i) the parameters that constitute the sampling space, with symbols, units, and short descriptions; (ii) any notable derived or fixed parameters; (iii) the joint prior distribution, including hard truncation bounds and constraint equations that define the hyperboundary of the support; (iv) one-dimensional (marginal) $68.3\%$ credible interval estimates symmetric in posterior mass about the \textit{median} ($\widehat{\textrm{CI}}_{68\%}$); and (v) KL-divergence estimates in \textit{bits} ($\widehat{D}_{\textrm{KL}}$) representing prior-to-posterior information gain (a scalar with an associated error calculated as the $68.3\%$ distributional interval about the median with respect to sampling process realizations; see Appendix~\ref{subsubsec:divergence} for high-level description of the divergence). Constraint equations in terms of two or more parameters result in \textit{marginal} distributions that are not equivalent to those inverse-sampled.}\label{table: STpPST}\\
\hline\hline
Parameter & Description & Prior (density and support) & $\widehat{\textrm{CI}}_{68\%}$ & $\widehat{D}_{\textrm{KL}}$\\
\hline
\endfirsthead
\multicolumn{5}{c}%
{\tablename\ \thetable\---\textit{Continued from previous page}} \\
\hline
\endhead
\hline \multicolumn{5}{r}{\textit{Continued on next page}} \\
\endfoot
\hline\hline
\endlastfoot
$P$ $[$ms$]$ &
coordinate spin period &
$P=4.87$,\footnote{\citet[][]{nanograv11}.} fixed &
$-$ &
$-$ \\
\hline
$M$ $[\textrm{M}_{\odot}]$ &
gravitational mass\footnote{Interpreted as a rotationally perturbed mass monopole moment \citep[e.g.,][]{Hartle1967}, but the perturbation is small for the spin frequency of PSR~J0030$+$0451 (see Section~\ref{sec:mrcontext}).} &
$M\sim U(1,3)$\footnote{Hard lower-bound based loosely on plausible astrophysical formation channels \citep[see, e.g.,][]{Strobel99}.} &
$1.34_{-0.16}^{+0.15}$ &
$1.26_{-0.02}^{+0.02}$ \\
$R_{\textrm{eq}}$ $[$km$]$ &
coordinate equatorial radius\footnote{An alternative two-dimensional space of $g(M,R_{\textrm{eq}})$ and $h(M,R_{\textrm{eq}})$---where $g\neq h$---on which we could choose to specify a joint flat prior density distribution is that of $g\coloneqq M$ and $h\coloneqq R_{\textrm{eq}}/r_{g}(M)$.} &
$R_{\textrm{eq}}\sim U[3r_{g}(1),16]$\footnote{The function $r_{g}(M)$ denotes the gravitational radius explicitly in dimensions of length.} &
$12.71_{-1.19}^{+1.14}$ &
$0.78_{-0.02}^{+0.02}$ \\
\hline
&compactness condition & $R_{\textrm{eq}}/r_{g}(M)>3$\\
&compactness condition\footnote{The coordinate polar radius of the source $2$-surface, $R_{\textrm{polar}}(M,R_{\textrm{eq}},\Omega)$, is a quasi-universal function adopted from \citet[][]{AlGendy2014}, where $\Omega\coloneqq2\pi/P$ is the coordinate angular rotation frequency. This compactness condition, together with the elliptical surface requirement below are unimportant for the $P\sim5$~ms spin period of PSR~J0030$+$0451, and therefore we ignore these constraints in Tables~\ref{table: ST+EST} through \ref{table: CDT-U}.}  &$R_{\textrm{polar}}(M,R_{\textrm{eq}},\Omega)\geq3r_{g}(M)$\\
&enforce elliptical $2$-surface cross-section  & function of $(M,R_{\textrm{eq}},\Omega)$\\
\hline
$\Theta_{p}$ $[$radians$]$ &
\TT{ST} region center colatitude\footnote{Note that for parameters where a lower-bound of zero would correspond to absence of pulsations, we use some small finite number as a lower-bound.} &
$\Theta_{p}\sim U(0,\pi)$ &
$2.23_{-0.10}^{+0.10}$ &
$2.98_{-0.02}^{+0.02}$\\
$\phi_{p}$ $[$cycles$]$ &
\TT{ST} region initial phase (from Earth) &
$\phi_{p}\sim U(a,a+0.2)$\footnote{Where $\phi_{p}=a$ is an arbitrary phase dependent on event data pre-processing. We set $a=0.35$.}&
$0.46_{-0.00}^{+0.00}$ &
$4.44_{-0.04}^{+0.03}$\\
$\zeta_{p}$ $[$radians$]$ &
\TT{ST} region angular radius &
$\zeta_{p}\sim U(0,\pi/2)$ &
$0.09_{-0.01}^{+0.01}$ &
$4.62_{-0.03}^{+0.03}$\\
\hline
$\Theta_{s}$ $[$radians$]$ &
\TT{PST} region superseding center colatitude &
$\Theta_{s}\sim U(0,\pi)$ &
$2.91_{-0.05}^{+0.05}$ &
$3.85_{-0.03}^{+0.03}$\\
$\phi_{s}$ $[$cycles$]$ &
\TT{PST} region initial phase (from Earth antipode) &
$\phi_{s}\sim U(-0.5,0.5)$, periodic\footnote{The periodic boundary is admitted and handled by \MultiNest. However, this is an unnecessary measure because we straightforwardly define the mapping from the native sampling space to the space of $\phi_{s}$ such that the likelihood function maxima are not in the vicinity of this boundary.} &
$-0.59_{-0.04}^{+0.04}$ &
$2.43_{-0.02}^{+0.02}$\\
$\psi_{s}^{+}$ $[$radians$]$ &
\TT{PST} region angular radii difference\footnote{The difference is defined as $\psi_{s}^{+}\coloneqq\zeta_{s}-\psi_{s}$, where $\zeta_{s}$ is the angular radius of the ceding member that radiates where it is not superseded.} &
$\xi_{s}\sim U(0,\pi/2)$, $\psi_{s}^{+}=\psi_{s}^{+}(\xi_{s},f_{s})\footnote{See Section~\ref{subsubsec:PST and PDT}.}$ &
$-0.01_{-0.02}^{+0.02}$ &
$3.16_{-0.03}^{+0.03}$\\
$\psi_{s}$ $[$radians$]$ &
\TT{PST} region superseding angular radius  &
$f_{s}\sim U(0,2)$, $\psi_{s}=\psi_{s}(\xi_{s},f_{s})$ &
$0.47_{-0.13}^{+0.13}$ &
$1.52_{-0.02}^{+0.03}$\\
$\varkappa_{s}$ &
\TT{PST} region ceding fractional angular offset &
$\varkappa_{s}\sim U(0,1)$ &
$0.10_{-0.03}^{+0.05}$ &
$2.67_{-0.05}^{+0.05}$\\
$\varphi_{s}$ $[$radians$]$ &
\TT{PST} region ceding azimuthal offset &
$\varphi_{s}\sim U(0,2\pi)$ &
$0.71_{-0.10}^{+0.10}$ &
$2.28_{-0.03}^{+0.03}$\\
\hline
&non-overlapping hot regions\footnote{Refer to Appendix~\ref{app:non-overlapping}.} & function of $\Theta_{p}$ through $\varphi_{s}$ above\\
$\log_{10}\left(\mathcal{T}_{p}\;[\textrm{K}]\right)$ &
\TT{ST} region \TT{NSX} effective temperature &
$\log_{10}\mathcal{T}_{p}\sim U(5.1,6.8)$, \TT{NSX} limits &
$6.11_{-0.01}^{+0.01}$ &
$5.29_{-0.03}^{+0.03}$\\
$\log_{10}\left(\mathcal{T}_{s}\;[\textrm{K}]\right)$ &
\TT{PST} region \TT{NSX} effective temperature &
$\log_{10}\mathcal{T}_{s}\sim U(5.1,6.8)$, \TT{NSX} limits &
$6.11_{-0.01}^{+0.01}$ &
$5.75_{-0.02}^{+0.03}$\\
$i$ $[$radians$]$ &
Earth inclination to rotation axis &
$i\sim U(0,\pi/2) $ &
$0.94_{-0.10}^{+0.11}$ &
$1.91_{-0.02}^{+0.02}$\\
$D$ $[$kpc$]$ &
Earth distance &
$D\sim N(0.325,0.009)$\footnote{Constructed to approximate the information in the measurement (statistical information) reported by \citet[][]{nanograv11}. The support $D\in[0.235,0.415]$ is equivalent to the interval $\mu\pm10\sigma$.} &
$0.33_{-0.01}^{+0.01}$ &
$0.07_{-0.01}^{+0.01}$\\
$N_{\textrm{H}}$ $[10^{20}$cm$^{-2}]$ &
interstellar neutral H column density &
$N_{\textrm{H}}\sim U(0,5)$ &
$0.78_{-0.16}^{+0.17}$ &
$2.86_{-0.03}^{+0.03}$\\
$\NICER\;\alpha$ &
calibrated matrix scaling &
$\alpha\sim N(1,0.1)$, $\alpha\in[0.5,1.5]$ &
$0.99_{-0.09}^{+0.08}$ &
$0.04_{-0.01}^{+0.01}$\\
$\NICER\;\beta$ &
reference-to-calibrated matrix weighting &
$\beta\sim U(0,1)$ &
$0.19_{-0.12}^{+0.17}$ &
$0.96_{-0.03}^{+0.02}$\\
$\NICER\;\gamma$ &
reference matrix scaling &
$\gamma\sim N(1,0.1)$, $\gamma\in[0.5,1.5]$ &
$0.98_{-0.08}^{+0.08}$ &
$0.08_{-0.01}^{+0.01}$\\
\pagebreak
&Sampling process information\footnote{Refer to Appendix~\ref{sec:posterior computation} for definitions.}&&& \\
\hline
&number of free parameters:\footnote{In the sampling space; the number of background count-rate parameters is equal to the number of channels defined by the data set.} $19$ &&& \\
&number of runs:\footnote{The mode-separation \MultiNest variant was deactivated, meaning that isolated modes are not evolved independently and nested sampling threads contact multiple modes.} $1$ &&& \\
&number of live points: $10^{3}$ &&& \\
&inverse hypervolume expansion factor:\footnote{For this sampling process, nor any such process reported in this work, we did \textit{not} activate constant-efficiency \MultiNest active-point bounding variant.} $0.8$ &&& \\
&termination condition: $10^{-1}$ &&& \\
&evidence:\footnote{Defined as the prior predictive probability $p(\boldsymbol{d}\,|\,\TT{ST+PST})$. We report the interval about the median containing $\pm45\%$ of $10^{3}$ joint bootstrap-weight replications for the combined run. Note, however, that in order to complete the reported evidence for comparison to models other than those defined in this work, upper-bounds for the background parameters need to be specified as described in Section~\ref{subsubsec:likelihood functions}.} $\widehat{\ln\mathcal{Z}}=-36368.28_{-0.46}^{+0.49}$ &&&\\
&global KL-divergence: $\widehat{D}_{\textrm{KL}}=68.9_{-0.8}^{+0.9}$ bits &&&\\
&number of core\footnote{Intel$\textsuperscript{\textregistered}$ Xeon E5-2697Av4 (2.60 GHz; Broadwell) processors on the SURFsara Cartesius supercomputer. Note that these are physical cores---i.e., hyper-threading technology is not invoked.} hours: $42453$ &&& \\
&likelihood evaluations: $78343018$ &&& \\
&nested replacements: $57972$ &&& \\
&weighted posterior samples:\footnote{Excludes samples with important weight smaller than $10^{-6}$ times the largest such weight amongst samples.} $20177$ &&& \\
\end{longtable}

\textbf{Marginalization.} For parameters that are shared between discrete models $\{\mathcal{M}_{m}\}$, where $\mathcal{M}_{m}\subset\mathscr{M}$, we could in principle marginalize over the discrete parameter $m$ \citep[see appendix~C of][for a formulation consistent with Section~\ref{subsec: generative model space}]{riley19b}, provided we accept a marginal prior mass distribution of $m$. If two or more models are competitive, the marginal joint posterior distribution of the shared parameters is not dominated by the information from a single model. As stated above, we cannot distinguish between four of the highest-evidence models (\TT{ST+PST}, \TT{ST+EST}, \TT{ST+CST}, and \TT{CDT-U}). Moreover, for \TT{ST+CST}, \TT{ST+EST}, and \TT{ST+PST}, the marginal joint posterior distribution of the shared spacetime parameters of interest is only mildly sensitive to model choice---i.e., the compactness constraints are commensurate, whilst the mass and radius are only weakly constrained individually, and the joint credible regions exhibit high partial overlap. For \TT{ST+PST}, the marginal joint posterior distribution of the shared spacetime parameters discernibly evolves, but the mass and radius remain weakly constrained individually. Given that $m$ labels models that are often (approximately) nested and which differ only in phenomenological complexity, we opt not to marginalize shared parameters over those models; instead, we report headline parameter estimates for \TT{ST+PST}, which exhibits the largest background-marginalized likelihood function values in any posterior typical set.

\section{Discussion}\label{sec:discussion}

In this section we highlight how our inferences may be sensitive to the modeling assumptions made, and discuss the implications of our inferences for both dense matter physics and NS astrophysics.

\subsection{Modeling assumptions}\label{sec:assump}

The inferences that we report are conditional upon a number of modeling assumptions. These assumptions were physically motivated, and reassuringly there are no obvious large discrepancies or structures in the pulse-profile residuals to indicate a major problem. Nonetheless, the sensitivity of our results to these assumptions should be explored in future work, given additional computational resource allocations. With regard to the MSP, the biggest model-dependencies are: (i) the atmosphere  (see Section~\ref{subsubsec: discuss atmosphere}); (ii) the treatment of phase-invariant components of the total signal (see Seciton~\ref{subsubsec: discuss phase-invariant treatment}); and (iii) the assumption there exist two disjoint hot regions, each of which is radiatively contiguous and has a temperature field that is adequately represented by one of the models described in Section~\ref{sec:models} (see Section~\ref{subsubsec: discuss phase-invariant treatment}). The instrument model  (see \ref{subsubsec: discuss instrument}) is of less concern.

\subsubsection{Atmosphere}\label{subsubsec: discuss atmosphere}
Two properties that could affect the atmosphere models used in our analyses are chemical composition (hydrogen as opposed to helium) and ionization state (fully versus partially ionized). In this Letter we have considered only a hydrogen composition: hydrogen would dominate the composition of matter accreted from the interstellar medium \citep{Blaes92}, whilst matter accreted from a binary companion star would be predominantly hydrogen or helium. There are other processes that may drive changes in composition. For instance: hydrogen would result from spallation \citep{Bildsten92}; diffusive nuclear burning could convert hydrogen to helium \citep{Chang03,Chang04}; and significant pulsar wind excavation could make visible an underlying heavy element layer \citep{Chang04}. If the atmosphere were in fact dominated by helium, we could expect changes because helium atmospheres radiate differently from those of hydrogen. For example, hydrogen and helium model specific intensities, at an atmosphere effective temperature of $10^6$ K (the approximate inferred temperature for the hot regions for this source), have fractional differences of at most $2$--$5\%$ at $0.5$--$1$~keV in the (maximal) forward direction \citep{bogdanov19c}. 

Regarding ionization state, our atmosphere models are constructed assuming the atmospheric plasma is
fully ionized, such that the dominant opacity in regimes of interest is that due to electron free-free absorption \citep{Ho01}. While opacity tables for partially ionized matter exist \citep{OPAL,OP,OPLIB}, they do not cover the full range of energies and temperatures needed for our analysis. However, the hydrogen neutral fraction in the atmosphere at $10^6$~K is low, and a comparison of our fully ionized hydrogen atmosphere model with that constructed using the OP \citep{OP} opacity table yields specific intensity fractional differences of at most $1$--$2\%$ at $0.5$--$1$~keV in the (maximal) forward direction at an atmosphere effective temperature of $10^6$ K \citep{bogdanov19c}. The importance of including partial ionization, in comparison to developing other aspects of the model (e.g., Section~\ref{subsubsec: discuss phase-invariant treatment}), is not clear; but partially-ionized models will be part of future re-analysis as updated opacity tables become available.

\subsubsection{Surface heating estimation \& treatment of phase-invariant components}\label{subsubsec: discuss phase-invariant treatment}

Predicting the MSP surface temperature field from \textit{ab initio}\footnote{Terminology adopted from the series of studies by \citet{Philippov15,Philippov15a} and \citet{Philippov18} on pulsar magnetospheric simulations.} calculations of energy deposition by magnetospheric currents is, as described in Section \ref{sec:models}, notoriously challenging. In this work we have assumed that there are only two distinct hot regions, motivated by the fact that there appear to be two pulsed components in the pulse profile. If the (surface) field structure involves higher-order multipoles,\footnote{With respect to some coordinate system that simplifies vector spherical harmonic field expansion; this coordinate system will generally be rotated and displaced from the system with stellar spin axis defined as the polar axis.} additional polar caps---and thus additional disjoint surface heating---may be possible, and/or more complicated polar cap topologies (such as ring-like) may be possible. We have also assumed specific forms for each hot region; the true temperature field is more complex, and both the physical complexity and our statistical sensitivity to such complexity should be investigated further. In lieu of a physical emission model for the stellar surface exterior of the hot regions, we subsumed the non-pulsed component of any such emission (which is expected \textit{a priori} to be dominant if hot regions with smaller angular extent are favored) within phase-invariant count-rate terms; however, this would be a minor concern on the premise that outside of the footpoints of open magnetic field lines at the polar caps, there is no energetically comparable heating to which we are sensitive when observing with \NICER.

All emission from sources other than the hot regions---a combination of astrophysical and instrumental---was left free in our models in the form of a set of phase-invariant count-rate terms (background parameters), one per channel, which we collected under the envelope of \textit{background contribution} (refer to Section~\ref{subsubsec:likelihood functions}). Moreover, an improper joint flat prior was implemented that was separable with respect to these background parameters; no upper-bounds (or lower-bounds) were defined for the prior support, in lieu of a physical (generative) model for the total contribution from the hot regions (and thus surface if emission exterior of the regions is considered unimportant). Such a model would need to account for the combined (phase-invariant) signal attributed to off-surface emission and any astrophysical backgrounds in the field of view, in the \NICER waveband, based on previous observations of PSR~J0030$+$0451 and/or theoretical modeling.

Remarkably, by neglecting any physical (generative) model for the total counts attributed to \textit{all} surface emission, the phase-invariant terms can even capture emission from the hot regions. One can reason that by permitting the phase-invariant terms to capture all or most of the phase-invariant signal components, there may then exist \textit{background-marginalized} likelihood function maxima corresponding to signals that are: (i) dominated by phase-invariant terms over a fraction of a rotational cycle; and (ii) elsewhere found to describe the pulse-profile adequately (in combination with the phase-invariant terms). Near pulse minima, the hot region contribution can be entirely dominated in linear combination with the phase-invariant terms, if the regions are both close to the visible limb of the star, or even partially or wholly non-visible. Such a heating configuration may exhibit systematic bias in the sense that: (i) it is not considered an adequate approximation of the configuration inferred when physical limits are modeled for the contribution from the stellar surface (or specifically the hot regions); or (ii) that a configuration alluded to in point (i) is not encompassed by a given posterior credible region boundary. On the other hand, it could be viewed that our treatment of the phase-invariant component of the total signal is in some aspects conservative: posterior credible regions may be appreciably larger than if such physical limits on surface contribution are imposed. In the \XPSI documentation \citep{riley19b} a simple parameter estimation workflow is demonstrated using the same default background treatment implemented in this present work; whilst there is no evidence for systematic bias nor credible region inflation in that specific case (where the true data-generating process is known), guarantees cannot be made universally.

Upon examination of Figure~\ref{fig:STpPST source} we see that the combined signal from the hot regions falls to near zero at its minimum---such that the fractional amplitude of the signal is near unity---and is thus \textit{not inconsistent} with a phase-invariant component of the combined signal being subsumed in the background. Further, examination of Figure~\ref{fig:STpPST spectrum}, in which we cross-check the \TT{ST+PST} region spectrum with a PSR~J0030$+$0451 spectrum inferred by \citet[][]{Bogdanov09}, further suggests that a fraction of the contribution from the hot regions is subsumed in the background. Interestingly, \citet[][]{Bogdanov09} inferred a spectrum with two thermal components of different temperatures.\footnote{Note that \citet{Bogdanov09} explicitly calculate the phase-resolved signal generated by rotating single-temperature circular hot spots, and then phase-average; it is thus not the case that the discrepancy is explained by \textit{our} rotating circular hot spots (refer to \TT{ST-U}), whose effective temperatures are commensurate, mimicking a dual-temperature \textit{incident} spectrum due to relativistic rotation.}

We attribute the discrepancy to a number of factors. First, the \project{XMM} photon event set had size $\mathcal{O}(10^{4})$, nearly two orders of magnitude smaller than the number of \NICER events used in this work. Second, it is known that \citet{Bogdanov09} did not fully resolve degeneracy between the thermal components and the neutral hydrogen column density $N_{\rm H}$. Lastly, \project{XMM} is an imaging telescope: the background signal was well-determined by imaging nearby source-free regions of the sky, and was subsequently used to impose that the surface hot spots generate the remaining signal from the imaged (point-source) MSP. Such a model is distinctly different from those we consider here for \NICER, a non-imaging telescope; we do not impose (e.g., via some informative background prior) the signal to be generated by the surface hot regions. As a result, we may inaccurately subsume a cooler contribution from the hot regions into the phase-invariant likelihood terms as suggested above; \citet{Bogdanov09}, on the other hand, may miss some non-diffuse radiative component(s) in the near vicinity of the surface of PSR~J0030$+$0451, and thus require their hot spots to generate additional cooler emission than they do in physical reality.

With these considerations in hand, we conclude that without further work, it is unclear which inferred signal is more physically accurate. Indeed, the existence and treatment of additional X-ray emission by PSR~J0030$+$0451---or from its circumstellar vicinity---in the \NICER waveband is considered an open question for future modeling. Moreover, the sensitivity of parameter estimation to requiring a certain phase-invariant contribution specifically from the hot regions could be investigated with newly allocated computing resources.

A (superficially) straightforward alternative to a full physical generative model of the non-pulsed surface and off-surface emission would be to define bounds---upper in particular---on the prior support of the channel-by-channel background count-rate parameters (refer to related discussion on limits in Section~\ref{subsubsec:likelihood functions}). One must then address the question of how. One option is to move the difficulty of defining a prior density for the background to another level in a Bayesian hierarchy via a hyperparameter (e.g., the upper-bound of a flat density function) and a corresponding hyperprior for each channel; each hyperparameter and background parameter pair must then be jointly numerically marginalized over to ensure sampling is tractable in a lower-dimensional parameter space. Another option would be to define a conditional prior distribution for each background parameter, where the upper-limit of the support is a function of the source parameters---e.g., some fraction of the phase-average source counts in a channel.

We did not consider at the outset of this work the possibility of jointly modeling \NICER and \project{XMM} event data---a strategy both tractable and arguably more rigorous given that we do not have likelihood function information nor posterior information from \citet[][]{Bogdanov09} that is compatible as a prior information to be updated conditional on the \NICER data. We could thus consider constructing a joint likelihood function over observations with both telescopes (requiring definition of at least one more nuisance parameter). On the other hand, the \NICER event data comprise far more photons than do the \project{XMM} data, and so inferences may be dominated by the (background-marginalized) \NICER likelihood function. There are, however, good prospects for improving our understanding of the \NICER background \cite[particle radiation and diffuse sky terms; see Section~\ref{subsubsec:background general} and][]{bogdanov19a}, even without new input from \project{XMM} observations or other imaging capabilities.

Observations with a future high--time resolution soft X-ray spectroscopic imaging telescope, such as the \project{Athena X-ray Observatory} \citep[][]{Athena13}, would be synergistic with archival \NICER data---and/or observations with a future large-area soft spectroscopic timing telescope such as the \project{enhanced X-ray Timing and Polarimetry mission} \citep[\project{eXTP},][]{Zhang19,Watts19} or the \project{Spectroscopic Time-Resolving Observatory for Broadband Energy X-rays} \citep[\project{STROBE-X},][]{Ray19}. The high-sensitivity (read large-area) imaging capabilities of \project{Athena} (with $\sim\!5$--$10$\arcsec\ angular resolution\footnote{Refer to \citet{bogdanov19a} for an \project{XMM} image of the PSR~J0030$+$0451 field; \project{Athena} promises to improve on the \project{XMM} point-spread function half-energy width by a factor of $\sim\!1.5$--$3$.}) and its two instruments, the \project{X-ray Integral Field Unit} (\project{X-IFU}, $10\,\mu{\rm s}$ time resolution\footnote{Similar to that expected for the \project{eXTP Spectroscopic Focusing Array} ($\sim\!10\,\mu{\rm s}$), but not as good as spectro-timing dedicated missions, like \project{NICER} or \project{STROBE-X} with $\sim\!0.1\,\mu{\rm s}$ time resolution. Note that event time-tagging resolution offered by \project{X-IFU} will still achieve a level of $\sim\!1/500$ of the spin period of PSR~J0030$+$0451.}) and the \project{Wide Field Imager} ($80\,\mu{\rm s}$ time resolution), will permit modeling of the background emission not originating from the line-of-sight of the PSR~J0030$+$0451\footnote{And other MSPs targeted by a mission such as \NICER.} system (as permitted by the point-spread function) in a similar manner to \citet[][]{Bogdanov09} for \project{XMM} \citep[and also][for PSR~J0437$-$4715]{Bogdanov13}. Furthermore, the microcalorimeter spectral resolution of \project{X-IFU} ($\sim\!2.5$~eV resolution for photons energies $<7$~keV) will permit precise measurements of various edges of the interstellar medium (e.g., oxygen K edge at $0.55$~keV, iron L edge at $0.72$~keV), thereby enabling derivation of independent tight constraints on the hydrogen column density $N_{\rm H}$ for a given MSP. Archival \NICER MSP data, on the other hand, will remain valuable far into the future because the principal \NICER mission science objectives focus purely on MSPs; \NICER will have compiled $\mathcal{O}(10^{6})$~s integrated exposures on the primary MSP targets, meaning that despite the smaller total effective area, the size of the event data set will be synergistic with an advanced multi-faceted mission such as \project{Athena} whose science objectives are broader. Consequently, observations with the \project{Athena} instruments will enable us to jointly model data sets, and make definitive progress on disentangling the signal generated by surface hot regions from the complicated phase-invariant emission detected by \NICER (or a future non-imaging telescope).

\subsubsection{Instrument}\label{subsubsec: discuss instrument}
We implemented a specific instrument response model (Section \ref{subsubsec:instrument general}): a simple ad hoc parameterization designed to combine several available calibration products and thus account for both energy-independent (absolute flux calibration) and energy-dependent uncertainty. Sensitivity to our choice of astrophysical calibration source, the Crab, also needs further study; it is known that operation in response to an incident radiation field is a function of its properties (see Section~\ref{subsubsec:instrument general}). Future study could explore whether using different calibration sources affects inferences. Another interesting question may be posed as to how the posterior information about this instrument model evolves as we update our knowledge via analysis of other sources (MSPs and otherwise). 

As our understanding of the \NICER instrument improves, it should be possible to improve on this model. One option might be to adopt a more sophisticated approach based on \citet{Lee11} and \citet{Xu14} to estimating the uncertainty in instrumental response for \project{Chandra}. At present such a sophisticated approach may be unjustified because computational expense would be amplified by increasing the complexity of the instrument model instead of---or in addition to---the complexity of model astrophysical sources whose nature is more uncertain. 

\subsection{Mass and radius constraints in context}\label{sec:mrcontext}

For \TT{ST+PST}, the inferred mass $M$ and equatorial radius $R_\mathrm{eq}$ are $1.34^{+0.15}_{-0.16}$~\msol~and $12.71^{+1.14}_{-1.19}$ km, where the credible interval bounds are approximately the $16\%$ and $84\%$ quantiles in marginal posterior mass, given relative to the median. The marginal credible intervals may thus be considered as $1\sigma$ intervals (containing $68.3\%$ of the posterior mass), where $\sigma_{M}/M\sim 11\%$ and $\sigma_{R}/R\sim 9.2\%$. For completeness, and to assist comparison with results derived using other methods, we also give the values for the $90\%$ credible interval ($M = 1.34 \pm 0.24$~\msol, $R_\mathrm{eq} = 12.71^{+1.83}_{-1.85}$ km) and the $95\%$ credible interval ($M = 1.34^{+0.28}_{-0.27}$~\msol, $R_\mathrm{eq} = 12.71^{+2.15}_{-2.14}$ km). The (equatorial) compactness $GM/R_{\mathrm{eq}}c^2 = 0.156_{-0.010}^{+0.008}$ is more tightly constrained than both $M$ and $R_\mathrm{eq}$ individually, at the $\sim\!6\%$ level; the $90\%$ credible interval is $GM/R_{\mathrm{eq}}c^2 = 0.156_{-0.017}^{+0.013}$, and the $95\%$ credible interval is $GM/R_{\mathrm{eq}}c^2 = 0.156_{-0.021}^{+0.015}$.

The effect of a rotation rate of $200$~Hz on NSs is a small deformation of the star into an oblate spheroid. The deformation enters in two ways. A star rotating at $200$~Hz will have an equatorial radius that is at most $2\%$ larger than a nonrotating star with the same mass for the stiffest equations of state, with smaller increases in radius for soft equations of state \citep{Cook94}.\footnote{\added{Note that a one-parameter sequence of nonrotating stars deforms into a sequence of stars rotating at some rate, but a unique deterministic map between the central densities that parameterize those sequences does not exist. This reflects the absence of a unique physical mode for stars to evolve in rotation rate. For comparative purposes, one often considers sequences---parameterized by rotation rate---that conserve a quantity such as total mass, total baryon number, or central density. In the accompanying Letter of \citet{raaijmakers19}, it is the central density that is explicitly defined as a model parameter (to be marginalized out), and which is held constant for comparison of nonrotating stars to rotating stars in the context of likelihood function evaluation.}} The polar radius will be smaller than the equatorial radius, an effect that depends on the equatorial radius, mass, and spin, with very little dependence on the EOS \citep[][and references therein]{AlGendy2014}; for the \textit{a posteriori} most probable exterior spacetimes inferred in this present work, the polar radius is $\sim\!1\%$ smaller than the equatorial radius. Rotation also increases the mass of a star compared to a nonrotating star with the same number of baryons, but this is at most a $0.2\%$ effect for the rotation-powered pulsars observed by \NICER. The effect of rotation on the location of the innermost stable circular orbit (ISCO) is larger \citep{vanDoesburgh18}: the (prograde) ISCO typically ranges from $5.6r_{g}$--$5.8r_{g}$ for stars spinning at $200$~Hz for most theoretical equations of state, compared to the Schwarzschild ISCO radius of $6r_{g}$. The $95\%$ compactness credible interval corresponds to equatorial radii in the range of $5.84r_{g}$--$7.41r_{g}$, meaning that the most probable stellar surfaces\footnote{Which in this present work are only embedded in the ambient spacetime whilst neglecting rotational metric deformation (see Section~\ref{subsubsec:source}) and are thus not self-consistently computed with global numerical solutions to the field equations given interior conditions (including an EOS).} have radii close to or larger than the radius of the ISCO, and thus that the innermost permitted stable orbit is typically at (or just exterior to) the surface.

In this subsection we proceed to discuss how the constraints derived in this work compare to existing constraints on mass, radius, and compactness derived using other methods and independent observations. We also consider prospects for improving \NICER constraints on PSR~J0030$+$0451. When comparing radii, it should be remembered that most of the published radius determinations using other methods have assumed that the star is spherical. For luminosity radius determinations of rapidly rotating NSs (spins of a few hundred Hz), systematic errors of $5\%$ could be introduced \citep{Baubock15}, although other systematic errors may well dominate over an inaccurate treatment of rotational surface and metric deformation.

The present constraints on the radius are consistent with the previous radius lower-limit for this pulsar, $R_{\rm eq}>10.7$~km ($95\%$ confidence, assuming a $1.4$~\msol~NS), obtained from early pulse-profile modeling of \project{XMM} data \citep{Bogdanov09}. It is also in agreement with the radii inferred from \project{XMM} observations of other \replaced{millisecond pulsars}{MSPs}: $R_{\rm eq}>11$~km ($3\sigma$ confidence) for PSR~J0437$-$4715 \citep{Bogdanov13}; and $R_{\rm eq}>7.8$~km ($68\%$ confidence) for PSR~J$2124-3358$ \citep{Bogdanov08}. However, these early light-curve models only considered uniform-temperature circular hot spots, neglected stellar oblateness, and we consider the statistical computation described in this Letter as more advanced.

Early constraints on the NS radius were obtained from the X-ray spectroscopic modeling of the thermal emission originating from the entire surface of isolated NS RX~J1856.5$-$3754 \citep{Burwitz01,Drake02,Pons02}. However, uncertainties due to the distance and due to the calculations of radiative transfer in the magnetized atmosphere \replaced{weakened the}{hindered an accurate} radius constraint for this NS \citep{Ho07}. More robust constraints, on the other hand, can be extracted from the X-ray spectra of quiescent low-mass X-ray binaries (qLMXBs) hosted in globular clusters. Not only are their distances known to better than $\sim\!10\%$ precision, but their purely thermal emission is thought to emerge from  non-magnetic NS atmosphere models (similar to the \TT{NSX} model used in the present work) in order to estimate the apparent radius $R_\infty=R_{\rm eq}\left(1+z\right)=R_{\rm eq}\left(1-2 G M /R_{\rm eq} c^{2}\right)^{-1/2}$.  However, the degeneracy between $R_{\rm eq}$ and $M$ in the estimation of $R_\infty$ precluded obtaining useful constraints on the EOS \cite[e.g.,][]{Webb07, Heinke06, Guillot11, Heinke14}, due to the typical shapes of the $M$--$R_{\rm eq}$ confidence contours that made them compatible with \replaced{most}{many} families of EOS. These results prompted the simultaneous analysis of a set of sources.

In these combined analyses, the degeneracy between $M$ and $R_{\rm eq}$ was lifted by assuming a parameterized shape for the EOS, either a toy-model (constant-radius EOS, as a simplistic representation of nucleonic EOS) or  an analytical representation using polytropes. A handful of qLMXBs in globular clusters (up to seven) and of Type-I X-ray bursters (four or five) have been combined to produce constraints on the EOS and/or the radii of NSs (given the assumed EOS shape). Early works using only qLMXBs produced rather small NS radii, $R_{\rm eq}\approx 9$--$10$~km with $\sim\!10$--$15\%$ uncertainties \citep[$90\%$ credible interval;][]{Guillot13, Guillot14, Guillot16a}, but the addition of new data and the use of more recent globular cluster distance measurements resulted in higher values, $R_{\rm eq}$ in the $9.9$--$11.2$~km range for a $1.5$~\msol~NS \citep[2$\sigma$ credible interval;][]{Bogdanov16}. Combining qLMXBs and Type-I X-ray bursts, other works have found radii in a wide range of values: $R_{\rm eq}\approx 10.4$--$12.9$~km \citep[$95\%$ credible interval;][]{Steiner10,Steiner13}, $R_{\rm eq}\approx 10.5$--$12.7$~km \citep[$90\%$ credible interval;][]{Lattimer14}, and $R_{\rm eq}\approx 9.8$--$11.4$~km \citep[$95\%$ credible interval, ][]{Ozel16}, for a $1.4$~\msol~NS. However, these analyses may have been affected by systematics, such as those due to the modeling of piled-up X-ray photons in the \project{Chandra} data\footnote{It was shown that an unmodeled pile-up fraction as low as $\sim\!1\%$ could affect the radius estimated via modeling of spectral data by as much as $10\%$ \citep{Bogdanov16}.}---which was not considered for all qLMXBs in these early analyses---or those due to the choice of atmospheric composition for qLMXBs (generally H versus He). The mass-radius constraints extracted from Type-I X-ray bursts are dependent on the color-correction factors used (between the measured blackbody temperatures and the modeled effective temperatures of the burning atmospheres), which have been debated in the literature \citep[e.g.,][]{Suleimanov11b,Guver12,Guver13,Kajava14}. These issues cast doubt on the robustness of the error intervals reported in these early works.

\citet{Nattila17} recently analyzed Type-I X-ray bursts from 4U~1702$-$429 by fitting bursting atmosphere models directly to spectra during the cooling tail of the bursts, hence avoiding the use of color-correction factors. These authors found $R_{\rm eq}= 12.4\pm0.4$~km and $M=1.9\pm0.3$~\msol~($68\%$ credible interval), although the posterior distributions also allow smaller radii $\sim\!10$~km for higher masses $\sim\!2.1$~\msol. 

\citet{Steiner18} considered the effects of pile-up on qLMXB spectra, the possibility of pure helium atmospheres (instead of pure hydrogen atmospheres), as well as non-uniform surface temperature distributions. They obtained constraints on polytropic EOS via Bayesian inference of mass-radius probability distributions of seven NS qLMXBs. These constraints translate to a NS radius in the $10.0$--$14.4$~km range ($95\%$ credible interval, assuming a $1.4$~\msol~NS), when considering all of the models tested.

More recently, a physically justified parameterization of the EOS was proposed as an alternative to polytropes \citep{Margueron18a,Margueron18b}. In that work, the EOS is a meta-model expressed as a Taylor expansion of nuclear physics parameters, and was applied to a combined spectral analysis of seven qLMXBs to directly extract values of nuclear physics parameters $L_{\rm sym}$, $K_{\rm sym}$, and $Q_{\rm sat}$ \citep{Baillot19}. Radius estimates were also derived: $R_{\rm eq}= 12.35\pm0.37$~km ($2\sigma$ credible interval) assuming a $1.45$~\msol~NS. 

The cold emission from the MSP PSR~J0437$-$4715 is detectable in the far ultraviolet \citep{Durant12} and in the soft X-ray band \citep[$0.1$--$0.3$~keV;][]{Guillot16b}, and its mass and distance are known precisely from radio timing \citep{Reardon16}. Applying NS atmosphere models\footnote{Similar but not identical to the atmosphere models used in our \NICER analysis---the models used in \citet{Gonzalez19} incorporated various effects that are important at lower temperatures such as  partial ionization and plasma frequency effects.} has permitted estimation of this pulsar's radius: $R_{\rm eq}=13.1\pm0.8$~km \citep[$68\%$ credible interval;][]{Gonzalez19}. We note that this was a phase-averaged spectral analysis, wherein the emission is assumed to originate from the $\sim\!10^{5}$~K stellar surface exterior to the (heated) polar cap regions.

An indirect method to constrain NS radii is to use emission features from an inner accretion disk in accreting LMXBs. Narrow emission lines, such as Fe K, arising from this rotating material are asymmetrically broadened---to which there is a strong relativistic contribution based upon proximity of the inner disk to the NS surface \citep{Fabian00}. The accretion disk must truncate at the stellar surface or at a larger radius: spectral modeling of these emission lines enables derivation of a statistical constraint on the inner radius of the (prograde) disk \textit{in units of} the gravitational radius or the spin-dependent ISCO radius \citep{Cackett08,Cackett10,MillerJM13,Degenaar15,Ludlam17}. To do so, an approximative ambient spacetime solution (e.g., Schwarzschild or Kerr) is typically invoked---as in this present work---but without the embedding of a NS surface.\footnote{The real surface could thus in principle enclose the ISCO associated with the ambient spacetime solution.} The constrained inner radius can then be translated into a lower-limit on the stellar compactness for that particular NS.

\citet{Ludlam17} inferred, for two LMXB systems, inner radii that are consistent with the disks extending down to the ISCOs of their respective ambient spacetimes. In the absence of a constraint on the NS mass that is independent\footnote{For example, via a binary mass function together with classification of the companion star.} of the inner radius of the disk, a mass merely has to be assumed to obtain an upper-limit on the stellar radius. For example, suppose that the 4U~1636$-$53 system contains a nonrotating\footnote{Or at least a small dimensionless spin, despite the $581.0$~Hz spin frequency \citep[][and references therein]{Ludlam17}.} minimally compact NS (i.e., disk truncation by surface): for a mass of $1.4$~\msol, the upper-limit on the stellar radius would lie close to the Schwarzschild ISCO, at $12.4$--$13.1$~km \citep[$1\sigma$ confidence interval;][]{Ludlam17}, which is not inconsistent with the constraint derived conditional on the \NICER data.\footnote{Where as discussed above the rotational deformation of the surface is small for PSR~J0030$+$0451.}\deleted{ even if a shared EOS is assumed} Lastly, if the surface of a NS in the 4U~1636$-$53 system does \textit{not} approximately truncate the disk---and is thus more compact than the ISCO---and/or the (prograde) ISCO is more compact due to NS rotation, the NS could only be viewed as inconsistent with our compactness estimate for PSR~J0030$+$0451 if the following are true: (i) the NS masses are both tightly constrained and happen to be highly commensurate; \deleted{(ii) an EOS is shared; }and (ii) the \replaced{masses}{NSs} do not \replaced{correspond to a branch or branches of stable spacetime solutions wherein the}{occupy a segment of the mass-radius sequence along which the} radius is highly sensitive to increasing mass due to EOS softening or phase transitions \citep[e.g.,][]{Drago14,Alford16,Alford17}. However, one mass is unconstrained (disk modeling) whilst the other is constrained at the $\sim\!10\%$-level (\NICER pulse-profile modeling).

Overall, these recent publications have estimated NS radii in the range of $12$--$14$~km, which is compatible with our PSR~J0030$+$0451 radius estimate. We note that \replaced{direct equation of}{equating} \replaced{the}{accurately measured} radii of distinct NSs \replaced{is strictly only relevant under the assumptions that}{should elicit agreement if}: (i) \replaced{a shared EOS is assumed}{the EOS is shared} \added{from core to crust, meaning that perturbative effects attributed to crust composition, temperature, and magnetic field strength are sufficiently small in the context of measurement precision}; (ii) the EOS is of a nucleonic composition that supports NSs with similar radii over a wide range in mass ($0.8$ to $\sim\!2.0$~\msol); and (iii) differences due to spin-dependent rotational deformations are accounted for or are small enough to justify neglecting \citep[see the discussion in the second and third paragraphs of Section~\ref{sec:mrcontext}\added{, and in}][]{raaijmakers19}. For other families of EOS, such as those involving quarks or hyperons \added{in hybrid stars \citep[e.g.,][]{Zdunik13} or baryon resonances \citep{Drago14}}, the radius may \added{be} (highly) sensitive to increasing mass. In such cases, \replaced{it would not be inconsistent}{we would expect} to find NSs \deleted{that share an EOS but} whose radii differ by several km \deleted{\citep{Drago14}}, reinforcing the importance of jointly estimating both the radius and the mass of each member of a population of NSs.

Constraints on NS masses and tidal \replaced{deformability}{deformabilities} are now also \replaced{flowing from}{being reported based on} the first binary NS merger gravitational wave event, GW170817 \citep{Abbott18,Abbott19}. These can be translated into constraints on mass and radius. The inferred values vary somewhat depending on modeling and prior assumptions. \citet{Abbott18}, for example, employed two methods: the first, which did not assume that both stars had the same EOS, yielded radii of $10.8_{-1.7}^{+2.0}$~km and $10.7_{-1.5}^{+2.1}$~km for the two stars, with masses in the range $1.16$--$1.62$~\msol; the second, which assumed a common EOS, yielded a radius of $11.9\pm1.4$~km for both stars and masses in the range $1.18$--$1.58$~\msol(all results $90\%$ credible intervals). \citet{De18}, who also assume a common EOS, report radii of $10.7^{+2.1}_{-1.6} \pm 0.2$~km and masses in the range $1.12$--$1.67$~\msol ($90\%$ credible interval). Constraints taking into account additional information tend to support slightly larger mean values for the radius, and smaller uncertainties, e.g., those derived from the electromagnetic counterpart ($12.4^{+1.1}_{-0.4}$~km, 2$\sigma$ confidence interval; \citealt{Most18}); and those invoking a theoretical minimum \citep{Tews17} for neutron matter pressure ($11.4^{+1.9}_{-0.8}$~km, 90\% credible interval marginalizing over mass; Zhao T. \& J. M. Lattimer, in preparation). The results of \citet{Abbott18} and \citet{De18}, which employ the assumption that both stars share the same EOS, suggest that the radii of the two stars are nearly equal despite the fact that the mass ratio of the stars could lie between $0.7$ and $1.0$ with almost uniform probability. Let us make the assumption that PSR~J0030$+$0415 also has the same radius as the binary members to well within the posterior uncertainty on each star: the radius reported in this Letter, inferred from \NICER data, is more consistent with values in the upper ranges emerging from the gravitational wave analysis of GW170817. Larger radii would only be consistent with mass ratios closer to unity. It follows that if the three stars have nearly the same radius, and if the mass ratio of GW170817 was near its lower limit of $0.7$, the common radius should be at the lower end of the \NICER range. Electromagnetic observations might suggest relatively large amounts of dynamical ejecta, which would favor mass ratios considerably less than unity \citep{Radice18}.

\textit{What are the prospects for improving constraints on mass and radius for PSR~J0030$+$0451?} Unfortunately we cannot obtain an independent constraint on the mass for PSR~J0030$+$0451 because it is not in a binary, unlike some of \NICER's other MSP targets. Our model of the \NICER background (particle radiation and diffuse sky terms) is, however, expected to improve without needing to wait for input from other telescope missions; understanding this background accurately will prove crucial, as discussed in Section~\ref{subsubsec: discuss phase-invariant treatment}, and may well impact mass-radius estimation. A longer total exposure time could also certainly be accumulated \added{(and indeed the \NICER team anticipate doing this)}. Previous studies that have examined how posterior estimation of mass and radius is sensitive to factors such as geometry, spin rate, and the number of source counts in the event data, indicate that constraining power increases as the square root of the number of counts \citep{Lo13,Psaltis14b}. However, those studies all assumed a single circular single-temperature hot spot, not a more complex hot region configuration such as those we have considered and inferred here. While it is likely that gathering more data will improve the \added{joint} constraint \added{on mass and radius (without reference to an EOS model)}, the precise observing time required \added{to achieve a given level of precision} cannot be estimated \added{robustly} without further study. 

Our report here is encouraging in terms of prospects for other \NICER targets such as PSR~J0437$-$4715: the pulsar mass is constrained independently to within a few percent via radio pulsar timing because it is in a binary system \citep{Reardon16}. If such a constraint had been available for PSR~J0030$+$0451, it is clear that we could have obtained a comparable posterior uncertainty on the inferred radius of a few percent.
 
\subsection{EOS implications}\label{sec:eoscontext}

One of the primary goals of mass-radius inference is to use posterior information\footnote{Strictly, likelihood information \citep[][]{Riley18}.} to infer the properties of the dense matter EOS, if such information is deemed sufficiently likelihood-dominated to warrant the study. Studies have utilized joint mass-radius posteriors inferred from X-ray spectral modeling of bursting and quiescent NSs \citep{Steiner10,Steiner13,Ozel16,Raithel17}. Moreover, studies have utilized mass and tidal deformability constraints derived from analysis of the NS binary merger event GW170817 \citep{Abbott18,Annala18,Most18,Tews18,Lim18,Malik18,Carson19,Li19,Montana19}, and consideration is already being given to combining constraints from electromagnetic and gravitational wave analysis \citep{Kumar19,McNeilForbes19,Weih19}.

Given a suitable model for the EOS \citep[see, e.g.,][]{Read09,Raithel16,Tews18b,Lindblom18} there are two approaches to EOS inference: one is to jointly infer the EOS parameters (and central densities) directly from the data (e.g., pulse-profile data); the other is to jointly infer EOS parameters from per-source nuisance-marginalized likelihood functions of exterior-spacetime parameters (e.g., gravitational mass and equatorial radius). The former approach is \textit{at least} as computationally intensive as the direct mass-radius inference reported in this paper; the latter approach is less computationally intensive \textit{given} archival likelihood function information about exterior spacetime parameters \citep{Riley18}. In any case, care is required in both overall approach and the selection of (interior source matter and exterior spacetime) model parameterization and priors \citep{Riley18,Raaijmakers18,Greif19,Carney18,Landry19}. For our analysis here we deliberately defined a joint flat prior density function for $M$ and $R_{\textrm{eq}}$, with the intention that the posterior density function can be invoked as a likelihood function marginalized over all nuisance parameters. We explore the dense matter EOS implications of the inferred mass, radius, and compactness for PSR~J0030$+$0451 in an accompanying Letter \citep{raaijmakers19}, following the approach to EOS inference outlined in \citet{Greif19}.

\subsection{Implications of the surface heating configuration}\label{sec:heatedregions}

We constructed a sequence of simple models for the properties of the two hot regions, nevertheless motivated by (numerical) pulsar theory. We considered models in which the regions were related via antipodal reflection symmetry with respect to the stellar origin, and models that do not impose such symmetry, meaning that their properties and location were described with distinct parameters (with the restriction that the regions cannot overlap). The models included simply-connected circular and crescent regions, and rings (whose hole and annulus are concentric or eccentric), each filled with single-temperature material. The models also included annular (ring) regions whose concentric hole is filled with material of finite temperature distinct from that of the material in the annulus.  

We were able to rule out the hot regions being antipodal and identical based on clear systematic structure in the residuals between data and model \textit{a posteriori}; moreover, a model wherein the regions are both assumed to be simply-connected circular single-temperature spots was strongly disfavored. We inferred that the regions are configured to exist in the same rotational hemisphere: one region subtends an angular extent of only a few degrees (in spherical coordinates with origin at the stellar center) but whose other structural details we are insensitive to; the other region is far more azimuthally extended,\footnote{In spherical coordinates with polar axis coincident with the stellar rotation axis.} in the form of a narrow hot crescent. The inferred effective temperature of the \TT{NSX} atmosphere was remarkably consistent across all models considered---for both regions---at $\sim\!1.3\times 10^6$ K. The \TT{ST+PST} model exhibited the largest background-marginalized likelihood function values in the typical set of a posterior mode. Figure \ref{fig:STpPST MML config} renders a representative configuration from the posterior mode, and Table \ref{table: STpPST} reports the marginal credible intervals for the hot-region parameters. Note that \TT{ST+PST} includes within prior support, configurations wherein the regions are similar or even congruent in shape, and \textit{a priori} favors (albeit weakly) smaller angular extents---the heating asymmetry is emergent in spite of this.

One of the principal astrophysical questions arising is how such a heating configuration can occur. It appears to be incompatible with magnetospheric current heating at the footpoints of a simple near-centered dipole magnetic field, and is likely to require some higher-order multipole structure \citep{Barnard82,Gralla17,Lockhart19}. We now need to determine the type of field configuration required, the magnitude of the different moments, and whether this is feasible on physical grounds. Consideration will also need to be given to how magnetospheric currents actually map to temperature fields on the stellar surface. We note that there are clear similarities between the inferred \TT{ST+PST} configuration and the current heating distribution contemporaneously derived by \citet[][see their figure 6 in particular]{Lockhart19} via quadrupolar extension of the magnetic field, considering that the heating ring is asymmetric with respect to the dipole axis, and closely resembles a large-scale arc-like hot region. There are, however, also some differences: \citet{Lockhart19} restricted their study to configurations where the center of the heated ring is antipodal to the \added{heated} spot \added{(both of which emit as approximate blackbodies)}, an assumption that would need to be relaxed to recover our preferred configuration.

There are also implications for pulsar emission in wavebands other than the X-ray \citep[see the reviews by][]{Grenier15,Cerutti17}. \textit{If a multipolar field structure is required to explain the surface temperature field, how does this affect radio and gamma-ray emission generated further out in the magnetosphere? Could the multipole structures persist out to the point where emission in these wavebands is thought to happen?} Quadrupole fields fall off faster than dipole fields, as the inverse fifth power of the radius, but the radius beyond which the field is predominantly dipolar would depend on the ratio of quadrupole to dipole components. Most current models of radio and gamma-ray emission assume that the field structure is a centered dipole \citep[e.g.][]{Radhakrishnan69,Gil84,Kijak03,Dyks04,Johnson14}, and this would need to be revisited.

There are also questions pertaining to stellar evolution. NSs are born with a field structure that could be quite complex as a result of the supernova process \citep{Ardeljan05,Obergaulinger17} but various diffusive evolutionary processes can subsequently modify field structure even for isolated NSs \citep{Reisenegger09, Vigano13, Mitchell15, Gourgouliatos18}. In addition, rotation-powered MSPs are thought to go through an extended period of accretion-induced spin-up to reach the observed spin rates. The accretion process may also act to modify the field structure \citep{Romani90,Melatos01,Payne04}. It remains to be determined whether a complex multipolar field structure could emerge and survive from birth, or be generated during the accretion process. If such a field structure is present or evolves during the accretion phase, there will also be implications for the spin-up process and for X-ray emission during that phase of NS evolution. If the magnetic field were to channel accreting material onto two magnetic polar caps on the same hemisphere, for example, this would certainly affect the emission from accreting \replaced{millisecond pulsars}{MSPs} \citep{Long07,Long08,Patruno12}. Whether the star is even visible as an accreting pulsar will depend not only on the geometry of the hot regions where accreting material impacts the star, but also on whether the observer views the hemisphere containing the polar caps or the other one. The flow at the inner edge of the accretion disk, a strong source of potentially variable X-ray emission, would also be affected by a multipolar field structure. Finally, extremely off-center dipoles or strong non-centered multipole fields, for instance, will produce asymmetries in the Poynting flux of low-frequency radiation parallel to the spin axis, with consequences that could include a large space velocity \citep{Harrison75}. \citet{Lommen06} found that PSR~J0030$+$0451 has a relatively low transverse space velocity---a property that is potentially in contention with a field far from that of a centered dipole---although since PSR~J0030$+$0451 is isolated its space velocity would also depend on how the binary was disrupted after the spin-up phase.

\replaced{Detailed discussion on these issues is reserved for an accompanying Letter}{Discussion on some of these issues is reserved for an accompanying Letter} \citep{bilous19}. However, further work on the implications of the inferred configuration for pulsar field structure, emission mechanisms, and stellar evolution is certainly required. It is clear that the mass and radius inferred for PSR~J0030$+$0451 depend strongly on the surface radiation field models (including prior support) that we have explored. Further study may show that our models are either not general enough or too general (for example with respect to the prior support). For example, we may find that temperature gradients in the hot regions cannot be neglected; or that field stability considerations impose a minimum angular separation between the polar caps---and possibly, by extension, the hot regions---larger than found in our analysis; or that no magnetospheric model can generate a heating configuration in which one region is a small-scale spot whilst the other an azimuthally extended crescent.

There are a number of computational aspects for pulsar theorists to consider when developing surface heating models suitable for statistical inference. We were only able to consider the configurations offered uniquely by \TT{ST+PST} as we approached the limit of our computational resource allocation; we can therefore provide stronger guarantees about the accuracy of the posterior computation for the lower-complexity models with more than one run (\TT{ST-U} and \TT{ST+CST}). In the future, we suggest that more resources be devoted to models at the \TT{ST+PST}-level of complexity, in particular for work on: (i) parameterization of hot regions with more complex topologies and/or boundaries; (ii) their efficient numerical resolution for likelihood function evaluation; and (iii) exploration of the associated parameter space via sampling or other methods.

We also suggest that additional resources be devoted to research avenues such as self-consistent theory and computation of surface heating by magnetospheric currents, for the purpose of statistical computation. A question may be posed as to the generation of highly non-dipolar surface temperature fields---e.g., arc- or ring-like heating distributions---which can be: (i) parameterized such that approximate representations find compromise between accuracy and complexity, capturing the facets considered most crucial to signal generation, perhaps as a sequence of models increasing in complexity; and (ii) built into efficient software implementations. Progress on such fronts should encourage a bridge to form between phenomenological efforts and more self-consistent theory for the purpose of efficient statistical computation, and may also offer a way in which to connect distinct theoretical models (in an approximative manner) on a continuous space.

\subsection{Pulse-profile modeling for other types of NS}\label{sec:ppmimp}

\NICER is the first mission designed specifically to use the pulse-profile modeling technique to infer the mass and radius of NSs. As a soft X-ray telescope with an effective area of less than a square meter, it is optimized for applying the technique to MSPs, which have soft, stable pulse-profiles meaning we can use multiple exposures taken over a long baseline to accumulate a sufficient number of events to statistically probe MSP physics. However, the technique can also be applied to other NSs with emission modulated by rapid rotation: accretion-powered pulsars and thermonuclear burst oscillation sources.

In accretion-powered pulsars \citep[see][for a review]{Patruno12}, accreting material is channeled by the magnetic field onto the magnetic polar caps and the pulsed emission has two main components: thermal emission from the heated vicinity of the accretion impact zones, and nonthermal emission from the shock in the accretion funnel \citep{Poutanen03}. A third pulsed component may arise due to reflection from the accretion disk \citep{Wilkinson11}. Thermonuclear burst oscillations \citep[see][for a review]{Watts12} are generated by rotational modulation of global asymmetries that form in a surface radiation field during thermonuclear (Type-I X-ray) bursts; such a burst occurs in the ocean of an accreting NS, driven by unstable burning of accreted hydrogen, helium, or carbon \citep[see, e.g.,][]{Galloway08}. The precise mechanism driving the \textit{detectable} asymmetry (oscillations) is not clear: possibilities include disrupted flame spread \citep{Spitkovsky02,Cavecchi13}, large-scale waves in the burning ocean \citep{Heyl04,Piro05,Chambers19}, or patterns triggered by convection \citep{Garcia18}.

Accretion-powered pulsations and thermonuclear burst oscillations are radiatively harder ($\sim\!1$--$30$\,keV) than the pulsations of MSPs. Accumulating the requisite number of photons for tight constraints, in a realistic observing time, also requires a telescope with an effective area of several square meters \citep{Watts16,Watts19b}. Several mission concepts are currently being developed for large-area broadband X-ray timing telescopes that would access a larger, fainter population of MSPs than we can observe with \NICER, and fuel pulse-profile modeling for accretion-powered pulsars and thermonuclear burst oscillators: these include the \project{eXTP}~\citep[][]{Zhang19,Watts19}, and the \project{STROBE-X}~\citep[][]{Ray19}. For an idea of the constraints that can be delivered by pulse-profile modeling and inference using existing data, see the \citet{Salmi18} analysis of \project{Rossi X-ray Timing Explorer} observations of the accretion-powered \replaced{millisecond pulsar}{MSP} SAX~J1808.4$-$3658.

Our uncertainty toward surface heating physics affects pulse-profile modeling not only for rotation-powered MSPs, but also accretion-powered pulsars and thermonuclear burst oscillators. Although the general mechanism that gives rise to the pulsed components in accretion-powered pulsars is clear, the surface \textit{and off-surface} temperature field and local comoving beaming function---particularly from the accretion-funnel shock---are \textit{a priori} highly uncertain for any given source. Thermonuclear burst emission has a well-understood (local comoving) beaming function due to the sub-surface thermal origin \citep{Suleimanov11}, but the mechanism for generating asymmetries in the global surface radiation field, which in turn generate such rotational oscillations, remains highly uncertain. Reducing the remaining theoretical uncertainties, and developing physically motivated parameterized models of the surface temperature field, will be important. However, our analysis of \NICER data provides an important real-world demonstration that pulse-profile modeling is a viable technique for constraining masses and radii of NSs, and that the analysis machinery can operate on somewhat flexible models, with weakly informative priors, for both source and background emission.

\section{Conclusion}

We reported on pulse-profile modeling efforts for the rotation-powered \replaced{MSP}{millisecond X-ray pulsar} PSR~J0030$+$0451, conditional on \NICER data. We focused on PSR~J0030$+$0451, a challenging source due to the absence of an independent constraint on the gravitational mass (compared to \NICER's other primary target PSR~J0437$-$4715). Nevertheless, PSR~J0030$+$0451 was selected as the optimal source to demonstrate simultaneous inference of gravitational mass and equatorial radius given weakly informative priors, and to develop our analysis procedures.

The mass and radius each have marginal posterior $68\%$ credible interval half-widths at the $\sim\!9$--$11\%$ level, conditional on the \NICER~\project{XTI} event data. These constraints are consistent with those emerging from both gravitational wave analysis and X-ray spectral modeling, and are expected to improve with further exposure. The compactness is constrained more tightly, at the $\sim\!6\%$ level. Prospects for \NICER delivering tight constraints for rotation-powered MSPs where the mass is known independently to uncertainties of a few percent are clearly excellent.

In addition to inferring properties of the spacetime (mass, radius, and compactness), we were also able to infer the properties of the thermally-emitting hot regions that we assume generate the pulsations. For the specific set of models that we considered, the inferred configuration has both hot regions in the same rotational hemisphere, with one hot region being a small spot and the other an azimuthally-extended narrow crescent. Models wherein the hot regions are antipodal are strongly disfavored, implying a complex \added{offset dipolar and} multipolar field structure that, if accurate, has major implications for both pulsar emission and stellar evolution.

\acknowledgments
We thank the anonymous referees for their suggested improvements to this work. This work was supported in part by NASA through the \NICER mission and the Astrophysics Explorers Program. T.E.R., A.L.W., and A.V.B. acknowledge support from ERC Starting Grant No.~639217 CSINEUTRONSTAR (PI: Watts). A.L.W. would also like to thank Ralph Wijers for environmental support. This work was sponsored by NWO Exact and Natural Sciences for the use of supercomputer facilities, and was carried out on the Dutch national e-infrastructure with the support of SURF Cooperative. This research has made extensive use of NASA's Astrophysics Data System Bibliographic Services (ADS) and the arXiv. R.M.L. acknowledges the support of NASA through Hubble Fellowship Program grant HST-HF2-51440.001. S.G. acknowledges the support of the Centre National d'\'{E}tudes Spatiales (CNES). W.C.G.H. appreciates use of computer facilities at the Kavli Institute for Particle Astrophysics and Cosmology. S.M.M. thanks NSERC for support. J.M.L. acknowledges support from NASA through Grant 80NSSC17K0554 and the U.S. DOE from Grant DE-FG02-87ER40317.

\facility{\NICER~\citep{Gendreau16}}

\software{Python/C~language~\citep{python2007}, GNU~Scientific~Library~\citep[GSL;][]{Gough:2009}, NumPy~\citep{Numpy2011}, Cython~\citep{cython2011}, SciPy~\citep{Scipy}, OpenMP~\citep{openmp}, MPI~\citep{MPI}, \project{MPI for Python}~\citep{mpi4py}, Matplotlib~\citep{Hunter:2007,matplotlibv2}, IPython~\citep{IPython2007}, Jupyter~\citep{Kluyver:2016aa}, \TEMPO~\citep[\TT{photons};][]{Hobbs06}, PINT~(\TT{photonphase}; \url{https://github.com/nanograv/PINT}), \MultiNest~\citep{MultiNest_2009}, \textsc{PyMultiNest}~\citep{PyMultiNest}, \project{GetDist}~(\url{https://github.com/cmbant/getdist}), \project{nestcheck}~\citep{higson2018nestcheck,higson2018sampling,higson2019diagnostic}, \project{fgivenx}~\citep{fgivenx}, \XPSI~\citep[\texttt{v0.1}; \url{https://github.com/ThomasEdwardRiley/xpsi};][]{riley19b}}

\bibliographystyle{aasjournal}
\bibliography{nicer_pulse_profile_modeling}

\appendix

\section{Posterior computation}\label{sec:posterior computation}

Here we describe how we derived posterior inferences conditional on the models defined in Section \ref{sec:models}. 

\subsection{Nested sampling}
We implemented nested sampling using the open-source software \textsc{MultiNest}\footnote{\MultiNest \TT{v3.11} can be located at \url{https://github.com/farhanferoz/MultiNest} with SHA1-hash~4b3709c. \textsc{PyMultiNest} \TT{v2.6} can be located at \url{https://github.com/JohannesBuchner/PyMultiNest} SHA1-hash~5d8c103.} \citep[][]{MultiNest_2008,MultiNest_2009,MultiNest_2013,PyMultiNest}. For a subset of model nodes in Figure~\ref{fig: discrete model space diagram} and Figure~\ref{fig: discrete model space diagram 2} we executed two production runs---i.e., we generated two realizations of a particular stochastic sampling process---with the following resolution settings: the number of \textit{active} (or \textit{live}) points was $10^{3}$; the bounded-hypervolume expansion factor was (at least\footnote{A \textit{lower-bound} for reasons pertaining to the implementation of the joint prior distribution; the lower-bound itself is numerically transformed into an appropriate \MultiNest setting in order to achieve at least this desired expansion factor. Further detail is beyond the scope of this description \citep[refer to][]{riley19b}.}) $0.3^{-1}$; the termination condition as a function of iteration number $i$ was, schematically,
\begin{equation}
    \ln\left(\widehat{\mathcal{Z}}_{i}+\widehat{\Delta\mathcal{Z}}_{i}\right)
    -
    \ln\left(\widehat{\mathcal{Z}}_{i}\right)
    <
    x
    \implies
    \frac{\widehat{\mathcal{Z}}_{i}}{\widehat{\mathcal{Z}}_{i}+\widehat{\Delta\mathcal{Z}}_{i}}
    >
    e^{-x},
\end{equation}
where $\widehat{\mathcal{Z}}_{i}$ is an estimator for the evidence integral up to iteration $i$ over estimated prior mass $(1-\widehat{X}_{i})$, and $\widehat{\Delta\mathcal{Z}}_{i}$ is an estimator for the \textit{maximum remaining} contribution to the evidence over the complementary estimated prior mass $\widehat{X}_{i}$. As $x\to0$, $e^{-x}\to1^{-}$ and the estimated amassed evidence entirely dominates the estimated remaining evidence; we generally set $x=10^{-1}$, which is five times smaller than suggested by \citet[][]{MultiNest_2009}.\footnote{The minimum tolerance used for any process was $x=10^{-3}$ to confirm that termination was not premature.} The number of active points is chosen to be a number that is larger than will be typically reported in the literature for similar dimensional problems (perhaps with simpler distributional structure), or recommended by the authors of \MultiNest, for a compromise between resource consumption and accuracy. The hypervolume expansion factor is, for all but one run,\footnote{The \TT{ST+PST} model was implemented as we were exhausting our computational resources and thus we lowered resolution---increased the nested-sample  acceptance fraction---to ensure completion of an exploratory run---see Table~\ref{table: STpPST}.} greater than the number recommended by the authors of \MultiNest for accurate evidence estimation. The combination of a number of active points---which is between $40$ and $80$ times larger than the dimensionality of the sampling space---and the recommended expansion factor, targets posterior computation with an implementation-specific error\footnote{The error due to not sampling from the entire prior hypervolume subject to a given likelihood function constraint.} that is smaller than the error due to the inherent stochasticity of Monte Carlo sampling.

The \textit{constant-efficiency} \MultiNest bounding variant should in general be avoided where tractability is not compromised to reduce risk of under-sampling when a hyper-ellipsoidal decomposition does not conform well to likelihood level hypersurfaces; we did not use this bounding variant. However, in the higher-dimensional contexts that we are approaching in this work, it could be useful to activate this bounding variant if integration is forecasted to consume too many computing resources (e.g., the acceptance fraction drops too low below $10^{-3}$ when using a high-performance system and an expensive likelihood function). Imposing a target acceptance fraction may be useful for initial exploratory runs in order to probe for configurations with high (marginal) likelihood---a weak indicator of potential model performance if future efforts achieve greater computational efficiency. When \textit{constant-efficiency} mode is activated the evidence estimates should be assumed to be positively biased unless importance nested sampling is also activated (which can consume an appreciable fraction of the memory on a typical supercomputer node for $\sim\!10^{8}$ likelihood function evaluations). The acceptance fraction decaying to such levels is indicative that the minimum-bounding hyper-ellipsoidal decomposition does not conform sufficiently well to the nested likelihood function level hypersurfaces, resulting in: (i) a large excess of hypervolume exterior to the likelihood surface but within the bounding union of ellipsoids; and/or (ii) the union of bounding ellipsoids exhibits a large fractional overlap. In such cases one could construct an alternative parameterization of the problem that is more tractable with the \MultiNest bounding algorithm (in the native sampling space), or use an alternative (nested) sampling algorithm.

The mode-isolation sampling variant was \textit{not} activated unless stated otherwise. This variant isolates the evolution of local modes whose hyper-ellipsoidal clustering decompositions are mutually non-overlapping; upon isolation, the active points of sampling threads constituting each mode cannot migrate between modes at subsequent iterations---they are locked in, which can alleviate premature mode deactivation. Sampling resolution, however, can be absorbed by unimportant modes, which is undesirable. The mode separation sampling variant generates distinct statistics for each mode and labels actives points according to mode association.

With the above settings, amongst ulterior numerical settings for marginal likelihood function evaluation, \XPSI typically executes $\mathcal{O}(10^{7})$ marginal likelihood function evaluations in $O(10^{4})$ core hours via \MultiNest.

\subsection{Summary of information presentation}\label{subsec:information presentation}

We now summarize how we opt to present probabilistic information in the form of figures and tables. We condition on a set of model variants that are treated in effectively the same manner, and the information presentation is consistent for these models; we thus here describe the information once. Figures and tables are displayed in the main body---see Section~\ref{sec:inferences}---for the model we deem to perform best considering both prior and posterior predictive measures and checks; we refer the reader to Table~\ref{table: STpPST} and Figure~\ref{fig:STpPST residuals} through Figure~\ref{fig:STpPST obs marginal posterior} for reference.

\subsubsection{Graphical posterior-predictive checking}\label{subsubsec:posterior checking figure}
In a figure such as Figure~\ref{fig:STpPST residuals} we display salient information for assessing performance of a model in isolation. Our generative modeling process is fundamentally built on the statement that we do not believe the true data-generating process exists within the model space considered (refer to Section~\ref{subsec: generative model space}); however, the models may, for the purpose of generating data, be deemed adequate approximations. We aim to graphically approximate an answer to the question: \textit{Does the model generate, \emph{a posteriori}, synthetic data that emulates structure in the real event data?}

Graphical posterior predictive checking here relies on the power of human identification of systematic structural differences, which if physically characterized can drive future model development toward better-performing approximations. Structural differences include spectro-temporal correlations between random variates assumed to be statistically independent (e.g., the \TT{ST-S} version of Figure~\ref{fig:STpPST residuals} in the online figure set), and inaccurate noise modeling leading to under- or over-estimation of the variance of random variables; the latter also may manifest due to excessive, non-physical predictive complexity (over-fitting). In the top panel of such a figure we display the data set for convenience. In the middle panel we choose to display the posterior-mean Poisson expected count numbers:
\begin{equation}
\lambda_{ij}
\coloneqq
\mathbb{E}_{\pi(\boldsymbol{\theta})}[c_{ij}(\boldsymbol{\theta})]
=
\mathop{\int}_{\mathcal{S}}
c_{ij}(\boldsymbol{\theta},B_{i})
\pi(\boldsymbol{\theta},B_{i}\,|\,\boldsymbol{d})
d\boldsymbol{\theta}
\approx
\mathop{\sum}_{k}w_{k}c_{ij}(\boldsymbol{\theta}_{k}),
\label{eqn:posterior-expected counts approximation}
\end{equation}
where $\boldsymbol{\theta}_{k}\sim\pi(\boldsymbol{\theta}\,|\,\boldsymbol{d})$ are samples with normalized importance weights $w_{k}$ drawn from the background-marginalized posterior density $\pi(\boldsymbol{\theta}\,|\,\boldsymbol{d})$, and $\mathcal{S}$ is the prior support. Crucially, we do not sample the joint posterior distribution $\pi(\boldsymbol{\theta},\boldsymbol{B}\,|\,\boldsymbol{d})$ because of prohibitive scaling of expense with dimensionality;\footnote{
If our set of samples was drawn from the joint posterior distribution $\pi(\boldsymbol{\theta},\boldsymbol{B}\,|\,\boldsymbol{d})$ we could better approximate the expectation integral in Equation~(\ref{eqn:posterior-expected counts approximation}), or at greater cost, approximate the posterior-predictive probability mass distribution in data space:
\begin{equation*}
p(x_{ij}\,|\,\boldsymbol{d})
=
\mathop{\int}
p(x_{ij}\,|\,\boldsymbol{\theta},B_{i})
\pi(\boldsymbol{\theta},\boldsymbol{B}\,|\,\boldsymbol{d})
d\boldsymbol{B}d\boldsymbol{\theta}
\approx
\mathop{\sum}_{k}
w_{k}
p(x_{ij}\,|\,\boldsymbol{\theta}_{k},B_{i,k})
,
\end{equation*}
whose expectation $\mathbb{E}[x_{ij}]$ and variance $\mathbb{V}[x_{ij}]$ are indicators of residual structure in the data relative to the model \textit{a posteriori}. The posterior predictive distribution may also be constructed by jointly generating samples from the Bayesian joint distribution: $\boldsymbol{x}_{k}\sim p(\boldsymbol{x}\,|\,\boldsymbol{\theta}_{k},\boldsymbol{B}_{k})$ and $(\boldsymbol{\theta}_{k},\boldsymbol{B}_{k})\sim\pi(\boldsymbol{\theta},\boldsymbol{B}\,|\,\boldsymbol{d})$.
}
therefore, in order to compute a data-space posterior-mean quantity, we opt to maximize the \textit{conditional} likelihood function $L(B_{i};\boldsymbol{\theta}_{k})$ with respect to each background count-rate parameter $B_{i}$, generating an estimator $\widehat{B}_{i}$, such that $c_{ij}(\boldsymbol{\theta}_{k})\coloneqq c_{ij}(\boldsymbol{\theta}_{k},\widehat{B}_{i})$.  In the \textit{bottom} panel we display standardized residuals between data count numbers and the quantities $\lambda_{ij}$, where we define Poisson-random variables $x_{ij}\sim p(x_{ij}\,|\,\lambda_{ij})$ \explain{$c_{ij}\to x_{ij}$, because $c_{ij}$ used already for a similar object}. We consider the set of figures associated with Figure~\ref{fig:STpPST residuals} as \textit{graphical posterior-checking} plots. If no clear systematic structure manifests during posterior predictive checking and the posterior-predictive distribution is well-approximated by a sampling distribution conditional on a parameter vector, we consider the model to be appropriate for predicting future observations against which the model may be falsified.

\begin{figure*}
\centering
\includegraphics[clip, trim=0cm 0cm 0cm 0cm, width=0.9\textwidth]{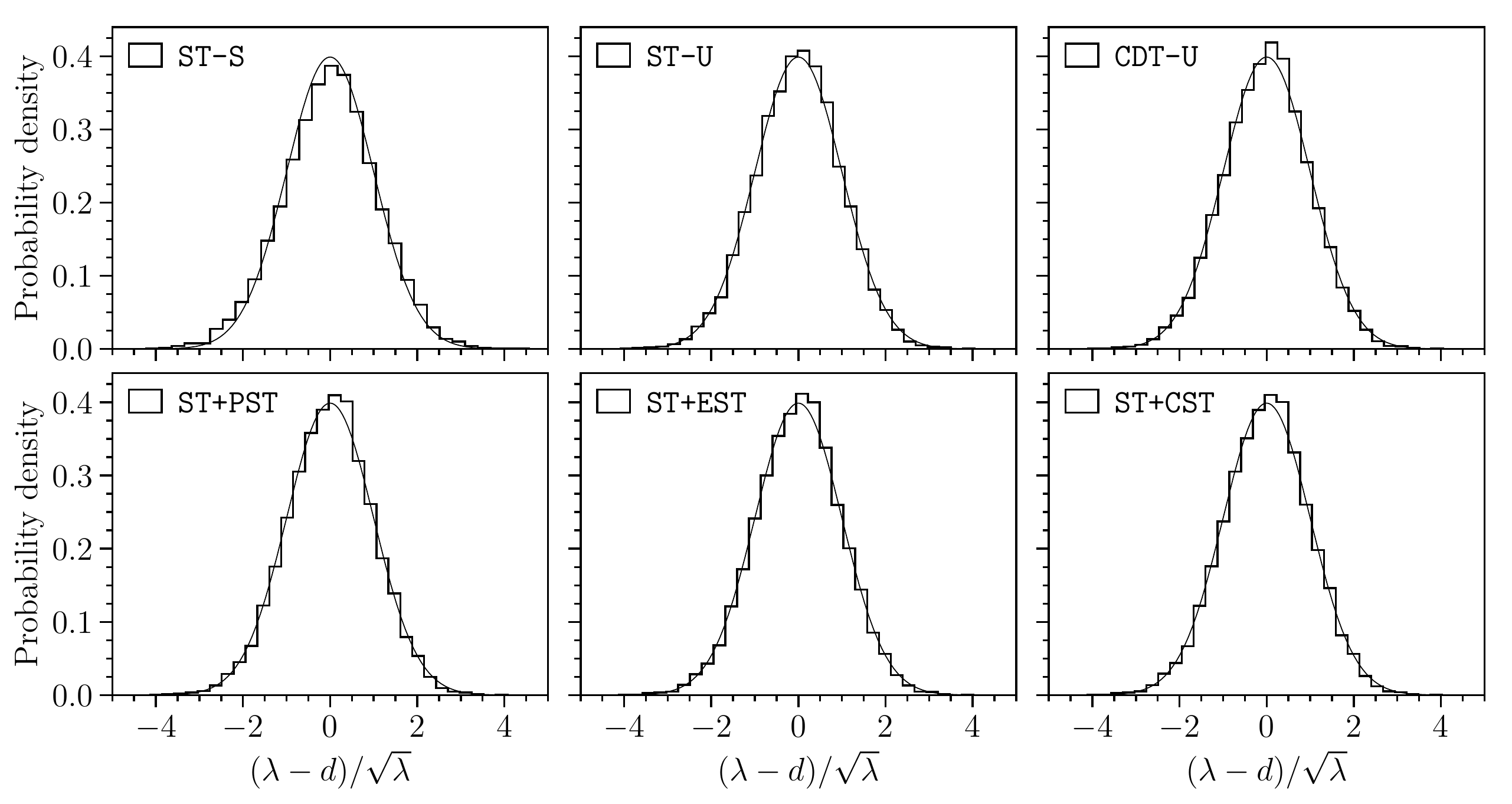}
\caption{\small{Summary of the two-dimensional residual panels in the online figure set associated with Figure~\ref{fig:STpPST residuals}. For each model we identify the sample (parameter vector) $\hat{\boldsymbol{\theta}}_{\rm BMML}$ that reported the highest \textit{background-marginalized} likelihood function value amongst identified posterior modes, together with the background count-rate vector $\hat{\boldsymbol{B}}_{\rm CML}$ that maximizes the \textit{conditional} likelihood function given $\hat{\boldsymbol{\theta}}_{\rm BMML}$. Given the expectations $\mathbb{E}[x_{ij}]=\lambda_{ij}(\hat{\boldsymbol{\theta}}_{\rm BMML},\hat{\boldsymbol{B}}_{\rm CML})$, we evaluate the residuals $(\lambda-d)/\sqrt{\lambda}$ for each phase-channel interval, where $\{d_{ij}\}$ are variates of independent Poisson-random variables $\{x_{ij}\}$. We display in each panel the distribution of these random variates (the real count numbers). The minimum number of expected counts in any phase-channel interval is $\sim\!30$, and thus the $\{x_{ij}\}$ drawn from the sampling-distribution are, approximately, identically and normally distributed if transformed as $y\coloneqq(\lambda-x)/\sqrt{\lambda}$. The smooth \textit{black} distribution in each panel is a normal distribution. The data sampling-distributions for the \TT{ST-U} model and models higher in complexity do not exhibit systematic structural inaccuracies in the context of the variates: there are no clear signs of noise-model inaccuracy or over-fitting, nor are there clear signs of residual correlations in, e.g., Figure~\ref{fig:STpPST residuals}. These models can thus perform adequately as point-measure predictors of structure in the real count-number data. The residuals conditional on the \TT{ST-S} model are distributed with larger variance than unity (normal): more weight is visible in the wings, and less in the near vicinity of zero.}}
\label{fig:residuals summary}
\end{figure*}

Bayesian global performance measures---such as the evidence---are useful for determining the utility of increments in the complexity of a generative model. However, whilst being a target of our posterior computation, prior predictive probabilities are known to not be universally robust and can be sensitive to prior choices and parameterization, especially in phenomenological contexts. Thus in order to assess performance and identify model features that conform well to data structure, it is necessary---and natural---to also visually inspect predictions \textit{a posteriori}. We consider other measures such as the expected posterior utility (i.e., power for future statistical falsification) beyond the scope of this work \citep{vehtari2012}.

In this work we refrain as much as possible from invoking \textit{point measures}---based on parameter vector point-estimates---to quantitatively summarize and compare models. The reason we do this is twofold: (i) a point estimate is usually the parameter vector that is estimated to globally optimize some quantity, and subsequent to estimation all information encoded by the posterior distribution is not explicitly regarded; and (ii) the target of our statistical computation (nested sampling) is not any particular point measure, but instead to draw samples from the posterior \textit{typical set} for the purpose of estimating posterior integrals. It follows that stronger guarantees can be made about the accuracy of statistical estimators that are posterior integrals than can be made about those which are point estimates; we therefore view integral estimators as generally more helpful and robust than point estimators.

As an example, consider the canonical estimation of the parameter vector that globally maximizes the likelihood function conditional on some model. In our case, our posterior computation does not target accurate maximization, and we are forced to marginalize over the phase-invariant (background) count-rate parameters in order to define a nested-sampling space whose dimensionality is not prohibitive. We thus do not guarantee that the sample (parameter vector) $\hat{\boldsymbol{\theta}}_{\rm BMML}$ that reported the highest \textit{background-marginalized} likelihood function value amongst identified posterior modes, together with the background count-rate vector $\hat{\boldsymbol{B}}_{\rm CML}$ that maximizes the \textit{conditional} likelihood function given $\hat{\boldsymbol{\theta}}_{\rm BMML}$, is an adequate estimator of the joint vector $(\boldsymbol{\theta},\boldsymbol{B})_{\rm ML}$ that globally maximizes the likelihood function. In Figure~\ref{fig:residuals summary} we consider the joint vector $(\hat{\boldsymbol{\theta}}_{\rm BMML},\hat{\boldsymbol{B}}_{\rm CML})$: empirically we see that all but one model (\TT{ST-S}) has, within prior support, parameter vectors that perform adequately as point-measure predictors of structure in the real count-number data.

\subsubsection{Parameter kernel density estimation and credible regions}\label{app:KDE}

We applied the post-processing module of the \XPSI package \citep[\added{\texttt{v0.1}};][]{riley19b}. \XPSI wraps---combines and adds functionality---to several other packages for statistical computation: relevant here for Gaussian kernel density estimation (KDE) is \project{GetDist}.\footnote{\url{https://github.com/cmbant/getdist}. Specifically, tag \TT{v0.3.1}, with some minor plotting customization locatable at \url{https://github.com/ThomasEdwardRiley/getdist/tree/customization}, with SHA1-hash~61f69d0. The technical KDE notes for \project{GetDist} are located at \url{https://cosmologist.info/notes/GetDist.pdf}.} We do not use \project{GetDist} to compute numerical one-dimensional credible intervals on each parameter, but \project{GetDist} is used whenever KDE is necessary for post-processing (posterior) samples into estimators: for calculation of the plotted one- and two-dimensional density functions; for calculation of the plotted two-dimensional (joint) credible regions; and for one-dimensional KL-divergence estimation, where both the marginal posterior and marginal prior density functions need to be estimated from samples.

\project{GetDist} can execute smoothing using either a manual Gaussian kernel bandwidth or an automatic optimized bandwidth. For all parameters and models, for simplicity and consistency, we invoke a manual bandwidth of $0.4$ times the estimated parameter standard deviations, based on the \project{GetDist} guidelines for choosing analysis settings. Moreover, \project{GetDist} performs various smoothing corrections to avoid smoothing bias as one attempts to mitigate finite-sample noise: (i) smoothed-density correction near hard one-dimensional parameter bounds defined through the prior support, with capability to estimate a finite local density function gradient; and (ii) iterative multiplicative bias-correction to the estimated density function to nullify over-smoothing. Density estimation near non-trivial prior support boundaries in two parameters does not account for the local boundary, but this is not problematic for our work here. We roll with the default bias-correction settings: a linear boundary kernel (in both one- and two-dimensional spaces) and zeroth-order multiplicative bias-correction (a single application with no iterations).

\subsubsection{Error analysis for statistical estimators}\label{subsubsec:estimators and errors}

Two frameworks now require distinction: (i) the parametric probabilistic framework of the generative model (a Bayesian context) for the data; and (ii) a non-parametric probabilistic framework that operates with realizations of a stochastic sampling process, which in turn operates on a deterministic \deleted{but unknown mathematical structure---a} target probability distribution defined in the parametric framework given data and a generative model \citep[e.g.,][]{skilling2006,higson2018sampling}. The sampling process \deleted{generally} has \added{a mixture of known and approximated properties, including} fixed non-physical settings. Estimators derived from the process output \replaced{exhibit distributions due to stochasticity which are estimated numerically, normalized, and interpreted probabilistically in a distinct context.}{are stochastic and we are interested in their distributions.}

The purpose of executing some number of computationally expensive \textit{repeats} is for \deleted{rigorous} error analysis---specifically \added{the} estimation of implementation-specific error \added{\citep{higson2019diagnostic}} pertaining to \deleted{the} sampling from the joint prior subject to a likelihood function constraint \added{\citep[a more thorough review of error analysis techniques may be found in][]{riley19b}}. \replaced{The aim}{One way to approach this problem} is to compute \deleted{parameter} posterior marginal density functions for many bootstrapped realizations of each stochastic run \replaced{to}{and then graphically} probe for the manifestation of implementation-specific error \citep{higson2019diagnostic}.

\XPSI wraps the package 
\project{nestcheck}\footnote{Specifically, tag \TT{v0.2.0}, with customization to support use of \project{GetDist} KDE, locatable at \url{https://github.com/ThomasEdwardRiley/nestcheck/tree/feature/getdist_KDE} with SHA1-hash~4555df0.}~\citep{higson2018nestcheck,higson2018sampling,higson2019diagnostic} to access existing error analysis routines. As an example of application in this work, consider Figure~\ref{fig:STpPST source marginal posterior}: in the associated online figure set, where we supply a higher-resolution version of each panel, we display the marginal posterior density distributions for each parameter as a set of shaded error bands. The shaded bands represent the distribution of posterior density, at each parameter value, based on simulated nested sampling process realizations; the posterior density function is estimated for each realization with \project{GetDist} as described above in Appendix~\ref{app:KDE}. The colorbar denotes the percentage of realizations spanned by a band with a given shade, where each band connects intervals (at each parameter value) containing the highest \textit{realization} density of \textit{posterior density}. Note that the colorbar is not associated with the shaded joint density distributions in the off-diagonal plots. The contours in the on-diagonal panels thus encode information on the variation of the parameter kernel density estimator due to the inherent stochasticity of each of sampling processes; the estimator distributions are connected as a function of each respective parameter to delineate the behavior of the probability mass. If the member processes are deemed to exhibit consistency under visual inspection, the combined process may be invoked to estimate distributions of estimators.

We apply \project{nestcheck} routines to bootstrap re-sample threads and simulate weights for the following: the one-dimensional quantiles in posterior mass for each parameter, which are in turn used to report the credible intervals; the global and parameter-by-parameter KL-divergences; and the evidence. The numerical values we report are in some cases (\TT{ST-S}, \TT{ST-U}, and \TT{ST+CST}) derived by combining two realizations into a single realization with $\mathcal{O}(10^{3})$ active points (sampling threads), if the runs are considered sufficiently consistent and exhibit sufficient resolution for our purposes here. However, due to computational expense we could only afford \textit{at most} two runs, and for both \TT{ST+EST} and \TT{ST+PST} we were limited to a single run with $\mathcal{O}(10^{3})$ active points. Moreover, note that processes executed by the mode-separation algorithm are incompatible with the notion of process combination: the theory nor software implementation exists at the time of writing, and therefore we only display such a run---when relevant and available---in the posterior figures, but do not use it to calculate numerical estimators.

\subsubsection{Estimating posterior information gain}\label{subsubsec:divergence}

In a Bayesian context, the KL-divergence \citep[][]{kullback1951} can be applied as a non-negative real scalar\footnote{And thus parameterization invariant.} measure of posterior information \textit{gain} about a parameter---or jointly about parameters---of a generative model, conditioned on the data set; it is also known via information-theoretic interpretation as \textit{relative entropy}. Equivalently it is the \textit{posterior-expected additional} number of bits necessary to encode the value of a parameter sample for lossless communication between agents, \textit{if} the (marginal) prior distribution is invoked to design an optimal encoding. Alternate interpretations exist to satiate a variety of readers---e.g., \citet[][]{Shlens14}.

KL-divergence maximization is central to an information-theoretic---but often in practice intractable---definition of a minimally-informative prior via reference to the generative model, but without reference to the data (via data-space marginalization). Whilst we cannot feasibly determine the maximal KL-divergence with respect to the space of all proper prior density functions, the number of bits of information gain is a useful indicator of the degree to which the likelihood function dominates the information encoded in the posterior. The KL-divergence has, as an example, recently been applied by \citet[][]{LIGO_161218_catalog} to probe posterior information gain and sensitivity to prior assumptions.

For continuous symbols, the KL-divergence is defined in the limit that a discrete symbol becomes a continuous subset of $\mathbb{R}^{n}$; in practice, of course, computer representation of the reals is discrete. Given the existence of a known optimal prior encoding, the number of additional bits for lossless communication of a posterior sample at a given precision is, in the context of our inference problem, far less than the number of bits required to store the sample. Mathematically, in units of \textit{bits}:
\begin{equation}
D_{\textrm{KL}}(\pi\;||\;p)
=
\mathop{\int}_{\mathcal{S}}
\pi(\boldsymbol{\theta}\,|\,\boldsymbol{d})
\log_{2}
\left[
\frac
{\pi(\boldsymbol{\theta}\,|\,\boldsymbol{d})}
{p(\boldsymbol{\theta})}
\right]
d\boldsymbol{\theta}
\approx
\mathop{\sum}_{k}
w_{k}
\log_{2}
\left[
\frac
{\pi(\boldsymbol{\theta}_{k}\,|\,\boldsymbol{d})}
{p(\boldsymbol{\theta}_{k})}
\right],
\end{equation}
where $\boldsymbol{\theta}_{k}\sim\pi(\boldsymbol{\theta}\,|\,\boldsymbol{d})$ are samples with normalized importance weights $w_{k}$ drawn from the background-marginalized density $\pi(\boldsymbol{\theta}\,|\,\boldsymbol{d})$. Computation of a single scalar divergence for the $n$-dimensional joint posterior is straightforwardly given by
\begin{equation}
\begin{aligned}
D_{\textrm{KL}}(\pi\;||\;p)
&=
\mathop{\int}_{\mathcal{S}}
\pi(\boldsymbol{\theta}\,|\,\boldsymbol{d})
\log_{2}
\left[
\frac
{\pi(\boldsymbol{\theta}\,|\,\boldsymbol{d})}
{p(\boldsymbol{\theta})}
\right]
d\boldsymbol{\theta}\\
&=
\mathop{\int}_{\mathcal{S}}
\pi(\boldsymbol{\theta}\,|\,\boldsymbol{d})
\log_{2}
\left[
\frac{1}{\mathcal{Z}}
\frac
{p(\boldsymbol{d}\,|\,\boldsymbol{\theta})\cancel{p(\boldsymbol{\theta})}}
{\cancel{p(\boldsymbol{\theta})}}
\right]
d\boldsymbol{\theta}\\
&=
-
\log_{2}\mathcal{Z}
+
\mathop{\int}_{\mathcal{S}}
\pi(\boldsymbol{\theta}\,|\,\boldsymbol{d})
\log_{2}
L(\boldsymbol{\theta})
d\boldsymbol{\theta}
\approx
-
\log_{2}\mathcal{Z}
+
\mathop{\sum}_{k}
w_{k}
\log_{2}
L(\boldsymbol{\theta}_{k}).
\end{aligned}
\end{equation}
It is more useful, however, to compute a marginal divergence for each parameter, yielding a handle on which parameters the (marginal) likelihood function is most insensitive to in the context of the (marginal) prior, and thus to which prior assumptions the global posterior inferences may be most sensitive. The parameters that exhibit the lowest marginal posterior information gain are those for which prior assumptions are generally more important to be aware of and should thus be an accurate representation of prior belief. Divergence estimation for each parameter (or jointly for $m<n$ parameters) requires more involved post-processing because kernel density estimation is performed for evaluation of the quotient of marginal densities appearing in the integrand. 

Note that if the marginal KL-divergence for some parameter $\theta$ is small (relative to the divergences of other parameters) but the divergence is close to the theoretical maximum expected divergence, the prior exists in the minimally-informative limit whilst being relatively informative. For example, if the Fisher information for $\theta$ is everywhere relatively small---meaning the experiment is at most a weak probe of the parameter---the marginal posterior for $\theta$ should be dominated even by a minimally-informative prior, but posterior inferences about other parameters should be insensitive to all information about $\theta$. If, on the other hand, the likelihood function is a useful probe of $\theta$ whilst the marginal divergence is relatively small, the global posterior may be sensitive to prior information about $\theta$. In practice, we determine that the parameters that typically exhibit the smallest divergences---pulsar distance $D$ and \NICER instrument parameters $\alpha$ and $\gamma$---are assigned marginal priors that are not weakly informative in the context of the likelihood function; moreover, we do not need to calculate Fisher information to understand that the likelihood function itself is a useful probe of these parameters, which exhibit degeneracies with other pulsar parameters. Therefore, we conclude that small divergences in these cases do indicate that our global posterior inferences are strongly conditional on these prior assumptions.

\section{Prior transforms}\label{app:prior transforms}

For nested sampling we aim to transform from a native sampling space---a unit hypercube $\mathcal{H}=[0,1]^{n}$---to a (physical) parameter space $\mathbb{R}^{n}$ according to an inverse transformation of the joint prior density distribution $p(\boldsymbol{\theta})$ defined on $\mathbb{R}^{n}$ (typically with some compact support). We thus require implementation of a mapping $\boldsymbol{x}\mapsto\boldsymbol{\theta}$ where $\boldsymbol{x}\in\mathcal{H}$ and $\boldsymbol{\theta}\in\mathbb{R}^{n}$. In this appendix we give the prior transforms implemented in order to facilitate reproduction of the sampling processes, and also to provide a demonstration of some of the necessary architectural work for parameter estimation via nested sampling.

In this work we provide a summary table for each model; as an example, refer to Table~\ref{table: STpPST}. Within each table the joint prior density function $q(\boldsymbol{\theta})$ and its support $\mathcal{S}$ are reported. Here we give a prescription for drawing a sample from a given prior: (i) draw a sample for each parameter according to the listed sampling distribution $p(\boldsymbol{\theta})$ with support $\mathcal{S}^{\dagger}\subset\mathbb{R}^{n}$---using one-dimensional inverse sampling---to generate a candidate vector $\widetilde{\boldsymbol{\theta}}\in\mathbb{R}^{n}$; (ii) systematically evaluate the constraint equations to determine whether $\widetilde{\boldsymbol{\theta}}\in\mathcal{S}\subset\mathbb{R}^{n}$ or whether $\mathcal{S}\not\ni\widetilde{\boldsymbol{\theta}}\in\mathbb{R}^{n}$; and (iii) accept the candidate sample if $\widetilde{\boldsymbol{\theta}}\in\mathcal{S}$, otherwise reject the sample. The form---i.e., the relative marginal density at two values of parameter $\theta$---is, for a subset of parameters, not given by the one-dimensional distribution $\theta\sim q(\theta)$ (whose definition in some cases, as the distance $D$, requires explicitly stated truncation bounds) explicitly written in the prior column: the set of constraint equations defining the joint compact support $\mathcal{S}$ often non-trivially modulate the density distribution $q(\theta)$ if $\theta$ appears in constraint equations jointly with some subset of the other parameters. If the joint prior is separable with respect $\theta$ to then by definition $p(\theta)\equiv q(\theta)$.

\subsection{Gravitational mass and equatorial radius}\label{app:mass-radius prior implementation}

We defined our joint prior distribution of gravitational mass and equatorial radius in Section~\ref{subsubsec:source}. We apply a technique from Appendix~E~of~\citet{riley19b}: inverse sampling of a joint flat density function $q(M,R_{\textrm{eq}})$ with a trivial rectangular boundary $M\in[M_{a},M_{b}]$ and $R_{\textrm{eq}}\in[R_{a},R_{b}]$, and subsequent rejection only if  $R_{\textrm{eq}}\notin[3r_{g},16]$ km.\footnote{Strictly speaking, we also impose several ulterior constraint equations. However, for the spin of PSR~J0030$+$0451 these constraint equations are unimportant for defining the prior support on the joint space of $M$ and $R_{\textrm{eq}}$.} In this case, let $\mathcal{H}=[0,1]\times[0,1]$ and let the support of $q(\boldsymbol{\theta})=q(M,R_\textrm{eq})$ be $\mathcal{S}^{\dagger}\subset\mathbb{R}^{2}$. The mapping is then $\mathcal{H}\to\mathcal{S}^{\dagger}, \boldsymbol{x}\mapsto\boldsymbol{\theta}$ \explain{$\mapsto$ change to $\to$ between sets}, where $\mathcal{S}^{\dagger}\supset\mathcal{S}$. For $(M,R_{\textrm{eq}})\in\mathcal{S}$, the joint prior density $p(M,R_{\textrm{eq}})\propto q(M,R_{\textrm{eq}})$ because the constraint equation in compactness is dependent only on $M$ and $R_{\textrm{eq}}$, and not on any ulterior source parameters. Such a procedure is also summarized in the preamble of Appendix~\ref{app:prior transforms} above.

A standard transform (in the context of the nested sampling software) on the other hand would take the form $\mathcal{H}\to\mathcal{S}, \boldsymbol{x}\mapsto\boldsymbol{\theta}$. To construct such a map, one may write $p(M,R_{\textrm{eq}})=p(R_{\textrm{eq}}\,|\,M)p(M)$ where
\begin{equation}
p(M)
=
\mathop{\int}_{3r_{g}(M)}^{R_{b}}
p(M,R_{\textrm{eq}})
dR_{\textrm{eq}}
\propto
R_{b}-3r_{g}(M).
\end{equation}
Then define $\boldsymbol{x}\coloneqq(x_{M},x_{R})$ where
\begin{equation}
x_{M}(M)
=
\mathop{\int}_{M_{a}}^{M}
p(M^{\prime})
dM^{\prime}
\end{equation}
and
\begin{equation}
x_{R}(R_{\textrm{eq}};M)
=
\mathop{\int}_{3r_{g}(M)}^{R_{\textrm{eq}}}
p(R_{\textrm{eq}}^{\prime}\,|\,M)
dR_{\textrm{eq}}^{\prime}
=
\mathop{\int}_{3r_{g}(M)}^{R_{\textrm{eq}}}
\frac{p(M,R_{\textrm{eq}}^{\prime})}{p(M)}
dR_{\textrm{eq}}^{\prime}
=
\frac{R_{\textrm{eq}}-3r_{g}(M)}{R_{b}-3r_{g}(M)}.
\end{equation}
Inverting, one has
\begin{equation}
R_{\textrm{eq}}(x_{R};M)
=
x_{R}R_{b}+
3(1-x_{R})r_{g}(M),
\end{equation}
and similarly one obtains a nonlinear function $M(x_{M})$.

The problem with such a standard transformation is clear from inspection of $x_{R}(R_{\textrm{eq}};M)$. The common $(M,R_{\textrm{eq}})$ degeneracy in pulse-profile modeling is linear due to sensitivity to compactness $M/R_{\textrm{eq}}$. Therefore, requiring a constant compactness $r_{g}(M)/R_{\textrm{eq}}=\textrm{const.}$ implies
\begin{equation}
x_{R}(R_{\textrm{eq}};M)
=
\frac{\textrm{const.} - 3}{R_{b}/r_{g}(M) - 3},
\end{equation}
where $M=M(x_{M})$, meaning that $x_{R}=x_{R}(x_{M};M/R_{\textrm{eq}})$ is generally a nonlinear function. An optimal mapping would preserve the linearity of the degeneracy, and thus we do not opt for a standard transformation where $\mathcal{H}\to\mathcal{S}$, $\boldsymbol{x}\mapsto\boldsymbol{\theta}$; instead we inverse sample with rejection as described above.

\subsection{Hot regions}\label{app:hotregions}

The joint prior distribution for the parameters of the members comprising both hot regions is non-trivial to implement, requiring a number of considerations. Moreover, the difficulty scales with the complexity of the hot regions. Here we break down the implementation into a series of steps. 

\subsubsection{Parameter space}

The joint parameter space for the members may be denoted $\boldsymbol{v}=(\Theta_{p},\phi_{p},\zeta_{p},\Theta_{s},\phi_{s},\zeta_{s},\ldots)$, where the six parameters explicitly written are inherent to every model wherein antipodal reflection symmetry is \textit{not} imposed, and any ulterior parameters depend on the model. For \TT{ST-U}, these six parameters are sufficient.

\subsubsection{Region-exchange degeneracy}

If the two hot regions sharing the stellar surface are of equivalent complexity, the prior support for the coordinates of the regions can be defined so as to avoid degeneracy of the likelihood function under exchange of the region positions.

For regions related via antipodal reflection symmetry, the prior support for the colatitude of the regions can be defined such that only at (or near to) the support boundary are there configurations wherein a pair of regions mutually map onto one another via a rotation about the stellar rotation axis. The same condition applies if the regions are \textit{not} related via antipodal reflection symmetry, but have equivalent complexities---e.g., \TT{ST-U} meaning two \TT{ST} regions. For instance, for \TT{ST-U}, we can impose that $\Theta_{p}\leq\Theta_{s}$, where $\Theta_{p}$ and $\Theta_{s}$ are, respectively, the colatitudes of the centers of the primary and secondary \TT{ST} regions.

If the regions are not related via antipodal reflection symmetry and do \textit{not} have equivalent complexities---e.g., \TT{ST+CST} or \TT{ST+EST}---then by definition it is not true that an arbitrary point in parameter space yields exactly the same system configuration as distinct point in parameter space due to region exchange. Indeed, region exchange degeneracy may only be exist for a subset of parameter space, and thus one need not impose a joint constraint on the prior support for the colatitudes of the regions.

\subsubsection{Ceding- and superseding-member radii}

As highlighted in Section~\ref{subsubsec:CDT-S and CDT-U}, it is advisable to define one's native nested-sampling space in order to linearize certain continuous degeneracies where possible.

The angular extent of the regions are remarkable in this respect for models constituted by at least one region at the \TT{CST} complexity level or beyond---i.e., when a hot region is constituted by a superseding member and a ceding member, each with an angular radius. Whilst it is not clear how to fully linearize degeneracy of type I illustrated in Figure~\ref{fig: continuous region degeneracies} (see Section~\ref{subsubsec:CST}), the important\footnote{In terms of prior mass in comparison to that associated with other degenerate structures.} degeneracy of type IV can be linearized by working with the joint space of $\psi$ and $\zeta$, the radii of the superseding and ceding members respectively (see Section~\ref{subsubsec:CDT-S and CDT-U}). Although such a choice may be appear obvious in isolation, it may not be the space on which one chooses to intuitively define a joint prior density distribution.

\textit{Transforms for a \TT{CST} or an \TT{EST} region.} If the superseding member subtends smaller angular extent than the ceding member, it is useful to consider $\psi\coloneqq f\zeta$ where $f\in[\epsilon_{f},1]$ for small (or zero) $\epsilon_{f}$. Moreover, it is common to invoke uniform prior density distributions for parameters with the intention of choosing a weakly informative prior but without rigorous proof. We therefore consider a flat separable density for $f$ and $\zeta$, such that $q(f,\zeta)=q(f)q(\zeta)$ where $f\sim U(\epsilon_{f},1)$ and $\zeta\sim U(\epsilon_{\zeta},\pi/2-\epsilon_{\zeta})$ for small (or zero) $\epsilon_{\zeta}$. More generally, one might choose $\zeta\in[\epsilon_{\zeta},b_{\zeta}]$; in our case $b_{\zeta}\coloneqq\pi/2-\epsilon_{\zeta}$. We then require the marginal density function $q(\psi)$ and the conditional density function $q(\zeta\,|\,\psi)$ in order to define a map $\mathcal{H}\to\mathcal{S}^{\dagger}$, where $\mathcal{H}=[0,1]\times[0,1]$ and where $\mathcal{S}^{\dagger}$ indicates that the support $\mathcal{S}$ of $p(\psi,\zeta)$, after all considerations in this appendix (see Appendix~\ref{app:non-overlapping}), will be such that $\mathcal{S}^{\dagger}\supset\mathcal{S}$.

We must now consider the size of $\epsilon_{f}$ and $\epsilon_{\zeta}$: these limits determine the boundary of the prior support in the joint space of $\psi$ and $\zeta$. For a single-temperature hot region, we simply choose $\epsilon_{f}=0$, such that either the superseding or ceding member can subtend zero angular extent at the boundary of the support; for a dual-temperature region one might choose a small finite value for $\epsilon_{f}$ given that it is filled with material of finite temperature. Given the choice $\epsilon_{f}=0$ the choice of $\epsilon_{\zeta}$ is unimportant for deriving the prior distributions of interest.

The joint density $q(\psi,\zeta)$ is given by
\begin{equation}
q(\psi,\zeta)
=
q(f,\zeta)
\biggr\lvert\frac{\partial f}{\partial\psi}\biggr\rvert
=
\zeta^{-1}q(\psi/\zeta)q(\zeta).
\end{equation}
The marginal density function $q(\psi)$ is thus given by
\begin{equation}
q(\psi)
=
\mathop{\int}
\zeta^{-1}
q(\psi/\zeta)q(\zeta)
d\zeta
\propto
\begin{cases}
\displaystyle{
\mathop{\int}_{\epsilon_{\zeta}}^{b_{\zeta}}
}
\zeta^{-1}
d\zeta
&
\text{if } \psi\leq\epsilon_{\zeta}
\\
\displaystyle{
\mathop{\int}_{\psi}^{b_{\zeta}}
}
\zeta^{-1}
d\zeta
&
\text{if } \epsilon_{\zeta}<\psi\leq b_{\zeta};
\end{cases}
\end{equation}
and further that
\begin{equation}
q(\psi)
\propto
\begin{cases}
\ln\left(b_{\zeta}/\epsilon_{\zeta}\right)
&
\text{if } \psi\leq\epsilon_{\zeta}
\\
\ln\left(b_{\zeta}/\psi\right)
&
\text{if } \epsilon_{\zeta}<\psi\leq b_{\zeta}.
\end{cases}
\end{equation}

Now define $\boldsymbol{x}\coloneqq(x_{\psi},x_{\zeta})$ where the mass
\begin{equation}
x_{\psi}(\psi)
\coloneqq
\mathop{\int}_{0}^{\psi}
q(\psi^{\prime})
d\psi^{\prime}
\propto
\begin{cases}
\psi\ln\left(b_{\zeta}/\epsilon_{\zeta}\right)
&
\text{if } \psi\leq\epsilon_{\zeta}
\\
\psi-\epsilon_{\zeta}
-
\psi\ln\left(\psi/b_{\zeta}\right)
&
\text{if } \epsilon_{\zeta}<\psi\leq b_{\zeta};
\end{cases}
\end{equation}
note that $x_{\psi}(\psi)$ is continuous at $\psi=\epsilon_{\zeta}$, and that $x_{\psi}\to1^{-}$ as $\psi\to b_{\zeta}^{-}$ because $q(\psi/\zeta)q(\zeta)=(b_{\zeta}-\epsilon_{\zeta})^{-1}$ where the joint density $q(f,\zeta)$ is finite. The function $\psi(x_{\psi})$ is not obviously obtainable in closed form for $\psi>\epsilon_{\zeta}$, and thus we interpolate to perform the transformation $x_{\psi}\mapsto\psi$.

We now require the conditional density $q(\zeta\,|\,\psi)$:
\begin{equation}
q(\zeta\,|\,\psi)
=
\frac{q(\psi,\zeta)}{q(\psi)}
=
\begin{cases}
\zeta^{-1}/\ln\left(b_{\zeta}/\epsilon_{\zeta}\right)
&
\text{if }
(\psi\leq\epsilon_{\zeta})\wedge(\epsilon_{\zeta}\leq\zeta\leq b_{\zeta})
\\
\zeta^{-1}/\ln\left(b_{\zeta}/\psi\right)
&
\text{if }
(\epsilon_{\zeta}<\psi\leq b_{\zeta})\wedge(\psi<\zeta\leq b_{\zeta}).
\end{cases}
\end{equation}
Then define the mass
\begin{equation}
x_{\zeta}(\zeta;\psi)
\coloneqq
\mathop{\int}_{a(\psi)}^{\zeta}
q(\zeta^{\prime}\,|\,\psi)
d\zeta^{\prime}
=
\begin{cases}
\displaystyle{
\ln\left(\zeta/\epsilon_{\zeta}\right) / \ln\left(b_{\zeta}/\epsilon_{\zeta}\right)
}
&
\text{if } 
(\psi\leq\epsilon_{\zeta})\wedge(\epsilon_{\zeta}\leq\zeta\leq b_{\zeta})
\\
\displaystyle{
\ln\left(\zeta/\psi\right)/\ln\left(b_{\zeta}/\psi\right)
}
&
\text{if } 
(\epsilon_{\zeta}<\psi\leq b_{\zeta})\wedge(\psi<\zeta\leq b_{\zeta}).
\end{cases}
\end{equation}
The function $\zeta(x_{\zeta};\psi)$ is written in closed form as
\begin{equation}
\zeta(\boldsymbol{x})
=
\begin{cases}
\displaystyle{
\epsilon_{\zeta}\exp\left(x_{\zeta}\ln\left(b_{\zeta}/\epsilon_{\zeta}\right)\right)
}
&
\text{if } 
(\psi\leq\epsilon_{\zeta})\wedge(\epsilon_{\zeta}\leq\zeta\leq b_{\zeta})
\\
\displaystyle{
\psi(x_{\psi})\exp\left(
x_{\zeta}\ln\left(b_{\zeta}/\psi\right)
\right)
}
&
\text{if } 
(\epsilon_{\zeta}<\psi\leq b_{\zeta})\wedge(\psi<\zeta\leq b_{\zeta}).
\end{cases}
\end{equation}

\textit{Transforms for a \TT{PST} region.} If the superseding member can subtend a \textit{larger} angular extent than the ceding member we consider the parameters $f\in[\epsilon_{f},2-\epsilon_{f}]$ for small (or zero) $\epsilon_{f}$ and $\xi\in[\epsilon_{\xi},b_{\xi}]$ for small $\epsilon_{\xi}$. If $f\leq1$, we define $\zeta=\xi$ and $\psi=f\zeta$, whilst if $f>1$ we define $\psi=\xi$ and $\zeta=(2-f)\psi$. We consider a flat separable joint density for $f$ and $\xi$: $f\sim U(\epsilon_{f},2-\epsilon_{f})$ and $\xi\sim U(\epsilon_{\xi},b_{\xi})$. We again require the marginal density function $q(\psi)$ and the conditional density function $q(\zeta\,|\,\psi)$ in order to define a map $\mathcal{H}\to\mathcal{S}^{\dagger}$, where $\mathcal{H}=[0,1]\times[0,1]$ and where $\mathcal{S}^{\dagger}$ indicates that the support $\mathcal{S}$ of $p(\psi,\zeta)$, after all considerations in this appendix (see Appendix~\ref{app:non-overlapping}), will be such that $\mathcal{S}^{\dagger}\supset\mathcal{S}$.

The joint density $q(\psi,\zeta)$ is piecewise with respect to $f$, given by
\begin{equation}
q(\psi,\zeta)
=
q(f,\xi)
\biggr\lvert\frac{\partial(f,\xi)}{\partial(\psi,\zeta)}\biggr\rvert
=
\begin{cases}
\zeta^{-1}q(f)q(\xi) &\text{if } f\leq1 \\
\psi^{-1}q(f)q(\xi) &\text{if } f>1.
\end{cases}
\end{equation}
We must now consider the size of $\epsilon_{f}$ and $\epsilon_{\xi}$: these limits determine the boundary of the prior support in the joint space of $\psi$ and $\zeta$. For a \TT{PST} region, we simply choose $\epsilon_{f}=0$, such that either the superseding or ceding member can subtend zero angular extent at the boundary of the support; for a \TT{PDT} region one might choose a small finite value for $\epsilon_{f}$ given that it is filled with material of finite temperature. Given the choice $\epsilon_{f}=0$ the choice of $\epsilon_{\xi}$ is unimportant for deriving the prior distributions of interest.

In order to construct a map $\mathcal{H}\to\mathcal{S}^{\dagger}$, we aim to obtain the joint density $p(\psi,\zeta)$ in the conditional form $p(\psi,\zeta)=p(\zeta\,|\,\psi)p(\psi)$. The marginal density function $p(\psi)$ is given by
\begin{equation}
q(\psi)
=
\mathop{\int}_{0}^{\psi}
\psi^{-1}
q(f)q(\xi)
d\zeta
+
\mathop{\int}_{\psi}^{b_{\xi}}
\zeta^{-1}
q(f)q(\xi)
d\zeta.
\end{equation}
If $\psi<\epsilon_{\xi}$, then: for $f>1$, $q(\xi=\psi)=0$; and for $f\leq1$, $q(\xi=\zeta)=0$ for $\zeta<\epsilon_{\xi}$. It follows that
\begin{equation}
q(\psi)
\propto
\begin{cases}
\displaystyle{
\mathop{\int}_{\epsilon_{\xi}}^{b_{\xi}}
}
\zeta^{-1}
d\zeta
&
\text{if } \psi<\epsilon_{\xi}
\\
\displaystyle{
\mathop{\int}_{0}^{\psi}
}
\psi^{-1}
d\zeta
+
\displaystyle{
\mathop{\int}_{\psi}^{b_{\xi}}
}
\zeta^{-1}
d\zeta
&
\text{if } \epsilon_{\xi}\leq\psi\leq b_{\xi},
\end{cases}
\end{equation}
and further that
\begin{equation}
q(\psi)
\propto
\begin{cases}
\ln\left(b_{\xi}/\epsilon_{\xi}\right)
&
\text{if } \psi<\epsilon_{\xi}
\\
1 + \ln\left(b_{\xi}/\psi\right)
&
\text{if } \epsilon_{\xi}\leq\psi\leq b_{\xi}.
\end{cases}
\end{equation}
Now define $\boldsymbol{x}\coloneqq(x_{\psi},x_{\zeta})$ where the mass
\begin{equation}
x_{\psi}(\psi)
\coloneqq
\mathop{\int}_{0}^{\psi}
q(\psi^{\prime})
d\psi^{\prime}
\propto
\begin{cases}
\psi\ln\left(b_{\xi}/\epsilon_{\xi}\right)
&
\text{if } \psi<\epsilon_{\xi}
\\
2(\psi-\epsilon_{\xi})
-
\psi\ln\left(\psi/b_{\xi}\right)
&
\text{if } \epsilon_{\xi}\leq\psi\leq b_{\xi};
\end{cases}
\end{equation}
note that $x_{\psi}(\psi)$ is continuous at $\psi=\epsilon_{\xi}$, and that $x_{\psi}\to1^{-}$ as $\psi\to b_{\xi}^{-}$ because $q(f)q(\xi)=(b_{\xi}-\epsilon_{\xi})^{-1}/2$ where the joint density $q(f,\xi)$ is finite. The function $\psi(x_{\psi})$ is not obviously obtainable in closed form for $\psi>\epsilon_{\xi}$, and thus we interpolate to perform the transformation $x_{\psi}\mapsto\psi$.

We now require the conditional density $q(\zeta\,|\,\psi)$:
\begin{equation}
q(\zeta\,|\,\psi)
=
\frac{q(\psi,\zeta)}{q(\psi)}
=
\begin{cases}
\zeta^{-1}/\ln\left(b_{\xi}/\epsilon_{\xi}\right)
&
\text{if }
(\psi<\epsilon_{\xi})\wedge(\epsilon_{\xi}\leq\zeta\leq b_{\xi})
\\
\psi^{-1}/\left[1+\ln\left(b_{\xi}/\psi\right)\right]
&
\text{if }
(\epsilon_{\xi}\leq\psi\leq b_{\xi})\wedge(0\leq\zeta\leq \psi)
\\
\zeta^{-1}/\left[1+\ln\left(b_{\xi}/\psi\right)\right]
&
\text{if }
(\epsilon_{\xi}\leq\psi\leq b_{\xi})\wedge(\psi<\zeta\leq b_{\xi}).
\end{cases}
\end{equation}
Then define the mass
\begin{equation}
x_{\zeta}(\zeta;\psi)
\coloneqq
\mathop{\int}_{a(\psi)}^{\zeta}
q(\zeta^{\prime}\,|\,\psi)
d\zeta^{\prime}
=
\begin{cases}
\displaystyle{
\ln\left(\zeta/\epsilon_{\xi}\right) / \ln\left(b_{\xi}/\epsilon_{\xi}\right)
}
&
\text{if } 
(\psi<\epsilon_{\xi})\wedge(\epsilon_{\xi}\leq\zeta\leq b_{\xi})
\\
\displaystyle{
\frac{\zeta}{\psi}
\left[1+\ln\left(b_{\xi}/\psi\right)\right]^{-1}
}
&
\text{if } 
(\epsilon_{\xi}\leq\psi\leq b_{\xi})\wedge(0\leq\zeta\leq \psi)
\\
\displaystyle{
\left[1 + \ln\left(\zeta/\psi\right)\right]
\left[1+\ln\left(b_{\xi}/\psi\right)\right]^{-1}

}
&
\text{if } 
(\epsilon_{\xi}\leq\psi\leq b_{\xi})\wedge(\psi<\zeta\leq b_{\xi}).
\end{cases}
\end{equation}
The function $\zeta(x_{\zeta};\psi)$ is written in closed-form as
\begin{equation}
\zeta(\boldsymbol{x})
=
\begin{cases}
\displaystyle{
\epsilon_{\xi}\exp\left(x_{\zeta}\ln\left(b_{\xi}/\epsilon_{\xi}\right)\right)
}
&
\text{if } 
(\psi<\epsilon_{\xi})\wedge(\epsilon_{\xi}\leq\zeta\leq b_{\xi})
\\
\displaystyle{
x_{\zeta}\psi(x_{\psi})\left[1+\ln\left(b_{\xi}/\psi\right)\right]
}
&
\text{if } 
(\epsilon_{\xi}\leq\psi\leq b_{\xi})\wedge(0\leq\zeta\leq \psi)
\\
\displaystyle{
\psi(x_{\psi})\exp\left(
x_{\zeta}\left[1+\ln\left(b_{\xi}/\psi\right)\right] - 1
\right)
}
&
\text{if } 
(\epsilon_{\xi}\leq\psi\leq b_{\xi})\wedge(\psi<\zeta\leq b_{\xi}).
\end{cases}
\end{equation}

\subsubsection{Non-overlapping hot regions}\label{app:non-overlapping}

We implicitly define the support $\mathcal{S}$ of the joint density $p(\boldsymbol{v})$, where  $\boldsymbol{v}=(\Theta_{p},\phi_{p},\zeta_{p},\Theta_{s},\phi_{s},\zeta_{s},\ldots)$ is the vector of parameters controlling both hot regions sharing the stellar surface, by imposing a constraint equation in terms of $\boldsymbol{v}$: we require that the regions are \textit{non-overlapping}. More explicitly: two radiating regions associated with distinct regions cannot overlap---if they were to overlap, additional logical conditions would be required to specify an order of precedence for intensity evaluation at spacetime events at the stellar surface.

For single-temperature regions with a single member---i.e., simply-connected and circular---one need only determine whether the ceding members, with their simple (circular) boundaries, overlap; the same condition is true if there exists a superseding member that is a \textit{hole} in a ceding member---i.e., \TT{ST}, \TT{CST}, or \TT{EST}. For dual-temperature hot regions, whose boundary is always constituted by a maximum of two simple (circular) boundaries, one need only determine whether any pair of members---from two distinct regions---overlap. However, for a \TT{PST} region the boundary of the radiating region is more unwieldy for evaluating whether or not the radiating region of the \TT{PST} region overlaps with another region sharing the stellar surface; in this case we simply define an overlap condition only in terms of the simple boundary of the ceding member, a subset of which is superseded by non-radiating surface.

To derive the joint prior density $p(\boldsymbol{v})$ and its support $\mathcal{S}$, we: (i) define $q(\boldsymbol{v})$ as a product of density functions; (ii) inverse sample as $\mathcal{H}\to\mathcal{S}^{\dagger}$, $\boldsymbol{x}\mapsto\boldsymbol{v}$; and (iii) accept the sample $\widetilde{\boldsymbol{v}}$ if the regions are determined to be non-overlapping. Algorithmically, we identify a set of pairs of members with simple boundaries whose center coordinates and angular radii are defined by vector $\widetilde{\boldsymbol{v}}$, and evaluate via a spherical coordinate transformation whether or not the angular separation of the member centers is at least equal to the sum of the angular radii. If any of the pairs of members overlap, then $\widetilde{\boldsymbol{v}}\notin\mathcal{S}$.

As an example, for \TT{ST+EST}, $\boldsymbol{v}=(\Theta_{p},\phi_{p},\zeta_{p},\Theta_{s},\phi_{s},\psi_{s},\zeta_{s},\varepsilon_{s},\varphi_{s})$ where the coordinates of the center of the ceding member of the \TT{EST} region is derived from the vector $(\Theta_{s},\phi_{s},\psi_{s},\zeta_{s},\varepsilon_{s},\varphi_{s})$. Overlap is then evaluated for this ceding member in relation to the \TT{ST} spot whose boundary is derived from the vector $(\Theta_{p},\phi_{p},\zeta_{p})$. The marginal density function for \textit{every} parameter constituting (the space of) vector $\boldsymbol{v}$ is thus modulated by excluding overlaps.

\section{Model summary tables}\label{app:model summaries}
In this appendix we provide posterior summary tables for all models applied to the PSR~J0030$+$0451 event data, other than \TT{ST+PST} (Table~\ref{table: STpPST}). For \TT{ST+EST} see Table~\ref{table: ST+EST}. For \TT{ST+CST} see Table~\ref{table: ST+CST}. For \TT{ST-U} see Table~\ref{table: ST-U}. For \TT{ST-S} see Table~\ref{table: ST-S}. For \TT{CDT-U} see Table~\ref{table: CDT-U}.

\section{Supplementary ideas for model extension}\label{app: supplementary models}

We now make note of models within the scope of this work that were either clearly (without need for explicit posterior computation due to the posterior properties of simpler models) not competitive for PSR~J0030$+$0451 or unhelpfully complex in phenomenologically describing the structure of a hot region. Excess complexity does not mean that the modeling has been optimized, but indicates that a particular extension to a model is not warranted because we are insensitive to a subset of parameters (or combinations of parameters), \textit{and} that the signals that maximize the likelihood function are signals that are effectively generated by a simpler (nested) model.

\begin{figure*}[t!]
\centering
\includegraphics[clip, trim=0cm 0cm 0cm 0cm, width=0.7\textwidth]{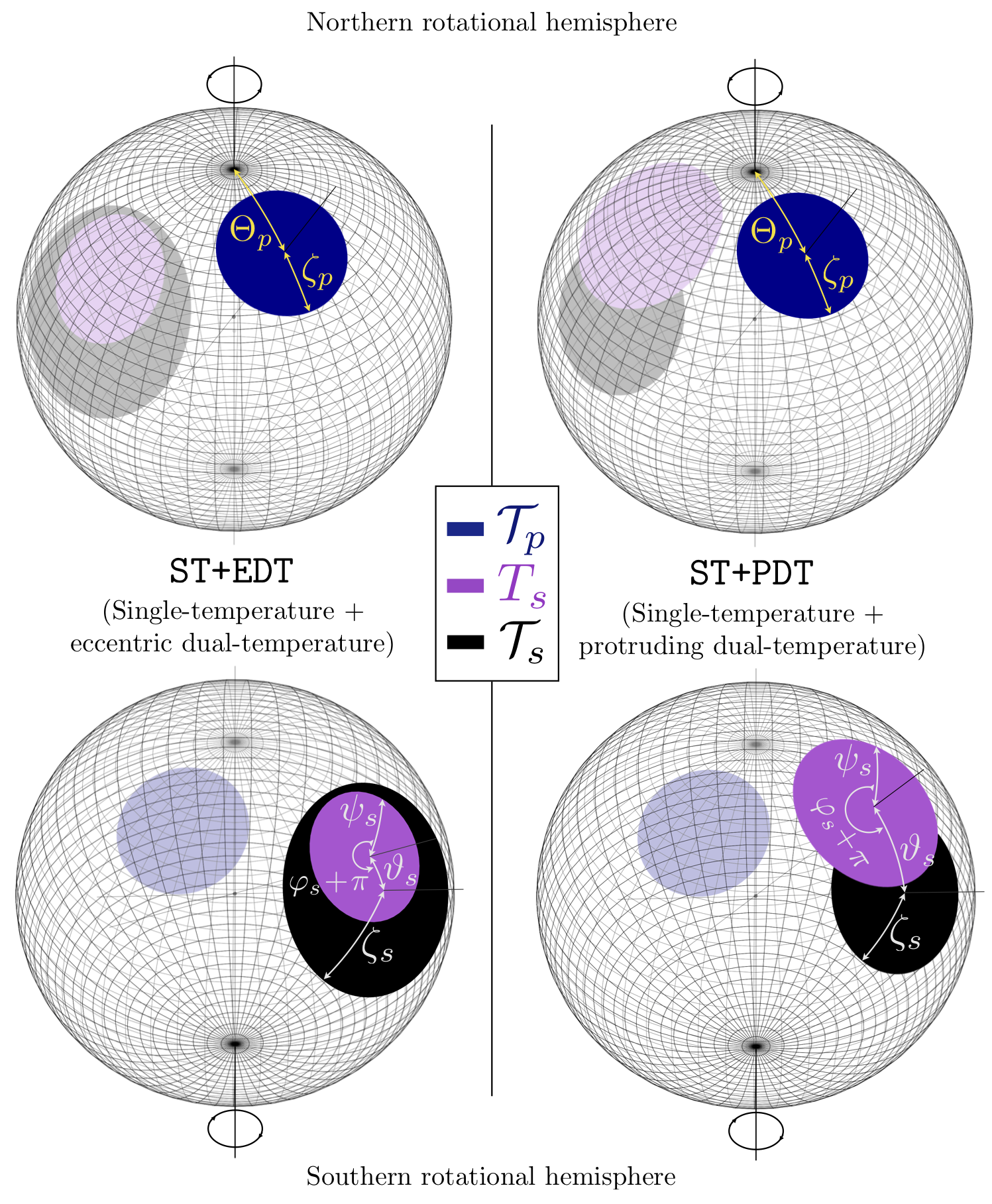}
\caption{\small{Schematic diagrams of models wherein an \TT{ST} region shares the stellar surface with a higher-complexity \TT{EDT} or \TT{PDT} region. The complete figure set ($3$ images) is available in the online journal.}}
\label{fig: schematic ST+EDT and ST+PDT diagrams}
\end{figure*}

\textit{\TT{ST+EDT} and \TT{ST+PDT}.} Obtained via simple extension of \TT{ST+EST} and \TT{ST+PST}: let the superseding member (the hole for \TT{EST}) contain radiating material. Refer to Figure~\ref{fig: schematic ST+EDT and ST+PDT diagrams}.


\figsetstart
\figsetnum{23}
\figsettitle{Extended models for the configuration of the surface hot regions.}

\figsetend

\textit{\TT{EDT-S} and \TT{EDT-U}.} Refer to the online figure set associated with Figure~\ref{fig: schematic ST+EDT and ST+PDT diagrams}. For \TT{EDT-S}, the surface radiation field associated with the secondary region is derived exactly by applying antipodal symmetry to the primary region: there are no free parameters associated with the secondary region. Conversely, for \TT{EDT-U}, the secondary region is endowed with distinct parameters---i.e., it is \textit{not} derived from the primary region under antipodal symmetry. However, the parameters of the secondary region have an otherwise equivalent meaning---in terms of surface radiation field specification---to their primary-region counterparts. As an example, the azimuth of the center of the secondary ceding member is defined relative to the meridian passing through the center of the secondary hole and through the rotational poles.

\textit{\TT{PDT-S} and \TT{PDT-U}.} Refer to the online figure set associated with Figure~\ref{fig: schematic ST+EDT and ST+PDT diagrams}. For \TT{PDT-S}, the surface radiation field associated with the secondary region is derived exactly by applying antipodal symmetry to the primary region: there are no free parameters associated with the secondary region. Conversely, for \TT{PDT-U}, the secondary region is endowed with distinct parameters---i.e., it is \textit{not} derived from the primary region under antipodal symmetry. However, the parameters of the secondary region have an otherwise equivalent meaning---in terms of surface radiation field specification---to their primary-region counterparts.

\begin{table}[t!]
\begin{ruledtabular}
\begin{tabular}{l|l|l|ll} 
Parameter & Description & Prior (density and support) & $\widehat{\textrm{CI}}_{68\%}$ & $\widehat{D}_{\textrm{KL}}$\\
\hline
$P$ $[$ms$]$ &
coordinate spin period &
$P=4.87$, fixed &
$-$ &
$-$ \\
\hline
$M$ $[\textrm{M}_{\odot}]$ &
gravitational mass &
$M\sim U(1,3)$ &
$1.46_{-0.18}^{+0.17}$ &
$1.15_{-0.02}^{+0.02}$ \\
$R_{\textrm{eq}}$ $[$km$]$ &
coordinate equatorial radius &
$R_{\textrm{eq}}\sim U[3r_{g}(1),16]$ &
$13.89_{-1.30}^{+1.14}$ &
$0.69_{-0.02}^{+0.02}$ \\
\hline
&compactness condition & $R_{\textrm{eq}}/r_{g}(M)>3$\\
\hline
$\Theta_{p}$ $[$radians$]$ &
\TT{ST} region center colatitude &
$\Theta_{p}\sim U(0,\pi)$ &
$2.22_{-0.10}^{+0.09}$ &
$3.01_{-0.03}^{+0.03}$\\
$\phi_{p}$ $[$cycles$]$ &
\TT{ST} region initial phase (from Earth) &
$\phi_{p}\sim U(a,a+0.2)$\footnote{Where $\phi_{p}=a$ is an arbitrary phase dependent on event data pre-processing. We set $a=0.35$.}&
$0.45_{-0.00}^{+0.00}$ &
$6.59_{-0.03}^{+0.02}$\\
$\zeta_{p}$ $[$radians$]$ &
\TT{ST} region angular radius &
$\zeta_{p}\sim U(0,\pi/2)$ &
$0.07_{-0.01}^{+0.01}$ &
$4.78_{-0.02}^{+0.03}$\\
\hline
$\Theta_{s}$ $[$radians$]$ &
\TT{EST} region hole center colatitude &
$\Theta_{s}\sim U(0,\pi)$ &
$2.66_{-0.09}^{+0.07}$ &
$3.36_{-0.03}^{+0.02}$\\
$\phi_{s}$ $[$cycles$]$ &
\TT{EST} region initial phase (from Earth antipode) &
$\phi_{s}\sim U(-0.5,0.5)$, periodic &
$-0.51_{-0.01}^{+0.01}$ &
$4.28_{-0.02}^{+0.03}$\\
$\psi_{s}^{+}$ $[$radians$]$ &
\TT{EST} region angular radii difference &
$\zeta_{s}\sim U(0,\pi/2)$, $\psi_{s}^{+}\coloneqq\zeta_{s}-\psi_{s}$ &
$0.03_{-0.01}^{+0.01}$ &
$2.93_{-0.03}^{+0.03}$\\
$\psi_{s}$ $[$radians$]$ &
\TT{EST} region hole angular radius  &
$f_{s}\sim U(0,1)$, $\psi_{s}\coloneqq f_{s}\zeta_{s}$ &
$0.25_{-0.04}^{+0.04}$ &
$2.29_{-0.03}^{+0.02}$\\
$\varepsilon_{s}$ &
\TT{EST} region annulus fractional angular offset &
$\varepsilon_{s}\sim U(0,1)$\footnote{If we were to parameterize the eccentricity in terms of the \textit{sum} of angular radii of the superseding (hole) and ceding regions, $\vartheta\coloneqq \varepsilon(1+f)\zeta$, a \textit{conditional} prior such as $\varepsilon\,|\,f\sim U(0,(1-f)/(1+f))$ would be necessary, where the upper-bound imposes that the radiating region is not simply-connected (i.e., is an annulus): $$\varepsilon(1+f)\zeta+f\zeta\leq\zeta\implies \varepsilon\leq(1-f)/(1+f).$$} &
$0.45_{-0.33}^{+0.45}$ &
$0.08_{-0.01}^{+0.01}$\\
$\varphi_{s}$ $[$radians$]$ &
\TT{EST} region annulus azimuthal offset &
$\varphi_{s}\sim U(0,2\pi)$ &
$0.79_{-0.26}^{+0.35}$ &
$0.64_{-0.06}^{+0.06}$\\
\hline
&non-overlapping hot-region annuli & function of $(\Theta_{p}, \Theta_{s}, \phi_{p}, \phi_{s}, \zeta_{p}, \zeta_{s})$\\
\hline
$\log_{10}\left(\mathcal{T}_{p}\;[\textrm{K}]\right)$ &
\TT{ST} region \TT{NSX} effective temperature &
$\log_{10}\mathcal{T}_{p}\sim U(5.1,6.8)$, \TT{NSX} limits &
$6.11_{-0.01}^{+0.01}$ &
$5.03_{-0.03}^{+0.03}$\\
$\log_{10}\left(\mathcal{T}_{s}\;[\textrm{K}]\right)$ &
\TT{EST} region \TT{NSX} effective temperature &
$\log_{10}\mathcal{T}_{s}\sim U(5.1,6.8)$, \TT{NSX} limits &
$6.11_{-0.01}^{+0.01}$ &
$5.60_{-0.03}^{+0.03}$\\
$i$ $[$radians$]$ &
Earth inclination to rotation axis &
$i\sim U(0,\pi/2) $ &
$1.01_{-0.07}^{+0.07}$ &
$2.40_{-0.02}^{+0.03}$\\
$D$ $[$kpc$]$ &
Earth distance &
$D\sim N(0.325,0.009)$ &
$0.33_{-0.01}^{+0.01}$ &
$0.02_{-0.01}^{+0.01}$\\
$N_{\textrm{H}}$ $[10^{20}$cm$^{-2}]$ &
interstellar neutral H column density &
$N_{\textrm{H}}\sim U(0,5)$ &
$0.61_{-0.16}^{+0.18}$ &
$2.84_{-0.03}^{+0.03}$\\
$\NICER\;\alpha$ &
calibrated matrix scaling &
$\alpha\sim N(1,0.1)$, $\alpha\in[0.5,1.5]$ &
$0.99_{-0.09}^{+0.09}$ &
$0.02_{-0.00}^{+0.01}$\\
$\NICER\;\beta$ &
reference-to-calibrated matrix weighting &
$\beta\sim U(0,1)$ &
$0.16_{-0.11}^{+0.17}$ &
$1.07_{-0.03}^{+0.03}$\\
$\NICER\;\gamma$ &
reference matrix scaling &
$\gamma\sim N(1,0.1)$, $\gamma\in[0.5,1.5]$ &
$0.99_{-0.09}^{+0.09}$ &
$0.04_{-0.01}^{+0.01}$\\
\hline
\hline
&Sampling process information&&& \\
\hline
&number of free parameters: $19$ &&& \\
&number of runs: $1$ &&& \\
&number of live points: $10^{3}$ &&& \\
&inverse hypervolume expansion factor: $0.3$ &&& \\
&termination condition: $10^{-1}$ &&& \\
&evidence:\footnote{Defined as the prior predictive probability $p(\boldsymbol{d}\,|\,\TT{ST+EST})$. We report the interval about the median containing $\pm45\%$ of $10^{3}$ joint bootstrap-weight replications for the combined run.} $\widehat{\ln\mathcal{Z}}= -36367.81_{-0.43}^{+0.48}$ &&&\\
&global KL-divergence: $\widehat{D}_{\textrm{KL}}=62.1_{-0.8}^{+0.8}$ bits &&&\\
&number of core\footnote{Intel$\textsuperscript{\textregistered}$ Xeon E5-2697Av4 (2.60 GHz; Broadwell) processors on the SURFsara Cartesius supercomputer.} hours: $61210$ &&& \\
&likelihood evaluations: $88965106$ &&& \\
&nested replacements: $53149$ &&& \\
&weighted posterior samples: $17671$ &&& \\
\end{tabular}
\end{ruledtabular}
\caption{Summary table for \TT{ST+EST}, introduced in Section~\ref{subsubsec:unequal complexities} and illustrated in Figure~\ref{fig: schematic ST+EST and ST+PST diagrams}.}
\label{table: ST+EST}
\end{table}

\begin{table}[t!]
\begin{ruledtabular}
\begin{tabular}{l|l|l|ll} 
Parameter & Description & Prior (density and support) & $\widehat{\textrm{CI}}_{68\%}$ & $\widehat{D}_{\textrm{KL}}$\\
\hline
$P$ $[$ms$]$ &
coordinate spin period &
$P=4.87$, fixed &
$-$ &
$-$ \\
\hline
$M$ $[\textrm{M}_{\odot}]$ &
gravitational mass &
$M\sim U(1,3)$ &
$1.44_{-0.19}^{+0.18}$ &
$1.07_{-0.02}^{+0.02}$ \\
$R_{\textrm{eq}}$ $[$km$]$ &
coordinate equatorial radius &
$R_{\textrm{eq}}\sim U[3r_{g}(1),16]$ &
$13.88_{-1.38}^{+1.23}$ &
$0.61_{-0.01}^{+0.01}$ \\
\hline
&compactness condition & $R_{\textrm{eq}}/r_{g}(M)>3$\\
\hline
$\Theta_{p}$ $[$radians$]$ &
\TT{ST} region center colatitude &
$\Theta_{p}\sim U(0,\pi)$ &
$2.24_{-0.09}^{+0.09}$ &
$5.14_{-0.02}^{+0.03}$\\
$\phi_{p}$ $[$cycles$]$ &
\TT{ST} region initial phase (from Earth) &
$\phi_{p}\sim U(-0.5,0.5)$, periodic &
$0.46_{-0.00}^{+0.00}$ &
$6.66_{-0.02}^{+0.03}$\\
$\zeta_{p}$ $[$radians$]$ &
\TT{ST} region angular radius &
$\zeta_{p}\sim U(0,\pi/2)$ &
$0.07_{-0.01}^{+0.01}$ &
$4.59_{-0.02}^{+0.02}$\\
\hline
$\Theta_{s}$ $[$radians$]$ &
\TT{CST} region center colatitude &
$\Theta_{s}\sim U(0,\pi)$ &
$2.60_{-0.06}^{+0.05}$ &
$2.79_{-0.02}^{+0.02}$\\
$\phi_{s}$ $[$cycles$]$ &
\TT{CST} region initial phase (from Earth antipode) &
$\phi_{s}\sim U(-0.5,0.5)$, periodic &
$-0.50_{-0.00}^{+0.00}$ &
$7.33_{-0.03}^{+0.02}$\\
$\psi_{s}^{+}$ $[$radians$]$ &
\TT{CST} region annulus angular width &
$\zeta_{s}\sim U(0,\pi/2)$, $\psi_{s}^{+}\coloneqq\zeta_{s}-\psi_{s}$ &
$0.04_{-0.01}^{+0.01}$ &
$2.71_{-0.02}^{+0.02}$\\
$\psi_{s}$ $[$radians$]$ &
\TT{CST} region hole angular radius  &
$f_{s}\sim U(0,1)$, $\psi_{s}\coloneqq f_{s}\zeta_{s}$ &
$0.23_{-0.03}^{+0.03}$ &
$2.52_{-0.02}^{+0.02}$\\
\hline
&enforce \TT{ST} and \TT{CST} colatitude order\footnote{Based on learning that additional complexity (of the form we consider in our model space) beyond \TT{ST} is not warranted for the one region, but is warranted for the other region.} & $\Theta_{s}\geq\Theta_{p}$\\
&non-overlapping hot regions & function of $(\Theta_{p}, \Theta_{s}, \phi_{p}, \phi_{s}, \zeta_{p}, \zeta_{s})$\\
\hline
$\log_{10}\left(\mathcal{T}_{p}\;[\textrm{K}]\right)$ &
\TT{ST} region \TT{NSX} effective temperature &
$\log_{10}\mathcal{T}_{p}\sim U(5.1,6.8)$, \TT{NSX} limits &
$6.10_{-0.01}^{+0.01}$ &
$4.99_{-0.02}^{+0.02}$\\
$\log_{10}\left(\mathcal{T}_{s}\;[\textrm{K}]\right)$ &
\TT{CST} region \TT{NSX} effective temperature &
$\log_{10}\mathcal{T}_{s}\sim U(5.1,6.8)$, \TT{NSX} limits &
$6.11_{-0.01}^{+0.01}$ &
$5.58_{-0.02}^{+0.02}$\\
$i$ $[$radians$]$ &
Earth inclination to rotation axis &
$i\sim U(0,\pi/2) $ &
$1.02_{-0.08}^{+0.07}$ &
$2.32_{-0.02}^{+0.02}$\\
$D$ $[$kpc$]$ &
Earth distance &
$D\sim N(0.325,0.009)$ &
$0.33_{-0.01}^{+0.01}$ &
$0.01_{-0.00}^{+0.00}$\\
$N_{\textrm{H}}$ $[10^{20}$cm$^{-2}]$ &
interstellar neutral H column density &
$N_{\textrm{H}}\sim U(0,5)$ &
$0.62_{-0.18}^{+0.19}$ &
$2.75_{-0.02}^{+0.02}$\\
$\NICER\;\alpha$ &
calibrated matrix scaling &
$\alpha\sim N(1,0.1)$, $\alpha\in[0.5,1.5]$ &
$0.99_{-0.10}^{+0.10}$ &
$0.01_{-0.00}^{+0.00}$\\
$\NICER\;\beta$ &
reference-to-calibrated matrix weighting &
$\beta\sim U(0,1)$ &
$0.16_{-0.11}^{+0.18}$ &
$0.98_{-0.02}^{+0.02}$\\
$\NICER\;\gamma$ &
reference matrix scaling &
$\gamma\sim N(1,0.1)$, $\gamma\in[0.5,1.5]$ &
$0.99_{-0.09}^{+0.09}$ &
$0.01_{-0.00}^{+0.00}$\\
\hline
\hline
&Sampling process information&&& \\
\hline
&number of free parameters: $17$ &&& \\
&number of runs: $2$ &&& \\
&number of live points \textit{per run}: $10^{3}$ &&& \\
&inverse hypervolume expansion factor: $0.3$ &&& \\
&termination condition: $10^{-1}$ &&& \\
&evidence:\footnote{Defined as the prior predictive probability $p(\boldsymbol{d}\,|\,\TT{ST+CST})$. We report the interval about the median containing $\pm45\%$ of $10^{3}$ joint bootstrap-weight replications for the combined run.} $\widehat{\ln\mathcal{Z}}= -36368.00_{-0.33}^{+0.34}$ &&&\\
&global KL-divergence: $\widehat{D}_{\textrm{KL}}=62.9_{-0.5}^{+0.6}$ bits &&&\\
&combined number of core\footnote{Approximate equal-partition between Intel$\textsuperscript{\textregistered}$ Xeon E5-2697Av4 (2.60 GHz; Broadwell) and E5-2690v3 (2.60 GHz; Haswell) processors on the SURFsara Cartesius supercomputer.} hours: $23010$ &&& \\
&combined likelihood evaluations: $39501475$ &&& \\
&combined nested replacements: $105264$ &&& \\
&combined weighted posterior samples: $32839$ &&& \\
\end{tabular}
\end{ruledtabular}
\caption{Summary table for \TT{ST+CST}, introduced in Section~\ref{subsubsec:unequal complexities}.}
\label{table: ST+CST}
\end{table}

\begin{table}[t!]
\begin{ruledtabular}
\begin{tabular}{l|l|l|ll} 
Parameter & Description & Prior (density and support) & $\widehat{\textrm{CI}}_{68\%}$ & $\widehat{D}_{\textrm{KL}}$\\
\hline
$P$ $[$ms$]$ &
coordinate spin period &
$P=4.87$, fixed &
$-$ &
$-$ \\
\hline
$M$ $[\textrm{M}_{\odot}]$ &
gravitational mass &
$M\sim U(1,3)$ &
$1.09_{-0.07}^{+0.11}$ &
$2.09_{-0.03}^{+0.02}$ \\
$R_{\textrm{eq}}$ $[$km$]$ &
coordinate equatorial radius &
$R_{\textrm{eq}}\sim U[3r_{g}(1),16]$ &
$10.44_{-0.86}^{+1.10}$ &
$1.44_{-0.03}^{+0.03}$ \\
\hline
&compactness condition & $R_{\textrm{eq}}/r_{g}(M)>3$\\
\hline
$\Theta_{p}$ $[$radians$]$ &
$p$ region center colatitude &
$\Theta_{p}\sim U(0,\pi)$ &
$2.48_{-0.06}^{+0.06}$ &
$6.92_{-0.02}^{+0.02}$\\
$\Theta_{s}$ $[$radians$]$ &
$s$ region center colatitude &
$\Theta_{s}\sim U(0,\pi)$ &
$2.78_{-0.02}^{+0.02}$ &
$4.08_{-0.02}^{+0.02}$\\
$\phi_{p}$ $[$cycles$]$ &
$p$ region initial phase (from Earth) &
$\phi_{p}\sim U(-0.5,0.5)$, periodic &
$0.46_{-0.00}^{+0.00}$ &
$7.51_{-0.02}^{+0.02}$\\
$\phi_{s}$ $[$cycles$]$ &
$s$ region initial phase (from Earth antipode) &
$\phi_{s}\sim U(-0.5,0.5)$, periodic &
$-0.50_{-0.00}^{+0.00}$ &
$8.05_{-0.02}^{+0.02}$\\
$\zeta_{p}$ $[$radians$]$ &
$p$ region angular radius &
$\zeta_{p}\sim U(0,\pi/2)$ &
$0.14_{-0.02}^{+0.02}$ &
$3.95_{-0.02}^{+0.02}$\\
$\zeta_{s}$ $[$radians$]$ &
$s$ region angular radius &
$\zeta_{s}\sim U(0,\pi/2)$ &
$0.29_{-0.03}^{+0.04}$ &
$2.98_{-0.02}^{+0.02}$\\
\hline
&eliminate region-exchange degeneracy & $\Theta_{s}\geq\Theta_{p}$\\
&non-overlapping hot regions & function of $(\Theta_{p}, \Theta_{s}, \phi_{p}, \phi_{s}, \zeta_{p}, \zeta_{s})$\\
\hline
$\log_{10}\left(\mathcal{T}_{p}\;[\textrm{K}]\right)$ &
$p$ region \TT{NSX} effective temperature &
$\log_{10}\mathcal{T}_{p}\sim U(5.1,6.8)$, \TT{NSX} limits &
$6.11_{-0.01}^{+0.01}$ &
$5.53_{-0.02}^{+0.03}$\\
$\log_{10}\left(\mathcal{T}_{s}\;[\textrm{K}]\right)$ &
$s$ region \TT{NSX} effective temperature &
$\log_{10}\mathcal{T}_{s}\sim U(5.1,6.8)$, \TT{NSX} limits &
$6.10_{-0.01}^{+0.01}$ &
$5.71_{-0.02}^{+0.02}$\\
$i$ $[$radians$]$ &
Earth inclination to rotation axis &
$i\sim U(0,\pi/2) $ &
$1.04_{-0.08}^{+0.07}$ &
$2.38_{-0.02}^{+0.02}$\\
$D$ $[$kpc$]$ &
Earth distance &
$D\sim N(0.325,0.009)$ &
$0.33_{-0.01}^{+0.01}$ &
$0.35_{-0.02}^{+0.02}$\\
$N_{\textrm{H}}$ $[10^{20}$cm$^{-2}]$ &
interstellar neutral H column density &
$N_{\textrm{H}}\sim U(0,5)$ &
$1.23_{-0.17}^{+0.17}$ &
$2.82_{-0.02}^{+0.02}$\\
$\NICER\;\alpha$ &
calibrated matrix scaling &
$\alpha\sim N(1,0.1)$, $\alpha\in[0.5,1.5]$ &
$0.96_{-0.10}^{+0.10}$ &
$0.11_{-0.01}^{+0.01}$\\
$\NICER\;\beta$ &
reference-to-calibrated matrix weighting &
$\beta\sim U(0,1)$ &
$0.23_{-0.16}^{+0.23}$ &
$0.57_{-0.02}^{+0.02}$\\
$\NICER\;\gamma$ &
reference matrix scaling &
$\gamma\sim N(1,0.1)$, $\gamma\in[0.5,1.5]$ &
$0.91_{-0.09}^{+0.10}$ &
$0.56_{-0.03}^{+0.04}$\\
\hline
\hline
&Sampling process information&&& \\
\hline
&number of free parameters: $16$ &&& \\
&number of runs:\footnote{The mode-separation \MultiNest variant was deactivated for these two runs that were combined to compute estimators. Mode separation means that modes are not evolved independently and nested sampling threads contact multiple modes; a mode-separation run was executed and is displayed in Figure~\ref{fig:STpPST spacetime marginal posterior} and in the figure sets available in the online corresponding to Figures~\ref{fig:STpPST source marginal posterior} and \ref{fig:STpPST obs marginal posterior}. The theory nor software implementation exists for combining this run with the two reported in the table.} $2$ &&& \\
&number of live points \textit{per run}: $10^{3}$ &&& \\
&inverse hypervolume expansion factor: $0.3$ &&& \\
&termination condition: $10^{-3}$ &&& \\
&evidence:\footnote{Defined as the prior predictive probability $p(\boldsymbol{d}\,|\,\TT{ST-U})$. We report the interval about the median containing $\pm45\%$ of $10^{3}$ joint bootstrap-weight replications for the combined run. See the footnote in Table~\ref{table: STpPST}.} $\widehat{\ln\mathcal{Z}}= -36377.60_{-0.35}^{+0.36}$ &&&\\
&global KL-divergence: $\widehat{D}_{\textrm{KL}}=63.7_{-0.5}^{+0.5}$ bits &&&\\
&combined number of core\footnote{Approximate equal-partition between Intel$\textsuperscript{\textregistered}$ Xeon E5-2697Av4 (2.60 GHz; Broadwell) and E5-2690v3 (2.60 GHz; Haswell) processors on the SURFsara Cartesius supercomputer.} hours: $9588$ &&& \\
&combined likelihood evaluations: $25346841$ &&& \\
&combined nested replacements: $121617$ &&& \\
&combined weighted posterior samples: $49481$ &&& \\
\end{tabular}
\end{ruledtabular}
\caption{Summary table for \TT{ST-U}, introduced in Section~\ref{subsubsec:ST-S and ST-U} and illustrated in Figure~\ref{fig: schematic ST diagrams}.}
\label{table: ST-U}
\end{table}

\begin{table}[t!]
\begin{ruledtabular}
\begin{tabular}{l|l|l|ll} 
Parameter & Description & Prior (density and support) & $\widehat{\textrm{CI}}_{68\%}$ & $\widehat{D}_{\textrm{KL}}$\\
\hline
$P$ $[$ms$]$ &
coordinate spin period &
$P=4.87$, fixed &
$-$ &
$-$ \\
\hline
$M$ $[\textrm{M}_{\odot}]$ &
gravitational mass &
$M\sim U(1,3)$ &
$2.93_{-0.01}^{+0.01}$ &
$6.75_{-0.03}^{+0.03}$ \\
$R_{\textrm{eq}}$ $[$km$]$ &
coordinate equatorial radius &
$R_{\textrm{eq}}\sim U[3r_{g}(1),16]$ &
$15.97_{-0.04}^{+0.02}$ &
$6.11_{-0.04}^{+0.04}$ \\
\hline
&compactness condition & $R_{\textrm{eq}}/r_{g}(M)>3$\\
\hline
$\Theta_{p}$ $[$radians$]$ &
$p$ region center colatitude &
$\Theta_{p}\sim U(0,\pi/2)$ &
$1.26_{-0.02}^{+0.02}$ &
$4.07_{-0.02}^{+0.03}$\\
$\phi_{p}$ $[$cycles$]$ &
$p$ region initial phase (from Earth) &
$\phi_{p}\sim U(-0.5,0.5)$, periodic &
$-0.09_{-0.00}^{+0.00}$ &
$8.57_{-0.03}^{+0.02}$\\
$\zeta$ $[$radians$]$ &
\TT{ST} region angular radius (shared) &
$\zeta\sim U(0,\pi/2)$ &
$0.09_{-0.00}^{+0.01}$ &
$6.55_{-0.02}^{+0.02}$\\
$\log_{10}\left(\mathcal{T}\;[\textrm{K}]\right)$ &
\TT{ST} region \TT{NSX} effective temperature (shared) &
$\log_{10}\mathcal{T}\sim U(5.1,6.8)$, \TT{NSX} limits &
$6.08_{-0.00}^{+0.00}$ &
$6.97_{-0.02}^{+0.02}$\\
\hline
$\Theta_{s}$ [radians]&
$s$ region center colatitude &
$\Theta_{s}=\pi-\Theta_{p}$, derived\\
$\phi_{s}$ [cycles]&
$s$ region initial phase &
$\phi_{s}=\phi_{p}+0.5$, derived\\
\hline
$i$ $[$radians$]$ &
Earth inclination to rotation axis &
$i\sim U(0,\pi/2) $ &
$1.23_{-0.03}^{+0.02}$ &
$3.98_{-0.02}^{+0.03}$\\
$D$ $[$kpc$]$ &
Earth distance &
$D\sim N(0.325,0.009)$ &
$0.32_{-0.01}^{+0.01}$ &
$0.02_{-0.00}^{+0.00}$\\
$N_{\textrm{H}}$ $[10^{20}$cm$^{-2}]$ &
interstellar neutral H column density &
$N_{\textrm{H}}\sim U(0,5)$ &
$0.02_{-0.02}^{+0.03}$ &
$5.93_{-0.04}^{+0.04}$\\
$\NICER\;\alpha$ &
calibrated matrix scaling &
$\alpha\sim N(1,0.1)$, $\alpha\in[0.5,1.5]$ &
$1.00_{-0.09}^{+0.09}$ &
$0.01_{-0.00}^{+0.00}$\\
$\NICER\;\beta$ &
reference-to-calibrated matrix weighting &
$\beta\sim U(0,1)$ &
$0.36_{-0.19}^{+0.22}$ &
$0.35_{-0.02}^{+0.02}$\\
$\NICER\;\gamma$ &
reference matrix scaling &
$\gamma\sim N(1,0.1)$, $\gamma\in[0.5,1.5]$ &
$1.01_{-0.09}^{+0.09}$ &
$0.03_{-0.00}^{+0.00}$\\
\hline
\hline
&Sampling process information&&& \\
\hline
&number of free parameters: 12 &&& \\
&number of runs: $2$ &&& \\
&number of live points \textit{per run}: $10^{3}$ &&& \\
&inverse hypervolume expansion factor: $0.3$ &&& \\
&termination condition: $10^{-3}$ &&& \\
&evidence:\footnote{Defined as the prior predictive probability $p(\boldsymbol{d}\,|\,\TT{ST-S})$. We report the interval about the median containing $\pm45\%$ of $10^{3}$ joint bootstrap-weight replications for the combined run. See the footnote in Table~\ref{table: STpPST}.} $\widehat{\ln\mathcal{Z}}= -37211.71_{-0.31}^{+0.29}$ &&&\\
&global KL-divergence: $\widehat{D}_{\textrm{KL}}=51.3_{-0.5}^{+0.5}$ bits &&&\\
&combined number of core\footnote{Intel$\textsuperscript{\textregistered}$ Xeon E5-2697Av4 (2.60 GHz; Broadwell) processors on the SURFsara Cartesius supercomputer.} hours: $1494$ &&& \\
&combined likelihood evaluations: $1666483$ &&& \\
&combined nested replacements: $87488$ &&& \\
&combined weighted posterior samples: $30202$ &&& \\
\end{tabular}
\end{ruledtabular}
\caption{Summary table for \TT{ST-S}, introduced in Section~\ref{subsubsec:ST-S and ST-U} and illustrated in Figure~\ref{fig: schematic ST diagrams}.}
\label{table: ST-S}
\end{table}

\afterpage{
\begin{longtable}{l|l|l|ll}
\caption{Summary table for \TT{CDT-U}, introduced in Section~\ref{subsubsec:CDT-S and CDT-U} and illustrated in Figure~\ref{fig: schematic CDT diagrams}. In this table we only give the numerical details for one completed run. However, we attempted to perform a higher-efficiency higher-resolution second run, without activation of the mode-separation \MultiNest sampling algorithm. Our attempt to improve the sampling efficiency by linearizing the degeneracy (type IV in Figures~\ref{fig: continuous region degeneracies}~and~\ref{fig: CDTU degeneracies primary}) observed in the first run; however, we failed to fully apply the necessary transformations (later applied to models \TT{ST+CST} and beyond, as described in Section~\ref{subsubsec:CDT-S and CDT-U} and Appendix~\ref{app:hotregions}), and the mapping from the native sampling space to the physical parameter space inadvertently preserved the nonlinearity of the degeneracy. We thus did not attain higher efficiency, which coupled with higher-resolution calculation, meant that this run was nearing---but did not reach---termination according to the standard criterion used for the other sampling processes reported in this work. Due to the low sampling efficiencies being reported, we ceased computation after consumption of $\sim\!160540$ core hours in order to preserve resources and redesign our modeling route as described in Sections~\ref{subsubsec:degeneracy and complexity}~and~\ref{subsubsec: model relationships}. The highest-likelihood points from the posterior mode in this second run were however utilized, in combination with those from the first run, to map out all of the degenerate posterior structure in Figures~\ref{fig: CDTU degeneracies primary} and \ref{fig: CDTU degeneracy secondary}. The inefficiency suffered during \TT{CDT-U} posterior computation served as a stark reminder that we should design our problems as carefully as possible in order to avoid resource wastage, and was the motivation behind considering in detail the question of \textit{``How much hot-region complexity is helpful?''} Whilst we could in principle conclude this adjourned run with additional computing resources, we have argued in this work that would not be fruitful do so.}\label{table: CDT-U}\\
\hline\hline
Parameter & Description & Prior (density and support) & $\widehat{\textrm{CI}}_{68\%}$ & $\widehat{D}_{\textrm{KL}}$\\
\hline
\endfirsthead
\multicolumn{5}{c}%
{\tablename\ \thetable\---\textit{Continued from previous page}} \\
\hline
\endhead
\hline \multicolumn{5}{r}{\textit{Continued on next page}} \\
\endfoot
\hline\hline
\endlastfoot
$P$ $[$ms$]$ &
coordinate spin period &
$P=4.87$, fixed &
$-$ &
$-$ \\
\hline
$M$ $[\textrm{M}_{\odot}]$ &
gravitational mass &
$M\sim U(1,3)$ &
$1.44_{-0.18}^{+0.17}$ &
$1.15$ \\
$R_{\textrm{eq}}$ $[$km$]$ &
coordinate equatorial radius &
$R_{\textrm{eq}}\sim U[3r_{g}(1),16]$ &
$13.86_{-1.26}^{+1.16}$ &
$0.68$ \\
\hline
&compactness condition & $R_{\textrm{eq}}/r_{g}(M)>3$\\
\hline
$\Theta_{p}$ $[$radians$]$ &
$p$ region center colatitude &
$\Theta_{p}\sim U(0,\pi)$ &
$2.24_{-0.08}^{+0.08}$ &
$4.68$\\
$\phi_{p}$ $[$cycles$]$ &
$p$ region initial phase (from Earth) &
$\phi_{p}\sim U(-0.5,0.5)$, periodic &
$0.46_{-0.00}^{+0.00}$ &
$6.75$\\
$\psi_{p}^{+}$ $[$radians$]$ &
$p$ region annulus angular width &
$\zeta_{p}\sim U(0,\pi/2)$, $\psi_{p}^{+}\coloneqq\zeta_{p}-\psi_{p}$ &
$0.08_{-0.06}^{+0.16}$ &
$2.59$\\
$\psi_{p}$ $[$radians$]$ &
$p$ region hole angular radius  &
$f_{p}\sim U(0,1)$, $\psi_{p}\coloneqq f_{p}\zeta_{p}$ &
$0.07_{-0.01}^{+0.01}$ &
$5.28$\\
\hline
$\Theta_{s}$ $[$radians$]$ &
$s$ region center colatitude &
$\Theta_{s}\sim U(0,\pi)$ &
$2.61_{-0.06}^{+0.05}$ &
$2.98$\\
$\phi_{s}$ $[$cycles$]$ &
$s$ region initial phase (from Earth antipode) &
$\phi_{s}\sim U(-0.5,0.5)$, periodic &
$-0.50_{-0.00}^{+0.00}$ &
$7.38$\\
$\psi_{s}^{+}$ $[$radians$]$ &
$s$ region annulus angular width &
$\zeta_{s}\sim U(0,\pi/2)$, $\psi_{s}^{+}\coloneqq\zeta_{s}-\psi_{s}$ &
$0.04_{-0.01}^{+0.01}$ &
$5.23$\\
$\psi_{s}$ $[$radians$]$ &
$s$ region hole angular radius  &
$f_{s}\sim U(0,1)$, $\psi_{s}\coloneqq f_{s}\zeta_{s}$ &
$0.23_{-0.03}^{+0.03}$ &
$3.66$\\
\hline
&eliminate region-exchange degeneracy & $\Theta_{s}\geq\Theta_{p}$\\
&non-overlapping hot-region annuli & function of $(\Theta_{p}, \Theta_{s}, \phi_{p}, \phi_{s}, \zeta_{p}, \zeta_{s})$\\
\hline
$\log_{10}\left(\mathcal{T}_{p}\;[\textrm{K}]\right)$ &
$p$ region annulus \TT{NSX} effective temperature &
$\log_{10}\mathcal{T}_{p}\sim U(5.1,6.8)$, \TT{NSX} limits &
$5.43_{-0.20}^{+0.23}$ &
$1.11$\\
$\log_{10}\left(T_{p}\;[\textrm{K}]\right)$ &
$p$ region hole \TT{NSX} effective temperature &
$\log_{10}T_{p}\sim U(5.1,6.8)$, \TT{NSX} limits &
$6.11_{-0.01}^{+0.01}$ &
$5.06$\\
$\log_{10}\left(\mathcal{T}_{s}\;[\textrm{K}]\right)$ &
$s$ region annulus \TT{NSX} effective temperature &
$\log_{10}\mathcal{T}_{s}\sim U(5.1,6.8)$, \TT{NSX} limits &
$6.11_{-0.01}^{+0.01}$ &
$5.56$\\
$\log_{10}\left(T_{s}\;[\textrm{K}]\right)$ &
$s$ region hole \TT{NSX} effective temperature &
$\log_{10}T_{s}\sim U(5.1,6.8)$, \TT{NSX} limits &
$5.47_{-0.23}^{+0.19}$ &
$1.30$\\
$i$ $[$radians$]$ &
Earth inclination to rotation axis &
$i\sim U(0,\pi/2) $ &
$1.02_{-0.07}^{+0.07}$ &
$2.37$\\
$D$ $[$kpc$]$ &
Earth distance &
$D\sim N(0.325,0.009)$ &
$0.33_{-0.01}^{+0.01}$ &
$0.02$\\
$N_{\textrm{H}}$ $[10^{20}$cm$^{-2}]$ &
interstellar neutral H column density &
$N_{\textrm{H}}\sim U(0,5)$ &
$0.70_{-0.18}^{+0.19}$ &
$2.70$\\
$\NICER\;\alpha$ &
calibrated matrix scaling &
$\alpha\sim N(1,0.1)$, $\alpha\in[0.5,1.5]$ &
$0.99_{-0.09}^{+0.09}$ &
$0.02$\\
$\NICER\;\beta$ &
reference-to-calibrated matrix weighting &
$\beta\sim U(0,1)$ &
$0.16_{-0.11}^{+0.17}$ &
$1.05$\\
$\NICER\;\gamma$ &
reference matrix scaling &
$\gamma\sim N(1,0.1)$, $\gamma\in[0.5,1.5]$ &
$0.99_{-0.09}^{+0.09}$ &
$0.03$\\
\pagebreak
&Sampling process information&&& \\
\hline
&number of free parameters: $20$ &&& \\
&number of runs: $1$ &&& \\
&number of live points:\footnote{The mode-separation \MultiNest variant was activated, meaning that isolated modes are evolved independently and nested sampling threads migrate between multiple modes. A local posterior mode was identified, corresponding to a weaker phase solution in which the primary and secondary hot regions transpose---relative to the global posterior mode---in their coupling to pulse components visible in the phase-folded event data. The number of live points locked into the mode with dominant posterior mass was $637$, a number assigned according to the prior mass distribution upon mode separation and under the influence of Monte Carlo noise. The posterior mass ratio (or ratio of local evidences) is estimated to be $\sim\!1100$. As stated in Appendix~\ref{subsubsec:estimators and errors}, because of activation of mode-separation sampling, we cannot perform error analysis via process bootstrapping, and we would not be able to combine with another run, supposing that one was available. We nevertheless have an error on the log-evidence reported natively by \MultiNest. Another consequence of the activation of mode-separation is that sampling resolution was absorbed by the local posterior mode with much lower mass---in other words, the active points were sparser in the dominant mode. In combination with sampling error due to likelihood isosurface nonlinearity, the consequence was that the dominant and degenerate posterior mode---which forms a large connected structure in parameter space as discussed in Section~\ref{subsubsec:degeneracy and complexity}---was not fully resolved. In particular, the type I degeneracy branch (refer to Figure~\ref{fig: continuous region degeneracies}~and~\ref{fig: CDTU degeneracies primary}) was not fully resolved, with sampling threads (active points) migrating to and densely populating the type IV branch (again refer to Figure~\ref{fig: CDTU degeneracies primary}) in an unbalanced manner. The second, albeit incomplete, run described in the caption of this table exhibited much improved resolution of the dominant mode (see the points in Figure~\ref{fig: CDTU degeneracies primary}); the resolution remained incomplete, however, which we attribute to sampling error due to the clear nonlinear degeneracy present in the mode.} $10^{3}$ &&& \\
&inverse hypervolume expansion factor:\footnote{We decreased the expansion factor to $0.8^{-1}$ when the acceptance rate decayed to below $5\times10^{-4}$, which slightly decreased the rate of decay. At this point the process was sampling from the typical set and nearing termination ($2\times10^{4}$ core hours remaining at low acceptance fraction).} $0.3$ &&& \\
&termination condition: $10^{-1}$ &&& \\
&evidence:\footnote{Defined as the prior predictive probability $p(\boldsymbol{d}\,|\,\TT{CDT-U})$.} $\widehat{\ln\mathcal{Z}}= -36366.76\pm0.21$ &&&\\
&global KL-divergence: $\widehat{D}_{\textrm{KL}}=64.0$ bits &&&\\
&combined number of core\footnote{Intel$\textsuperscript{\textregistered}$ Xeon E5-2697Av4 (2.60 GHz; Broadwell) processors on the SURFsara Cartesius supercomputer.} hours: $101917$ &&& \\
&likelihood evaluations: $156707329$ &&& \\
&nested replacements: $54610$ &&& \\
&weighted posterior samples: $17503$ &&& \\
\end{longtable}
}

\clearpage
\listofchanges
\end{document}